\documentclass[rmp,aps,twocolumn,nofootinbib,floatfix]{revtex4} 

\usepackage[final]{graphicx}    
\graphicspath{{figures/}}       
          
\usepackage{dcolumn} 
\usepackage{bm} 
\usepackage{amsmath}
\usepackage{amsfonts} 
\usepackage{amssymb}
 
\newcommand{\bs}{\boldsymbol}

\usepackage[utf8]{inputenc}    
\usepackage[T1]{fontenc}       

\newcommand{\rvec}{\mathbf{r}}    

\newcommand{\xvec}{\mathbf{x}}
\newcommand{\Rvec}{\mathbf{R}} 
\newcommand{\dvec}{\mathbf{d}}     
\newcommand{\kvec}{\mathbf{k}}

\newcommand{\ahatdag}{\hat{a}^\dagger}
\newcommand{\ahat}{\hat{a}}      
\newcommand{\nhat}{\hat{n}}      
\newcommand{\chatdag}{\hat{c}^\dagger}
\newcommand{\chat}{\hat{c}} 
\newcommand{\spinupstate}{\vert\!\!\uparrow\rangle}
\newcommand{\spindownstate}{\vert\!\!\downarrow\rangle}

\begin{document}
\title{Many-Body Physics with Ultracold Gases}

\author{Immanuel Bloch}
\email{bloch@uni-mainz.de}
\affiliation{Institut f\"{u}r Physik,
Johannes Gutenberg-Universit\"at, D-55099 Mainz, Germany}
\author{Jean Dalibard}
\email{jean.dalibard@lkb.ens.fr}
\affiliation{Laboratoire Kastler Brossel, CNRS, Ecole Normale Superieure, 
24 rue Lhomond, F-75005 Paris, France}
\author{Wilhelm Zwerger}
\email{zwerger@ph.tum.de}
\affiliation{Physik-Department,
Technische Universit\"at M\"unchen, D-85748 Garching, Germany}

\begin{abstract}
This article reviews recent experimental and theoretical progress
on many-body phenomena in dilute, ultracold gases. Its focus are
effects beyond standard weak-coupling descriptions, like 
the Mott-Hubbard-transition in optical lattices, 
strongly interacting gases in one and two dimensions  or 
lowest Landau level physics in quasi two-dimensional 
gases in fast rotation. Strong correlations in fermionic gases are 
discussed in optical lattices or near Feshbach resonances in 
the BCS-BEC crossover. 
\end{abstract}

\date{7 October, 2007}


\maketitle
\tableofcontents

\section{INTRODUCTION}
\label{sec:intro}

The achievement of Bose-Einstein-Condensation (BEC)
\cite{Anderson:1995,Davis:1995c,Bradley:1995} and of Fermi degeneracy
\cite{DeMarco:1999b,Truscott:2001,Schreck:2001} in ultracold,
dilute gases has opened a new chapter in
atomic and molecular physics in which the particle
statistics and their interactions,
rather than the study of single atoms or photons, are at center stage.
For a number of years, a main focus in this field has been
to explore the wealth of phenomena associated with the
existence of coherent matter waves.  Major examples include
the observation of interference of two overlapping condensates
 \cite{Andrews:1997a}, of long range phase coherence 
 \cite{Bloch:2000} or of quantized vortices and vortex lattices \cite{Madison:2000a,
AboShaeer:2001} and molecular condensates with bound pairs of fermions
\cite{Greiner:2003b, Jochim:2003b, Zwierlein:2003a}.
Common to all of these phenomena is the existence of a coherent,
macroscopic matter wave in an interacting many-body system,
a concept familiar from
the classic area of superconductivity and superfluidity.
It was the basic insight of \textcite{Ginzburg:1950} that,
quite independent of a detailed microscopic understanding,
an effective description of the coherent many-body state is  provided by
a complex, macroscopic wave function $\psi(\mathbf{x})=|\psi(\mathbf{x})|
\exp{i\phi(\mathbf{x})}$. Its  magnitude
squared gives the superfluid density, while the  phase $\phi(\mathbf{x})$
determines the superfluid velocity via $\mathbf{v}_s=\hbar/M\cdot\nabla\phi(\mathbf{x})$
(see the Appendix for a discussion of these concepts and their subtle connection
with the microscopic criterion for BEC).
As emphasized by \textcite{Cummings:1966} and by
\textcite{Langer:1968}, this picture is similar to the description of laser light 
by a coherent state \cite{Glauber:1963}. It applies both to the standard
condensates of bosonic atoms and also to weakly bound fermion pairs
which are the building blocks of the BCS-picture of superfluidity in Fermi
systems. In contrast to conventional superfluids like $^4\,$He
or superconductors, where
the macroscopic wave function only provides a phenomenological
description of the superfluid degrees of freedom, the situation in dilute gases
is considerably simpler. In fact, as a result of the weak interactions,
the dilute BEC's  are essentially pure condensates sufficiently below the transition.
The macroscopic wave function is thus directly
connected with the microscopic degrees of freedom,  providing
a complete and quantitative description of both static and
time-dependent phenomena in terms
of a reversible, nonlinear Schr\"odinger-equation,
the famous Gross-Pitaevskii equation \cite{Gross:1961, Pitaevskii:1961}.
In dilute gases therefore, the many-body
aspect of a BEC is reduced to an effective single-particle description,
where interactions just give rise to an additional potential proportional to
the local particle density. Adding small
fluctuations around this zeroth-order picture leads to the well known
Bogoliubov theory of weakly interacting Bose gases. Similar to the
closely related BCS-superfluid of weakly interacting fermions, the
many-body problem is then completely soluble in terms of
a set of non-interacting quasiparticles. 
Dilute, ultracold gases provide a concrete realization of these
basic models of many-body physics and many of their
characteristic properties have indeed been
verified quantitatively. Excellent reviews of this
remarkably rich area of research have been given by 
\textcite{Dalfovo:1999} and by \textcite{Leggett:2001} and
- more recently -  in the comprehensive books by \textcite{Pethick:2002} 
and by \textcite{Pitaevskii:2003}.

In the past several years, two major new developments
have considerably  enlarged the range of physics which
is accessible  with ultracold gases. They are associated with
\begin{itemize}
\item  the ability to tune the interaction strength in cold gases
by Feshbach resonances \cite{Inouye:1998,Courteille:1998} and 
\item the possibility to change
the dimensionality with optical potentials and, in particular, to
generate strong periodic potentials 
for cold atoms through optical lattices \cite{Greiner:2002a}. 
\end{itemize}
Both developments, either individually or in combination, 
allow to enter a regime, in which the interactions even in 
extremely dilute gases can no longer
be described by a picture based on
non-interacting quasi-particles. The appearance of such phenomena
is characteristic for the physics of strongly correlated systems.
For a long time, this area of research was confined to the dense
and strongly interacting quantum liquids of condensed
matter or nuclear physics. By contrast, gases - almost by definition -
were never thought to exhibit strong correlations. 

The use of Feshbach resonances and optical potentials
for exploring strong correlations in ultracold gases
was crucially influenced by earlier ideas from theory. 
In particular, \textcite{Stoof:1996}
suggested that Feshbach resonances in a degenerate gas of
$^6$Li, which exhibits a tunable attractive interaction between two
different hyperfine states, may be used to realize BCS-pairing
of fermions in ultracold gases.  A remarkable idea in a
rather unusual direction in the context of atomic physics  
was the proposal by \textcite{Jaksch:1998} to realize a quantum
phase transition from a superfluid to a Mott-insulating state
by loading a BEC into an optical lattice and simply
raising its depth. Further directions into
the regime of strong correlations were opened with the suggestions
by \textcite{Olshanii:1998a} and \textcite{Petrov:2000b},
to realize a Tonks-Girardeau gas with BEC's confined in one dimension
and by \textcite{Wilkin:2000} to explore quantum Hall effect physics 
in fast rotating gases.

Experimentally, the strong coupling regime in dilute gases was first 
reached by \textcite{Cornish:2000}, using Feshbach 
resonances for bosonic atoms.
Unfortunately, in this case, increasing the scattering length $a$ 
leads to a strong decrease in the condensate 
lifetime due to three-body losses, whose rate on average
varies like $a^4$ \cite{Fedichev:1996b,Petrov:2004c}. A quite different approach 
to the regime of strong correlations, which does not suffer from 
problems with the condensate lifetime,
was taken by \textcite{Greiner:2002a}. Loading 
BEC's into an optical lattice, they observed 
a quantum phase transition from a superfluid
to a  Mott-insulating phase even in the standard regime 
where the average interparticle spacing is much larger 
than the scattering length. Subsequently,
the strong confinement available with optical lattices made possible
the achievement of low dimensional systems where new phases can emerge.
In fact,  the first example of a bosonic Luttinger liquid was obtained with the 
observation of a (Tonks-Girardeau) hard-core Bose gas 
in one dimension by \textcite{Paredes:2004} and \textcite{Kinoshita:2004}.
In two dimensions a Kosterlitz-Thouless crossover between a normal phase and
one with quasi-long range order was observed by \textcite{Hadzibabic:2006}. The physics of strongly interacting bosons in the lowest 
Landau level is accessible with fast rotating BEC's \cite{Schweikhard:2004a, Bretin:2004},
where the vortex lattice is eventually predicted to melt by quantum fluctuations. 
Using atoms like $^{52}$Cr, which have a larger permanent magnetic moment,
BEC's with strong dipolar interactions have been realized by \textcite{Griesmaier:2005}. 
In combination with Feshbach resonances, this opens the way tune the 
nature and range of the interaction \cite{Lahaye:2007}, which might, for instance, 
be used to reach novel many-body states that are not accessible 
in the context of the fractional Quantum Hall Effect. 

In Fermi gases, the  Pauli-principle strongly suppresses three-body losses, 
whose rate in fact {\it de}creases with increasing
values of the scattering length \cite{Petrov:2004b}. Feshbach resonances
therefore allow to enter the strong coupling regime $k_F\vert a\vert\gg 1$
in ultracold Fermi gases \cite{OHara:2002a,Bourdel:2003}. In particular, there exist 
stable molecular states of weakly bound fermion
pairs, in highly excited ro-vibrational states \cite{Strecker:2003,Cubizolles:2003}. 
The remarkable stability of fermions near Feshbach resonances
allows to explore the crossover from a molecular BEC 
to a BCS-superfluid of weakly bound Cooper-pairs
\cite{Regal:2004a, Zwierlein:2004, Bartenstein:2004b, Bourdel:2004}.
In particular, the presence of pairing due to many-body effects 
has been probed by spectroscopy of the gap \cite{Chin:2004},
or the closed channel fraction \cite{Partridge:2005} while superfluidity 
has been verified by the observation
of quantized vortices \cite{Zwierlein:2005}.
Recently, these studies have been extended to Fermi gases with unequal densities 
for the spin-up and spin-down components
\cite{Zwierlein:2006, Partridge:2006}, 
where pairing is suppressed  by the mismatch of the respective Fermi energies.
 
Repulsive fermions in an optical lattice
allow to realize an ideal and tunable version
of the Hubbard model,  a paradigm for the multitude of strong
correlation problems in condensed matter physics.
 Experimentally, some of the basic properties of degenerate fermions 
in periodic potentials like the existence of a Fermi surface and
the appearance of a band insulator at unit filling have 
been observed by \textcite{Kohl:2005a}. 
While it is difficult to cool fermions to temperatures 
much below the bandwidth in a deep optical lattice
these experiments 
give rise to the hope that eventually magnetically 
ordered or unconventional superconducting phases of the fermionic 
Hubbard model will be accessible with cold gases. 
The perfect control and tunability of the interactions in these systems
provide a completely novel approach to study basic  
problems in many-body physics and, in particular, to enter regimes
which have never been accessible in condensed matter or
nuclear physics.

The present review aims to give an overview of this rapidly
evolving field, covering both theoretical concepts and
their experimental realization. It provides an introduction 
to the strong correlation aspects of cold gases, that is,  
phenomena which are not captured by 
weak-coupling descriptions like the Gross-Pitaevskii or Bogoliubov theory.
The focus of this review   
is on examples which have already been realized 
experimentally. Even within this limitation, 
however, the rapid development of the field in recent years
makes it impossible to give a complete survey. In particular, 
important subjects like spinor gases, Bose-Fermi mixtures, quantum spin systems in optical lattices 
or dipolar gases will not be discussed here (see e.g. \textcite{Lewenstein:2007}). Also,
applications of cold atoms in optical lattices for
quantum information are omitted completely, 
for a recent introduction see \textcite{Jaksch:2005}.

\subsection{Scattering of ultracold atoms}

For an understanding of the interactions between neutral atoms, 
first at the two-body level, it is instructive to use a toy model 
\cite{Gribakin:1993}, in which the van der Waals attraction 
at large distances is cutoff by a hard core at some 
distance $r_c$ on the order of an atomic dimension. The 
resulting spherically symmetric potential
\begin{equation}
\label{eq:vdwpot}
V(r)=\begin{cases} -C_6/r^6 & \text{if} \qquad r>r_c\\
\infty & \text{if} \qquad r\leq r_c
\end{cases}
\end{equation}
is, of course, not a realistic description of the short
range interaction of atoms, however it captures the main features
of scattering at low energies. The asymptotic behavior of the
interaction potential is fixed by 
the van der Waals coefficient $C_6$. It defines a characteristic length
\begin{equation}
\label{eq:vdwlength}
a_c=\left(\frac{2M_rC_6}{\hbar^2}\right)^{1/4}
\end{equation}
at which the kinetic energy of the relative motion of two
atoms with reduced mass $M_r$ equals their interaction energy.
For alkali atoms, this length is typically on the order of several 
nano-meters. It is much larger than the atomic scale $r_c$
because alkalis are strongly polarizable, 
resulting in a large $C_6$ coefficient.
The attractive well of the van der Waals potential thus
supports many bound states (of order $100$ in $^{87}$Rb!).
Their number $N_b$ may be determined from the WKB-phase 
\begin{equation}
\label{eq:vdwphase}
\Phi=\int_{r_c}^{\infty}dr\sqrt{2M_r\vert V(r)\vert}/\hbar
=a_c^2/2r_c^2\gg 1
\end{equation}
at zero energy, via $N_b=[\Phi/\pi+1/8]$ where $[\,\, ]$ means taking the integer part\footnote{This result follows from Eq.~(\ref{eq:s-length}) below
by noting that a new bound state is pulled in from the
continuum each time the scattering length diverges (Levinson's theorem).}.
The number of bound states in this model
therefore crucially depends on the precise value of the short range
scale $r_c$. By contrast, the low energy scattering properties
are essentially determined by the van der Waals length $a_c$, which
is only sensitive to the asymptotic behavior of the potential. Consider
the scattering in states with angular momentum $l=0,1,2\ldots$ in the 
relative motion (for identical bosons or fermions only 
even or odd values of $l$ are possible, respectively). The effective potential 
for states with $l\ne 0$ contains a centrifugal barrier whose 
height is of order $E_c=\hbar^2/M_ra_c^2$. Converting this energy 
into an equivalent temperature, one obtains temperatures 
around $1\/$~mK for typical atomic masses. At temperatures below that, 
the energy $\hbar^2k^2/2M_r$ in the relative motion of two atoms 
is typically below the centrifugal barrier. Scattering 
in states with $l\ne 0$ is therefore frozen out, unless there exist 
so-called shape resonances,
i.e. bound states with $l\ne 0$ behind the centrifugal barrier, which may be
in resonance with the incoming energy, see \textcite{Boesten:1997,Durr:2005}.
For gases in the sub-mK regime, therefore, usually lowest angular momentum collisions dominate ($s$-wave for bosons, $p$-wave for fermions), which in fact {\it defines} the regime of ultracold atoms.
In the  $s$-wave case, the scattering amplitude is determined by the 
corresponding phase shift $\delta_0(k)$ via \cite{Landau:1987}
\begin{equation}
\label{eq:s-ampl}
f(k)=\frac{1}{k\cot{\delta_0(k)}-ik}\,\to\,
\frac{1}{-1/a+r_ek^2/2-ik}\, .
\end{equation}
At low energies, it is characterized by the scattering length $a$ and the 
effective range $r_e$ as the two single parameters. For the 
truncated van der Waals potential (\ref{eq:vdwpot}), the scattering length
can be calculated analytically as \cite{Gribakin:1993}
\begin{equation}
\label{eq:s-length}
a=\bar{a}\left[ 1-\tan\left(\Phi-3\pi/8\right)\right]
\end{equation}
where $\Phi$ is the WKB-phase (\ref{eq:vdwphase}) and 
$\bar{a}=
0.478\, a_c$
the so-called mean scattering length. The expression (\ref{eq:s-length}) shows that
the characteristic magnitude of the scattering length is the van der Waals length. 
Its detailed value, however, depends on the short range physics via 
the WKB-phase $\Phi$, which is sensitive to the hard core scale $r_c$. 
Since the detailed behavior of the potential is typically not known 
precisely, in many cases neither the sign of the scattering length  
nor the number of bound states can be determined from ab initio
calculations. The toy-model result, however, is useful beyond
the identification of $a_c$ as the characteristic scale for the scattering length. 
Indeed, if the ignorance about the short range physics is replaced 
by the (maximum likelihood) assumption of a uniform distribution of
$\Phi$ in the relevant interval $[0,\pi]$, the probability for finding
a positive scattering length, i.e. $\tan{\Phi}<1$ is $3/4$.  A 
repulsive interaction at low energy, which is 
connected with a positive scattering length,  is therefore three 
times more likely than an attractive one, where $a<0$ \cite{Pethick:2002}.
Concerning the effective range $r_e$ in Eq.~(\ref{eq:s-ampl}), it turns out that also 
$r_e$ is on the order of the van der Waals or the 
mean scattering length $\bar{a}$ rather than the short range
scale $r_c$ as might have been expected naively
\footnote{This is a general result for deep potentials with a power law
decay at large distances, as long as the 
scattering energy is much smaller than the depth of the potential well.} \cite{Flambaum:1999}.
Since $ka_c\ll 1$ in the regime of ultracold collisions, this implies that the 
$k^2\,$-contribution in the denominator of the scattering amplitude is 
negligible. In the low-energy limit, the two-body collision problem is thus
completely specified by the scattering length $a$ as the single parameter 
and a corresponding scattering amplitude
\begin{equation}
\label{eq:p-potampl}
f(k)=\frac{-a}{1+ika}\, .
\end{equation}  
As noted by Fermi in the context of scattering of slow
neutrons and by Lee, Huang and Yang for the low temperature
thermodynamics of weakly interacting quantum gases, 
Eq.~(\ref{eq:p-potampl}) is the exact scattering amplitude at \emph{arbitrary} 
values of $k$ for the pseudopotential
\footnote{Due to the delta-function, the last term involving the partial derivative with 
respect to $r=\vert\mathbf{x}\vert$ can be omitted when the potential acts on a function which
is regular at $r=0$.}   
\begin{equation}
\label{eq:p-pot}V(\mathbf{x})(\ldots)=\frac{4\pi\hbar^2 a}{2M_r}\cdot\delta(\mathbf{x})\,\frac{\partial}{\partial r}\left( r\ldots\right)\, .
\end{equation}
At temperatures such that $k_BT<E_c$, two-body interactions in 
ultracold gases may thus be 
described by a pseudopotential, with the scattering length
usually taken as an experimentally determined parameter.  This 
approximation is valid in a wide range of situations, provided 
no longer range contributions come into play as e.g. in the case 
of dipolar gases. The interaction is 
repulsive for positive and attractive for negative scattering length.
Now, as shown above, the
true interaction potential has many bound states, irrespective of the 
sign of $a$. For the low energy scattering of atoms, however, these bound
states are irrelevant as long as no molecule formation occurs via
three-body collisions. The scattering amplitude in the limit
$k\to 0$ is only sensitive to bound (or virtual for $a<0$) states near zero
energy. In particular, within the pseudopotential approximation  
the amplitude (\ref{eq:p-potampl}) has a single pole $k=i\kappa$, 
with $\kappa=1/a>0$  if the scattering length is positive. 
Quite generally, poles of the scattering amplitude in the upper 
complex $k\/$-plane are connected with bound states
with binding energy $\varepsilon_b=\hbar^2\kappa^2/2M_r$ \cite{Landau:1987}. 
In the pseudopotential approximation,
only a single pole is captured; the energy of the associated bound state
is just below the continuum threshold. A repulsive pseudopotential
thus describes a situation where the full potential has a bound state
with a binding energy  $\varepsilon_b=\hbar^2/2M_ra^2$ on the order of or smaller than the characteristic energy $E_c$
introduced above. The associated positive scattering length 
is then identical with the decay length of 
the wave function $\sim\exp{-r/a}$ of the 
highest bound state. In the attractive case $a<0$, in turn, there is
no bound state within a range $E_c$ below the continuum 
threshold, however there is a virtual state just above it. 
  
\subsection{Weak interactions} 

For a qualitative discussion of what defines the weak interaction
regime in dilute, ultracold gases, it is useful to start with the idealization
of no interactions at all.
Depending on the two fundamental possibilities for the statistics
of indistinguishable particles, Bose or Fermi, the ground state of a gas
of $N$ non-interacting particles is either a perfect BEC or a Fermi sea.
In the case of an ideal BEC, all
particles occupy the lowest available single particle level, consistent with
a fully symmetric many-body wave function. For fermions, in turn, the
particles fill the $N$ lowest single particle levels up to the Fermi-energy
$\epsilon_F(N)$, as required by the Pauli-principle.  
At finite temperatures, the discontinuity in the 
Fermi-Dirac distribution at $T=0$ is smeared out, giving rise to
a continuous evolution from the degenerate gas at $k_BT\ll\epsilon_F$
to a classical gas at high temperatures $k_BT\gtrsim\epsilon_F$.
By contrast, bosons exhibit in 3D a phase transition at finite temperature,
where the macroscopic occupancy of the ground state
is lost. In the homogeneous gas, this transition occurs when the thermal
de Broglie wavelength $\lambda_T=h/\sqrt{2\pi M k_BT}$
reaches the average interparticle
distance $n^{-1/3}$. The surprising fact that a phase 
transition appears even
in an ideal Bose gas is a consequence of the correlations
imposed by the particle statistics alone, as noted already
in Einstein's fundamental paper \cite{Einstein:1925a}.
For trapped gases, with geometrical mean trap frequency
$\bar{\omega}$, the transition to a BEC
is in principle smooth\footnote{A Bose gas in a trap exhibits a sharp transition only
in the limit $N\to\infty,\,\bar\omega\to 0$ with $N{\bar\omega}^3=\text{const}$,
i.e. when the critical temperature approaches a finite value in the 
thermodynamic limit.}. 
Yet, for typical particle numbers in the
range $N\approx 10^4-10^7$, there is a rather sharply defined
temperature $k_BT_c^{(0)}=\hbar\bar{\omega}\cdot\left(N/\zeta(3)\right)^{1/3}\/$,
above which the occupation of the oscillator ground state
is no longer of order $N$. This temperature is again 
determined by the condition that the thermal de Broglie wavelength 
reaches the average interparticle distance at the center of the trap
(see Eq.~(\ref{eq:n_2semiclassical}) and below).

\begin{figure}
\begin{center}
\includegraphics[width=0.9\linewidth]{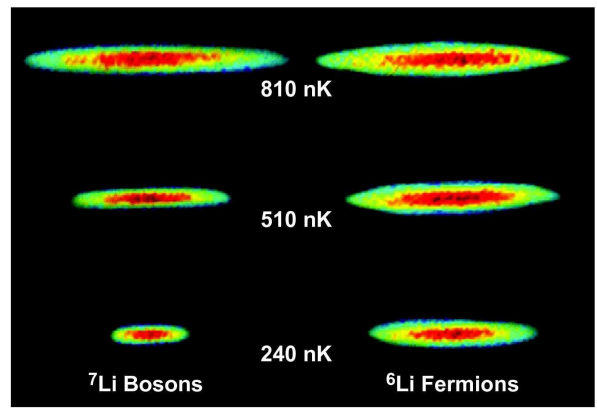}
\caption{Simultaneous cooling of a bosonic and fermionic quantum gas of $^7$Li and $^6$Li to quantum degeneracy. In the case of the Fermi gas, the Fermi pressure prohibits the atom cloud to shrink further in space as quantum degeneracy is approached. Reprinted with permission from \textcite{Truscott:2001}.\label{fig:FermionsBosonsHulet}}
\end{center}
\end{figure}

As discussed above, interactions between ultracold atoms
are described by a pseudopotential (\ref{eq:p-pot}),
whose strength $g=4\pi\hbar^2a/2M_r$ is fixed by the exact s-wave 
scattering length $a$. Now,
for identical fermions, there is no $s$-wave scattering due to the
Pauli-principle. In the regime $ka_c\ll 1$, where all
higher momenta $l\ne 0$ are frozen out, a single component Fermi gas thus
approaches an ideal, non-interacting quantum gas. To reach the
necessary temperatures, however, requires thermalization
by elastic collisions. For identical fermions, $p$-wave collisions 
dominate at low temperatures, whose cross section $\sigma_p\sim E^2$
leads to a vanishing of the scattering rates $\sim T^2$ \cite{DeMarco:1999a}. 
Evaporative cooling therefore does not work for a single component
Fermi gas in the degenerate regime. This 
problem may be circumvented by cooling in the presence
of a different spin state which is then removed, or by 
sympathetic cooling with a another atomic species. 
In this manner, an ideal Fermi gas, which is one of the paradigms of statistical physics, 
has first been realized by \textcite{DeMarco:1999b}, 
\textcite{Truscott:2001} and \textcite{Schreck:2001}
(see Fig.~\ref{fig:FermionsBosonsHulet}).

In the case of mixtures of 
fermions in different internal states or for bosons, there is 
in general a finite scattering length $a\ne 0$, which is typically of 
the order of the van der Waals length Eq.~(\ref{eq:vdwlength}). 
By a simple dimensional argument, interactions are expected to be 
weak when the scattering length is much smaller than the average 
interparticle spacing.  Since ultracold alkali gases have 
densities between $10^{12}\/$ and $10^{15}\/$ particles per cm$^{-3}$,
the average interparticle spacing $n^{-1/3}$ typically is in the range
$0.1\, -\, 1\,\mu$m.              
As shown above, the scattering length, in turn,  is usually only in the few 
nm range. Interaction effects are thus expected to be very small, unless
the scattering length happens to be large near a zero energy resonance of
Eq.~(\ref{eq:s-length}).  
In the attractive case $a<0$, however, 
even small interactions can lead to instabilities. 
In particular, attractive bosons are unstable 
towards collapse. However, in a trap, a metastable gaseous state arises for sufficiently small atom numbers \cite{Pethick:2002}. 
For mixtures of fermions in different internal states, 
an arbitrary weak attraction leads to the BCS-instability, where 
the ground state is essentially a BEC of Cooper pairs (see section VIII). 
In the case of repulsive interactions, in turn, perturbation theory
works in the limit $n^{1/3}a\ll 1$
\footnote{We neglect the possibility of a Kohn-Luttinger instability \cite{Kohn:1965} 
of repulsive fermions to a (typically) p-wave superfluid state,
which usually only appears at temperatures very far below $T_F$, see
\textcite{Baranov:1996}.}. For fermions 
with two different internal states, an appropriate description
is provided by the dilute gas version of Landau's theory of Fermi liquids.
The associated ground state chemical potential is given by \cite{Lifshitz:1980}
\begin{equation}
\label{eq:mu_F}
\mu_{\rm Fermi}=\frac{\hbar^2k_F^2}{2M}\Bigl(1+\frac{4}{3\pi}k_Fa+
\frac{4(11-2\ln{2})}{15\pi^2}(k_Fa)^2+\ldots\Bigr) \, ,
\end{equation}
where the Fermi wavevector $k_F=(3\pi^2 n)^{1/3}$ is determined 
by the total density $n$ in precisely the same manner as
in the non-interacting case. 
Weakly interacting  Bose gases, in turn, are described by the 
Bogoliubov theory, which has $\sqrt{na^3}$ as the relevant small parameter.
For example, the chemical potential at zero temperature for 
a homogeneous gas is given by \cite{Lifshitz:1980} 
\begin{equation}
\label{eq:mu_B}
\mu_{\rm Bose}=\frac{4\pi\hbar^2 a}{M}n\Bigl(1+
\frac{32}{3}\left(\frac{na^3}{\pi}\right)^{1/2} +\ldots \Bigr)\, .
\end{equation}
Moreover, interactions lead to a depletion
\begin{equation}
\label{eq:depletion}
n_0=n\Bigl(1-\frac{8}{3}\left(\frac{na^3}{\pi}\right)^{1/2} +\ldots\Bigr)
\end{equation}
of the density $n_0$ of particles at zero momentum compared to the 
perfect condensate of an ideal Bose gas.
The finite value of the chemical potential at zero temperature defines a characteristic
length $\xi$ by the relation $\hbar^2/2M\xi^2=\mu_{\rm Bose}$. This is the
so-called healing length \cite{Pitaevskii:2003}
which is the scale, over which the macroscopic
wave function $\psi(\mathbf{x})$ varies near a boundary (or a vortex core, see
section VII) where BEC is suppressed.
To lowest order in $\sqrt{na^3}$, this length is given by
$\xi=(8\pi na)^{-1/2}$. In the limit $na^3\ll 1$, the healing length
is therefore much larger than the average interparticle spacing $n^{-1/3}$.
In practice, the dependence on the gas parameter $na^3$ is
so weak, that the ratio $\xi n^{1/3}\sim(na^3)^{-1/6}$
is never very large. On a microscopic level, $\xi$ is the length
associated with the ground state energy per particle by the
uncertainty principle.  It can thus be
identified with the scale, over which bosons may be considered to
be localized spatially. For weak coupling BEC's, the
atoms are therefore smeared out over
distances much larger than the average interparticle spacing.
 
Interactions also shift the critical temperature for BEC away from
its value $T_c^{(0)}$ in the ideal Bose gas. To lowest order in the
interactions, the shift is positive and linear in the scattering length,
\cite{Baym:1999}
\begin{equation}
\label{eq:Tc-shift}T_c/T_c^{(0)}=1+c n^{1/3}a+\ldots
\end{equation}
with a numerical constant $c\approx 1.32$ \cite{Arnold:2001,Kashurnikov:2001}. 
The unexpected {\it in}crease of the BEC condensation temperature
with interactions is due to a reduction of the critical 
density. While a quantitative derivation of Eq.~(\ref{eq:Tc-shift})
requires quite sophisticated techniques \cite{Holzmann:2004}, the 
result can be recovered by a simple argument. To leading order,
the interaction induced change in $T_c$ only depends on the scattering length.
Compared with the non-interacting case, the finite scattering length may be 
thought of as effectively increasing the quantum mechanical uncertainty in
the position of each atom due to thermal motion from 
$\lambda_T$ to $\bar{\lambda}_T=\lambda_T+a$.
To lowest order in $a$, the modified ideal gas criterion 
$n\bar{\lambda}_{T_c}^3=\zeta(3/2)$ then
gives rise to the linear and positive shift of the 
critical temperature in Eq.~(\ref{eq:Tc-shift}) with a 
coefficient $\bar{c}\approx 1.45$, which is not far from the numerically 
exact value.

In the standard situation of a gas confined in a harmonic
trap with characteristic frequency $\bar{\omega}$, the influence of 
weak interactions is quantitatively quite different for temperatures near
$T=0$ or near the critical temperature $T_c$. At zero temperature,
the non-interacting Bose gas has a density distribution
$n^{(0)}(\mathbf{x})=N\cdot\vert\phi_0(\mathbf{x})\vert^2$, 
which just reflects the harmonic oscillator 
ground state wave function $\phi_0(\mathbf{x})$. 
Its characteristic width is the oscillator length $\ell_0=\sqrt{\hbar/M\bar{\omega}}$
which is on the order of one $\mu$m for typical confinement 
frequencies. Adding even small repulsive interactions changes 
the distribution quite strongly.  Indeed, in the experimentally relevant 
limit $Na\gg\ell_0$, 
the density profile $n(\mathbf{x})$ in the presence of an 
external trap potential $U(\mathbf{x})$ can be obtained from
a local density approximation (LDA)
\begin{equation}
\label{eq:LDA}\mu\left[ n(\mathbf{x})\right]+U(\mathbf{x})=
\mu\left[ n(0)\right]\, .
\end{equation}
For weakly interacting bosons in an isotropic harmonic trap, 
the linear dependence $\mu_{\rm Bose}=gn$ of the chemical potential
on the density in the homogeneous case then leads to a 
so-called Thomas-Fermi profile
$n(\mathbf{x})=n(0)\left(1-(r/R_{TF})^2\right)$.  
Using the condition $\int n(\mathbf{x})=N$, the associated radius
$R_{TF}=\zeta\ell_0$ considerably exceeds 
the oscillator length since the dimensionless 
parameter $\zeta=\left(15 Na/\ell_0\right)^{1/5}$
is typically much larger than one  
\cite{Giorgini:1997b} 
\footnote{For fermions, the validity of the LDA,
which is in fact just a semiclassical 
approximation (see e.g. \textcite{Brack:1997}), does not
require interactions. The leading term $\mu_{\rm Fermi}\sim n^{2/3}$
of Eq.~(\ref{eq:mu_F}) leads to a density profile
$n(\mathbf{x})=n(0)\left(1-(r/R_{TF})^2\right)^{3/2}$
with a radius $R_{TF}=\zeta\ell_0$. Here
$\zeta=k_F(N)\ell_0=(24\, N)^{1/6}\gg 1$ and the Fermi 
wavevector $k_F(N)$ in a trap is defined by
$\epsilon_F(N)=\hbar^2k_F^2(N)/2M$}. 
This broadening leads to a 
significant decrease in the density $n(\mathbf{0})$ at the trap
center by a factor $\zeta^{-3}$ compared 
with the non-interacting case. The strong effect of even weak interactions
on the ground state in a trap may be understood from the fact 
that the chemical potential $\mu=\hbar\bar{\omega}\cdot\zeta^{2}/2$
is much larger than the oscillator ground state energy. Interactions 
are thus able to mix in many
single particle levels beyond the harmonic trap ground state.  
Near the critical temperature, in turn, the ratio 
$\mu/k_BT_c\simeq \bigl(n(\mathbf{0})a^3\bigr)^{1/6}$
is small. Interaction corrections to the
condensation temperature, which dominate finite size corrections 
for particle numbers much larger than $N\simeq 10^4$, 
are therefore accessible perturbatively \cite{Giorgini:1997b}.  In contrast to the 
homogeneous case, where the density is fixed and $T_c$ is shifted
upwards, the dominant effect in a trap arises from 
the reduced density at the trap center. Near $T_c$ this 
effect is small and the corresponding shift of may be expressed in the form
$\Delta T_c/T_c=-\text{const}\, a/\lambda_{T_c}$ 
\cite{Giorgini:1997b, Holzmann:2004, Davis:2006}. 
A precise measurement of this shift has been 
performed by \textcite{Gerbier:2004}. Their results are in quantitative 
agreement with mean-field theory, with no observable contribution of
critical fluctuations at their level of sensitivity. Quite recently,
evidence for critical fluctuations has been inferred from
measurements of the correlation length $\xi\sim (T-T_c)^{-\nu}$ 
very close to ${T_c}$. The observed value
$\nu=0.67\pm 0.13$ \cite{Donner:2007} agrees
well with the expected critical exponent  of the 3D XY-model.

In spite of the strong deviations in the density distribution 
compared to the non-interacting case, the one- and 
two-particle correlations of weakly interacting bosons 
are well described by approximating the many-body
ground state of $N$ bosons by a product 
\begin{equation}
\label{eq:GP-wfct}\Psi_{GP}\left( \mathbf{x}_1, \mathbf{x}_2, \ldots \mathbf{x}_N\right)\,
=\,\prod_{i=1}^N\,\phi_1(\mathbf{x}_i)
\end{equation}
in which all atoms are in the identical single particle state
$\phi_1(\mathbf{x})$. Taking Eq.~(\ref{eq:GP-wfct}) as a variational ansatz, 
the optimal macroscopic
wave function $\phi_1(\mathbf{x})$ is found to obey the well known Gross-Pitaevskii
equation. More generally, it turns out that for trapped BEC's the Gross-Pitaevskii 
theory can be derived in a mathematically rigorous manner by taking
the limits $N\to\infty$ and $a\to 0$ in such a way that the 
ratio $Na/\ell_0$ is fixed \cite{Lieb:2000}.
A highly nontrivial aspect of
these derivations is that they show explicitly that in the dilute limit 
interactions enter only via the scattering length.  The Gross-Pitaevskii
equation thus remains valid e.g. for a dilute gas of hard spheres.
Since all the interaction energy is of kinetic origin in this case,
the standard mean field derivation of the Gross-Pitaevskii equation 
via the replacement of the field operators by a classical c-number $\hat{\Psi}(\mathbf{x})\to\sqrt{N}\phi_1(\mathbf{x})$ is thus incorrect 
in general.  
From a many-body point of view, the Ansatz Eq.~(\ref{eq:GP-wfct}), where
the ground state is written as a product of optimized single-particle
wave functions,  is just the standard Hartree-approximation.
It is the simplest possible approximation to account for interactions, 
however it clearly contains no interaction induced correlations between
\emph{different} atoms at all. A first step to go beyond that 
is the well known Bogoliubov theory. This is usually
introduced by considering small fluctuations around the Gross-Pitaevskii
equation in a systematic expansion in the number of noncondensed
particles \cite{Castin:1998}. 
It is more instructive from a many-body point of view, however,
to formulate Bogoliubov theory
in such a way that the boson ground state is
approximated by an optimized product \cite{Lieb:1963a, Leggett:2001}
\begin{equation}
\label{eq:Bog-wfct}
\Psi_{Bog.}\left( \mathbf{x}_1, \mathbf{x}_2, \ldots \mathbf{x}_N\right)\,=\,\prod_{i<j}
\phi_2(\mathbf{x}_i,\mathbf{x}_j)
\end{equation}
of identical, symmetric {\it two}-particle wave functions $\phi_2$. This allows
to include interaction effects beyond the Hartree potential of the 
Gross-Pitaevskii theory by suppressing
configurations in which two particles are close together. The
many-body state thus incorporates two-particle correlations which are
important e.g. to obtain the standard sound modes and the related
coherent superposition of 'particle' and 'hole' excitations. This structure,
which has been experimentally verified by \textcite{Vogels:2002},  is
expected to apply in a qualitative form even for strongly interacting BEC's,
whose low energy excitations are exhausted by harmonic phonons (see Appendix).

Quantitatively, however, the Bogoliubov theory is restricted to the regime $\sqrt{na^3}\ll 1$,
where interactions lead only to a small depletion (\ref{eq:depletion}) of 
the condensate at zero temperature. Going beyond that requires to  
specify the detailed form of the interaction 
potential $V(r)$ and not only the associated scattering length $a$. 
The ground state of a gas of hard sphere bosons, for instance, looses BEC 
already for $na^3\gtrsim0.24$ by a first order transition to a solid state \cite{Kalos:1974}. 
On a variational level, choosing the
two-particle wave functions in (\ref{eq:Bog-wfct}) 
of the form $\phi_2(\mathbf{x}_i,\mathbf{x}_j)\sim\exp{-u(|\mathbf{x}_i-\mathbf{x}_j|)}$
with an effective two-body potential $u(r)$, 
describes so-called Jastrow wave functions. They allow taking into 
account strong short range correlations, however they still
exhibit BEC even in a regime, where the associated
one-particle density describes a periodic crystal rather than a uniform liquid, 
as shown by \textcite{Chester:1970}.  Crystalline order
may thus coexist with BEC! For a recent discussion of this issue 
in the context of a possible supersolid phase of $^4$He, see \textcite{Clark:2006}.

For weakly interacting fermions at $k_Fa\ll 1$,
the variational ground state which is analogous to 
Eq.~(\ref{eq:GP-wfct}), is a Slater determinant
\begin{equation}
\label{eq:HF-wfct}
\Psi_{HF}\left( \mathbf{x}_1, \mathbf{x}_2, \ldots \mathbf{x}_N\right)\,=\,\text{Det}
\left[\phi_{1,i}(\mathbf{x}_j)\right]\, ,
\end{equation}   
of optimized single particle states $\phi_{1,i}(\mathbf{x}_j)$.
In the translational invariant case, they are plane waves
$\phi_{1,i}(\mathbf{x})=V^{-1/2}\exp{i\mathbf{k}_i\mathbf{x}}$, where 
the momenta $\mathbf{k}_i$ are filled up to the Fermi-momentum
$k_F$.
Although both the Bose and the Fermi groundstate wave functions
consist of symmetrized or anti-symmetrized single-particle 
states, they describe - of course - fundamentally different physics. 
In the Bose case, the one particle density matrix $g^{(1)}(\infty)=n_0/n$ 
approaches a finite constant at infinite separation, which is the 
basic criterion for BEC (see Appendix). The many-body wave function is 
thus sensitive to changes of the phase at points separated by 
distances $r$ which are large compared to the 
interparticle spacing. By contrast, the Hartree-Fock state 
(\ref{eq:HF-wfct}) for fermions shows 
no long range phase coherence and indeed, the one particle 
density matrix decays exponentially  $g^{(1)}(r)\sim
\exp{-\gamma r}$ at any finite temperature \cite{Ismail-Beigi:1999}.
The presence of $N$ \emph{distinct} eigenstates in Eq.~(\ref{eq:HF-wfct}), 
which is a necessary consequence 
of the Pauli-principle, thus leads to a many-body wave function
which may be characterized as {\it nearsighted}. 
The notion of nearsightedness, depends on the observable, however. 
As defined originally by \textcite{Kohn:1996}, it means that  
a localized external potential around some point $\mathbf{x}'$
is not felt at a point $\mathbf{x}$ at a distance much 
larger than the average interparticle spacing . This requires
the density response function $\chi(\mathbf{x},\mathbf{x}')$ to be 
short ranged in position space. In this respect, weakly interacting 
bosons, where $\chi(\mathbf{x},\mathbf{x}')\sim
(\exp{-\vert\mathbf{x}-\mathbf{x}'\vert/\xi})/
\vert\mathbf{x}-\mathbf{x}'\vert$ decays exponentially on  
the scale of the healing length $\xi$ are more nearsighted than fermions
at zero temperature, where $\chi(\mathbf{x},\mathbf{x}')\sim\sin(2k_F\vert\mathbf{x}-\mathbf{x}'\vert)/
\vert\mathbf{x}-\mathbf{x}'\vert^3$ exhibits an algebraic decay with Friedel
oscillations at twice the Fermi wave vector $2k_F$. The characterization
of many-body wave functions in terms of the associated correlation functions 
draws attention to another basic point emphasized by \textcite{Kohn:1999}: in situations 
with a large number of particles  the many-body wave function itself is 
not a meaningful quantity because it cannot be reliably calculated for $N\gtrsim 100$.
Moreover, physically accessible observables are only sensitive to the 
resulting one- or two-particle correlations.
Cold gases provide a concrete example for the latter statement:
the standard time-of-flight technique of measuring the 
absorption image after a given free expansion time $t$ 
provides the one-particle density matrix in Fourier space, 
while the two-particle density matrix is revealed in the noise 
correlations of the absorption images (see section III).  

\subsection{Feshbach resonances}

The most direct way of reaching the strong interaction regime in
dilute, ultracold gases are Feshbach resonances, which allow to
increase the scattering length to values beyond the average
interparticle spacing. In practice, this method works best for
fermions because for them the lifetime due to three-body
collisions becomes very large near a Feshbach resonance, in stark
contrast to bosons, where it goes to zero. The concept of Feshbach
resonances was first introduced in nuclear physics in the context
of reactions forming a compound nucleus \cite{Feshbach:1958}.
Quite generally, a Feshbach resonance in a two-particle collision
appears whenever a bound state in a closed channel is coupled
resonantly with the scattering continuum of an open channel. 
The two channels may correspond, for example, to different spin
configurations for the atoms. The
scattered particles are then temporarily captured in the
quasi-bound state and the associated long time delay gives rise to
a Breit-Wigner type resonance in the scattering cross-section.
What makes Feshbach resonances in the scattering of cold atoms
particularly useful, is the ability to tune the scattering length simply by 
changing the magnetic field \cite{Tiesinga:1993}. This
tunability relies on the difference in the magnetic moments of the
closed and open channels, which allows to change the position of
closed channel bound states relative to the open channel threshold
by varying the external, uniform magnetic field. Note that
Feshbach resonances can alternatively be induced optically via
one- or two-photon transitions,  
\cite{Fedichev:1996a, Bohn:1999} as realized by
\textcite{Theis:2004}. The control parameter is
then the detuning of the light from atomic resonance. Although
more flexible in principle, this method suffers, however, from
 heating problems for typical atomic transitions, associated with the
spontaneous emission processes created by the light
irradiation.

\begin{figure}
\begin{center}
\includegraphics[width=0.9\linewidth]{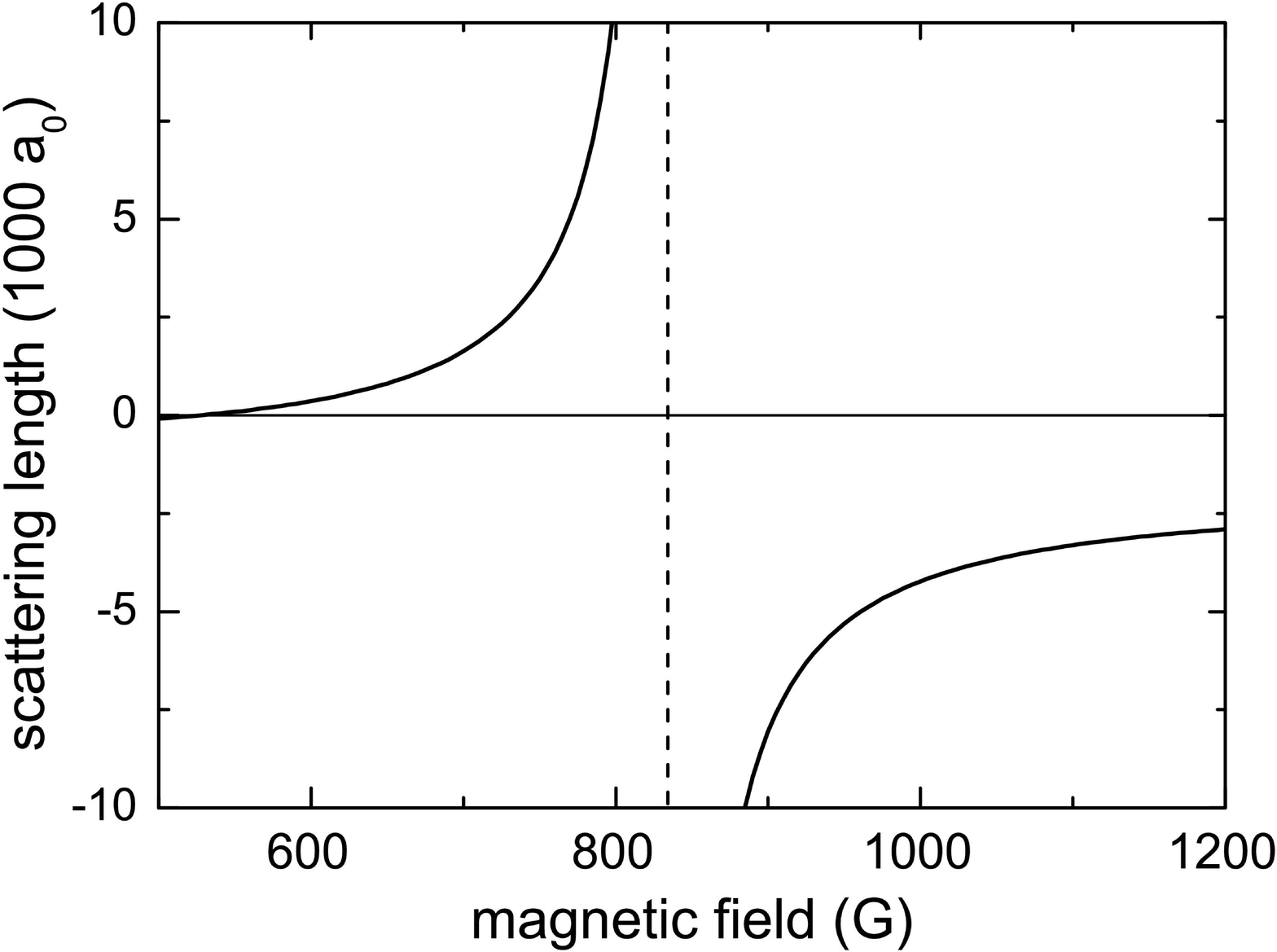}
\caption{Magnetic field dependence of the scattering length 
between the two lowest magnetic sub-states of $^6$Li with a Feshbach 
resonance at $B_0=834\,$G and a zero crossing at 
$B_0+\Delta B=534\,$G. The background scattering length $a_{\rm bg}=-1405\, a_B$ 
is exceptionally large in this case ($a_B$ being the Bohr radius).
\label{fig:FeshbachLi6}}
\end{center}
\end{figure}

On a phenomenological level, Feshbach resonances are described by
an effective pseudopotential between atoms in the open channel
with scattering length
\begin{equation}
\label{eq:FBs-length}
a(B)=a_{\rm bg}\left(1-\frac{\Delta B}{B-B_0}\right)\, .
\end{equation}
Here $a_{\rm bg}$ is the off-resonant background scattering length
in the absence of the coupling to the closed channel while $\Delta
B$ and $B_0$ describe the width and position of the resonance
expressed in magnetic field units (see Fig.~\ref{fig:FeshbachLi6}). 
In this section we outline the basic
physics of magnetically tunable Feshbach resonances, providing a 
connection of the parameters in Eq.~(\ref{eq:FBs-length}) with 
the interatomic potentials. Of course, our
discussion only covers the basic background
for understanding the origin of large and tunable scattering
lengths. A much more detailed presentation of Feshbach resonances 
can be found in the reviews by
\textcite{Timmermans:2001}, \textcite{Duine:2004} and by \textcite{Koehler:2006}.    

\paragraph*{Open and closed channels}
We start with the specific example of fermionic $^6$Li atoms,
which have electronic spin $S=1/2$ and nuclear spin $I=1$. In the
presence of a magnetic field $\mathbf{B}$ along the
$z\/$-direction, the hyperfine coupling and Zeeman energy lead for
each atom to the Hamiltonian
\begin{equation}
\label{eq:hf-int}\hat{H}'=a_{\rm hf}
\mathbf{\hat{S}}\cdot\mathbf{\hat{I}} + \left(2\mu_B\hat{S_z}
-\mu_n \hat{I_z}\right)B\, .
\end{equation}
Here $\mu_B$ is the standard Bohr magneton and $\mu_n$ ($\ll
\mu_B$) the magnetic moment of the nucleus. This hyperfine-Zeeman
Hamiltonian actually holds for any alkali atom, with a single
valence electron with zero orbital angular momentum. If $B\to 0$
the eigenstates of this Hamiltonian are labeled by the quantum
numbers $f$ and $m_f$, giving the total spin angular momentum and
its projection along the $z$ axis, respectively. In the opposite
Paschen-Back regime of large magnetic fields ($B \gg \hbar a_{\rm
hf}/\mu_B \simeq 30$~G in Lithium), the eigenstates are labeled by
the quantum numbers $m_s$ and $m_I$, giving the projection on the
$z$ axis of the electron and nuclear spins, respectively. The
projection $m_f=m_s+m_I$ of the total spin along the $z$ axis
remains a good quantum number for any value of the magnetic field.

Consider a collision between two lithium atoms, prepared in the
two lowest eigenstates $|a\rangle$ and $|b\rangle$ of the
Hamiltonian (\ref{eq:hf-int}) in a large magnetic field. The
lowest state $|a\rangle$ (with $m_{fa}=1/2$) is $\approx\vert m_s=
-1/2, m_I=1\rangle$ with a small admixture of $\vert m_s=1/2,
m_I=0\rangle$, whereas $|b\rangle$ (with $m_{fb}=-1/2$) is
$\approx \vert m_s= -1/2, m_I=0\rangle$ with a small admixture of
$\vert m_s=1/2, m_I=-1\rangle$. Two atoms in these two lowest states 
thus predominantly scatter in their triplet state
\footnote{The fact that there is a nonvanishing $s$-wave 
scattering length for these states is thus connected with 
the different {\it nuclear} and not electronic spin
in this case !}.
Quite generally, the interaction potential during the collision 
can be written as a sum 
 \begin{equation}
 \label{eq:singlet-triplet}
 V(r)=\frac{1}{4}(3V_t(r)+V_s(r))+\hat{\mathbf
S}_1\cdot\hat{\mathbf S}_2\;(V_t(r)-V_s(r))\ ,
 \end{equation}
of projections onto the singlet $V_s(r)$ and triplet $V_t(r)$
molecular potentials, where the $\hat{\mathbf S}_i$'s ($i=1,2$) are the spin operators
for the valence electron of each atom.
These potentials have the same van der Waals
attractive behavior at long distances, but they differ
considerably at short distances, with a much deeper attractive
well for the singlet than for the triplet potential.  Now, in a large but finite
magnetic field the initial state $|a,b\rangle$ is not a purely
triplet state. Because of the tensorial nature of $V(r)$, this
spin state will thus evolve during the collision. More precisely, 
since the second term in Eq.~(\ref{eq:singlet-triplet})
is not diagonal in the basis $|a,b\rangle$, 
the spin state $|a,b\rangle$ may be coupled to other
scattering channels $|c,d\rangle$,
provided the $z$ projection of the total spin is conserved
($m_{fc}+m_{fd}=m_{fa}+m_{fb}$). When the atoms are far apart the
Zeeman+hyperfine energy of $|c,d\rangle$ exceeds the initial
kinetic energy of the pair of atoms prepared in $|a,b\rangle$
by an energy on the order of the hyperfine energy. Since 
the thermal energy is much smaller than that for ultracold collisions,
the channel $|c,d\rangle$ is closed and the atoms always
emerge from the collision in the open channel state $|a,b\rangle$.
However, due to the strong coupling of $(a,b)$ to $(c,d)$ via 
the second term in Eq.~(\ref{eq:singlet-triplet}), which is 
typically on the order of eV, the effective scattering amplitude
in the open channel can be strongly modified.

\begin{figure} 
\includegraphics[width=0.9\columnwidth]{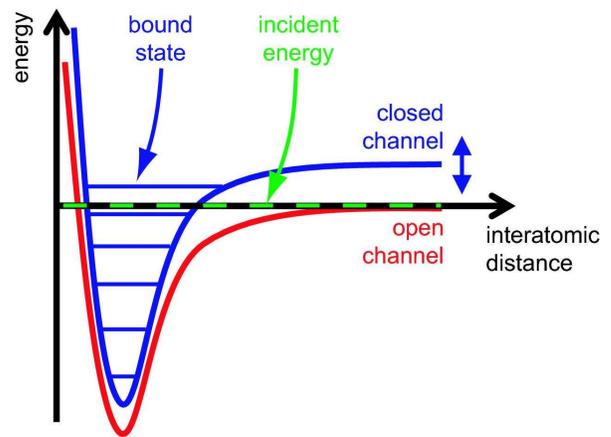}
\caption{The two-channel model for a Feshbach resonance. Atoms
 prepared in the open channel, corresponding to the interaction potential
 $V_{\rm op}(r)$ (in red), undergo a collision at low incident energy.
In the course of the collision the open channel is coupled to the
closed channel $V_{\rm cl}(r)$ (in blue). When a bound state 
of the closed channel has an energy close to zero, a
scattering resonance occurs. The position of the closed channel
can be tuned with respect to the open one e.g. by varying the
magnetic field $B$.
\label{fig:FeshbachSchematic}} 
\end{figure}

\paragraph*{Two-channel model}

We now present a simple two-channel model which captures the main
features of a Feshbach resonance (see Fig. \ref{fig:FeshbachSchematic}). 
We consider a collision between two atoms with reduced mass $M_r$,
and model the system in the vicinity of the resonance by the
Hamiltonian \cite{Nygaard:2006}
\begin{equation}
\label{eq:2channel} 
\hat{H}=\begin{pmatrix}
 -\frac{\hbar^2}{2M_r}\mathbf{\nabla}^2+V_{\rm op}(r) & W(r) \\
W(r) & -\frac{\hbar^2}{2M_r}\mathbf{\nabla}^2+V_{\rm cl}(r)
\end{pmatrix}\ .
\end{equation}
Before collision the atoms are
prepared in the \emph{open channel}, whose potential $V_{\rm
op}(r)$ gives rise to the background scattering length $a_{\rm
bg}$. Here the zero of energy is chosen such that $V_{\rm
op}(\infty)=0$. In the course of the collision a coupling to the
\emph{closed channel} with potential $V_{\rm cl}(r)$ ($V_{\rm
cl}(\infty)>0$) occurs via the matrix element $W(r)$, whose range is
on the order of the atomic scale $r_c$. For
simplicity, we consider here only a {\it single} closed channel, which is appropriate for an isolated resonance. We also assume that the value of $a_{\rm bg}$ is on the order of the van der Waals length (\ref{eq:vdwlength}).
If $a_{\rm bg}$ is anomalously large, as occurs e.g. 
for the $^{6}$Li resonance shown in Fig.~\ref{fig:FeshbachLi6}, an 
additional open channel resonance has to be included in the model, 
as discussed by \textcite{Marcelis:2004}.

We assume that the magnetic moments of the colliding states differ
for the open and closed channels, and we denote their difference
by $\mu$. Varying the magnetic field by $\delta B$ therefore
amounts to shifting the closed channel energy by $\mu\, \delta B$
with respect to the open channel. In the following we are
interested in the magnetic field region close to $B_{\rm res}$
such that one (normalized) bound state $\phi_{\rm res}(r)$ of the
closed channel potential $V_{\rm cl}(r)$ has an energy $E_{\rm
res}(B)=\mu (B-B_{\rm res})$ close to $0$. It can thus be
resonantly coupled to the collision state where two atoms in the
open channel have a small positive kinetic energy. 
In the vicinity of the Feshbach resonance, the situation 
is now very similar to the well know
Breit-Wigner problem (see e.g. \textcite{Landau:1987}, section
134). A particle undergoes a scattering process
in a (single channel) potential with a quasi or true bound state
at an energy $\nu$, which is nearly resonant with the 
incoming energy $E(k)=\hbar^2k^2/(2M_r)$.  
According to Breit and Wigner, this 
leads to a resonant contribution 
 \begin{equation}
\delta_{\rm res}(k)=-\arctan \left[\frac{\Gamma(k)/2}{E(k)-\nu}\right]
 \label{eq:Breit-Wigner}
 \end{equation}
to the scattering phase shift, where $\nu=\mu(B-B_0)$ is conventionally 
called the {\it detuning}  in this context (for the difference 
between $B_{\rm res}$ and $B_0$ see below).The associated 
resonance width $\Gamma(k)$ vanishes near zero energy,  
with a  threshold behavior linear in $k=\sqrt{2M_rE}/\hbar$ 
due to the free particle density of states. It is convenient to 
define a characteristic length $r^{\star}>0$ by 
\begin{equation}
\label{eq:rstar}
\Gamma(k\to 0)/2=\frac{\hbar^2}{2M_r r^{\star}}\, k\, .
\end{equation} 
The scattering length $a=-\lim_{k\to 0}\tan{(\delta_{bg}+\delta_{res})}/k$
then has the simple form
\begin{equation}
\label{eq:FBs-length2}
a=a_{bg}-\frac{\hbar^2}{2M_r r^{\star}\nu}\, .
\end{equation}
This agrees precisely with Eq.~(\ref{eq:FBs-length}) provided the 
width parameter $\Delta B$ is identified with the combination 
$\mu\Delta B a_{bg}=\hbar^2/(2M_r r^{\star})$ of the two 
characteristic lengths $a_{bg}$ and $r^{\star}$.

On a microscopic level, these parameters may be 
obtained from the two channel Hamiltonian ~(\ref{eq:2channel})
by the standard Green
function formalism. In absence of coupling $W(r)$ the scattering
properties of the open channel are characterized by $G_{\rm
op}(E)=\left(E-H_{\rm op}\right)^{-1}$ with $H_{\rm
op}=P^2/(2M_r)+V_{\rm op}(r)$. We denote by $|\phi_0\rangle$ the
eigenstate of $H_{\rm op}$ associated with the energy $0$
 which behaves as $\phi_0(r)\sim (1-a_{\rm bg}/r)$ for large $r$. 
In the vicinity of the resonance the closed channel contributes
essentially through the state $\phi_{\rm res}$, and its Green
function reads
 \begin{equation}
G_{\rm cl}(E,B)\simeq \frac{|\phi_{\rm res}\rangle \langle
\phi_{\rm res}|}{E-E_{\rm res}(B)}\ .
 \end{equation}
With this approximation one can project the eigenvalue equation
for the Hamiltonian $\hat H$ onto the background and closed
channels. One can then derive the scattering length $a(B)$ of the
coupled channel problem and write it in the form of Eq.~(\ref{eq:FBs-length}). The position of the zero energy resonance
$B_0$ is shifted with respect to the `bare' resonance value
$B_{\rm res}$ by
 \begin{equation}
\mu (B_0-B_{\rm res})=- \langle \phi_{\rm res}| WG_{\rm op}(0)W
|\phi_{\rm res}\rangle \, .
 \end{equation}
The physical origin of this 
resonance shift is that an infinite scattering length  
requires that the contributions to $k\cot\delta(k)$ in the total 
scattering amplitude from
the open and the closed channel precisely cancel. In a 
situation where the background scattering length 
deviates considerably from its typical value $\bar{a}$
and where the off-diagonal coupling measured by $\Delta B$ 
is strong, this cancellation appears already when the bare
closed channel bound state is far away from the continuum
threshold. A simple analytical estimate
for this shift has been given by \textcite{Julienne:2004}
\begin{equation}
\label{eq:res-shift}B_0=B_{res}+\Delta B\cdot\frac{x(1-x)}{1+(1-x)^2}\, ,
\end{equation} 
where $x=a_{\rm bg}/\bar a$. The characteristic length $r^*$ defined in (\ref{eq:rstar})
is determined by the off-diagonal coupling via  
 \begin{equation}
 \langle \phi_{\rm res}| W |\phi_{0}\rangle
 =\frac{\hbar^2}{2M_r}\sqrt{\frac{4\pi}{r^{\star}}}\ .
 \label{eq:rstar2}
 \end{equation}

We shall be mostly interested in the \emph{wide resonance} case,
which corresponds to the situation where $r^*\ll |a_{\rm bg}|$.
Since the background scattering length
is generically on the order of the van der Waals length Eq.~(\ref{eq:vdwlength}),
this implies that the width $\mu|\Delta B|$ in the detuning over which
the scattering length deviates considerably from its background value is much 
larger than the characteristic energy $E_c$ below which 
collisions are in the ultracold regime.  For a
quantitative estimate, consider the specific resonances in
fermionic $^6$Li and $^{40}$K at $B_0=834\,$G and $B_0=202\,$G
respectively, which have been used to study the BCS-BEC crossover
with cold atoms (see section VIII). They are characterized by the
experimentally determined parameters $a_{\rm bg}=-1405\, a_B$,
$\Delta B=-300\,$G,  $\mu=2\mu_B$ and $a_{\rm bg}=174\, a_B$,
$\Delta B=7.8\,$G,  $\mu=1.68\mu_B$ respectively, where $a_B$ and
$\mu_B$ are the Bohr radius and Bohr magneton. From these
parameters, the characteristic length associated with the two
resonances turns out to be $r^{\star}=0.5\, a_B$ and
$r^{\star}=28\, a_B$, both obeying the wide resonance condition
$r^{\star}\ll\vert a_{\rm bg}\vert $.

\paragraph*{Weakly bound states close to the resonance}

In addition to the control of scattering properties, an important
feature of the Feshbach resonance concerns the possibility to
resonantly form weakly bound dimers in the region $a>0$. We
briefly present below some key properties of these dimers,
restricting for simplicity to the wide resonance case and to the
vicinity of the resonance $|B-B_0| \ll |\Delta B|$, so that the
scattering length $a(B)$ deviates substantially from its
background value $a_{\rm bg}$.

To determine the bound state for the two-channel Hamiltonian 
(\ref{eq:2channel}), one considers the
Green function $G(E)=(E-\hat H)^{-1}$ and looks for the low energy
pole at $E=-\varepsilon_{\rm b}<0$ of this function. The corresponding bound state
can be written
 \begin{equation}\langle\mathbf{x}|\Psi^{\rm (b)}\rangle =
 \begin{pmatrix}
 \sqrt{1-Z}\ \psi_{\rm bg}(r) \\
 \sqrt{Z}\; \phi_{\rm res}(r)
 \end{pmatrix}\ ,
 \end{equation}
where the coefficient $Z$ characterizes the closed channel
admixture. The values of $\varepsilon_{\rm b}$ and $Z$ can be calculated
explicitly by projecting the eigenvalue equation for $\hat H$ on
each channel. The binding energy
\begin{equation}
\label{eq:bindingenergy}
\varepsilon_b=\bigl(\mu(B-B_0)\bigr)^2/\varepsilon^{\star}\, +\ldots
\end{equation}
of the weakly bound state vanishes quadratically near the resonance, 
with characteristic energy
$\varepsilon^{\star}=\hbar^2/2M_r(r^{\star})^2$. The length $r^{\star}$ 
thus provides a direct measure of the curvature in the energy of
the weakly bound state as a function of
the detuning $\mu(B-B_0)$. In an experimental situation
which starts from the atom continuum,
it is precisely this weakly bound state which is reached upon
varying the detuning by an adiabatic change in the magnetic field
around $B_0$. The closed channel admixture $Z$ can be written as
 \begin{equation}
Z\simeq 2\frac{r^*}{|a_{\rm bg}|}\; \frac{|B-B_0|}{|\Delta B|}
=2\frac{|\nu|}{\epsilon^*} \, .
 \end{equation}
For a wide resonance, where $r^{\star}\ll\vert a_{\rm bg}\vert $,
this admixture is always much smaller than one 
over the magnetic field range $|B-B_0|\lesssim |\Delta B|$.
The bound state near the Feshbach resonance can thus be described
in a single-channel picture, without explicitly taking into
account the closed channel state.

The bound state $|\Psi^{\rm (b)}\rangle$ that we just presented
should not be confused with the bound state $|\Phi_{\rm
op}^{\rm (b)}\rangle$, that  exists for $a_{\rm bg}>0$ in the
open channel, for a vanishing coupling $W(r)$. The bound state
$|\Phi_{\rm op}^{\rm (b)}\rangle$ has a binding energy $\sim
\hbar^2/(2M_r a_{\rm bg}^2)$, that is much larger than that of 
Eq.~(\ref{eq:bindingenergy}) when $|B-B_0| \ll |\Delta B|$. For $|B-B_0| \sim |\Delta B|$
the states $|\Psi^{\rm (b)}\rangle$ and $|\Phi_{\rm op}^{\rm
(b)}\rangle$ have comparable energies and undergo an avoided
crossing. The universal character of the above results is then
lost and one has to turn to a specific study of the eigenvalue
problem.

To conclude, Feshbach resonances provide a flexible tool to change
the interaction strength between ultracold atoms over a wide
range. To realize a proper many-body  Hamiltonian with tunable
two-body interactions, however, an additional requirement is that
the relaxation rate into deep bound states due to three-body
collisions is negligible. As will be discussed in section VIII.A,
this is possible for fermions, where the relaxation rate is small
near Feshbach resonances \cite{Petrov:2004b,Petrov:2005}. 


\section{OPTICAL LATTICES}
\label{sec:lattices}

In the following, we will discuss how to confine cold 
atoms by laser light into configurations of a reduced dimensionality or 
in periodic lattices, thus generating situations in which
the effects of interactions are strongly enhanced.

\subsection{Optical potentials}
The physical origin of the confinement of cold 
atoms with laser light is the dipole force
\begin{equation}
\label{eq:dipoleforce}
\mathbf{F}=\frac{1}{2}\alpha(\omega_L)\nabla (|\mathbf{E}(\rvec)|^2) 
\end{equation}
due to a spatially varying ac-Stark shift
which atoms experience in an off-resonant light field \cite{Grimm:2000}.
Since the time scale for the center-of-mass motion of the atoms
is much slower that the inverse laser frequency $\omega_L$, only the 
time averaged intensity $|\mathbf{E}(\rvec)|^2$ enters.
The direction of the force depends on
the sign of the polarizability $\alpha(\omega_L)$.  In the vicinity
of an atomic resonance from the ground $\vert g\rangle$ to an 
excited state $\vert e\rangle$ at frequency
$\omega_0$, the polarizability has the standard form $\alpha(\omega_L)
\approx \vert\langle e\vert \hat{d}_{\mathbf{E}}\vert g\rangle\vert^2/\hbar
(\omega_0-\omega_L)$, with $\hat{d}_{\mathbf{E}}$ the dipole operator
in the direction of the field. Atoms are thus 
attracted to the nodes or to the anti-nodes of the laser intensity
for blue detuned $(\omega_L>\omega_0)$ or red detuned
$(\omega_L<\omega_0)$ laser light respectively.  
A spatially dependent intensity profile $I(\rvec)$ therefore
creates a trapping potential for neutral atoms.  Within a 
two level model, an explicit form of the dipole potential may
be derived by using the rotating wave approximation, which is a
reasonable approximation provided that the detuning
$\Delta=\omega_L-\omega_0$ of the laser field is small compared to the
transition frequency itself $|\Delta| \ll \omega_0$. With $\Gamma$ as 
the decay rate of the excited state, one obtains for $|\Delta| \gg \Gamma$ 
\cite{Grimm:2000}

\begin{equation}
\label{eq:dippot}
V_{\rm dip}(\rvec)= \frac{3\pi c^2}{2\omega_0^3} \frac{\Gamma}{\Delta}
I(\rvec),
\end{equation}
which is attractive or repulsive for red ($\Delta<0$) or blue 
($\Delta>0$) detuning, respectively. 
Atoms are thus attracted or repelled from
an intensity maximum in space. It is important to note that,
in contrast to the form suggested in Eq.~(\ref{eq:dipoleforce}), the light force
is not fully conservative.  Indeed, spontaneous emission gives rise
to an imaginary part of the polarizability. Within a two level
approximation, the related scattering rate  
$\Gamma_{sc}(\rvec)$ leads to 
an absorptive contribution $\hbar\Gamma_{sc}(\rvec)$ 
to the conservative dipole potential (\ref{eq:dippot}), which 
can be estimated as \cite{Grimm:2000} 
\begin{equation}
\label{eq:scattrate}
\Gamma_{sc}(\rvec) = \frac{3\pi
c^2}{2\hbar\omega_0^3}\left(\frac{\Gamma}{\Delta}\right)^2 I(\rvec)\, .
\end{equation}
As the eqs.~(\ref{eq:dippot},\ref{eq:scattrate}) show, the
ratio of scattering rate to the optical potential depth 
vanishes in the limit $|\Delta| \gg \Gamma$. A strictly conservative
potential can thus be reached in principle 
by increasing the detuning of the laser field. In practice
however, such an approach is limited by the maximum available laser
power. For experiments with ultracold quantum gases of alkali atoms,
the detuning is typically chosen to be large compared to the excited
state hyperfine structure splitting and in most cases even large
compared to the fine structure splitting in order to sufficiently
suppress spontaneous scattering events. 

The intensity profile $I(r,z)$ of a gaussian laser beam
propagating along the $z\,$-direction has the form
\begin{equation}
\label{eq:gaussianbeam}
    I(r,z)=\frac{2P}{\pi w^2(z)} e^{-2 r^2/w^2(z)}\, .
\end{equation}
Here $P$ is the total power of the laser beam,
$r$ is the distance from the center and
$w(z)=w_0\sqrt{1+z^2/z_R^2}$ is the $1/e^2$ radius.
This radius is characterized by a beam waist $w_0$
which is typically around $100\,\mu$m. 
Due to the finite beam divergence, the beam width
increases linearly with $z$ on a scale $z_R=\pi w_0^2/\lambda$
which is called the Rayleigh length.  Typical 
values for $z_R$ are in the mm range. Around the
intensity maximum a potential depth minimum occurs for a red detuned
laser beam, leading to an approximately harmonic potential 
\begin{equation}
V_{\rm dip}(r,z)\approx - V_{trap}\left\{ 1-2 \left( \frac{r}{w_0}\right)^2
- \left( \frac{z}{z_R}\right)^2\right\}\, .
\end{equation}
The trap depth $V_{trap}$ is linearly proportional to the laser
power and typically ranges from a few kHz up to a MHz (from the Nanokelvin to the Microkelvin regime). 
The harmonic confinement is characterized by radial $\omega_r$ and
axial $\omega_{z}$ trapping frequencies $\omega_r= (4 V_{trap}/M w_0^2
)^{1/2}$ and $\omega_z=(2V_{trap}/M z_R^2)$. Optical traps for 
neutral atoms have a wide range of applications \cite{Grimm:2000}.
In particular, they are inevitable in situations where magnetic trapping does 
not work for the atomic states under consideration. This is often the case 
when the interactions are manipulated via Feshbach resonances, 
involving high field seeking atomic states.
\bigskip

\noindent {\it Optical Lattices} A periodic potential is simply generated by overlapping two
counterpropagating laser beams. Due to the interference between the
two laser beams an optical standing wave with period $\lambda/2$ is
formed, in which the atoms can be trapped.
More generally, by choosing the two laser beams to 
interfere under an angle less than 180$^\circ$, one can also realize
periodic potentials with a larger period \cite{Peil:2003,Hadzibabic:2004}.
The simplest possible periodic optical potential is formed by
overlapping two counterpropagating beams. For a gaussian profile,
this results in a trapping potential of the form
\begin{equation}\label{eq:simple1Dlat}
V(r,z)= -V_0\cdot e^{-2r^2/w^2(z)}\cdot \sin^2(kz)
\end{equation}
where $k=2\pi/\lambda$ is the wave
vector of the laser light and $V_0$ the maximum depth of the
lattice potential. Note that due to the interference of the two
laser beams $V_0$ is four times larger than $V_{\rm trap}$ if the laser
power and beam parameters of the two interfering lasers are equal.

Periodic potentials in two dimensions can be formed by overlapping
two optical standing waves along different, usually orthogonal,  directions. 
For orthogonal polarization vectors of the two laser fields no interference
terms appear. The resulting optical potential in the center of the trap is then a simple sum of a purely sinusoidal potential in both directions.  

\begin{figure}
\begin{center}
\includegraphics[width=0.9\linewidth]{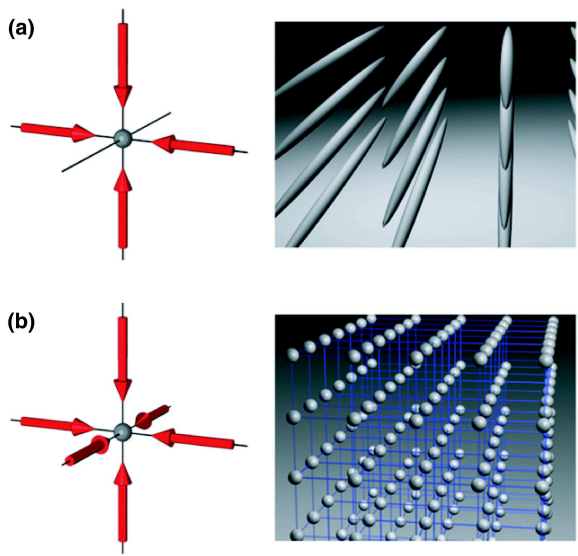}
\caption{Two-dimensional {\bf (a)} and three-dimensional {\bf (b)}
optical lattice potentials formed by superimposing two or three
orthogonal standing waves. For a two-dimensional optical lattice,
the atoms are confined to an array of tightly confining
one-dimensional potential tubes, whereas in the three-dimensional
case the optical lattice an be approximated by a three dimensional
simple cubic array of tightly confining harmonic oscillator
potentials at each lattice site.\label{fig:lattices2d3d}}
\end{center}
\end{figure}

In such a two-dimensional optical lattice potential, the atoms are
confined to arrays of tightly confining one-dimensional tubes (see
Fig.~\ref{fig:lattices2d3d}a). For typical experimental parameters
the harmonic trapping frequencies along the tube are very weak and
on the order of 10-200\,Hz, while in the radial direction the
trapping frequencies can become as high as up to 100\,kHz. 
For sufficiently deep lattice depths, atoms 
can thus move only axially along the tube. In this manner, 
it is possible to realize quantum wires
with neutral atoms, which allow to study strongly correlated gases
in one dimension, as discussed in Section~\ref{sec:onedim}. Arrays of 
such quantum wires have been realized by several groups  
\cite{Greiner:2001b,Moritz:2003,Tolra:2004,Paredes:2004,Kinoshita:2004}.

For the creation of a three-dimensional lattice potential, three
orthogonal optical standing waves have to be overlapped. 
The simplest case of independent standing waves, with no cross
interference between laser beams of different standing waves
can be realized by choosing orthogonal polarization
vectors and also by
using slightly different wavelengths for the three standing waves. 
The resulting optical potential is then simply given by the sum of
three standing waves. In the center of the trap, for
distances much smaller than the beam waist, the trapping potential
can be approximated as the sum of a homogeneous periodic lattice
potential 
\begin{equation}
\label{eq:V_p}
V_p(x,y,z)=V_{0}\left(\sin^2{kx}+\sin^2{ky}
+\sin^2{kz}\right)
\end{equation}
and an additional external harmonic confinement due to the
gaussian laser beam profiles. In addition to this harmonic confinement, a confinement due to the magnetic trapping is often used, which has to be taken into account as well for the total harmonic confinement of the atom cloud.

For deep optical lattice potentials, the confinement on
a single lattice site is approximately harmonic. The atoms
are then tightly confined at a single lattice site, with trapping frequencies
$\omega_{0}$ of up to $100$\,kHz. 
The energy $\hbar\omega_0=2E_r\left(V_0/E_r\right)^{1/2}$ of
local oscillations in the well is on the order of several recoil energies $E_r=\hbar^2 k^2/2m$, which is a natural measure of energy scales in optical lattice potentials. Typical values of $E_r$ are in the range of several kHz for $^{87}$Rb. 
\bigskip                                                      

\noindent {\it\label{sec:spindepoptlat} Spin Dependent Optical Lattice Potentials}
For large detunings of the laser light forming the optical lattices compared
to the fine structure splitting of a typical alkali atom, the resulting optical lattice potentials are almost the same for all magnetic sublevels in the ground state manifold of the atom. However, for more near-resonant light fields, situations can be created for which the different magnetic sublevels can be exposed to vastly different optical potentials \cite{Jessen:1996}. Such spin dependent lattice potentials can e.g. be created in a standing
wave configuration formed by two counterpropagating laser beams with
linear polarization vectors enclosing an angle $\theta$ \cite{Jessen:1996,Brennen:1999,Jaksch:1999,Mandel:2003a}.
 The resulting standing wave light field can be decomposed into a
superposition of a $\sigma^+$ and $\sigma^-$ polarized standing wave
laser field, giving rise to lattice potentials $V_+(x,\theta)=V_0
\cos^2(kx+\theta/2)$ and $V_-(x,\theta)=V_0 \cos^2(kx-\theta/2)$.
By changing the polarization angle $\theta$ one can thereby control the
relative separation between the two potentials $\Delta x =\theta/\pi
\cdot \lambda_x/2$. When increasing $\theta$, both potentials shift
in opposite directions and overlap again when
$\theta=n\cdot \pi$, with $n$ being an integer. Such a configuration has been used to coherently move atoms across lattices and
realize quantum gates between them \cite{Jaksch:1999,Mandel:2003a,Mandel:2003b}. Spin dependent lattice potentials, furthermore offer a convenient way to tune the interactions between two atoms in different spin states. By shifting the spin dependent lattices relative to each other, the overlap of the on-site spatial wave function can be tuned between zero and its maximum value, thus controlling the interspecies interaction strength within a restricted range. 
Recently, \textcite{Sebby:2006} have also been able to demonstrate a novel spin dependent lattice geometry, in which 2D arrays of double well potentials could be realized. Such 'superlattice' structures allow for versatile intra- and inter-well manipulation possibilities \cite{Sebby:2007,Lee:2007,Foelling:2007}. A variety of lattice structures can be obtained by interfering laser beams under different angles, see e.g. \cite{Jessen:1996,Grynberg:2001}.

\subsection{Band Structure}

 We consider in this section the single particle eigenstates in an infinite periodic potential.
Any additional potential, that could originate from the intensity profile of the laser beams, or from some magnetic confinement is neglected (for the single particle spectrum in the presence of an 
additional harmonic confinement see \cite{Hooley:2004}). In a simple cubic lattice, the potential is given by Eq.~(\ref{eq:V_p}), with a tunable amplitude $V_0$ and lattice constant 
$d=\pi/k$. In the limit $V_0\gg E_r$, each well supports a number of 
vibrational levels, separated by an energy $\hbar\omega_0\gg E_r$.
At low temperatures, the atoms are restricted to the lowest vibrational level at each site.
Their kinetic energy is then frozen, except for the small tunneling amplitude to neighboring sites. 
The associated single-particle eigenstates in the lowest band are Bloch waves with
quasi-momentum $\mathbf{q}$ and energy 
\begin{equation}
\label{eq:Blochband}
\varepsilon_0(\mathbf{q})=\frac{3}{2}\hbar\omega_0 -2J\left(\cos{q_xd}+\cos{q_yd}
+\cos{q_zd}\right) +\ldots
\end{equation}
The parameter $J>0$ is the gain in kinetic energy
due to nearest neighbor tunneling. In the limit $V_0\gg E_r$, it can be
obtained from the width $W\to 4J$ of the lowest band in the
1D Mathieu-equation
\begin{equation}
\label{eq:Jmathieu}
J=\frac{4}{\sqrt\pi}E_r\left(\frac{V_0}{E_r}\right)^{3/4}
\exp {-2\left(\frac{V_0}{E_r}\right)^{1/2}} \, .
\end{equation}
For lattice depths larger than $V_0>15 E_r$ this approximation yields $J$ to better than $10\%$ accuracy (see Table I). More generally, for any periodic potential $V_p(\rvec+\mathbf{R})
=V_p(\rvec)$ which is not necessarily deep and separable,
the exact eigenstates are Bloch functions $\psi_{n,\mathbf{q}}(\rvec)$. They are
characterized by a discrete band index $n$ and a quasi-momentum $\mathbf{q}$ 
within the first Brillouin-zone of the reciprocal lattice \cite{Ashcroft:1976}.
Since Bloch functions are multiplied by a pure phase factor 
$\exp{i\mathbf{q}\cdot\mathbf{R}}$ upon translation by one of the lattice vectors $\mathbf{R}$, 
they are extended over the whole lattice. An alternative single-particle basis, 
which is more useful for describing the hopping of particles among the 
discrete lattice sites $\mathbf{R}$, are 
the Wannier functions $w_{n,\mathbf{R}}(\rvec)$. They are connected 
with the Bloch functions by a Fourier transform
\begin{equation}
\label{eq:Wannier}
\psi_{n,\mathbf{q}}(\rvec)=\sum_{\mathbf{R}}
w_{n,\mathbf{R}}(\rvec)e^{i\mathbf{q}\cdot\mathbf{R}}
\end{equation}
on the lattice. The Wannier functions 
depend only on the relative distance $\rvec-\mathbf{R}$ and,
at least for the lowest bands, they are centered around the lattice 
sites $\mathbf{R}$ (see below). By choosing a convenient normalization, they
obey the orthonormality relation 
\begin{equation}
\label{eq:orthonorm}
\int d^3r\, w_n^*(\rvec-\mathbf{R})w_{n'}(\rvec-\mathbf{R'})=
\delta_{n,n'}\,\delta_{\mathbf{R},\mathbf{R'}}
\end{equation}
for different bands $n$ and sites $\mathbf{R}$. Since the 
Wannier functions for all bands $n$ and sites $\mathbf{R}$ 
form a complete basis, the operator $\hat{\psi}(\rvec)$ which 
destroys a particle at an arbitrary point $\rvec$ can be 
expanded in the form
\begin{equation}
\label{eq:fieldop}
\hat{\psi}(\rvec)=\sum_{\mathbf{R},n}w_n(\rvec-\mathbf{R})\,\hat{a}_{\mathbf{R},n}\, . 
\end{equation}
Here, $\hat{a}_{\mathbf{R},n}$ is the annihilation operator for particles in the 
corresponding Wannier states, which are not necessarily well localized at site $\mathbf{R}$. 
The Hamiltonian for free motion on a periodic lattice then has the form 
\begin{equation}
\label{eq:H_0,p}
\hat{H}_0=\sum_{\mathbf{R},\mathbf{R'},n}J_n(\mathbf{R}-\mathbf{R'})\,
\hat{a}^{\dagger}_{\mathbf{R},n}\hat{a}_{\mathbf{R'},n}\, .
\end{equation}
It describes the hopping in a given band $n$ with matrix elements $J_n(\mathbf{R})$,
which in general connect lattice sites at arbitrary distance $\mathbf{R}$.
The diagonalization of this Hamiltonian by Bloch states (\ref{eq:Wannier})
shows that the hopping matrix elements $J_n(\mathbf{R})$ are uniquely determined by
the Bloch band energies $\varepsilon_n(\mathbf{q})$ via
\begin{equation}
\label{eq:J_n}
\sum_{\mathbf{R}}J_n(\mathbf{R})\exp{i\mathbf{q}\cdot\mathbf{R}}=
\varepsilon_n(\mathbf{q})\, .
\end{equation}
In the case of seperable periodic potentials
$V_p(\rvec)=V(x)+V(y)+V(z)$, generated by three orthogonal optical lattices,
the single particle problem is one-dimensional, and a complete analysis 
of Wannier functions has been given by \textcite{Kohn:1959}.
Choosing appropriate phases for the Bloch functions,
there is a unique Wannier function
for each band, which is real and exponentially localized. The decay 
$\sim\exp{-h_n\vert x\vert}$ is characterized by 
a decay constant $h_n$, which is a decreasing function of the band index $n$.
For the lowest band $n=0$, where the Bloch function at $q=0$ is finite at the origin, 
the Wannier function $w(x)$ can be choosen to be symmetric around $x=0$
(and correspondingly it is antisymmetric for the first excited band). 
More precisely,  the asymptotic behavior of the 1D Wannier functions
and the hopping matrix elements is
$\vert w_n(x)\vert\sim \vert x\vert^{-3/4}\exp{-h_n\vert x\vert}$ 
and $J_n(R)\sim\vert R\vert^{-3/2}\exp{-h_n\vert R\vert}$, 
respectively \cite{He:2001}. In the particular case of a purely 
sinusoidal potential $V_0\sin^2(kx)$ with lattice constant $d=\lambda/2$,
the decay constant $h_0$ increases monotonically with $V_0/E_r$. 
In the deep lattice limit $V_0\gg E_r$,
it approaches $h_0d=\pi\sqrt{V_0/E_r}/2$.
It is important to realize, that even in
this limit, the Wannier function does not uniformly converge to the local harmonic 
oscillator ground state $\phi$ of each well: the $w_n(x)$ decay exponentially
rather than in a Gaussian manner and they always have nodes in order to 
guarantee the orthogonality relation (\ref{eq:orthonorm}). Yet, as shown in
Table~\ref{tab:hoppingelements}, the overlap is near one even for shallow optical lattices.
\begin{table}
\begin{tabular}{|c|c|c|c|c|c||} \hline 
$V_0/E_r$  &  $~~~4J/E_r~~~$  &  $~~~W/E_r~~~$  &  $~~~J(2d)/J~~~$  
	   &  $~|\langle w|\phi\rangle|^2~$		\\ \hline
3   &  0.444109  &  0.451894  &  0.101075  &  0.9719	\\
5   &	 0.263069  &  0.264211  &  0.051641  &  0.9836	\\
10  &	 0.076730  &  0.076747  &  0.011846  &  0.9938	\\
15  &	 0.026075  &  0.026076  &  0.003459  &  0.9964  \\
20  &	 0.009965  &  0.009965  &  0.001184  &  0.9975	\\	\hline
\end{tabular}
\caption{Hopping matrix elements to nearest ($J$) and 
next nearest neighbors ($J(2d)$), bandwidth $W$ and overlap between the Wannier
function and the local gaussian ground state in 1D optical lattices. Table courtesy of M.~Holthaus.}
\label{tab:hoppingelements}
\end{table}

\subsection{Time-of-flight and adiabatic mapping}
\label{sec:TOF}

\begin{figure}
\begin{center}
\includegraphics[width=\columnwidth]{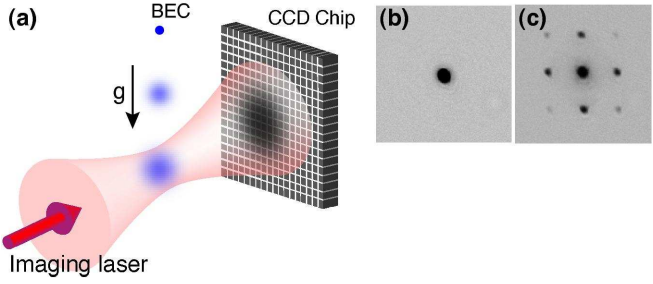}
\caption{Schematic setup for absorption imaging after a time-of-flight period {\bf (a)}. Absorption image for a BEC released from a harmonic trap {\bf (b)}. Absorption image for a BEC released from a shallow optical lattice ($V_0=6\,E_r$){\bf (c)}. Note the clearly visible interference peaks in the image. \label{fig:Imaging}}
\end{center}
\end{figure}

\noindent {\it Sudden release} When releasing ultracold quantum gases from an optical lattice, two possible release methods can be chosen. If the lattice potential is turned off abruptly and 
interaction effects can be neglected, a given Bloch state with quasi-momentum $q$ will expand according to its momentum distribution as a superposition of plane waves with momenta $p_n=\hbar q \pm n\times2\hbar k$. This is a direct consequence of the fact that Bloch waves can be expressed as a superposition of plane wave states $\exp{i(\mathbf{q}+\mathbf{G})\cdot\mathbf{r}}$ 
with momenta $\mathbf{q}+\mathbf{G}$, which include arbitrary reciprocal lattice vectors 
$\mathbf{G}$. In a simple cubic lattice with lattice spacing $d=\pi/k$, the vectors $\mathbf G$ 
are integer multiples of the fundamental reciprocal lattice vector $2k$. 
After a certain time-of-flight time, this momentum distribution can be imaged using standard absorption imaging methods. If only a single Bloch state is populated, as is the case for a Bose-Einstein condensate with quasi-momentum $q=0$, this results in a series of interference maxima that can be observed after a time-of-flight period $t$ (see Fig.~\ref{fig:Imaging}).
As will be shown in section III.A below, the density distribution observed after a fixed time-of-flight at position $\mathbf{x}$, is nothing but the momentum distribution of the particles trapped in the lattice

\begin{equation}
    n(\mathbf{x})=\left(\frac{M}{\hbar t}\right)^3 |\tilde{w}(\mathbf{k})|^2 \mathcal{G}(\mathbf{k}).  
\label{eq:momentumdistribution}             
\end{equation}                                                          

\noindent Here $\mathbf{k}$ is related to $\mathbf{x}$ by $\mathbf{k}=M\mathbf{x}/\hbar t$
due to the assumption of ballistic expansion while $\tilde{w}(\mathbf{k})$ is the Fourier transform 
of the Wannier function. The coherence properties of the many-body state are 
characterized by the Fourier transform 

\begin{equation}
	\mathcal{G}(\mathbf{k})= \sum_{\Rvec,\Rvec'}e^{i \mathbf{k} \cdot (\Rvec-\Rvec')} 
	G^{(1)}\!\left(\Rvec, \Rvec'\right) 
	\label{eq:structurefunctionlattice}
\end{equation}

\noindent of the one-particle density matrix $G^{(1)}\!\left(\Rvec, \Rvec'\right)\!=\!\langle\ahatdag_\Rvec \ahat_{\Rvec'} \rangle$. 

In a BEC, the long range order in the amplitudes leads to a constant value of 
the first order coherence function $G^{(1)}\left(\Rvec, \Rvec'\right)$ at 
large separations $|\mathbf{R}-\mathbf{R}'|$ (see Sec.~\ref{sec:appendix}). The resulting momentum distribution coincides with the standard multiple wave interference pattern obtained with light diffracting off a material grating (see Fig.~\ref{fig:Imaging}c and section IV.B below). The atomic density distribution observed after a fixed time-of-flight time, thus yields information on the coherence properties of the many-body system. It should be noted, however, that the observed density distribution after time-of-flight can deviate from the in-trap momentum distribution, if interaction effects during the expansion occur, or the expansion time is not so long that the initial size of the atom cloud can be neglected ("far-field approximation") \cite{Pedri:2001,Gerbier:2007a}.  It is important to be aware of these discrepancies and take them into account for an interpretation of the experimental data.           
\bigskip

\begin{figure}
\begin{center}
\includegraphics[width=\linewidth]{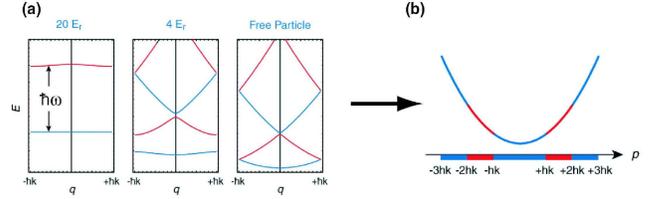}
\caption{{\bf (a)} Bloch bands for different potential depths.
During an adiabatic ramp down the quasi momentum is conserved and
{\bf (b)} a Bloch wave with quasi momentum $q$ in the $n$th energy
band is mapped onto a free particle with momentum $p$ in the $n$th
Brillouin zone of the lattice. Reprinted with permission from \textcite{Greiner:2001b}.}\label{fig:AdiabaticRampDown}
\end{center}
\end{figure}

\noindent {\it Adiabatic mapping} One of the advantages of using optical lattice potentials is
that the lattice depth can be dynamically controlled by
simply tuning the laser power. This opens another possibility for releasing the atoms from the lattice potential e.g. by adiabatically
converting a deep optical lattice into a shallow one and eventually
completely turning off the lattice potential. Under adiabatic
transformation of the lattice depth, the quasi-momentum $\mathbf q$ is preserved
and during the turn off process a Bloch wave in the $n$th energy
band is mapped onto a corresponding free particle momentum $\mathbf p$ in
the $n$th Brillouin zone (see Fig.~\ref{fig:AdiabaticRampDown})
\cite{Kastberg:1995,Greiner:2001b,Kohl:2005a}.

\begin{figure}
\begin{center}
\includegraphics[width=\linewidth]{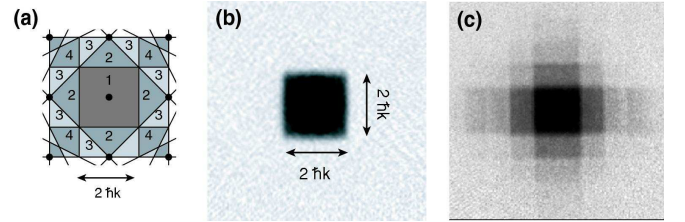}
\caption{{\bf (a)} Brillouin zones of a 2D simple cubic optical
lattice. For a homogeneously filled lowest Bloch band, an adiabatic
shut off of the lattice potential leads to a homogeneously populated
first Brillouin zone, which can be observed through absorption
imaging after a time-of-flight expansion {\bf (b)}. If in addition
higher Bloch bands were populated, higher Brillouin zones become
populated as well {\bf (c)}. Reprinted with permission of \textcite{Greiner:2001b}.\label{fig:BrillouinExp}}
\end{center}
\end{figure}

The adiabatic mapping technique has been used with both bosonic
\cite{Greiner:2001b} and fermionic \cite{Kohl:2005a} atoms. For
the situation of a homogeneous filled lowest energy band, an adiabatic ramp
down of the lattice potential leaves the central Brillouin zone - a
square of width $2\hbar k$ - fully occupied (see
Fig.~\ref{fig:BrillouinExp}b). If on the other hand higher energy
bands are populated, one also observes populations in higher
Brillouin zones (see Fig.~\ref{fig:BrillouinExp}c). As in this method each Bloch wave is mapped onto a specific free-particle momentum state, it can
be used to efficiently probe the distribution of the particles over Bloch states in different energy bands.

\subsection{Interactions and two-particle effects}

So far we have only discussed single particle behavior of ultracold atoms in optical lattices. However, the short-ranged $s$-wave interactions between the particles give rise to an on-site interaction energy, when two or more atoms occupy a single lattice site. Within the pseudopotential approximation, the interaction between 
bosons has the form
\begin{equation}
\hat{H}^{'}=\frac{g}{2} \int d^3r \, \hat{\psi}^{\dagger}(\rvec)\hat{\psi}^{\dagger}(\rvec)
 \hat{\psi}(\rvec)\hat{\psi}(\rvec)\, .
\label{eq:BoseH}
\end{equation}
Inserting the expansion Eq.~(\ref{eq:fieldop}) leads to interactions 
involving Wannier states both in different bands and different lattice sites. 
The situation simplifies, however, for a deep optical lattice and the 
assumption that only the lowest band is occupied.
The overlap integrals are then dominated by the 
on-site term $U\hat n_{\mathbf{R}}(\hat n_{\mathbf{R}}-1)/2$, which is 
non-zero, if two or more atoms are in the same Wannier state. 
At the two-particle level, the 
interaction between atoms in Wannier states localized around $\mathbf{R}$ and 
$\mathbf{R}'$ is thus reduced to a local form $U\cdot\delta_{\mathbf{R},\mathbf{R}'}$ with 
\begin{equation}
\label{eq:U}
U=g\int d^3r\, \vert w(\rvec) \vert^4
 =\sqrt{\frac{8}{\pi}} ka\, E_{r}
\left(\frac{V_{0}}{E_{r}}\right)^{3/4}
\end{equation}
(for simplicity, the band index $n=0$ is omitted for the lowest band). 
The explicit result for the on-site interaction $U$ is obtained by taking $w(\rvec)$ as the Gaussian
ground state in the local oscillator potential. As mentioned above, this is {\it not}
the exact Wannier wave function of the lowest band. In the
deep lattice limit $V_0\gg E_r$, however , the result (\ref{eq:U})
provides the asymptotically correct behavior. Note that the
strength $|U|$ of the on-site interaction {\it in}creases with $V_0$, which is
due to the squeezing of the Wannier wave function $w(\rvec)$. 
\bigskip

\noindent {\it Repulsively bound pairs} Consider now an optical lattice at very low filling. 
An occasional pair of atoms at the same site has an energy $U$
above or below the center of the lowest band. In the attractive case
$U<0$, a two-particle bound state will form for sufficiently large 
values of $|U|$. In the repulsive case, in turn, the pair is expected 
to be unstable with respect to breakup into two separate 
atoms at different lattice sites, to save repulsive interaction. 
This process, however, is forbidden if the repulsion is {\it above} 
a critical value $U>U_c$. The physical origin for this surprising result is that momentum 
and energy conservation do not allow the two particles to separate.
There are simply no free states available if the energy lies more than 
$zJ$ above the band center, which is the upper edge of the tight binding band. Here $z$ denotes the number of nearest neighbours on a lattice.
Two bosons at the same lattice site will thus stay together if 
their interaction is sufficiently repulsive. 
In fact, the two-particle bound state above the band
for a repulsive interaction is the precise analog of the 
standard bound state below the band for 
attractive interactions, and there is a perfect 
symmetry around the band center.

\begin{figure}
\begin{center}
\includegraphics[width=\columnwidth]{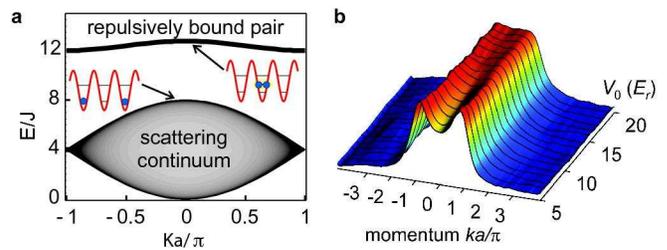}
\caption{Repulsively bound atom pairs. {\bf (a)} Spectrum of energy $E$ of the 1D Bose-Hubbard hamiltonian for $U/J=8$ as a function of the center of mass quasi-momentum $K$. The Bloch band for respulsively bound pairs is located above the continuum of unbound states. {\bf (b)} Experimentally measured quasi-momentum distribution of repulsively bound pairs vs. lattice depth $V_0$. Reprinted with permission from \textcite{Winkler:2006}. }\label{fig:RepulsivePairs}
\end{center}
\end{figure}

Such 'repulsively bound pairs' have been observed in a recent 
experiment by \textcite{Winkler:2006}.  A dilute gas of
$^{87}$Rb$_2$ Feshbach molecules was prepared in 
the vibrational ground state of an optical lattice. 
Ramping the magnetic field across a Feshbach resonance
to negative $a$, these molecules can be adiabatically dissociated and then
brought back again to positive $a$ as repulsive pairs. Since the  
bound state above the lowest band is built from states in which the 
relative momentum of the two particles is near the edge of the Brillouin zone, 
the presence of repulsively bound pairs can be inferred
from corresponding peaks in the quasi-momentum distribution observed
in a time-of-flight experiment (see Fig.~\ref{fig:RepulsivePairs}) \cite{Winkler:2006}. The 
energy and dispersion relation of these pairs follows from 
solving the equation $UG_{K}(E,0)=1$ for a bound state of two particles
with center-of-mass momentum $K$.  
In close analogy to Eq.~(\ref{eq:boundstate}) below,  
$G_{K}(E,0)$ is the local Green function 
for free motion on the lattice with hopping matrix element $2J$. 
Experimentally, the optical lattice was strongly anisotropic such that tunneling is possible only 
along one direction. The corresponding bound state equation in one 
dimension can be solved explicitely, giving \cite{Winkler:2006}

\begin{equation}
\label{eq:pairs}
E(K,U_1)=2J\sqrt{(2\cos{Kd/2})^2+(U_1/2J)^2}-4J
\end{equation}
for the energy with respect to the upper band edge.  
Since $E(K=0,U_1)>0$ for arbitrary small values of $U_1>0$,
there is always a bound state in one dimension. 
By contrast, in 3D there is 
a finite critical value, which is $U_c=7.9136\, J$ for a simple cubic lattice.
The relevant on-site interaction $U_1$ in one dimension is obtained from
Eq.~(\ref{eq:g_1}) below for the associated pseudopotential.
With $\ell_0$ the oscillator length for motion along the direction of hopping,
it is given by
\begin{equation}
\label{eq:U_1}
U_1=g_1\int dx\, |w(x)|^4=\sqrt{\frac{2}{\pi}}\,\hbar\omega_{\perp}\cdot\frac{a}{\ell_0}\, .
\end{equation}
Evidently, $U_1$ 
has the transverse confinement energy $\hbar\omega_{\perp}$ as the characteristic scale, 
rather than the recoil energy $E_r$ of Eq.~(\ref{eq:U}) in the 3D case.  
\bigskip

\noindent {\it Tightly confined atom pairs} The truncation to the lowest band requires that both the thermal and the on-site 
interaction energy $U$ are much smaller than $\hbar\omega_0$. In the deep lattice 
limit $V_0\gg E_r$, the condition $U\ll\hbar\omega_0$ leads to
$ka(V_0/E_r)^{1/4}\ll 1$ using (\ref{eq:U}).  This is equivalent to $a\ll\ell_0$, where
$\ell_0=\sqrt{\hbar/M\omega_0}=(E_r/V_0)^{1/4}d/\pi$ is the
oscillator length associated with the local harmonic motion in the deep wells
of the optical lattice. The assumption of staying in the lowest band in the 
presence of a repulsive interaction,  thus requires the scattering 
length to be much smaller than $\ell_0$ which is itself smaller, but of the 
same order, than the lattice spacing $d$.  For standard values 
$a\approx 5\,$nm and  $d\approx 0.5\,\mu$m, this condition is very well 
justified. In the vicinity of Feshbach resonances, however, the 
scattering lengths become comparable to the lattice spacing. 
A solution of the two-particle problem in the presence of 
an optical lattice for arbitrary values of the ratio $a/\ell_0$ has been given by
\textcite{Fedichev:2004}. Neglecting interaction induced couplings to higher  bands, they have shown that the effective interaction at energies 
smaller than the bandwidth
is again described by a pseudopotential. For repulsive interactions $a>0$, 
the associated effective scattering length reaches a bound 
$a_{\rm eff}\approx d$ on the order of the lattice spacing, 
even if $a\to\infty$ near a Feshbach resonance. In the case where the 
free space scattering length is negative,  $a_{\rm eff}$ exhibits
a geometric resonance which precisely describes the formation of a 
two-particle bound state at $|U|=7.9136 J$ discussed above.

This analysis is based on the assumption that the particles remain in 
a given band even in the presence of strong interactions. Near Feshbach 
resonances, however, this is usually not the case.  
In order to adress the question of interaction induced transitions between 
different bands, it is useful to consider the simple
problem of two interacting particles in a harmonic well \cite{Busch:1998}. 
Provided the range of the 
interaction is much smaller than the oscillator length $\ell_0$, the interaction
of two particles in a single well is still described by a pseudopotential.
The ratio of the scattering length $a$ to $\ell_0$, however, may be arbitrary. 
The corresponding energy levels $E=\hbar\omega_0(3/2\, -\Omega)$ as a function 
of the ratio $\ell_0/a$ follow from the transcendental equation 
\begin{equation}
\label{eq:Busch}
\frac{\ell_0}{a}=\frac{\sqrt{2}\,\Gamma(\Omega/2)}{\Gamma((\Omega -1)/2)}
=f_3(\Omega)\, .
\end{equation}
where $\Gamma(z)$ is the standard Gamma function.
In fact, this is the analytical continuation to an arbitrary sign of 
the dimensionless binding energy $\Omega$ in Eq.~(\ref{eq:f_n}) 
below, for the case $n=3$, since
a harmonic confinement is present in all three spatial directions.

\begin{figure}
\begin{center}
\includegraphics[width=0.7\columnwidth]{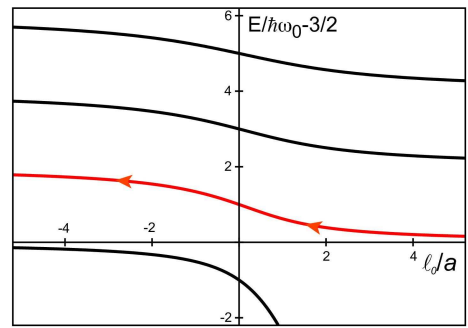}
\caption{Energy spectrum of two interacting particles in a 3D harmonic
oscillator potential from Eq.~(\ref{eq:Busch}). The arrows indicate the transfer of a pair
in the ground state to the first excited level by sweeping across a
Feshbach resonance. There is a single bound state below the lowest 
oscillator level, whose energy has been measured by \textcite{Stoferle:2006}.
\label{fig:FeshbachTightConfinement}.}
\end{center}
\end{figure}

\begin{figure}
\begin{center}
\includegraphics[width=\columnwidth]{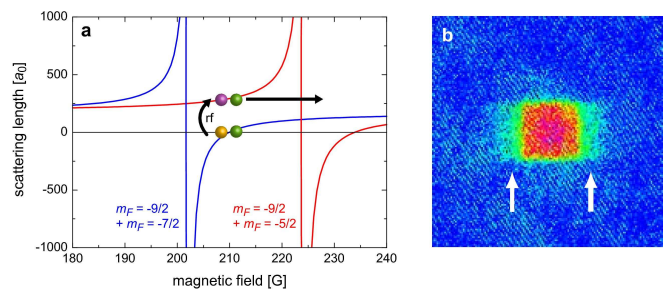}
\caption{Experimentally observed interaction induced transitions between
Bloch bands. {\bf a} Two Feshbach resonances between the $|F=9/2,m_F=-9/2\rangle$ and $|F=9/2,m_F=-7/2\rangle$ states (left) and the $|F=9/2,m_F=-9/2\rangle$ and $|F=9/2,m_F=-5/2\rangle$ states (right) are exploited to tune the interactions in the gas. {\bf b} Quasi-momentum distribution for a final magnetic field of B = 233G. Arrows indicate the atoms in the higher bands. Reprinted with permission from \textcite{Kohl:2005a}. \label{fig:HigherBands}.}
\end{center}
\end{figure}

As shown in Fig.~\ref{fig:FeshbachTightConfinement}, 
the discrete levels for the {\it relative} motion of two particles 
form a sequence which, at infinite scattering length, is
shifted upwards by precisely $\hbar\omega_0$ compared to the non-interacting
levels at zero angular momentum $E^{(0)}(n_r)=\hbar\omega_0(2n_r+3/2)$. 
In particular, a change of the scattering length from small positive to small negative values, through a Feshbach resonance where $a$ diverges, increases the energy by $2\hbar\omega_0$
while the particles are transferred to the next higher level $n_r=1$. 
Feshbach resonances can thus be used to switch pairs of particles
in individual wells of a deep optical lattice, where tunneling 
is negligible, to higher bands. Experimentally, this has been 
studied by \textcite{Kohl:2005a}. Starting from a two-component 
gas of fermionic $^{40}$K at $a\approx 0$ and unit filling, i.e. with two fermions
at each lattice site in the center of the trap,  the atoms were transferred to a different 
hyperfine state and the magnetic field was then increased
beyond the associated Feshbach resonance at $B_0=224\,$G
\footnote{Changing $a$ from positive to negative values avoids creation of 
molecules in an adiabatic ramp.}.
The resulting transfer of particles into higher bands is then revealed
by observing the quasi-momentum distribution in time-of-flight images
after adiabatically turning off the optical lattice,
see Fig.~\ref{fig:HigherBands}. It was pointed out by \textcite{Ho:2006}
that such Feshbach sweeps open novel possibilities to create fermionic 
Mott insulating states.

\section{DETECTION OF CORRELATIONS}
\label{sec:correlations}
In order to probe interacting many-body quantum states with strong correlations, it is essential to use detection methods that are sensitive to higher order correlations. Here, recent proposals for using analogues of quantum optical detection techniques have proven to be novel tools for analyzing strongly interacting quantum matter \cite{Altman:2004,Polkovnikov:2006a,Gritsev:2006,Zhang:2007,Duan:2006,Niu:2006}. Most of these techniques make use of the fact that the quantum fluctuations in many observables, such as e.g. the visibility of the interference pattern between two released quantum gases or the fluctuations in the momentum distribution after release from the trap, contain information of the initial correlated quantum state. Whereas in the usual time-of-flight momentum distributions one essentially probes first order coherence properties of the system, the noise-correlation techniques introduced below will yield information on the second (or higher) order correlation properties and therefore possible long range order in real space. Such correlation techniques in expanding atom clouds have begun to be successfully employed in recent experiments, probing the momentum correlations between atomic fragments emerging from a dissociated molecule \cite{Greiner:2005a}, revealing the quantum statistics of bosonic or fermionic atoms in an optical lattice \cite{Foelling:2005,Rom:2006}, or in exploring the Kosterlitz-Thouless transition in two-dimensional Bose-Einstein condensates \cite{Hadzibabic:2006}. All the correlation techniques for strongly correlated quantum gases can also greatly benefit from efficient single atom detectors that have recently begun to be used in the context of cold quantum gases \cite{Oettl:2005,Schellekens:2005,Jeltes:2007}. 

\subsection{Time-of-flight versus noise correlations}
Let us begin by considering a quantum gas released from a trapping potential. After a finite time-of-flight time $t$,
the resulting density distribution yields a three-dimensional density distribution $n_{3D}(\xvec)$ \footnote{In this section, we denote in-trap spatial coordinates by $\rvec$ and spatial coordinates after time-of-flight by $\xvec$ for clarity.}. If interactions can be neglected during time-of-flight, the average density distribution is related to the in-trap quantum state via:

\begin{eqnarray}
	        \langle \hat{n}_{3D}(\xvec)\rangle_{\rm tof} &=& \langle \ahatdag_{tof}(\xvec) \ahat_{\rm tof}(\xvec)\rangle_{\rm tof} \label{eq:tof_two}  \\
	&\approx& \langle \ahatdag(\kvec) \ahat(\kvec) \rangle_{\rm trap} = \langle \hat{n}_{3D}(\kvec)\rangle_{\rm trap}, \nonumber    
\end{eqnarray}
                                         
\noindent where  $\kvec$ and $\xvec$ are related by the ballistic expansion 
condition $\kvec = M \xvec/\hbar t$ (a factor $(M/\hbar t)^3$ from the transformation 
of the volume elements $d^3x\to d^3k$ is omitted, see Eq.~(\ref{eq:momentumdistribution})).
Here we have used the fact that for long time-of-flight 
times, the initial size of the atom cloud in the trap can be neglected. 
It is important to realize, that in each 
experimental image, a single realization of the density is observed, not an average. Moreover, each 
pixel in the image records on average a substantial number $N_{\sigma}$ of atoms. For each of those pixels, however, the number of atoms recorded in a {\it single realization} of an experiment will exhibit shot noise fluctuations of relative order $1/\sqrt{N_{\sigma}}$ which will be discussed below.                            
As shown in Eq.~(\ref{eq:tof_two}), the density distribution after time-of-flight represents a momentum distribution reflecting the first order coherence properties of the in-trap quantum state.
This assumption is however only correct, if during the expansion process interactions between the atoms do not modify the initial momentum distribution, which we will assume throughout the text. When the interactions between the atoms have been enhanced, e.g. by a Feshbach resonance, or a high density sample is prepared, such an assumption is not always valid. Near Feshbach resonances one therefore often ramps back to the zero crossing of the scattering length before expansion.

\bigskip

\noindent {\it Density-density correlations in time-of-flight images} Let us now turn to the observation of density-density correlations in the expanding atom clouds \cite{Altman:2004}. These are characterized by the density-density correlation function
\begin{equation}
    \langle\hat{n}(\xvec)\hat{n}(\xvec')\rangle= \langle\hat{n}(\xvec)\rangle\!\langle\hat{n}(\xvec')\rangle
    g^{(2)}(\xvec,\xvec')+\delta(\xvec - \xvec')\langle\hat{n}(\xvec)\rangle 
    \label{eq:g_2} 
\end{equation} 
which contains the normalized pair distribution $g^{(2)}(\xvec,\xvec')$ and a self correlation term.  
Relating the operators after time-of-flight expansion to the in-trap momentum operators, using Eq.~(\ref{eq:tof_two}), one obtains:
\begin{eqnarray}
	  & \langle \hat{n}_{3D}(\xvec)\hat{n}_{3D}(\xvec')\rangle_{\rm tof} \approx 
	  \langle \ahatdag(\kvec) \ahat(\kvec) \ahatdag(\kvec')\ahat(\kvec') \rangle_{\rm trap}=&\nonumber \\
  & \langle \ahatdag(\kvec) \ahatdag(\kvec')\ahat(\kvec')\ahat(\kvec)\rangle_{\rm trap}
+ \delta_{\kvec \kvec'} \langle \ahatdag(\kvec) \ahat(\kvec)\rangle_{\rm trap}&. \label{eq:corr1}
\end{eqnarray}

The last term on the rhs of the above equation is the autocorrelation term and will be dropped in the subsequent discussion, as it only contributes to the signal for $\xvec=\xvec'$ and contains no more information about the initial quantum state, than the momentum distribution itself. The first term, however, shows that for $\xvec \neq \xvec'$, subtle momentum-momentum correlations of the in-trap quantum states are present in the noise-correlation signal of the expanding atom clouds. 

Let us discuss the obtained results for two cases that have been analyzed in the experiment: (1) Ultracold atoms in a Mott insulating state or a fermionic band insulating state released from a 3D optical lattice and (2) two interfering one-dimensional quantum gases separated by a distance $\dvec$.\bigskip   
\par                                                            

\subsection{Noise correlations in bosonic Mott and fermionic band insulators}
Consider a bosonic Mott insulating state or a fermionic band insulator in a three-dimensional simple cubic lattice. In both cases, each lattice site $\Rvec$ is occupied by a fixed atom number $n_\Rvec$. Such a quantum gas is released from the lattice potential and the resulting density distribution is detected after a time-of-flight $t$. In a deep optical lattice, the (in-trap) field operator $ \hat{\psi}(\rvec) $ can be expressed as a sum over destruction operators $\ahat_{\Rvec}$ of localized Wannier states,
by using the expansion (\ref{eq:fieldop}) and neglecting all but the lowest band. The field 
operator for destroying a particle with momentum $\kvec$ is therefore given by
\begin{equation}
     \ahat(\kvec)= \int e^{-i \kvec \rvec} \hat{\psi}(\rvec) d^3r 
\simeq\tilde{w}(\kvec)\sum_{\Rvec} e^{-i \kvec \Rvec} \ahat_\Rvec,  
\end{equation}                                               
where $\tilde{w}(\kvec)$ denotes the Wannier function in momentum space.

\begin{figure}
\includegraphics[width=1\columnwidth]{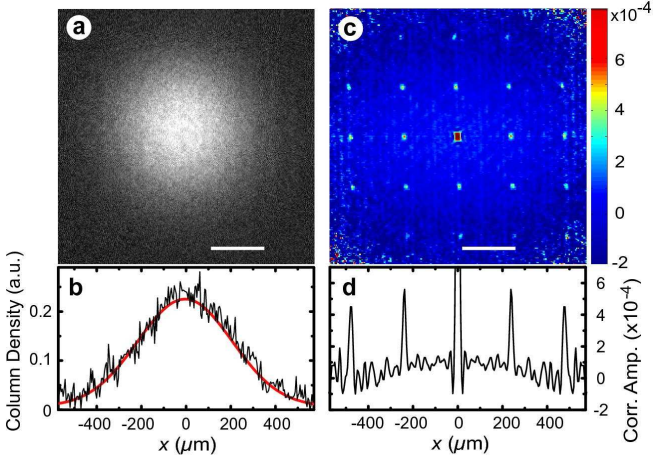}
\caption[]{Noise correlations of a Mott insulator released from a 3D optical lattice. {\bf (a)} Single shot absorption image of a Mott insulator released from an optical lattice and associated cut through the image {\bf (b)}. A statistical correlation analysis over several independent images such as the one in {\bf (a)} yields the correlation function {\bf (c)}. A cut through this two-dimensional correlation function reveals a Hanbury-Brown \& Twiss type bunching of the bosonic atoms {\bf (d)}. Reprinted with permission from \textcite{Foelling:2005}.}
\label{fig:NoiseCorrelations}
\end{figure}
For the two states considered here, the expectation value in Eq.~(\ref{eq:corr1}) 
factorizes into one-particle density matrices 
$\langle\ahatdag_\Rvec \ahat_{\Rvec'} \rangle=n_{\mathbf{R}}\,\delta_{\Rvec,\Rvec'}$
with vanishing off-diagonal order. The density-density correlation function after a 
time-of-flight is then given by (omitting the autocorrelation term of order $1/N$) 

\begin{eqnarray}
   && \langle \hat{n}_{3D}(\xvec)\hat{n}_{3D}(\xvec')\rangle = |\tilde{w}(M\xvec/\hbar t)|^2 |\tilde{w}(M\xvec'/\hbar t)|^2 N^2 \nonumber \\  
   && \times \left[ 1 \pm \frac{1}{N^2} \left| \sum_\Rvec e^{i(\xvec-\xvec') \cdot\Rvec (M/\hbar t)} n_\Rvec\right|^2 \right].\label{eq:noisecorr}     
\end{eqnarray}

The plus sign in the above equation corresponds to the case of bosonic particles and the minus sign to the case of fermionic particles in a lattice. 
Both in a Mott state of bosons and in a filled band of fermions, 
the local occupation numbers $n_{\mathbf{R}}$ are fixed integers. The above equation then shows that correlations or anticorrelations in the density-density expectation value appear for bosons or 
fermions, whenever the difference $\kvec-\kvec'$ is equal to a reciprocal lattice 
vector $\mathbf{G}$ of the underlying lattice. In real space, where the images are 
actually taken, this corresponds to spatial separations for which

\begin{equation}
	|\xvec-\xvec'|=\ell = \frac{2ht}{\lambda M}.
\end{equation}

Such spatial correlations or anticorrelations in the quantum noise of the density distribution of expanding atom clouds can in fact be traced back to the famous Hanbury Brown \& Twiss effect \cite{HanburyBrownTwiss:1956a,HanburyBrownTwiss:1956b,Baym:1998} and its analogue for fermionic particles \cite{Henny:1999,Oliver:1999,Kiesel:2002,Iannuzzi:2006,Rom:2006,Jeltes:2007}. For the case of two atoms localized at two lattice sites this can be readily understood in the following way: there are two possible ways for the particles to reach two detectors at positions $\xvec$ and $\xvec'$
which differ by exchange. A constructive interference for the case of bosons or a destructive interference for the case of fermions then leads to correlated or anticorrelated quantum fluctuations that are registered in the density-density correlation function \cite{Altman:2004,Baym:1998}.  

\begin{figure}
\includegraphics[width=1\columnwidth]{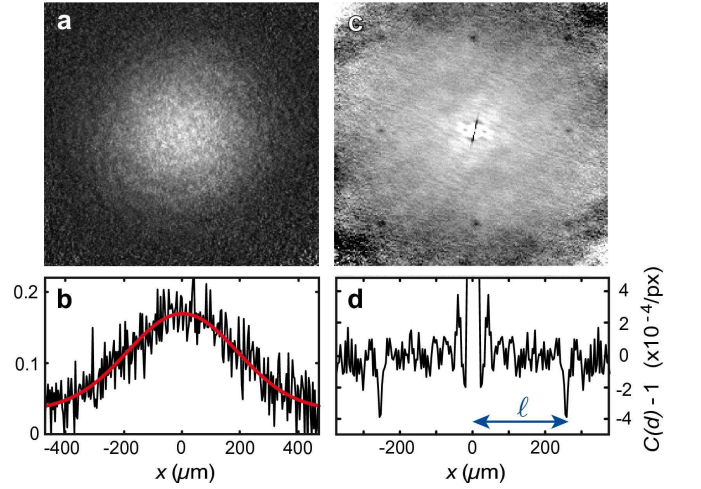}
\caption{Noise correlations of a band insulating Fermi gas. Instead of the correlation ''bunching'' peaks observed in Fig.~\ref{fig:NoiseCorrelations} the fermionic quantum gas shows an HBT type antibunching effect, with dips in the observed correlation function. Reprinted with permission from \textcite{Rom:2006}}
\label{fig:Antibunching}
\end{figure}                  

The correlations for the case of a bosonic Mott insulating state and anticorrelations for the case of a fermionic band insulating state have recently been observed experimentally 
\cite{Foelling:2005,Rom:2006,Spielman:2007}. 
In these experiments several single images of the desired quantum state are recorded after releasing the atoms from the optical trapping potential and observing them after a finite time-of-flight time (for a single of these images see e.g. Fig.~\ref{fig:NoiseCorrelations}a or Fig.~\ref{fig:Antibunching}a). These individually recorded images only differ in the atomic shot noise from each other. A set of such absorption images is then processed to yield the spatially averaged second order correlation function $g^{(2)}_{\rm exp}(\mathbf{b})$:

\begin{equation}
   g^{(2)}_{\rm exp}(\mathbf{b}) = \frac{\int \langle n(\xvec+\mathbf{b}/2) \cdot n(\xvec-\mathbf{b}/2)\rangle\, d^2\xvec}{\int \langle n(\xvec+\mathbf{b}/2)\rangle \langle n(\xvec-\mathbf{b}/2)\rangle\,d^2\xvec}\label{eq:corrfuncexp}.	
\end{equation}

As shown in Fig.~\ref{fig:NoiseCorrelations}, the Mott insulating state exhibits
long range order in the pair correlation function $g^{(2)}(\mathbf{b})$. This order is not connected with the trivial periodic modulation of the average density imposed by the optical lattice after time-of-flight,
which is factored out in $g^{(2)}(\xvec,\xvec')$ (see Eq.~(\ref{eq:g_2})). Therefore, in the superfluid regime, 
one expects $g^{(2)}(\xvec,\xvec')\equiv 1$ despite the periodic density modulation in the interference pattern after time-of-flight. 
It is interesting to note that the correlations or anticorrelations can also be traced back to the enhanced fluctuations in the population of the Bloch waves with quasi momentum $q$ for the case of the bosonic particles and the vanishing fluctuations in the population of Bloch waves with quasi momentum $q$ for the case of fermionic particles \cite{Rom:2006}.

Note that in general the signal amplitude obtained in the experiments for the correlation function deviates significantly from the theoretically expected value of 1. In fact, one typically observes signal levels of $10^{-4}-10^{-3}$ (see Figs.~(\ref{fig:NoiseCorrelations},\ref{fig:Antibunching})). This can be explained by the finite optical resolution when imaging the expanding atomic clouds, thus leading to a broadening of the detected correlation peaks and thereby a decreased amplitude, as the signal weight in each correlation peak is preserved in the detection process. Using single atom detectors with higher spatial and temporal resolution such as the ones used in \textcite{Schellekens:2005} and \textcite{Jeltes:2007}, one can overcome such limitations and thereby also evaluate higher order correlation functions.
\bigskip    
\par

\subsection{Statistics of interference amplitudes for low-dimensional quantum gases} 
As a second example, we consider two bosonic one-dimensional quantum gases oriented along the $z-$direction and separated by a distance $d$ along the $x-$direction. The density-density correlation function at positions $\xvec=(x,y=0,z)$ and $\xvec'=(x',y'=0,z')$ after time-of-flight is then given by formula:
\begin{eqnarray}
 \langle \hat{n}_{3D}(\xvec)\hat{n}_{3D}(\xvec') \rangle &=& \langle \ahatdag_{tof}(\xvec) \ahatdag_{tof}(\xvec') \ahat_{tof}(\xvec) \ahat_{tof}(\xvec')\rangle \nonumber \\
&&+ \delta_{\xvec \xvec'} n(\xvec)n(\xvec').
\label{eq:corrfunc_1}   
\end{eqnarray}                                                            
The operators for the creation $\ahatdag_{tof}(\xvec)$ and destruction $\ahat_{tof}(\xvec)$ of a particle after a time-of-flight period at position $\xvec$ can be related to the in-trap operators describing the trapped quantum gases 1 and 2. Since the expansion mostly occurs along the initially strongly confined directions $x$ and $y$, we can neglect for simplicity the expansion along the axial direction $z$ and obtain for $y=0$ 
\begin{equation}
   \ahat_{tof}(\xvec)=\ahat_1(z) e^{i k_1 x} + \ahat_2(z) e^{i k_2 x},  
\end{equation}
with $k_{1,2} = M (x \pm d/2) /\hbar t$. 
The interference part 
\begin{equation}
 \langle \ahatdag_2(z_1)\ahat_1(z_1)\ahatdag_1(z_2) \ahat_2(z_2) \rangle \times \left [ e^{i k (x_1-x_2)} + c.c. \right ],
\end{equation}
of the correlation function in Eq.~(\ref{eq:corrfunc_1}) is 
an oscillatory function, with wavevector $k=Md/\hbar t$.  
In a standard absorption image, with the propagation direction of the imaging beam pointing along the $z-$direction, one has to additionally take into account an integration along this direction over a length $L$ from which a signal is recorded. Using the above equation, one obtains for this case: 
                                                                
\begin{equation}
   \langle \nhat(x)\nhat(x') \rangle_{\rm int} =  \langle |\hat{A}_k|^2 \rangle \left [ e^{i k (x_1-x_2)} + c.c. \right ],                              
\end{equation}                          
with the observable

\begin{equation}
   \hat{A}_k=\int_{-L/2}^{L/2} dz\,\ahatdag_1(z)\ahat_2(z).\label{eq:observable}
\end{equation}

The above observable characterizes the visibility of an interference pattern obtained in a single run of the experiment. Note that the density-density correlations in the expanding atom clouds are however determined by the expectation value of                                  

\begin{equation}
\langle |\hat{A}_k|^2 \rangle = \int dz_1\!\int\!dz_2 \, \langle \ahatdag_2(z_1)\ahat_1(z_1)\ahatdag_1(z_2) \ahat_2(z_2) \rangle,	    
\label{eq:quantcorramp}
\end{equation}
which can be obtained by a statistical analysis of the visibility of the interference patterns obtained in several runs of the experiment. The basic example in this context is the observation of 
a pronounced interference pattern in a single realization of two overlapping but independent condensates with fixed particle numbers by \textcite{Andrews:1997a}. As discussed e.g. by 
\textcite{Castin:1997}, the detection of particles at certain positions entails a non-vanishing
interference amplitude in a single realization, whose typical visibility is determined by 
$\langle |\hat{A}_k|^2 \rangle\neq0$. Averaging over many realizations, in turn,
completely eliminates the interference because $\langle \hat{A}_k \rangle=0$
\cite{Leggett:2001}. 

For the case of identical (but still independent) quantum gases, one can simplify Eq.~(\ref{eq:quantcorramp}), to yield: 

\begin{equation}
 \langle |\hat{A}_k|^2 \rangle \approx  L \int_{-L/2}^{L/2}  dz \langle \ahatdag(z)\ahat(0)\rangle^2 = L \int |G^{(1)}(z)|^2 dz. 
\end{equation}                                        
The fluctuations in the interference pattern are thus directly linked to the coherence properties of the one-dimensional quantum systems. For the case of Luttinger liquids, the one-particle density matrix 
$G^{(1)}(z)\sim z^{-1/(2K)}$ at zero temperature decays algebraically with an exponent determined by the Luttinger parameter $K$ (see section~\ref{subsec:LiebLiniger}). As a result, the interference 
amplitudes exhibit an anomalous scaling 
$\langle |\hat{A}_k|^2 \rangle \propto L^{-1/K}$  \cite{Polkovnikov:2006a}. 
By determining higher moments $\langle |\hat{A}_k|^{2n} \rangle$ of arbitrary order $n$ of the visibility in an interference experiment, one can characterize the full distribution function of the normalized random variable $|\hat{A}_k|^2/\langle |\hat{A}_k|^2 \rangle$. Full knowledge of the distribution function in fact amounts to a complete characterization of the correlations in the many-body systems, as has been recently shown by \textcite{Gritsev:2006}. For the case of 1D Bose-Einstein condensates, this has recently been tested by \textcite{Hofferbeth:2007}.

The above analysis for one-dimensional quantum systems can be readily extended to the case of two-dimensional systems \cite{Polkovnikov:2006a} and has been directly used to detect a Berezinskii-Kosterlitz-Thouless-transition \cite{Hadzibabic:2006} with ultracold quantum gases (see sec.~\ref{sec:twodim}). For the case of lattice based systems it has been shown that noise correlations can be a powerful way to reveal e.g. an antiferromagnetically ordered phase of two-component bosonic or fermionic quantum gases \cite{Altman:2004,Werner:2005}, to characterize Bose-Fermi mixtures \cite{Wang:2005,Ahufinger:2005} and quantum phases with disorder \cite{Rey:2006}, as well as to detect supersolid phases \cite{Scarola:2006}.

\section{MANY-BODY EFFECTS IN OPTICAL LATTICES}
\label{sec:bosons}

As a first example,
illustrating how cold atoms in optical lattices can be used to study
genuine many-body phenomena in dilute gases, we discuss
the Mott-Hubbard transition for bosonic atoms. Following the 
original idea by \textcite{Jaksch:1998}, this transition was
first observed experimentally by \textcite{Greiner:2002a}.
The theory of the underlying quantum phase transition is based on 
the Bose-Hubbard model, originally introduced by 
\textcite{Fisher:1989} to describe the destruction of
superfluidity due to strong interactions and disorder. 

\subsection{Bose-Hubbard model}

A conceptually simple model to describe cold atoms in 
an optical lattice at finite density is obtained by 
combining the kinetic energy (\ref{eq:H_0,p}) in the 
lowest band with the on-site repulsion arising from
(\ref{eq:BoseH}) in the limit of
a sufficiently deep optical lattice. More precisely,
the Bose-Hubbard model (BHM) is obtained from 
a general many-body Hamiltonian with a pseudoptential 
interaction under the assumptions
\begin{itemize}

\item
both the thermal and the mean interaction energies at a single site
are much smaller than the separation $\hbar\omega_0$ to the first 
excited band.
\item
the Wannier functions decay essentially within a single lattice constant.
\end{itemize}\bigskip

Under these assumptions, only the lowest band needs to be taken 
into account in Eq.~(\ref{eq:fieldop}).  Moreover, the hopping matrix  elements 
$J(\mathbf{R})$ are non-negligible only for $\mathbf{R}=0$ or to nearest 
neighbors (NN) in Eq.~(\ref{eq:H_0,p}) and the interaction constants
are dominated by the on-site contribution ~(\ref{eq:U}). 
This leads to the Bose-Hubbard model (BHM) 
\begin{equation}
\label{eq:BHM}
\hat H=-J\sum_{\langle\mathbf{R},\mathbf{R'}\rangle}\,\hat a_{\mathbf{R}}^{\dagger}\hat a_{\mathbf{R'}}\, +
\,\frac{U}{2}\sum_{\mathbf{R}}\,\hat n_{\mathbf{R}}(\hat n_{\mathbf{R}}-1)\, +\,
\sum_{\mathbf{R}}\,\epsilon_{\mathbf{R}}\hat n_{\mathbf{R}}\, .
\end{equation}
($\langle\mathbf{R},\mathbf{R'}\rangle$ denotes a sum over all lattice sites
$\mathbf{R}$ and its nearest neighbors at $\mathbf{R}'=\mathbf{R}+\mathbf{d}$, where
$\mathbf{d}$ runs through the possible nearest neighbor vectors). 
The hopping matrix element $J(\mathbf{d})=-J<0$ to nearest neighbors is
always negative in the lowest band, because the ground state must have 
zero momentum $\mathbf{q}=0$ in a time-reversal invariant situation. 
For a separable lattice and in the limit $V_0\gg E_r$, it is given by
Eq.~(\ref{eq:Jmathieu}). More generally, the hopping matrix elements are 
determined by the exact band energy using Eq.~(\ref{eq:J_n}). An alternative,
but more indirect, expression is $J(\mathbf{R})=\langle w(\mathbf{R})|\hat{H}_0| 
w(0)\rangle$ \cite{Jaksch:1998}.

Since the standard BHM includes next neighbor hopping only, a convenient approximation for $J$ in Eq.~(\ref{eq:BHM}) is obtained by simply adjusting it to the given bandwidth. Concerning  the on-site repulsion $U$, which disfavors 
configurations with more than one boson at a given site, its precise value
as determined by Eq.~(\ref{eq:U}) requires the exact Wannier function.
In the low filling $\bar n\sim 1$ regime, it follows from the single-particle
Bloch states via Eq.~(\ref{eq:Wannier}). For higher fillings, the mean-field
repulsion on each lattice site leads to an admixture of excited states
in each well and eventually to a description, where for $\bar n\gg 1$ one has
a lattice of coupled Josephson junctions with a Josephson
coupling $E_J=2{\bar n}J$ and an effective 'charging 
energy' $U$ \cite{Fisher:1989, Cataliotti:2001}. 
For intermediate fillings, the Wannier functions
entering both the effective hopping matrix element $J$ and on-site
repulsion $U$ have to be adjusted to account for the mean-field interaction
\cite{Niu:2006b}. The change in the on-site interaction energy
with filling has been 
observed experimentally by \textcite{Campbell:2006}. 
In a more detailed description, the effects of interactions at higher 
filling can be accounted for by a multi-orbital generalization of the 
Gross-Pitaevskii ansatz \cite{Alon:2005}. This leads to
effective 'dressed' Wannier states which include higher bands and 
coupling between different sites. The last term with a variable on-site energy
$\epsilon_{\mathbf{R}}=\tilde{V}(\mathbf{R})$ describes the effect of 
the smooth trapping potential $\tilde{V}(\mathbf{r})$. It includes 
the constant band center energy, arising from the $J(\mathbf{R}\!=\! 0)\,$-term of the 
hopping contribution (\ref{eq:H_0,p}) and acts like a spatially varying chemical potential.

The BHM describes the competition between the kinetic energy
$J$ which is gained by delocalizing particles over the lattice 
sites in an extended Bloch state and the repulsive on-site 
interaction $U$, which disfavors having more than one 
particle at any given site.  In an optical lattice loaded with 
cold atoms, the ratio $U/J$ between these two energies can 
be changed easily by varying the dimensionless depth 
$V_0/E_r$ of the optical lattice. Indeed, from Eqs.~(\ref{eq:Jmathieu}) 
and ~(\ref{eq:U}), the ratio 
$U/J\sim (a/d)\cdot\exp{(2\sqrt{V_0/E_r})}$
increases exponentially with the lattice depth. Of course, to see
strong interaction effects, 
the average site occupation $\langle\hat{n}_{\mathbf{R}}\rangle$ 
needs to be on the order of one, 
otherwise the atoms never see each other. This was the situation 
for cold atoms in optical lattices in the 90's, studied e.g. by
\textcite{Westbrook:1990,Grynberg:1993,Hemmerich:1993,Kastberg:1995}.

\subsection{Superfluid-Mott-Insulator transition}
\label{subsec:SF-MI}

The BHM Eq.~(\ref{eq:BHM}) is not an exactly soluble model, not even in one dimension,
despite the fact that the corresponding continuum model in 1D, the Lieb-Liniger 
model, is exactly soluble. Nevertheless, the essential physics of the model 
and, in particular, the existence and properties of the 
quantum phase transition which the BHM exhibits as a function of
$U/J$ are rather well understood \cite{Fisher:1989}. In fact, for the 3D case and effectively unit filling, the existence of a quantum 
phase transition from a homogeneous BEC to a MI with a nonzero gap has 
been proven rigorously in a model of hard core bosons in the presence of
a staggered field by \textcite{Aizenman:2004}. Let us first discuss the limiting cases, which describe the two possible phases within the BHM. 

\bigskip

\noindent {\it Superfluid phase} In the trivial limit $U=0$,
the many-body ground state is simply an ideal
BEC where all $N$ atoms are in the $\mathbf{q}=0$
Bloch-state of the lowest band. Including the normalization factor in a
lattice with $N_L$ sites, this state can be written in the
form
\begin{equation}
\label{eq:BEClattice}
\vert\Psi_{N}\rangle (U=0)=\frac{1}{\sqrt{N!}}
\left(\frac{1}{\sqrt{N_L}}\sum_{\mathbf{R}}\, \hat a_{\mathbf{R}}^{\dagger}\right)^N
\,\vert 0\rangle\, .
\end{equation}
In the limit $U/J\to 0$ therefore, the ground state of the BHM is a  Gross-Pitaevskii 
type state with a condensate fraction which is trivially equal to one.
The critical temperature of the ideal Bose gas in an optical lattice 
at filling $\bar{n}=1$ can be obtained from the condition $\int d\varepsilon\,
g(\varepsilon)n_{\rm B}(\beta_c\varepsilon)=1$, where $g(\varepsilon)$ is 
the density of states in the lowest band and $n_{\rm B}(x)=(\exp{(x)}-1)^{-1}$
the Bose-Einstein distribution.  This gives $k_{\rm B}T_c=5.59\, J$.
In the presence of an optical lattice, therefore, the critical temperature for BEC is significantly reduced compared with the free space situation, essentially due to the increased effective mass $M^*$ of the particles in the lattice. 
The relevant parameter, however, is not the temperature but the entropy.
Indeed, by starting with a deeply degenerate gas and adiabatically switching on the 
optical lattice, the degeneracy parameter stays constant and the temperature
is essentially reduced by a factor $M/M^*$ \cite{Olshanii:2002b,Hofstetter:2002,Blakie:2004}.

For a sufficiently large system $N,N_L\to\infty$ at fixed (not necessarily integer) density $N/N_L$
(in the experiment \cite{Greiner:2002a}, the total number of occupied lattice sites was
about $10^5$),
the perfect condensate Eq.~(\ref{eq:BEClattice})  becomes indistinguishable 
in practice from a coherent state
\begin{equation}
\exp \left( \sqrt{N}\hat a_{q=0}^{\dagger}\right)\vert 0\rangle 
= \prod_{\mathbf{R}} \left(\exp\left [ \sqrt{\frac{N}{N_L}}\hat a_{\mathbf{R}}^{\dagger}\right ]
\vert 0\rangle_{\mathbf{R}}\right). \label{eq:coherentstate}
\end{equation}
It factorizes into a product of {\it local} coherent  states at every lattice site $\mathbf{R}$
with average $\bar{n}=\langle\hat n\rangle=N/N_L$ because boson operators at different 
sites commute. The probability distribution
for the number of atoms at any given site for a perfect BEC
in an optical lattice is therefore Poissonian with a standard deviation given by $\sigma(\bar{n})=\sqrt{\bar{n}}$. Taking $N=N_L$, i.e. an average density
such that there is one atom for each lattice site, there is a 
$1-2/e=0.27$ probability that any given site is occupied with more than one atom. 
The kinetic energy minimization requirement that every atom 
wants to be at all lattice sites with equal amplitude thus
necessarily leads to a substantial probability of finding more than one atom 
on a given site. At finite repulsion $U>0$, such configurations
are, of course, disfavoured. \bigskip

\noindent {\it Mott insulating phase} To understand the behavior in the opposite limit $U\gg J$, 
it is useful to consider the case of unit filling, i.e. the number $N$ of atoms is precisely
equal to the number $N_L$ of lattice sites. In the limit $U\gg J$, 
hopping of the atoms is negligible and the obvious ground state
\begin{equation}
\label{eq:MIstate}
\vert\Psi_{N=N_L} \rangle(J=0)=\bigl(
\prod_{\mathbf{R}}\hat a_{\mathbf{R}}^{\dagger} \bigr) \vert 0\rangle
\end{equation}
is a simple product of local Fock-states with precisely one atom 
per site. With increasing $J$, the atoms start to hop around, which
necessarily involves double occupancy, increasing the energy by $U$. Now as
long as the gain $J$ in kinetic energy due to hopping is smaller than $U$,
the atoms remain localized. For any $J\not= 0$, however, the ground state is no longer a simple
product state as in Eq.~(\ref{eq:MIstate}). Once $J$ becomes of order or larger than $U$,
the gain in kinetic energy outweighs the repulsion due to double
occupancies.  The atoms then undergo a transition to a superfluid,
in which they are delocalized over the whole lattice. This 
is a sharp quantum phase transition in the thermodynamic limit, 
because the state (\ref{eq:BEClattice}), in contrast to (\ref{eq:MIstate}),
exhibits off-diagonal long range order, which cannot disappear in 
a continuous manner. By contrast, the evolution between these 
two states is completely smooth for say two particles in two wells,
where a simple crossover occurs from a state with a well defined
relative phase at $J\gg U$ to one with a well defined particle number 
in each well at $J\ll U$.   
\bigskip
 
\begin{figure}
\includegraphics[width=.9\columnwidth]{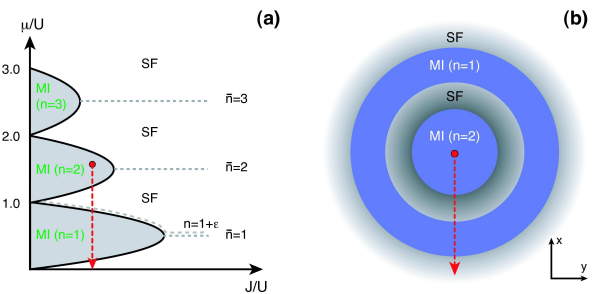}
\caption{Schematic zero temperature phase diagram of the Bose-Hubbard model.
The dashed lines of constant integer density $\langle\hat n\rangle=1,2,3$ in the SF hit the corresponding MI phases at the tips of the
lobes at a critical value of $J/U$, which decreases with density $\bar n$. For
$\langle\hat n\rangle=1+\varepsilon$ the line of constant density stays outside the $\bar n=1$ MI because a fraction $\varepsilon$ of the particles remains superfluid down to the lowest values of $J$. In an external trap with a $\bar n=2$ MI phase in the center, a series of MI and SF regions appear by going towards the edge of the cloud, where the local chemical potential has dropped to zero}
\label{fig:ShellStructureMI}
\end{figure}

\noindent {\it Phase diagram} The zero temperature phase diagram of the homogeneous BHM
is shown schematically in Fig.~\ref{fig:ShellStructureMI}a as a function of $J/U$, 
with the density controlled by a chemical potential $\mu$. At 
$U/J\to 0$, the kinetic
energy dominates and the ground state is a delocalized superfluid,
described by Eq.~(\ref{eq:coherentstate}) to lowest order. 
At large values of $U/J$, interactions dominate and one obtains a series of
Mott-insulating (MI) phases with fixed integer filling
$\bar n=1,2,\ldots$. These states are {\it incompressible}, implying that 
their density remains unchanged upon varying the chemical 
potential. In fact, it is the property $\partial n/\partial\mu=0$, 
which is {\it the} defining property of a MI,
and not the existence of local Fock states which only
exist at $J=0$. The transition between the SF and MI phases 
is associated with the loss of long range order in the one-particle 
density matrix $g^{(1)}(\mathbf{x})$. In the 3D case, the order parameter
of the SF-MI transition is therefore the condensate fraction $n_0/n$, which 
drops continuously from one at $U/J\ll 1$ to zero at $(U/J)_c$. The 
continuous nature of the SF-MI quantum phase transition in any 
dimension follows from the fact that the effective field theory for the 
complex order parameter $\psi$ is of that of a $d+1$-dimensional 
XY-model \cite{Fisher:1989, Sachdev:1999}. 
More precisely, this is valid only for the special transition at 
integer density, which is driven by phase fluctuations only.  
By contrast, crossing the SF-MI phase boundary 
by a change in the chemical potential, the associated change in the 
density gives rise to a different critical behavior \cite{Fisher:1989}. 
For instance, the excitation gap in the MI phase
vanishes linearly with the distance from the boundary of the Mott lobe
in this more generic case. 

Within a mean-field approximation, the critical value for the transition from a MI to a SF in a three dimensional optical lattice, is given by $(U/J)_c=5.8z$ for $\bar n=1$ and
$(U/J)_c=4\bar nz$ for $\bar n\gg 1$
\cite{Fisher:1989,Sheshadri:1993,Oosten:2001}. Here $z$ is the number of nearest
neighbors and thus $2zJ$ is the total bandwidth of the lowest Bloch band,
which is the relevant parameter which has to be compared with $U$. Recently, precise Quantum Monte-Carlo simulations by \textcite{Capogrosso:2007}
have determined the critical value for the $\bar n=1$ transition in a  simple cubic lattice 
to be at $(U/J)_c=29.36$ with an accuracy of about $0.1$\%. In one dimension, 
the SF-MI transition is of the Kosterlitz-Thouless type, with a finite
jump of the superfluid density at the transition. Precise values 
for the critical coupling are available from DMRG calculations,
giving  $(U_1/J)_c=3.37$ \cite{Kuehner:2000, Kollath:2004} for the 
$\bar n=1$ transition. For $\bar n\gg 1$, the BHM is equivalent to a 
chain of Josephson junctions with coupling energy $E_J=2{\bar n}J$.
The SF-MI transition is then desribed by the $1+1$-dimensional 
$O(2)$-model, which gives $(U_1/J)_c=2.2\,\bar n$ 
\cite{Hamer:1979, Roomany:1980}.

From Eqs.~(\ref{eq:Jmathieu}) and ~(\ref{eq:U}), the critical
value of the dimensionless lattice depth $V_0/E_r$ for rather deep
lattices is obtained from
\begin{equation}
\label{eq:U/Jcrit}
\left(V_0/E_r\right)_c=\frac{1}{4}
\ln^2{\Bigl(\frac{\sqrt{2}d}{\pi a}\cdot\left( U/J\right)_{c}\Bigr)}
\end{equation}
Using the experimental parameters $d=426\,$nm 
and $a=5.7\,$nm \cite{Greiner:2002a}, the precise result 
for $\left( U/J\right)_{c}$ in a simple cubic lattice gives
a critical value $V_0/E_r\vert_c=11.89$ for the SF-MI transition with $\bar n=1$.
Given that Eq.~(\ref{eq:Jmathieu}), on which the above estimate for the 
critical lattice depth is based, is not very precise in this regime, this 
result is in reasonable agreement with the lattice depth of $V_0=12-13 E_r$, where the transition is observed
experimentally \cite{Greiner:2002a,Gerbier:2005a}.

Consider now a filling with $\langle\hat n\rangle=1+\varepsilon$ which is slightly larger
than one. For large $J/U$ the ground state has all the atoms delocalized
over the whole lattice and the situation is hardly different from the
case of unit filling. Upon lowering $J/U$, however, the line of constant
density remains slightly above the $\bar n=1$ 'Mott-lobe',
and stays in the SF regime down to the lowest $J/U$ (see Fig.~\ref{fig:ShellStructureMI}). For any
noninteger filling, therefore, the ground state remains SF as long as the
atoms can hop at all. This is a consequence of the fact, that even for
$J\ll U$ there is a small fraction $\varepsilon$ of atoms which remain SF
on top of a frozen MI-phase with $\bar n=1$. Indeed this fraction
can still gain kinetic energy by delocalizing over the whole lattice
without being blocked by the repulsive interaction $U$ because two
of those particles will never be at the same place. The same argument applies to
holes when $\varepsilon$ is negative. As a result, in the homogeneous system,
the quantum phase transition from a SF to a MI  only appears 
if the density is equal to a commensurate, integer value. 
\bigskip

\noindent {\it In-trap density distribution} Fortunately,
the situation is much less restrictive in the presence of a harmonic trap.
Indeed, within a local density approximation, the inhomogeneous 
situation in a harmonic trap is described by a spatially varying chemical potential
$\mu_{\mathbf{R}}=\mu(0)-\epsilon_{\mathbf{R}}$ with $\epsilon_{\mathbf{R}}=0$ at the trap
center. Assuming e.g. that the chemical potential $\mu(0)$ at trap center
falls into the $\bar n=2$ 'Mott-lobe', one obtains a series of
MI domains separated by a SF by moving to the boundary of the trap
where $\mu_{\mathbf{R}}$ vanishes (see Fig.~\ref{fig:ShellStructureMI}b). 
In this manner, all the different phases which exist for given $J/U$ 
below $\mu(0)$ are present simultaneously ! 
The SF phase has a finite compressiblity 
$\kappa=\partial n/\partial\mu$ and a gapless excitation spectrum of the
form $\omega(q)=cq$ because there is a finite superfluid density $n_s$. By contrast, 
in the MI-phase both $n_s$ and $\kappa$ vanish. 
As predicted by \textcite{Jaksch:1998}, the incompressibility 
of the MI phase allows to distinguish it from the SF
by observing the local density distribution in a trap. 
Since $\kappa=0$ in the MI, the density stays
constant in the Mott phases, even though the external trapping potential is
rising. In the limit of $J\rightarrow 0$ the SF regions vanish and one obtains a 'wedding cake' type density profile, with radii $R_n$ of the different Mott insulating regions, given by 
$R_n=\sqrt{(2[\mu(0)-nU]/M\omega^2)}$ \cite{DeMarco:2005}.

The existence of such wedding-cake like density profiles of a Mott insulator has been supported by Monte-Carlo \cite{Kashurnikov:2002,Batrouni:2002,Wessel:2004,Rigol:2006} and DMRG \cite{Kollath:2004} calculations in one, two, and three dimensions. Very recently number state resolved, in-trap density profiles have been detected experimentally by \textcite{Campbell:2006} and \textcite{Foelling:2006}. In the latter case it has been possible to directly observe the wedding cake density profiles and thus confirm the incompressibility of the Mott insulating regions of the atomic gas in the trapping potential. A sharp drop in the radii of the $n=2$ occupied regions has been observed when the crossing the transition point \cite{Foelling:2006}. It should be noted that the in-trap density profiles can be used as a sensitive thermometer for the strongly interacting quantum gas. For typical experimental parameters, one finds that for  temperatures around $T^*\gtrsim 0.2 U/k_B$, the wedding cake profiles become completely washed out \cite{Gerbier:2007b}. Within the strongly interacting regime, the superfluid shells accomodate most of the entropy of the system and can turn already into a normal thermal gas at a lower temperature $T_c\sim zJ$ with the Mott insulating shells still intact \cite{Capogrosso:2007,Gerbier:2007b,Ho:2007}. In order to reach the lowest temperatures in this regime, it is advantageous to keep the external harmonic confinement as low as possible, or even decrease it during an increase of the lattice depth \cite{Gerbier:2007b,Ho:2007}.
\bigskip

\begin{figure}
\begin{center}
\includegraphics[width=0.9\columnwidth]{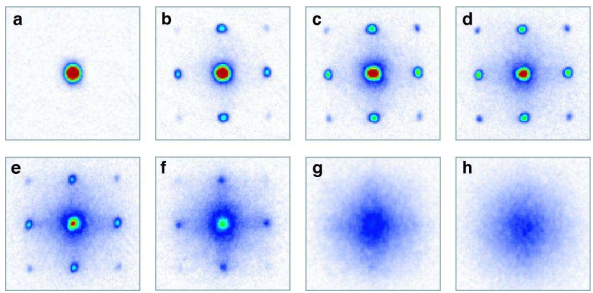}
\caption{\small Absorption images of multiple matter wave
interference patterns after releasing the atoms from an optical
lattice potential with a potential depth of {\bf a} $0\,E_r$, {\bf c} $3\,E_r$
{\bf c} $7\,E_r$ {\bf d} $10\,E_r$ {\bf e} $13\,E_r$ {\bf f} $14\,E_r$, {\bf g} $16\,E_r$  and
{\bf b} $20\,E_r$. The ballistic expansion time was
15\,ms. Reprinted with permission from  \textcite{Greiner:2002a}. \label{fig:reciproclattice}}
\end{center}
\end{figure}

\noindent {\it Phase coherence across the SF-MI transition} The disappearance of superfluidity 
(or better of BEC) at the SF-MI transition was initially observed experimentally by a time-of-flight 
method 
\cite{Greiner:2002a}.  The corresponding series of images is shown in Fig.~\ref{fig:reciproclattice} for different values of $V_0$, ranging between $V_0=0$ (a) and
$V_0=20E_r$ (h). One observes a series of interference peaks around the
characteristic 'zero-momentum' peak of a condensate in the absence
of an optical lattice.
With increasing $V_0$ these peaks become more
pronounced. Beyond a critical lattice depth around $V_0\approx12-13E_r$ (e),
which agrees very well with the above estimate for the SF-MI transition
for one Boson per site, this trend is suddenly reversed, however, and the interference peaks
eventually disappear completely. In order to understand why these
pictures indeed provide a direct evidence for a SF
to MI transition predicted by the Bose-Hubbard model, it is useful to 
consider the idealized situation of a perfect periodic lattice in the 
absence of any trapping potential. From Eq.~(\ref{eq:momentumdistribution})
the observed density at position $\mathbf{x}$ reflects 
the momentum distribution at  $\mathbf{k}=M\mathbf{x}/\hbar t$.
Factoring out the number of lattice sites, it is proportional to
the lattice Fourier transform
\begin{equation}
\label{eq:n(k)}
n(\mathbf{k})\sim \vert \tilde{w}(\mathbf{k})\vert^2\,\sum_{\Rvec}e^{i\mathbf{k}\cdot\mathbf{R}}
G^{(1)}(\Rvec).
\end{equation}
of the one-particle density matrix $G^{(1)}(\mathbf{R})$ at separation
$\mathbf{R}$. For optical lattice depths below the critical value, the ground state 
in a 3D situation is a true BEC, where $G^{(1)}(|\mathbf{R}|\to\infty)=n_0$ approaches 
a finite value at large separation.
 For the MI phase,
in turn, $G^{(1)}(\mathbf{R})$ decays to zero exponentially. The
SF phase of cold atoms in a homogeneous  optical lattice is thus
characterized by a momentum distribution which exhibits sharp
peaks at the reciprocal lattice vectors $\mathbf{k}=\mathbf{G}$
(defined by $\mathbf{G}\cdot\mathbf{R}=2\pi$ times an integer,
see e.g. \cite{Ashcroft:1976}) plus a smooth background from the short range correlations. The fact that the
peaks in the momentum distribution at $\mathbf{k}=\mathbf{G}$ initially grow
with increasing depth of the lattice potential is a result of the
strong decrease in spatial extent of the Wannier function $w(\mathbf{r})$, which entails a corresponding increase in its Fourier transform $\tilde{w}(\mathbf{k})$
at higher momenta. In the MI regime, where $G^{(1)}(\mathbf{R})$ decays to
zero, remnants of the interference peaks still remain (see e.g. Fig.~\ref{fig:reciproclattice}f)
as long as $G^{(1)}(\mathbf{R})$ extends over several lattice spacings, because
the series in Eq.~(\ref{eq:n(k)}) adds up constructively at $\mathbf{k}=\mathbf{G}$. 
A more detailed picture for the residual short range coherence features beyond the 
SF-MI transition is obtained by considering perturbations deep in the Mott insulating 
regime at $J=0$. There, $G^{(1)}(\mathbf{R})$ vanishes beyond $\mathbf{R}=0$ 
and the momentum distribution is a structureless Gaussian, reflecting the Fourier transform of the
Wannier wave function (see Fig.~\ref{fig:reciproclattice}h).  With increasing  
tunneling $J$, the Mott state at $J/U \rightarrow 0$ is modified by a coherent admixture of particle-hole pairs. However due the presence of a gapped excitation spectrum, such particle hole pairs cannot spread out and are rather tightly bound to close distances. They do, however, give rise
to a significant degree of short range coherence. Using first order perturbation theory with the tunneling operator as a perturbation on the dominating interaction term, one finds that the amplitude of the coherent particle hole admixtures in a Mott insulating state is proportional to $J/U$:
\begin{equation}
   |\Psi\rangle_{U/J}\approx  |\Psi\rangle_{U/J\rightarrow\infty} + \frac{J}{U} \sum_{\langle \Rvec,\Rvec' \rangle} \ahatdag_\Rvec \ahat_{\Rvec'} |\Psi\rangle_{U/J\rightarrow\infty}.
\end{equation}

Close to the  transition point, higher order perturbation theory or a Green function analysis can account for coherence beyond nearest neighbors and the complete liberation of the particle-hole pairs,  which eventually leads to the formation of long range coherence in the superfluid regime. The coherent particle hole admixture and its consequence on the short range coherence of the system have been investigated theoretically and experimentally in \cite{Gerbier:2005a,Gerbier:2005b,Sengupta:2005}.

The SF-MI quantum phase transition therefore shows up directly
in the interference pattern. For the homogeneous system it 
reveals the existence or not of off-diagonal long range order in the
one-particle density matrix. The relevant order parameter is
the condensate fraction. Of course the actual system is
not homogeneous and a numerical computation of the interference pattern is
necessary for a quantitative comparison with experiment. This has been
done e.g. for the 3D case in \cite{Kashurnikov:2002}  and for
1D case in \cite{Batrouni:2002,Kollath:2004}. Due to the
finite size and the fact that different MI phases are involved, the pattern
evolves continuously from the SF to the MI regime. 
While the critical values for $J/U$ are different
for the MI phases with $\bar n=1$ and $\bar n=2$ which are present
in the experiment \cite{Greiner:2002a}, the transition seen in the 
time-of-flight images occurs rather rapidly with increasing lattice depth. Indeed, 
from Eq.~(\ref{eq:U/Jcrit}), 
the experimental control parameter $V_0/E_r$ depends only logarithmically 
on the relevant parameter $U/J$ of the BHM. The small change from $V_0=13E_r$ 
in (e) to $V_0=14E_r$ in (f) thus covers a range in $J/U$ wider than that, 
which would be required to distinguish the $\bar n=1$ from the $\bar n=2$ transition. For a quantitative evaluation of the interference patterns, one must  also take into account the broadening 
mechanism during time-of-flight expansion, as discussed in Sec.~\ref{sec:TOF}. 

When approaching the SF-MI transition from the superfluid regime, the increasing interactions tend to increase the depletion of the condensate and thereby reduce the long range phase coherent component with increasing $U/J$ \cite{Orzel:2001,Hadzibabic:2004,Schori:2004,Xu:2006}. For increasing lattice depth, the condensate density as a measure of the long range coherent fraction, then decreases continuously and vanishes at the transition point. The visibility of the interference pattern in general, however, evolves smoothly across the SF-MI transition, due to the presence of a strong short range coherent fraction in the MI just across the transition point (see discussion above). Above the transition point the visibility of the momentum distribution can also show kinks as the lattice depth is increased, which have been attributed to the beginning formation of shell structures in the MI state \cite{Gerbier:2005a,Gerbier:2005b,Sengupta:2005}.\bigskip

\noindent {\it Excitation spectrum} A second signature of the SF-MI transition is the appearance of a finite
excitation gap $\Delta\not= 0$ in the Mott insulator.  Deep in the MI phase, this gap
has size $U$, which is just the increase in energy if an atom tunnels to
an already occupied adjacent site (note that $U$ is much smaller than
the gap $\hbar\omega_0$ for the excitation of the next vibrational state).
The existence of a gap has been
observed experimentally by applying a potential gradient in the MI \cite{Greiner:2002a} or by using a modulation spectroscopy method \cite{Stoferle:2004} and
measuring the resulting excitations. Recent calculations indicate that such measurements simultaneously probe global \cite{Iucci:2006,Huber:2007} and local properties of the system. In particular, e.g. a peaked excitation spectrum can also appear in a strongly interacting superfluid regime, where $U>J$ \cite{Kollath:2006a}. A way to probe global features of the many-body excitation spectrum, also close to the transition point, might be achieved by employing Bragg spectroscopy techniques as proposed in \cite{Oosten:2005,Rey:2005,Pupillo:2006}.
 
In the SF regime, there is no excitation gap. Instead, the homogeneous system exhibits a sound like mode with frequency $\omega(q)=cq$. As shown in the appendix, the associated sound velocity $c$ is determined by $Mc^2=n_s/\kappa$
and thus gives information about the superfluid
density $n_s$. The existence of a sound like excitation even in the
presence of an underlying lattice which explicitely breaks translation
invariance is a consequence of long range phase coherence in the SF. Its
observation would therefore directly probe superfluidity, in contrast to the peaks in the interference pattern, which measure BEC.   
   
\bigskip

\noindent {\it Number statistics} Associated with the transition from a superfluid to a Mott insulating state is a profound change in the atom number statistics per lattice site. As noted above, 
in the homogeneous system the ground state
in the extreme MI limit ($J/U \rightarrow 0$) is a product of Fock states with an integer
number $\bar n$ of particles at each site. At finite hopping $J\not= 0$,
this simple picture breaks down because the atoms have a finite amplitude
to be at different sites. The many-body ground state can then no longer
be written as a simple product state. In the opposite limit
$U\to 0$, the ground state is a condensate of zero
quasi-momentum Bloch states.
In the limit $N,N_L\to\infty$ at fixed (not necessarily integer) density $\bar{n}=N/N_L$,
the associated perfect condensate is a product of coherent states on each lattice site       
\begin{equation} 
	|\alpha\rangle = e^{-|\alpha|^2/2} \sum_n \frac{\alpha^n}{\sqrt{n!}} |n\rangle 
\end{equation}
with $\alpha$ describing the amplitude and phase of the coherent matter wave field. This corresponds to a Poissonian atom number distribution on each lattice site with 
average $|\alpha|^2=\bar n$. 

A remarkable consequence of the representation ~(\ref{eq:coherentstate}) 
is that, at least for integer densities
$\bar n=1,2,\ldots$, the many-body ground state may be factorized into a
product over single sites 
\begin{equation}
\label{eq:Gutzwiller}
\vert\Psi_{GW}\rangle\, =
\prod_{\mathbf{R}}\left(\sum_{n=0}^{\infty}c_{n}\vert n\rangle_{\mathbf{R}}\right)
\end{equation}
in both limits $J\to 0$ and $U\to 0$. The associated atom number
probability distribution $p_n=|c_n|^2$ is either a pure Fock or a full
Poissonian distribution. It is now very plausible to use the factorized
form in Eq.~(\ref{eq:Gutzwiller}) as an {\it approximation} for arbitrary $J/U$, taking the coefficients
$c_n$ as variational parameters which are determined by minimizing the
ground state energy \cite{Rokhsar:1991, Sheshadri:1993}. As pointed out by 
\textcite{Rokhsar:1991},
this is effectively a Gutzwiller ansatz for bosons. Beyond being
very simple computationally, this ansatz describes the SF to MI
transition in a mean-field sense, becoming exact in infinite
dimensions. In addition, it provides
one with a very intuitive picture of the transition to a MI state,
which occurs precisely at the point, where the local number distribution
becomes a pure Fock distribution. Indeed, within
the Gutzwiller approximation, the expectation value 
\begin{equation}
\langle\Psi_{GW}\vert\hat a_{\mathbf{R}}\vert\Psi_{GW}\rangle=\sum_{n=1}^{\infty}\,
\sqrt{n}c_{n-1}^*c_n
\end{equation}
of the local matter wave field vanishes if and only if the probability 
for finding different particle numbers at any given site is equal to zero. 
It is important, however, to emphasize
that the Gutzwiller ansatz fails to account for the nontrivial correlations
between different sites present at any finite $J$. These correlations
imply that the one particle density matrix $G^{(1)}(\mathbf{R})$ is
different from zero at finite distance $|\mathbf{R}|\not= 0$,
becoming long ranged at the transition to a SF. By contrast, in the
Gutzwiller approximation, the one particle density matrix has no
spatial dependence at all: it is zero at any $|\mathbf{R}|\not= 0$ in the
MI and is completely independent of $\mathbf{R}$ in the SF. Moreover, in the
Gutzwiller approximation, the phase transition is directly reflected in the
local number fluctuations, with the variance of $n_{\mathbf{R}}$ vanishing
throughout the MI phase. In reality, however, local variables
like the on-site number distribution will change in a smooth manner near the
transition and the variance of the local particle number will only
vanish in the limit $J\to 0$.

Crossing the SF-MI transition, therefore, the number statistics evolves rather smoothly from a Poissonian distribution to Fock states on each lattice site. Recent experimental progress has allowed measurements of the number distribution in the optical lattice via microwave spectroscopy exploiting collisional frequency shifts \cite{Campbell:2006} or spin changing collisions \cite{Gerbier:2006b}. When crossing the SF-MI transition, \textcite{Campbell:2006} were able to observe the emergence of a discrete excitation spectrum with Hz resolution. In the second experiment, the change in atom number statistics from Poissonian to Fock states could be revealed \cite{Gerbier:2006b}. Another possibility to observe the number squeezing of the initially Poissonian atom number distribution in the weakly interacting regime due to increasing interatomic interactions has been to use a matter wave beam splitter and observe the timescale of the collapse in the ensuing phase diffusion dynamics \cite{Greiner:2002b,Sebby:2007,Jo:2007}, which is discussed in the following paragraph.   

\subsection{Dynamics near quantum phase transitions} 
\label{subsec:dynamics}

One of the major advantages of cold atoms in studying many-body 
phenomena is the possibility to change the parameters characterizing
the relative strength of the kinetic and interaction energy dynamically.
This opens the possibility to study the real time dynamics of strongly
correlated systems in a controlled manner. As a 
simple example, we discuss the quench of the system from the superfluid into the Mott insulating regime.
This issue has been investigated
in an experiment, observing collapses and revivals
of the matter wave due to the coherent superposition of states with
different atom numbers in the SF \cite{Greiner:2002b}. 
In the weakly interacting regime of a BEC in an optical lattice potential,  the ground state (\ref{eq:coherentstate}) is a product of coherent states on each lattice site  
with a Poissonian atom number distribution. If the lattice depth is now suddenly increased to a parameter regime, where the ground state of the system is a Mott insulating state, the initial atom number fluctuations of the coherent state will be frozen out, as the system is not given enough time to redistribute towards the novel many-body ground state. The evolution with time of such a coherent state can be evaluated by taking into account the time evolution of the different
Fock states forming the coherent state:

\begin{equation}
|\alpha\rangle(t)=e^{-|\alpha|^2/2}\sum_n \frac{\alpha^n}{\sqrt{n!}}
e^{-i\frac{1}{2}Un(n-1)t/\hbar} |n\rangle.
\end{equation}

The coherent matter wave field $\psi$ on each lattice site can
then simply be evaluated through $\psi = \langle \alpha(t)|\hat
a|\alpha(t)\rangle$, which exhibits an intriguing dynamical
evolution
\cite{Yurke:1986,Sols:1994,Wright:1996,Imamoglu:1997,Castin:1997,Dunningham:1998}.
At first, the different phase evolutions of the atom number states
lead to a collapse of $\psi$. However, at integer multiples in time
of $h/U$ all phase factors in the above equation re-phase modulo
$2\pi$ and thus lead to a revival of the initial coherent state . 
In fact, precise revivals appear as long as 
the initial state can be written in the factorized form of 
Eq.~(\ref{eq:Gutzwiller}). Since the time evolution operator
$\exp{-i\hat{H}t/\hbar}$ factorizes into a product of on-site
terms $\exp{-in(n-1)Ut/2\hbar}$, the time dependence is perfectly
periodic with period $t_{\rm rev}=h/U_f$, where $U_f$ is the value of 
the on-site repulsion after the quench. Clearly the period is independent of
the precise form of the initial number distribution $|c_n|^2$. 
The collapse time $t_c\approx t_{\rm rev}/\sigma_n$ in turn, depends on the variance
$\sigma_n^2=\langle n^2\rangle-\langle n\rangle^2$ of the local 
number distribution. Its measurement thus provides information 
about how the coherent superposition of different particle numbers 
in the SF state is eventually destroyed by approaching the 
MI regime \cite{Greiner:2002b}.

The collapse and revival of
the coherent matter wave field of a BEC is reminiscent to the
collapse and revival of the Rabi oscillations in the interaction of
a single atom with a single mode electromagnetic field in cavity
quantum electrodynamics \cite{Rempe:1987,Brune:1996}. There, the
nonlinear atom-field interaction induces the collapse and revival of
the Rabi oscillations whereas here the nonlinearity due to the
interactions between the atoms themselves leads to the series of
collapse and revivals of the matter wave field. It should be pointed
out that such a behavior has also been theoretically predicted to
occur for a coherent light field propagating in a nonlinear medium
\cite{Yurke:1986} but to our knowledge has never been observed experimentally. 
Such a dynamical evolution of the atomic quantum state due to the nonlinear interactions between the particles is also known as quantum phase diffusion and has been detected in \cite{Greiner:2002b} for low atom numbers on each site. For larger atom numbers, the initial time evolution of the quantum phase diffusion could be recently observed by \textcite{Jo:2007} in a double well scenario.

The simple single site description is valid only in the limits 
$U_i\ll J$ of a nearly perfect SF in the initial state and 
$U_f\gg J$ of negligible tunneling in the final state. To determine
the dynamics in a more general situation, is a complicated 
non-equilibrium many-body problem. Numerical 
results for arbitrary values of $U_i$ and $U_f$ have been obtained
for the 1D BHM by \textcite{Kollath:2007}, using the time-dependent 
density matrix renormalization group \cite{Schollwoeck:2005}.                                                        

In a related scenario it has be proposed that when jumping from an initial Mott insulating state into the superfluid regime, one should observe oscillations of the superfluid order parameter \cite{Altman:2002,Polkovnikov:2002}. For large filling factors, oscillating coherence has been observed after a quench from a deep to a shallow lattice by \textcite{Tuchman:2006}. The formation of a superfluid from an initial Mott insulating phase poses a general problem of interest in the context of the dynamics of strongly correlated quantum systems. Both experiment \cite{Greiner:2002a} and theory \cite{Clark:2004} have confirmed that the emergence of coherence in the system can occur rather rapidly on timescales of a few tunneling times $\hbar/J$. It is an an open question, however, whether off-diagonal long range order in the one-particle density matrix indeed sets in within such a short time and what length scales are relevant over which order has established in order to observe coherence in a time-of-flight picture.

\noindent \subsection{Bose-Hubbard model with finite current}
The SF-MI transition discussed in \ref{subsec:SF-MI} above
is a continuous phase transition in the ground state of a
many-body Hamiltonian. The observation from the 
time-of-flight images, that long range phase coherence is lost
beyond a critical value of $U/J$, provides 
a signature for the disappearance of BEC.
The expected simultaneous loss of {\it superfluidity} across this 
transition may be studied by considering  the phase boundary, where
stationary states with a finite current loose their stability.
Such stationary out-of-equilibrium states
may be created experimentally by boosting the condensate to a finite 
momentum state \cite{Fallani:2004},
or by inducing a center-of-mass oscillation in the trap \cite{Fertig:2005}.
The question, what happens to the equilibrium  SF-MI transition
in a situation with a finite current has been addressed by \textcite{Polkovnikov:2005}.
For a given number $\bar{n}$ of bosons per site, the kinetic energy term
in the BHM (\ref{eq:BHM}) gives rise to a Josephson coupling energy $E_J=2\bar{n}J$
due to next neighbor-tunneling which favors a vanishing relative phase between
adjacent lattice sites. In the limit $E_J\gg U$, there is non-vanishing matter 
wave field $\psi_{\mathbf{R}}=\langle\hat{a}_{\mathbf{R}}\rangle$. In the ground state,
all bosons have zero momentum and $\psi_{\mathbf{R}}$ is uniform. 
States with a finite current,  in turn, are BEC's in which single particle states 
with non-zero momentum $\mathbf{q}$ are macroscopically occupied. 
To zeroth order in $U/E_J$, their energy 
is the Bloch band energy Eq.~(\ref{eq:Blochband}). The associated current 
per particle ${\cal J}=(2J/\hbar)\,\sin{q_xd}$ for motion along the $x\,$-direction has 
a maximum at $p=q_xd=\pi/2$. States with a larger momentum are unstable
in a linear stability analysis \cite{Polkovnikov:2005}. 
This instability was observed  experimentally by \textcite{Fallani:2004}. 
A moving optical lattice is created by two counterpropagating beams
at frequencies which differ by a small detuning $\delta\nu$. Averaged over 
the optical frequencies, this gives rise to an interference pattern which is
a standing wave moving at velocity $v=\lambda\delta\nu/2$. Adiabatically 
switching on such a lattice in an existing BEC then leads to a condensate 
in a state with quasi-momentum $q=Mv/\hbar$. Its lifetime 
shows a very rapid decrease for momenta
near the critical value $q_c$. 

\begin{figure}
\begin{center}
\includegraphics[width=0.8\columnwidth]{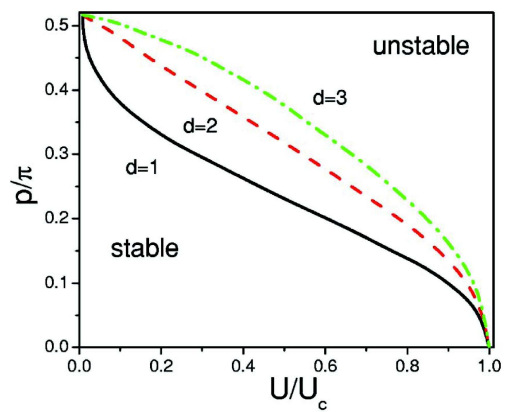}
\caption{Mean-field phase diagram separating stable and unstable motion of condensate regions. The vertical axis denotes the condensate momentum in inverse lattice units and the horizontal axes denotes the normalized interaction. Reprinted with permission from  \cite{Polkovnikov:2005}.\label{fig:PolkCurrent}}
\end{center}
\end{figure}

In the strongly 
interacting regime near the SF-MI transition, such a single particle picture is
no longer valid. At the mean-field level, the problem may be solved by using
the field theoretical description of the SF-MI transition. The SF phase is then
characterized by a nonzero complex order parameter $\psi$, whose 
equilibrium value $|\psi|\sim\xi^{-1}$ vanishes like the inverse of the correlation
length $\xi$ (this relation holds in the mean-field approximation, which is
appropriate for the transition at integer densities in 3D). The stationary   
solutions of the dimensionless order parameter equation $\nabla^2\psi+\xi^{-2}\psi=|\psi |^2\psi $
with {\it finite} momentum are of the form $\psi(x)=\sqrt{\xi^{-2}-p^2}\exp{(ipx)}$. 
Evidently, such solutions exist only if $ |p|<1/\xi $. The critical value of the dimensionless 
momentum $p$, where current currying states become unstable, thus approaches zero 
continuously at the SF-MI transition \cite{Polkovnikov:2005}. In fact, the same argument 
can be used to discuss the vanishing of the critical current in 
superconducting wires near $T_c$, see \textcite{Tinkham:1996}.
The complete mean field phase diagram, shown in Fig.~\ref{fig:PolkCurrent} 
interpolates smoothly between the classical instability at $p_c=\pi/2$ and
$p_c\to 0$ in the limits $U\to 0$ and $U\to U_c$ respectively. In contrast 
to the equilibrium transition at $p=0$ which is continuous, the dynamical transition
is of {\it first order}. Crossing the phase boundary at any nonzero current is therefore
connected with an irreversible decay of the current to zero. Experimentally, the decrease of the critical momentum near the SF-MI transition has been observed by \textcite{Mun:2007}. Their results are in good agreement with the phase diagram shown in Fig.~\ref{fig:PolkCurrent}.
       
\begin{figure}
\begin{center}
\includegraphics[width=0.7\columnwidth]{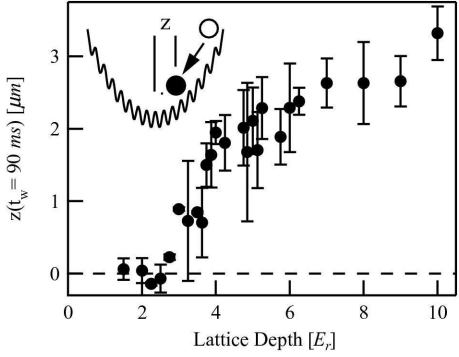}
\caption{Inhibition of transport in a one-dimensional bosonic quantum system with an axial optical lattice. For lattice depths above approx. $2\,E_r$, an atom cloud displaced to the side of the potential minimum (see inset) is stuck at this position and does not relax back to the minimum. Reprinted with permission from \textcite{Fertig:2005}.\label{fig:PortoTransport}}
\end{center}
\end{figure}

In the mean-field picture, states of a SF with non-zero 
momentum have an infinite lifetime.
More precisely, however, such states can only
be {\it meta}stable, because the ground state of any time-reversal 
invariant Hamiltonian necessarily has zero current.  
The crucial requirement for SF in practice, therefore, is that
current carrying states have lifetimes which by far exceed
experimentally relevant scales. This requires these states 
to be separated from the state with vanishing current by energy barriers,
which are much larger than the thermal or relevant zero point energy
\footnote{This is different from the well known Landau 
criterion of superfluid flow below a finite critical velocity
\cite{Pitaevskii:2003}.  Indeed, the existence of 
phase slips implies that the critical velocity is always zero in a  strict  sense.}. 
The rate for phase slips near the critical line in Fig.~\ref{fig:PolkCurrent} has been calculated
by \textcite{Polkovnikov:2005}. It turns out, that the mean-field
transition survives fluctuations in 3D, so in principle it is 
possible to locate the equilibrium  SF-MI transition by 
extrapolating the dynamical transition line to zero momentum. 
In the experiments of \textcite{Fertig:2005}, the 
system still showed sharp interference peaks even in the 
'overdamped' regime where the condensate motion was 
locked by the optical lattice (see Fig.~\ref{fig:PortoTransport}). This may be due to localized 
atoms at the sample edges, which block the dipole oscillation 
even though the atoms in the center of the trap are still in 
the SF regime.A theoretical study of the damped oscillations of 1D bosons has 
been given by \cite{Banacloche:2006}.

A different method to drive a SF-MI transition dynamically has been
suggested by \textcite{Eckardt:2005}.
Instead of a uniformly moving optical lattice, it employs an oscillating
linear potential $K\cos{(\omega t)}\cdot\hat{x}$ along one of the 
lattice directions (in a 1D BHM $\hat{x}=\sum_j j\hat{n}_j$ is the 
dimensionless position operator). For modulation frequencies such that $\hbar\omega$ is much larger 
than the characteristic scales $J$ and $U$ of the unperturbed BHM,
the driven system behaves like the undriven one, however with a
renormalized tunneling matrix element $J_{\rm eff}=J\cdot{\cal{J}}_0(K/(\hbar\omega))$,
where ${\cal{J}}_0(x)$ is the standard Bessel function. Since $U$ is unchanged
in this limit, the external perturbation completely supresses the tunneling
at the zero's of the Bessel function. Moreover, it allows to invert the sign of 
$J_{\rm eff}$ to negative values, where e.g. the superfluid phase corresponds
to a condensate at finite momentum $q=\pi$. It has been shown numerically \cite{Eckardt:2005}
that by a slow variation of the driving amplitude $K$ from zero to 
$K=2.4\,\hbar\omega$ (where  $J_{\rm eff}\approx 0$) and back to zero 
allows to adiabatically transform a superfluid into a Mott insulator and 
then back to a superfluid.  In recent experiments by \textcite{Lignier:2007}, 
the dynamical supression of tunneling with increasing driving $K$ has been 
observed through a measurement of the expansion velocity along the 
direction of the optical lattice after switching off the axial confinement.

\subsection{Fermions in optical lattices}
\label{sec:fermions}
In this section we would like to focus on fermions in 3D optical lattice potentials and experimental results that have been obtained in these systems (see also \textcite{Giorgini:2007}). 
Interacting fermions in an periodic potential can be
described by the Hubbard hamiltonian. Let us for now restrict the discussion to the case of atoms confined to the lowest energy band and to two possible spin states $\spinupstate,\spindownstate$ for the fermionic particles. The single band Hubbard hamiltonian thus reads:

\begin{eqnarray}
   H&=&-J\sum_{\langle \Rvec,\Rvec' \rangle,\sigma} \left( \chatdag_{\Rvec,\sigma} \chat_{\Rvec',\sigma} + h.c.\right) + U \sum_\Rvec \hat{n}_{\Rvec\uparrow} \hat{n}_{\Rvec\downarrow} \nonumber \\ && + \frac{1}{2}M\omega^2 \sum_{\Rvec,\sigma} \Rvec^2 \hat{n}_{\Rvec,\sigma}                   
\label{eq:fermionichubbard}
\end{eqnarray}

As in the case of bosonic particles, the phase diagram depends strongly on the interaction strength vs. kinetic energy of the atoms in the optical lattices. An important difference between the bosonic and fermionic Hubbard hamiltonian can also be seen in the form of the interaction term, where only two particles of different spin states are allowed to occupy the same lattice site, giving rise to an interaction energy $U$ between the atoms. 
\bigskip

\begin{figure}
\begin{center}
\includegraphics[width=0.8\columnwidth]{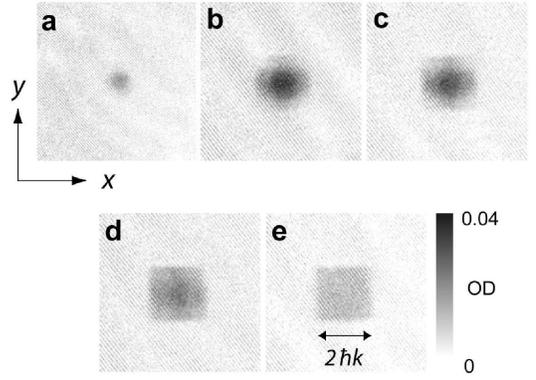}
\caption{Fermi surfaces vs band-filling for ultracold fermionic $^{40}$K atoms
in a three-dimensional simple cubic lattice potential. From {\bf (a)} to {\bf (e)} the filling factor has been continuously increased, bringing the system from a conducting to a band insulating state. Reprinted with permission from \textcite{Kohl:2005a}.\label{fig:FermiSurfaces}}
\end{center}
\end{figure}

\noindent {\it Filling factor and Fermi surfaces}  A crucial parameter in the fermionic Hubbard model is the filling factor of the atoms in the lattice. Due to the overall harmonic confinement of the atoms (last term in Eq.~(\ref{eq:fermionichubbard})), this filling fraction changes over the cloud of trapped atoms. One can however specify an average characteristic filling factor,
\begin{equation}
	\rho_c = \frac{N_F d^3}{\zeta^{3}}, 
\end{equation}                                                
with $\zeta=\sqrt{2J/M\omega^2}$ describing the typical delocalization length of the single particle wave-functions in the combined periodic lattice and external harmonic trapping potential \cite{Kohl:2005a,Rigol:2004a}.
The characteristic filling factor can be controlled experimentally, by either increasing the total number of fermionic atoms $N_F$, by reducing $J$ via an increase of the lattice depth or via an increase of the overall harmonic confinement. The latter case however has the disadvantage that a strong harmonic confinement will lead to a decoupling of the independent lattice sites, as the tunnel coupling $J$ will not be large enough to overcome the potential energy offset due to the overall harmonic confinement. One is then left with an array of uncoupled, independent harmonic oscillators. The characteristic filling factor of the system can be revealed experimentally by observing the population of the different Bloch states via adiabatic band mapping, introduced in Sec.~\ref{sec:TOF}. By changing the atom number or the lattice depth, \textcite{Kohl:2005a} could thus observe a change from a low filling to a band insulating state, where the single species particles are completely localized to individual lattice sites (see Fig.~\ref{fig:FermiSurfaces}).

\noindent {\it Thermometry and pair correlations}
An important question regarding fermionic quantum gases in optical lattices is the temperature of the many-body system. It has been shown that for non-interacting 50/50 spin mixtures of fermions, the number of doubly occupied spin-states can be used to determine the temperature of the system. For zero temperature and deep optical lattices, one would expect all lattice sites to be occupied by a spin-up and spin-down atom equally. For finite temperatures, however, atoms could be thermally excited to higher lying lattice sites at the border of the system, thus reducing the number of doubly occupied lattice sites. By converting doubly occupied sites into fermionic molecules, it has been possible to determine the number of doubly vs. singly occupied sites and obtain an estimate for the temperature of the spin-mixture \cite{Stoferle:2006,Kohl:2006}. Another possibility to determine the temperature of the system even for a single species fermionic quantum gas in the lattice has been provided by the use of quantum noise correlations, as introduced in Sec.~\ref{sec:correlations} . For higher temperatures of the quantum gas, the atoms tend to spread out in the harmonic confinement and thus increase the spatial size of the trapped atom cloud. As the shape of each noise-correlation peak essentially represents the Fourier transform of the in-trap density distribution, but with fixed amplitude of 1 (see Eq.~(\ref{eq:noisecorr})), an increase in the size of the fermionic atom cloud by temperature will lead to a decrease in the observed correlation signal, as has recently been detected in \cite{Rom:2006}.

When increasing a red-detuned optical lattice on top of a fermionic atom cloud, this usually also leads to an increased overall harmonic confinement of the system. It has been shown that for such a case, the density of states of the system can be significantly modified, thus leading to an adiabatic heating of the fermionic system by up to a factor of two for a strong overall harmonic confinement \cite{Kohl:2006}. In order to reach low temperatures for the fermionic system, it would thus be advantageous to keep the harmonic confinement as low as possible or even decrease it as the optical lattice depth is increased. Such a configuration is e.g. possible with a blue-detuned optical lattice in conjunction with a red-detuned optical dipole trap.


\section{COLD GASES IN ONE DIMENSION}
\label{sec:onedim}

In the following it is shown that the confinement of
cold atoms in a quantum wire geometry which can
be achieved via strong optical
lattices \cite{Paredes:2004, Kinoshita:2004} 
provides a means of reaching the strong interaction
regime in dilute gases. This opens the
possibility to realize both bosonic and fermionic
Luttinger liquids and a number of exactly soluble
models in many-body physics.

\subsection{Scattering and bound states}
\label{subsec:1Dscattering}

Consider cold atoms subject to a strong 2D optical lattice
in the $y,z\,$-plane which confines
the motion to the (axial) $x\,$-direction. In the
regime, where the excitation energy $\hbar\omega_{\perp}$ for motion
in the radial $y,z\,$-directions is much larger than the
chemical potential, only the lowest transverse eigenmode
is accessible. In terms of the oscillator length
$\ell_{\perp}=\sqrt{\hbar/M\omega_{\perp}}$
for the transverse motion, this requires the
1D density $n_1=n\pi\ell_{\perp}^2$ to obey $n_1a\ll 1$. The
effective interaction of atoms confined in such a geometry
has first been discussed by \textcite{Olshanii:1998a}.
In the realistic case, where $\ell_{\perp}$ is much larger than the
effective range $r_e$ of the atom-atom interaction
\footnote{For a typical frequency
$\omega_{\perp}=2\pi\cdot 67\,$kHz, the transverse
oscillator length is equal to $41.5\,$nm for
$^{87}$Rb},
the two-particle scattering problem at low energies can again be
described by a (3D) pseudopotential. However the
asymptotic scattering states are now of the form
$\phi_0(y,z)\exp{\pm ikx}$, where $\phi_0(y,z)$ is the Gaussian
ground state wave function for the transverse motion
and $k$ the wavevector for motion along the axial direction.
Now, away from Feshbach resonances, the 3D scattering length
is much smaller than the transverse oscillator length. In this
weak confinement limit $|a|\ll\ell_{\perp}$, the effective 1D
interaction is simply obtained by integrating the 3D pseudopotential $g\int d^3x\,\vert\phi_0(y,z)\vert^2\delta(\mathbf{x})$ over the ground state density
of the transverse motion. The resulting effective 1D interaction is then of the
form $V(x)=g_1\,\delta(x)$ with
\begin{equation}
\label{eq:g_1}
g_1\bigl(|a|\ll\ell_{\perp}\bigr)
=g\vert\phi_0(0,0)\vert^2=2\hbar\omega_{\perp}\cdot a\, .
\end{equation}
Trivially, an attractive or repulsive pseudopotential lead to
a corresponding sign of the 1D interaction.
In order to discuss what happens near Feshbach resonances,
where the weak confinement assumption breaks down,
we consider the question of possible bound states
of two particles in a strong transverse confinement.
Starting with an attractive 3D pseudopotential $a<0$,
the effective 1D potential  (\ref{eq:g_1}) has a bound state
with binding energy $\varepsilon_b=M_rg_1^2/2\hbar^2$. With increasing
magnitude of $a$, this binding energy increases, finally diverging
at a Feshbach resonance. In turn, upon crossing the resonance
to the side were $a>0$, the effective potential (\ref{eq:g_1})
becomes repulsive. The bound state, therefore, has disappeared
even though there is one in the 3D pseudopotential for $a>0$. Obviously, this
cannot be correct. As pointed out above, the result (\ref{eq:g_1})
only applies in the weak confinement limit $|a|\ll\ell_{\perp}$. In the following we present a treatment of the scattering properties of confined particles for arbitrary values of the ratio
$|a|/\ell_{\perp}$. To this end, we will first consider the issue of two-particle bound states
with a pseudopotential interaction in quite general terms by mapping the problem to a random walk process. Subsequently, we will derive the low energy scattering amplitudes by analytic continuation.

\paragraph*{Confinement induced bound states}

Quite generally, two-particle bound states may be determined
from the condition $V\hat{G}=1$ of a pole in the exact T-matrix.
In the case of a pseudopotential with scattering length $a$,
the matrix elements $\langle\mathbf{x}|\ldots | 0\rangle$ of this
equation lead to
\begin{equation}
\label{eq:boundstate}
\frac{1}{4\pi a}=\frac{\partial}{\partial r}\left( r\, G(E, \mathbf{x})\right)_{r=0}
=\lim_{r\to 0}\left( G(E,\mathbf{x})+\frac{1}{4\pi r}\right)\, ,
\end{equation}
in units, where $\hbar=2M_r=1$.
Here, $G(E, \mathbf{x})=\langle\mathbf{x}\vert (E-\hat{H}_0)^{-1}\vert 0\rangle$
is the Green function of the free Schr\"odinger equation and $\hat{H}_0$ includes
both the kinetic energy and the harmonic confining potential. As $r\to 0$, the Green function
diverges like $-(4\pi r)^{-1}$, thus providing the regularization for the
pseudopotential through the second term of Eq.~(\ref{eq:boundstate}). For energies
$E$ below the bottom of the spectrum of $\hat{H}_0$,
the resolvent $(E-\hat{H}_0)^{-1}=-\int_0^{\infty}dt\,\exp{(E-\hat{H}_0)t}$
can be written as a time integral. Moreover, by the Feynman-Kac formulation of
quantum mechanics, the (imaginary time) propagator $P(\mathbf{x},t)=\langle\mathbf{x}\vert
\exp{-\hat{H}_0t}\vert 0\rangle$ can be interpreted as the sum over
Brownian motion trajectories from $\mathbf{x}=0$ to $\mathbf{x}$ in time $t$
weighted with  $\exp{-\int_0^t U[\mathbf{x}(t')]}$, where $U(\mathbf{x})$ is the
external potential in the free Hamiltonian $\hat{H}_0$. Similarly, the
contribution $(4\pi r)^{-1}=\int_0^{\infty}dt\, P^{(0)}(\mathbf{x},t)$ can be written
in terms of the probability density of free Brownian motion with, again,
diffusion constant $D=1$ in units where $\hbar=2M_r=1$.
Taking the limit $r\to 0$, Eq.~(\ref{eq:boundstate}) finally leads to the exact equation
\begin{equation}
\label{eq:diffusion}
\frac{1}{4\pi a}=\int_0^{\infty}dt\,\left[ P^{(0)}(t)-e^{Et}P(t)\right]
\end{equation}
for the bound state energies $E$ of two particles with a pseudopotential
interaction and scattering length $a$. Here $P^{(0)}(t)=(4\pi t)^{-3/2}$ is
the probability density at the origin of a free random walk after
 time $t$, starting and ending at $\mathbf{x}=0$,
 while $P(t)$ is the same quantity in the presence of an additional
confining potential. Note that in the formulation of Eq.~(\ref{eq:diffusion}),
the regularization of the $1/r\,$-singularity in $G(E,\mathbf{x})$
is accounted for by the cancellation of the short time divergence due
to $P(t\to 0)=(4\pi t)^{-3/2}$, because the random walk does not feel
the confinement as $t\to 0$. For two particles in free space, where
$P(t)=P^{(0)}(t)$ at all times, Eq.~(\ref{eq:diffusion}) gives the
standard result that a bound state at $E=-\varepsilon_b$ below
the continuum at $E=0$ exists only for $a>0$, with the
standard value $\varepsilon_b=\hbar^2/2M_ra^2$. In the presence of an
additional confinement however, there is a bound state for arbitrary
sign of the scattering length. The quasi-bound state in the
continuum at $a<0$ thus becomes
a true bound state,  i.e. it is shifted upwards by {\it less}
than the continuum threshold $E_c$. The physics behind this
is the fact that the average time $\sim\int dt\, P(t)\exp{E_ct}$ spent near the origin
is infinite for the {\it confined} random walk.  This provides
an intuitive understanding for why an infinitesimally small
(regularized) delta potential is able to bind a state.

For harmonic confinement, the probability density
$P(t)=\bigl((4\pi)^3\,\text{det}\,\hat{J}(t)\bigr)^{-1/2}$ can be calculated
from the determinant  $\text{det}\,\hat{J}(t)$ for small fluctuations
around the trivial Brownian path $\mathbf{x}(t')\equiv 0$.
Here, $\hat{J}(t)$ obeys the simple $3$ by $3$ matrix equation \cite{Schulman:1981}
\begin{equation}
\label{eq:Jacobi}
\left( -\partial_t^2 +\hat{\omega}^2\right)\hat{J}(t)=0\;\text{with}\;\hat{J}(0)=0\;\;  \partial_t\hat{J}(t)\vert_{t=0}=1\, ,
\end{equation}
where $\hat{\omega}^2$ is the (diagonal) matrix of the trap frequencies.
The fluctuation determinant thus is equal to $\text{det}\,\hat{J}(t)=t(\sinh(\omega_{\perp}t)/
\omega_{\perp})^2$ for the situation with two confining directions and to
$\text{det}\,\hat{J}(t)=t^2\sinh(\omega_{z}t)/
\omega_{z}$ in the case of a pancake geometry.
Since the continua start at $\hbar\omega_{\perp}$ and
$\hbar\omega_{z}/2$, respectively, we write the bound state
energies as $E=\hbar\omega_{\perp}-\varepsilon_b$ or $E=\hbar\omega_{z}/2-\varepsilon_b$.
The dimensionless binding energy $\Omega=\varepsilon_b/\hbar\omega_{{\perp},z}$
in the presence of confinement then follows from the transcendental equation
\begin{equation}
\label{eq:f_n}
\frac{\ell_{{\perp},z}}{a}\!=\!\int_0^{\infty}\!\!\!\frac{du}{\sqrt{4\pi u^3}}\left( 1-
\frac{e^{-\Omega u}}{\left[ (1-\exp{-2u})/2u\right]^{n/2}}\right)\!
=f_n(\Omega)
\end{equation}
where $n=1,2$ is the number of confined directions.

\begin{figure}
\includegraphics[width=0.9\columnwidth]{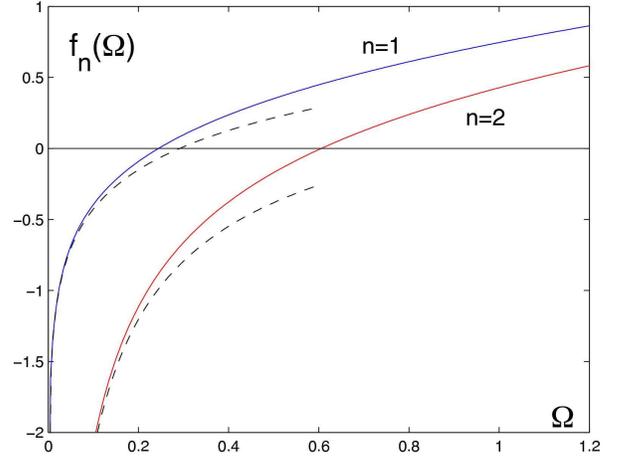}
\caption{The functions $f_n(\Omega)$ defined in Eq.~(\ref{eq:f_n})
for $n=1,2$. The limiting dependence $f_1(\Omega)= \ln{(\pi\Omega/B)}/\sqrt{2\pi}$ and
$f_2(\Omega)= -1/\sqrt{\Omega}+A$ on the
dimensionless binding energy $\Omega$ for $\Omega\ll 1$
is indicated by the dashed lines. \label{fig:fOmega}}
\end{figure}
The functions $f_{1,2}$ are shown in Fig.~\ref{fig:fOmega}. For small binding energies
$\Omega\ll 1$, their asymptotic behavior is
$f_1(\Omega)= \ln{(\pi\Omega/B)}/\sqrt{2\pi}$ and
$f_2(\Omega)= -1/\sqrt{\Omega}+A$
with numerical constants $A=1.036$ and $B=0.905$.
In the range $\Omega\lesssim 0.1$, where these expansions are
quantitatively valid, the resulting bound state energies are
\begin{equation}
\label{eq:2Dboundstate}
\varepsilon_b=\hbar\omega_{z}\cdot\frac{B}{\pi}\,\exp\left(-\sqrt{2\pi}\ell_{z}/|a|\right)
\end{equation}
in a 2D pancake geometry or
\begin{equation}
\label{eq:1Dboundstate}
\varepsilon_b=\frac{\hbar\omega_{\perp}}{\left(\ell_{\perp}/|a|+A\right)^2}
\end{equation}
in a 1D waveguide. These results were obtained, using different methods,
by \textcite{Petrov:2001} and by \textcite{Bergeman:2003}, respectively.
With increasing values of $|a|$
the binding energy increases,  reaching finite, universal values
$\varepsilon_b=0.244\,\hbar\omega_{z}$
or $\varepsilon_b=0.606\,\hbar\omega_{\perp}$ precisely at the
Feshbach resonance for one or two confined directions.
Going beyond the Feshbach resonance, where $a>0$, the binding energy
increases further, finally reaching the standard 3D result in the
weak confinement limit $a\ll\ell_{\perp}$.
Here the binding is unaffected by the nature of confinement
and $f_n(\Omega\gg 1)=\sqrt{\Omega}=f_0(\Omega)$.

Experimentally, confinement induced bound states have
been observed by rf spectroscopy in an array of 1D quantum wires using
a mixture of fermionic $^{40}$K atoms
in their two lowest hyperfine states $m_F=-9/2$ and
$m_F=-7/2$ \cite{Moritz:2005}. The different states
allow for a finite s-wave scattering length which may
be tuned using a Feshbach resonance at $B_0=202\,$G.
In the absence of the optical lattice a finite binding energy appears
only for magnetic fields below $B_0$, where $a>0$. In the
situation with a strong transverse confinement, however,
there is a nonzero binding energy both below and above $B_0$.
Using Eq.~(\ref{eq:FBs-length}) for the magnetic field dependence of the
scattering length, its magnitude is in perfect agreement with the
result obtained from Eq.~(\ref{eq:f_n}). In particular, the prediction
of a universal value $\varepsilon_b=0.606\,\hbar\omega_{\perp}$
of the binding energy right at the Feshbach resonance has
been verified by changing the confinement frequency
$\omega_{\perp}$. 

\paragraph*{Scattering amplitudes for confined particles}
Consider now two-particle scattering in the continuum,
i.e. for energies $E=\hbar\omega_{\perp}+\hbar^2 k^2/2M_r$
{\it above} the transverse ground state energy. Quite generally,
the effective 1D scattering problem is described by a unitary
S-matrix $\hat S=\bigl(\begin{smallmatrix}  t & r \\  r & t
\end{smallmatrix}\bigr)$,
with reflection and transmission
amplitudes $r$ and $t$.  They are related to the even and odd scattering
phase shifts $\delta_{e,o}(k)$ by the eigenvalues
$\exp\left(2i\delta_{e,o}(k)\right)=t\pm r$ of the S-matrix. In analogy to the 3D case,
the corresponding scattering amplitudes are  defined as $f_{e,o}(k)=
\left(\exp\left(2i\delta_{e,o}(k)\right)-1\right)/2$. For the particular case of a
delta-function interaction, the odd scattering amplitude and phase shift
vanish identically. The relevant dimensionless scattering amplitude
$f_e(k)=r(k)=t(k)-1$ is thus simply the standard reflection amplitude.
In the low energy limit, a representation analogous to Eq.~(\ref{eq:s-ampl})
allows to define a 1D scattering length $a_1$  by
\footnote{As in the 3D case, the scattering length is finite only for potentials
decaying faster than $1/x^3$. Dipolar interactions are therefore marginal
even in one dimension.}
\begin{equation}
\label{eq:1Ds-ampl}f(k)=\frac{-1}{1+i\cot{\delta(k)}}\,\to\,
\frac{-1}{1+ika_1}
\end{equation}
with corrections of order $k^3$ because $\cot{\delta(k)}$ is
odd in $k$. The universal limit $f(k=0)=-1$ reflects the fact
that finite 1D potentials become impenetrable at zero energy.
For a delta-function potential $V(x)=g_1\,\delta(x)$, the low energy form
of Eq.~(\ref{eq:1Ds-ampl}) holds for arbitrary $k$, with
a scattering length $a_1\equiv -\hbar^2/M_rg_1$, which approaches
$a_1\to-\ell_{\perp}^2/a$ in the weak confinement limit.
More generally,  the exact value of $a_1$ for an arbitrary ratio $a/\ell_{\perp}$
may be determined by using the connection $\varepsilon_b=\hbar^2\kappa^2/2M_r$
between the bound state energy and a pole at $k=i\kappa=i/a_1$ of the
1D scattering amplitude (\ref{eq:1Ds-ampl}) in the upper complex $k\,$-plane.
Using Eq.~(\ref{eq:1Dboundstate}) for the
bound state energy in the regime $a<0$ and the fact that $a_1$ must be
positive for a bound state, one finds
$a_1(a)=-\ell_{\perp}^2/a+A\ell_{\perp}$. Since the effective 1D pseudopotential
is completely determined by the scattering length $a_1$, two-particle scattering
is described by an interaction of the form
\begin{equation}
\label{eq:g_1exact}
V_{1D}(x)=g_1\,\delta(x)\;\;\;\text{with}\;\;\; g_1(a)=
\frac{2\hbar\omega_{\perp}\cdot a}{1-A\,a/\ell_{\perp}}\, .
\end{equation}
From its derivation, this result appears to be valid only for $g_1$ small and negative,
such that the resulting dimensionless binding energy $\Omega\lesssim 0.1$ is
within the range of validity of Eq.~(\ref{eq:1Dboundstate}). Remarkably,
however, the result (\ref{eq:g_1exact}) is exact, as far as scattering of
two particles above the continuum is concerned, at {\it arbitrary} values of $a/\ell_{\perp}$.
This is a consequence of the fact that Eq.~(\ref{eq:g_1exact})
is the {\it unique} pseudopotential consistent with the behavior of
$f(k)$ up to order $k^3$. The exact  scattering length thus is fixed by the
binding energy (\ref{eq:1Dboundstate}) at small values of $\Omega$. Moreover, the
result uniquely extends into the regime where $a_1$ becomes negative,
i.e. the pseudopotential ceases to support a bound state. The associated
change of sign in $g_1$ at $a=\ell_{\perp}/A$ is called a confinement induced
resonance \cite{Olshanii:1998a}. It allows to change the sign of the
interaction between atoms by purely geometrical means !  As shown by \textcite{Bergeman:2003},
it can be understood as a Feshbach resonance where a bound state in the
closed channel drops below the continuum of the ground state. Now,
in a 1D waveguide, the separation between energy levels of successive
transverse eigenstates with zero angular momentum is exactly
$2\hbar\omega_{\perp}$, independent of the value of $a$ \cite{Bergeman:2003}.
The confinement induced resonance thus appears precisely when the exact bound state
energy $\varepsilon_b$ - which {\it cannot} be determined from
the pseudopotential (\ref{eq:g_1exact}) unless $\Omega<0.1$ -
reaches $2\hbar\omega_{\perp}$. An analogous confinement induced resonance
appears in a 2D pancake geometry, see \textcite{Petrov:2000a}.

\subsection{Bosonic Luttinger-liquids, Tonks-Girardeau gas}
\label{subsec:LiebLiniger}

To describe the many-body problem of bosons
confined to an effectively 1D situation, the basic microscopic starting point
is a model due to \textcite{Lieb:1963b}
\begin{equation}
\label{eq:LiebLiniger}
H=-\frac{\hbar^2}{2M}\sum_{i=1}^{N}\frac{\partial^2}{\partial x_i^2} +
g_1 \sum_{i<j}\delta(x_i-x_j)\, .
\end{equation}
It is based on pairwise interactions with
a pseudopotential $\sim g_1\delta(x)$, as given in Eq.~(\ref{eq:g_1exact}). 
This is a valid description of the actual interatomic potential provided that 
the two-body scattering amplitude has the low
energy form of Eq.~(\ref{eq:1Ds-ampl} for all relevant 
momenta $k$.  In the limit $\mu\ll\hbar\omega_{\perp}$ of a
single transverse mode, they obey $k\ell_{\perp}\ll 1$. Remarkably,
in this regime, the pseudopotential approximation is always applicable. 
Indeed,  it follows from the leading correction
$\sim\Omega$ in the low binding energy expansion 
$f_2(\Omega)= -1/\sqrt{\Omega}+A+b\Omega+\ldots$ of the function 
defined in Eq.~(\ref{eq:f_n}), that the denominator
of the 1D scattering amplitude (\ref{eq:1Ds-ampl}) has the form
$1+ika_1-ib(k\ell_{\perp})^3+\ldots$ at low energies with
a numerical coefficient $b=0.462$. Since $a_1\approx -\ell_{\perp}^2/a$
in typical situations where $a\ll\ell_{\perp}$, the condition 
$k^2<1/(a\ell_{\perp})$ for a negligible cubic term $\sim k^3$
in the scattering amplitude
is always obeyed in the regime $k\ell_{\perp}\ll 1$ of a single transverse mode.
In this limit, the interaction between cold atoms in a 1D tube is therefore 
generically described by the integrable Hamiltonian (\ref{eq:LiebLiniger}).
In the homogeneous case, stability requires $g_1$ to be positive
\footnote{For a possible extension to a metastable 'super-Tonks' regime
at $g_1<0$ see \textcite{Astrakharchik:2005a}},
i.e. $\!$ the 3D scattering length
obeys $\ell_{\perp}>Aa>0$. At  a given 1D density $n_1=N/L$
the strength of the interactions in Eq.~(\ref{eq:LiebLiniger}) is characterized by a single
dimensionless parameter
\begin{equation}
\label{eq:gamma_1}
\gamma=\frac{g_1n_1}{\hbar^2n_1^2/M}\, = \,\frac{2}{n_1|a_1|}\approx\, \frac{2a}{n_1\ell_{\perp}^2}
\;\;\;\text{if}\;\;\;\ell{_\perp}\gg a\, .
\end{equation}
In marked contrast to the 3D situation,  the
dimensionless interaction strength $\gamma$
scales inversely with the 1D density $n_1$ \cite{Petrov:2000b}.
In one dimension, therefore
it is the {\it low} density limit where interactions dominate. This
rather counterintuitive result can be understood physically
by noting that the scattering amplitude Eq.~(\ref{eq:1Ds-ampl})
approaches $-1$ as $k\to 0$. Since, at a given interaction
strength $g_1$, the low energy limit is reached at low densities,
the atoms in this regime are perfectly reflected by the repulsive potential of the surrounding
particles.  For $\gamma\gg 1$, therefore, the system approaches a gas
of impenetrable bosons where all the energy is kinetic
\footnote{In fact, it follows from the Lieb-Liniger solution discussed below,
that the ratio of the interaction and kinetic energy per particle diverges like
$1/\sqrt{\gamma}$ for $\gamma\ll 1$ and decreases monotonically to zero 
as $1/\gamma$ for $\gamma\gg 1$}. 
In particular, as was shown by \textcite{Girardeau:1960},
at $\gamma=\infty$, the hard-core
condition of a vanishing wave function whenever two particle coordinates
coincide is fulfilled by  a wave function
\begin{equation}
\label{eq:Girardeau}
\Psi_B(x_1\ldots x_N)=
\prod_{i<j}\,\vert\sin\left[ \pi(x_j-x_i)/L\right] \vert
\end{equation}
which coincides with the absolute
value of the wave function of a non-interacting spinless Fermi gas.
Strongly interacting bosons in 1D thus acquire a fermionic character, a fact
well known from the exact solution of a hard core Bose or spin $1/2$ system
on a 1D lattice in terms of non-interacting fermions by the Jordan-Wigner
transformation, see e.g. \textcite{Wen:2004}.

\paragraph*{Crossover diagram in a harmonic trap}
In the presence of an additional harmonic confinement $V(x)=M\omega_0^2x^2/2$
along the axial direction, the relative interaction strength depends,
in addition to the parameter $\gamma$ introduced above, also on the ratio
$\alpha=\ell_0/|a_1|\approx 2a\ell_0/\ell_{\perp}^2$
between the oscillator length $\ell_0=\sqrt{\hbar/M\omega_0}$ and the
magnitude of the 1D scattering length. For a tight radial confinement
$\ell_{\perp}\approx 40$nm and typical values $\ell_0\approx 2\,\mu$m
for the axial oscillator length, one obtains $\alpha\approx 12$ for $^{87}$Rb.
This is in fact the interesting regime, since for $\alpha\ll 1$
the typical relative momenta $k\approx 1/\ell_0$ of two particles are so large
that the strong interaction limit $k|a_1|\ll 1$ cannot be reached at all.
The conditions for realizing the Tonks-Girardeau (TG) limit in a trap have been
discussed by \textcite{Petrov:2000b} using a quantum hydrodynamic description
(see also \textcite{Ho:1999a} and the review by \textcite{Petrov:2004a} on trapped gases in low dimensions).
Using Eq.~(\ref{eq:phasefluct}) in the appendix, the phase fluctuations in a 1D Bose gas
behave like $\delta\phi^2(x)=\bigl(\ln{(|x|/\xi)}\bigl)/K$ at zero
temperature. From the Lieb-Liniger solution discussed below, both the
characteristic healing length $\xi\approx K(\gamma)/n_1$ and the dimensionless
so-called Luttinger parameter $K=\pi\hbar\kappa c$ are monotonically decreasing functions
of the interaction strength. In particular, $K(\gamma)\to\sqrt{\pi/\gamma}$
is much larger than one in the limit $\gamma\ll 1$. In this limit, the 1D Bose gas
looses its phase coherence on a scale $\ell_{\phi}(T=0)=\xi\exp{K}$
(defined by $\delta\phi^2(x=\ell_{\phi})\approx 1$), which
exceeds the healing length by an exponentially large factor. The gas
behaves like a true condensate as long as its size $R$ is smaller than $\ell_{\phi}$.
Applying the Gross-Pitaevskii equation plus the local
density approximation, the radius of the associated Thomas-Fermi profile
$n_1(x)$  is $R_{\rm TF}\simeq (N\alpha)^{1/3}\ell_0$ \cite{Petrov:2000b}. The
condition $R_{\rm TF}\gg \ell_0$ for the validity of the Thomas-Fermi
approximation is thus always obeyed if $\alpha\gtrsim 1$. In contrast to the
analogous situation in 3D, however, the weak coupling regime requires
{\it high} densities. The local value $n_1(0)\xi\approx K(\gamma(0))$
of the Luttinger parameter at the trap center must thus be large compared to one.
This requires $\gamma(0)\ll 1$ or, equivalently,
$n_1(0)\ell_0\approx N\ell_0/R_{\rm TF}\gg\alpha$. As a result, the
Thomas-Fermi profile becomes invalid if $N<N_{\star}=\alpha^2\gg 1$.
For particle numbers below $N_{\star}$, the trapped gas
reaches the TG-regime. The density distribution is
eventually that of a free Fermi gas, with a chemical potential $\mu=N\hbar\omega_0$
and a cloud size $R_{\rm TG}=\sqrt{2N}\ell_0$. The continuous evolution of
the density profile and cloud size between the weak coupling and the TG-limit
has been discussed by \textcite{Dunjko:2001}.

\begin{figure}
\includegraphics[width=0.9\columnwidth]{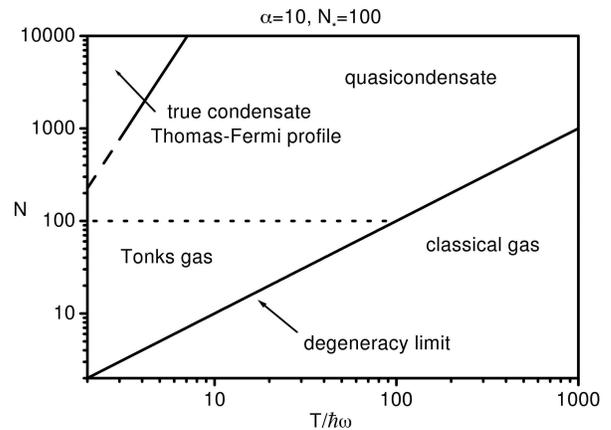}
\caption{Phase diagram for a 1D Bose gas in a harmonic trap with
$\alpha=10$. The Tonks-Girardeau regime is reached
for small particle numbers $N<N_{\star}=\alpha^2$ and temperatures
below the degeneracy limit $N\hbar\omega_0$. Reprinted with permission from \textcite{Petrov:2000b}.
\label{fig:Tonks-phasediagram}}
\end{figure}

At finite temperatures, the dominant phase fluctuations are {\it thermal}
and give rise to a linear increase $\delta\phi^2(x)=|x|/\ell_{\phi}(T)$
with distance on a scale $\ell_{\phi}(T)=\hbar^2n_s/Mk_BT$,
which only depends on the 1D superfluid density $n_s$
\footnote{To simplify the notation, the dimensionality is not
indicated in the superfluid or quasi-condensate densities
$n_s$ and $\tilde{n}_0$.}. In a trap of size $R$, these
fluctuations are negligible if $\ell_{\phi}>R$. Using $n_s\approx N/R$
and the zero temperature result for the Thomas-Fermi radius at $N>N_{\star}$,
this translates into $k_BT\lesssim\hbar\omega_0\, (N/N_{\star})^{1/3}$.
In this range, the trapped gas is effectively a true BEC with a Thomas-Fermi
density profile and phase coherence extending over the full cloud size.
With increasing temperature, phase fluctuations are nongegligible,
however density fluctuations become only relevant if $T$ exceeds
the degeneracy temperature $T_d=N\hbar\omega_0$ \cite{Petrov:2000b}.
Defining a characteristic temperature $T_{\phi}=T_d/(N\alpha)^{2/3}\ll T_d$
below which phase fluctuations are irrelevant over the system size,
there is a wide range $T_{\phi}<T<T_d$ in which the density profile
is still that of a BEC, however phase coherence is lost.
The system can be thought of as a collection
of independently fluctuating local BEC's and is called a {\it quasi-condensate}
\cite{Petrov:2000b}. At higher temperatures $k_BT\gtrsim N\hbar\omega_0$ the gas
eventually evolves into a non-degenerate regime of a
Boltzmann gas. The complete crossover diagram is
shown in Fig.~\ref{fig:Tonks-phasediagram}.

Experimentally, the presence of strong phase fluctuations
in a 1D situation already shows up in very elongated 3D condensates
which still have a Thomas-Fermi density profile in the radial direction
(i.e. $\mu\gg\hbar\omega_{\perp}$ such that many transverse modes are
involved).  In a trap, the strong interaction prevents that local velocity fields due to phase
fluctuations show up in the density profile. Switching off
the trap, however, the interaction becomes negligible after
a certain expansion time and then phase fluctuations are indeed
converted into density fluctuations. These have been seen
as stripes in absorption images of highly elongated BEC's
by \textcite{Dettmer:2001}. The linear increase of 
the phase fluctuations $\delta\phi^2(x)$ at finite temperature leads
to an exponential decay of the first order coherence function.
The resulting Lorentzian momentum distribution was observed 
experimentally by \textcite{Richard:2003} and by \textcite{Hellweg:2003}, 
using Bragg spectroscopy. This enabled a
quantitative measurement of $\ell_{\phi}(T)$. Moreover, 
at very low temperatures, no significant density fluctuations 
were present, thus confirming the quasi-BEC picture in which
$\langle n(\mathbf{x})^2\rangle \approx \langle n(\mathbf{x})\rangle^2$.

\begin{figure}
\includegraphics[width=0.95\columnwidth]{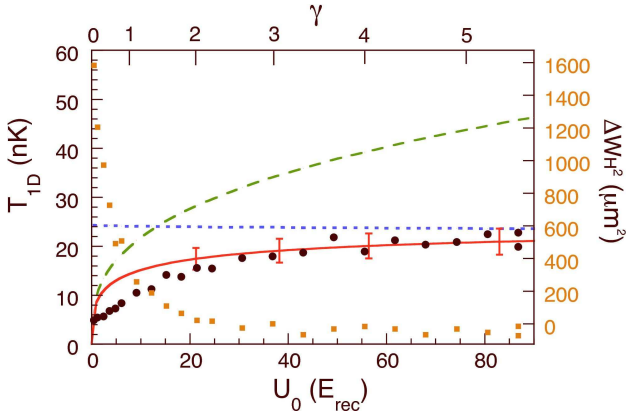}
\caption{Average axial energy per particle and equivalent temperature
$T_{1D}$ as a function of the transverse confinement $U_0$. With
increasing values of $U_0$, the energy crosses over from a weakly
interacting Bose gas (long-dashed line) to a Tonks-Girardeau gas
(short-dashed line), where $T_{1D}$ is independent of $U_0$.
Reprinted with permission from \textcite{Kinoshita:2004}. \label{fig:Tonks-energy}}
\end{figure}

\paragraph*{Lieb-Liniger solution}
As shown by \textcite{Lieb:1963b} the model Eq.~(\ref{eq:LiebLiniger}) can be solved by the Bethe ansatz. The essential physical property which lies behind the possibility of this exact solution is the fact that in one dimension interactions only give rise to scattering but not to diffraction. All eigenstates of the many-body problem
in the domain $0\leq x_1\leq x_2\ldots\leq x_N\leq L$ can thus be written
as a sum
\begin{equation}
\label{eq:Bethe}
\Psi_B(x_1\ldots x_N)=
\sum_{P}\, a(P)\, \exp{i\sum_{l=1}^{N}\, k_{P(l)}x_l}
\end{equation}
of plane waves with $N$ distinct wave vectors $k_l$. They are
combined with the coordinates $x_l$ in all $N!$ possible permutations
$k_{P(l)}$ of the $k_l$. The associated amplitude $a(P)=\prod_{(ij)}\, e^{i\theta_{ij}}$
factorizes into two-particle scattering phase shifts
$\theta_{ij}=\pi+2\arctan{(k_i-k_j)a_1/2}$, with $a_1$ the 1D scattering
length associated with the pseudopotential $g_1\delta(x)$. Here, the product
$(ij)$ runs over all permutations of two wavenumbers which are needed to
generate a given permutation $P$ of the $k_l$ from the identity. Due to
$a_1\sim -1/g_1$, the two-particle phase shifts $\theta_{ij}\sim g_1/(k_i-k_j)$ are
singular as a function of the momenta in the limit $g_1\to 0$
of an eventually ideal Bose gas.
In the limit $\gamma\gg 1$, the phase shifts approach
$\theta_{ij}=\pi$ for all momenta.
Thus $a(P)_{\gamma=\infty}=(-1)^{|P|}$ is just the parity of the
permutation $P$ and the Bethe ansatz wave function Eq.~(\ref{eq:Bethe})
is reduced to the free fermion type wave function Eq.~(\ref{eq:Girardeau})
of a Tonks-Girardeau gas.  For arbitrary coupling, both the ground state energy
per particle $E_0/N=n_1^2\hbar^2/2M\cdot e(\gamma)$ and the chemical
potential $\mu=\partial E_0/\partial N$ are monotonically
increasing functions of $\gamma$ at fixed density $n_1$. For weak
interactions $\gamma\ll 1$, the chemical potential $\mu=g_1n_1$ follows
the behavior expected from a mean field approach, which is valid here
for {\it high} densities $n_1|a_1|\gg 1$. In the low density regime $\gamma\gg 1$,
in turn, $\mu=\hbar^2(\pi n_1)^2/2M$ approaches a coupling independent value
which is just the Fermi energy associated with $k_F=\pi n_1$.  The energy per
particle in this regime is of completely kinetic origin, independent of the interaction
strength $\gamma$. This remarkable property of the Tonks-Girardeau
gas has been observed experimentally by \textcite{Kinoshita:2004} (see Fig.~\ref{fig:Tonks-energy}). They have measured the axial expansion energy of an array of 1D Bose gases
as a function of the strength $U_0$ of the transverse confinement.
Since $\omega_{\perp}\sim\sqrt{U_0}$, the dimensionless coupling $\gamma$
is monotonically increasing with $U_0$ at fixed density $n_1$. In the
weak confinement limit, the expansion energy scales linearly with $\sqrt{U_0}$,
reflecting the mean field behavior $e(\gamma)=\gamma(1-4\sqrt{\gamma}/3\pi +\ldots)$
of the average energy per particle. By contrast, for large values of $\sqrt{U_0}$,
where $|a_1|$ becomes shorter than the average interparticle spacing,
the energy $e(\gamma)=\pi^2(1-4/\gamma+\ldots)/3$  saturates at a value
which is fixed by the density.

The low lying excitations of the Lieb-Liniger model
have been obtained exactly by \textcite{Lieb:1963c}.
Surprisingly, it turned out that there are {\it two} types of excitations,
both with a linear spectrum $\omega=cq$ at low momenta. One of them
has a Bogoliubov like dispersion, linear at $q\xi\ll 1$ and
quadratic at $q\xi\gg 1$. The crossover from
collective to single-particle behavior occurs at a characteristic
length $\xi$, which is related to the chemical potential
 in the ground state via $\mu=\hbar^2/(2M\xi^2)$. In the limit $\gamma\ll 1$,
the crossover length can be expressed in the form $\xi n_1=\gamma^{-1/2}\gg 1$. Similar to the situation in three dimensions,  the weak coupling regime can therefore be characterized by the fact that the healing length is much larger
than the average interparticle spacing $n_1^{-1}$, By contrast, for
strong coupling $\gamma\gg 1$, where the chemical potential approaches the
Fermi energy  of a spinless, non-interacting Fermi gas
at density $n_1$, the healing length $\xi$ is essentially identical with the
average interparticle distance. The sound velocity $c$ turns out to coincide
with the simple thermodynamic formula $Mc^2=\partial\mu/\partial n$ which
is obtained from the quantum hydrodynamic Hamiltonian Eq.~(\ref{eq:QHD})
under the assumption that the superfluid density at $T=0$ coincides with
the full density. The ground state of the Lieb-Liniger gas is
in fact fully superfluid at arbitrary values of $\gamma$
in the sense of Eq.~(\ref{eq:n_s}), despite the fact that phase fluctuations
destroy plain BEC even at zero temperature. The sound velocity increases
monotonically with $\gamma$, approaching the finite value $c(\infty)=v_F=\hbar\pi n_1/M$
of an ideal Fermi gas in the Tonks-Girardeau limit. The second type of excitations
found by \textcite{Lieb:1963c} also has a linear dispersion at small $q$ with
the same velocity. However, in contrast to the Bogoliubov-like spectrum discussed
before, it is restricted to a finite range $|q|\leq\pi n_1$ and terminates
with a vanishing group velocity. It turns out,
that these excitations are precisely the solitons of the nonlinear Schr\"odinger
equation \cite{Ishikawa:1980}, for a discussion in the cold gas
context see \textcite{Jackson:2002a}.

\paragraph*{Momentum distribution in the Luttinger liquid regime}
To obtain the momentum distribution of a strongly correlated 
1D Bose gas, it is convenient to
start from a quantum hydrodynamic description of the
one-particle density matrix. At zero temperature,
the logarithmic increase $\delta\phi^2(x)=\bigl(\ln{|x|/\xi}\bigr)/K$
of the phase fluctuations with distance (see Eq.~(\ref{eq:phasefluct})),
leads to an algebraic decay of $g^{(1)}(x)$ at scales beyond the
healing length $\xi$. The associated exponent $1/(2K)$
is rather small in the weak coupling regime and approaches its
limiting value $1/2$ in the TG-limit. The resulting momentum distribution
thus exhibits a power-law divergence $\tilde{n}(k)\sim k^{-(1-(1/2K))}$
for $k\xi\ll 1$. At any finite temperature, however, this divergence is cutoff
due to thermal phase fluctuations.
Indeed, these fluctuations increase linearly with distance $\delta\phi^2(x)=|x|/\ell_{\phi}(T)$
on a scale $\ell_{\phi}(T)$, implying an exponential decay of
the one-particle density matrix for $|x|> \ell_{\phi}$.
 This leads to a rounding of the momentum distribution at small $k\simeq 1/\ell_{\phi}$.            

\begin{figure}
\begin{center}
\includegraphics[width=0.9\linewidth]{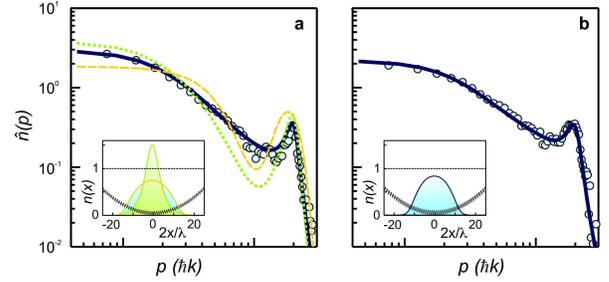}
\caption{Axial momentum distribution of a lattice based one-dimensional bosonic quantum gas for {\bf (a)} $\gamma_L \approx 14$ and {\bf (b)} $\gamma_L \approx 24$. The solid curve is the theoretical momentum distribution based on fermionization and the short and long-dashed curves in {\bf (a)} denote the expected values for a non-interacting Bose gas and a non-interacting Fermi gas, respectively. The insets show the corresponding in-trap density distributions. Reprinted with permission from  \textcite{Paredes:2004}.\label{fig:Tonksmomentum}}
\end{center}
\end{figure}

Experimentally, the momentum distribution of a Tonks-Girardeau gas has been observed
for ultracold atoms in a 2D optical lattice by \textcite{Paredes:2004}. There a weak optical
lattice along the one-dimensional quantum gases was applied in order to tune the system
into the strongly interacting regime, where $K$ is close to one. Indeed,
for the low filling factors $f\ll 1$ used in the experiment, there is no
Mott insulating phase. The 1D Bose-Hubbard model at $f\ll 1$
describes a bosonic Luttinger liquid with $K\approx 1+4\pi f/\gamma_L$
\cite{Cazalilla:2004}, where $\gamma_L=U/J$ is the effective coupling parameter on a lattice.
Increasing the axial lattice depth, this ratio becomes very large, of order
$\gamma_L=5-200$. As shown in Fig.~\ref{fig:Tonksmomentum}, the resulting momentum
distributions exhibit a power-law decay over a wide momentum range.
They are in very good agreement with a fermionization-based calculation of a
hard-core Bose gas in a harmonic confinement. The momentum distributions 
at finite values of $U$ have been studied by Quantum Monte Carlo calculations, noting that although the final hardcore limit is only reached for large values of $\gamma_L \rightarrow \infty$ \cite{Pollet:2004}, the deviations in the momentum distribution compared to fermionized bosons are less than a few percent already for $\gamma_L>5$ \cite{Wessel:2005a}.
In the experiment, a significant deviation from the limiting value of $1-1/(2K)\to 1/2$
of the exponent in the momentum distribution at small values of $k$ is found, which is
masked by finite temperature effects and finite size cut-offs. For larger momenta, the momentum distribution of the quantum gas is determined by short-range correlations between the particles, which can increase the coefficient of the power-law decay in the momentum distribution above $1/2$ in the experiment. In fact, it has been shown by \textcite{Olshanii:2003}, that the momentum distribution
of a homogeneous 1D Bose gas at large momenta $k\xi\gg 1$ should behave like $1/k^4$ as long as
$kr_e\ll 1$. 

The parameter $K=\pi\hbar\kappa c$ determines the asymptotic behavior
not only of the one-particle density matrix but in fact of {\it all}
correlation functions. This may be understood from
Haldane's description of 1D Bose liquids
\footnote{In the context of cold gases the notion of a quantum {\it liquid}
is - of course - purely conventional. These systems are stable only in the
gaseous phase, yet may be strongly interacting.}
in terms of their long wavelength density oscillations \cite{Haldane:1981a}.
In its most elementary form, this is just the 1D version of the
quantum hydrodynamic Hamiltonian Eq.~(\ref{eq:QHD}).
Introducing a field $\hat\theta(x)$ which is related to the
small fluctuations around the average
density by $\delta\hat n_1(x)=\partial_x\hat\theta(x)/\pi$, the effective Hamiltonian
describing the low lying excitations is of the form
\begin{equation}
\label{eq:1DQHD}
\hat H=\frac{\hbar c}{2\pi}\int dx
\left[K(\partial_{x}\hat\phi)^2+\frac{1}{K}(\partial_{x}\hat\theta)^2\right]
\end{equation}
with sound velocity $c$. The low energy physics of a 1D Bose liquid is thus
completely determined by the velocity $c$ and the dimensionless
parameter $K$. In the translation invariant case,
$K=\pi\hbar n_1/Mc$ is fixed by $c$ and the average density.
Moreover, for interactions which may be described by a 1D pseudopotential,
$K$ may be expressed in terms of the microscopic coupling
constant $\gamma$ by using the Lieb-Liniger solution. The resulting
value of $K=v_F/c$ decreases monotonically from
$K(\gamma)=\pi/\sqrt{\gamma}\gg 1$ in the weak interaction, high density limit
to $K(\gamma)= 1+4/\gamma+\ldots$ in the Tonks-Girardeau limit (see \textcite{Cazalilla:2004c}).
The property $K>1$ for repulsive bosons is valid quite generally for
interactions which decay faster than $1/x^3$ such that the 1D scattering
length $a_1$ is finite.
Formally, the algebraic decay of $g^{(1)}(x)$ in a 1D gas at {\it zero} temperature
is very similar to the situation in 2D at {\it finite} temperatures, where $g^{(1)}(r)$
exhibits a power law decay with exponent $\eta$ (see Eq.~(\ref{eq:powerlawdecay})).
Apart from the different nature of the phase fluctuations (quantum versus thermal),
there is, however, an important difference between both situations. In the 2D
case, superfluidity is lost via the BKT-transition once $\eta>1/4$
(see Eq.~(\ref{eq:BKT})). By contrast, in one dimension, there is
no such restriction on the exponent and superfluidity still persists if $K<2$. 
The origin of this difference is related to the fact that phase slips in one
dimension require a non-zero modulation of the potential, e.g. by a weak optical lattice 
\cite{Buchler:2003a}. Formally, it is related to a Berry phase term beyond 
Eq.~(\ref{eq:1DQHD}), which confines vortex-antivortex pairs in this case, 
see \textcite{Wen:2004}.

\begin{figure}
\includegraphics[width=0.8\columnwidth]{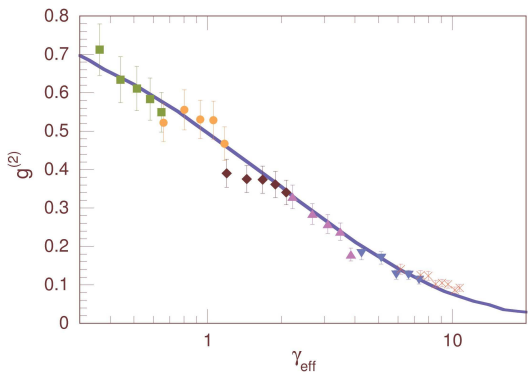}
\caption{Local pair correlation function from photo-association
measurements as a function of the interaction parameter $\gamma_{\rm eff}$
averaged over an ensemble of 1D Bose gases. The theoretical prediction
is shown as a solid line. Reprinted with permission from \textcite{Kinoshita:2005}. \label{fig:g2gamma}}
\end{figure}

\paragraph*{Two-and three-particle correlations}
An intuitive understanding of
the evolution from a weakly interacting quasi-condensate
to a Tonks-Girardeau gas with increasing values of $\gamma$
is provided by considering the pair distribution function $g^{(2)}(x)$.
It is defined by the density correlation function  $\langle\hat{n}(x)\hat{n}(0)\rangle=n_1\delta(x)+n_1^2\, g^{(2)}(x)$ and
is a measure of the probability to find two particles to be separated by a distance $x$.
For an {\it ideal} BEC in three dimensions, the pair distribution function
is equal to $g^{(2)}(\mathbf{x})\equiv 1$ at arbitrary distances. Above $T_c$, it drops
monotonically from $g^{(2)}(0)=2$ to the trivial limit $g^{(2)}(\infty)=1$ of any
homogeneous system  on the scale of the thermal wavelength $\lambda_T$.
For cold atoms in 3D, these basic results on bosonic two-particle
correlations have been verified experimentally by \textcite{Oettl:2005}
and \textcite{Schellekens:2005}.
For the Lieb-Liniger gas, the local value of the pair correlation
$g^{(2)}(0)\!=\! de(\gamma)/d\gamma$ can be obtained directly from
the derivative of the dimensionless ground state energy \cite{Gangardt:2003}.
The exact result for $e(\gamma)$ then gives $g^{(2)}(0)=1-2\sqrt{\gamma}/\pi+\ldots$
and $g^{(2)}(0)=\bigl(2\pi/\sqrt{3}\gamma\bigr)^2\to 0$ in the limits $\gamma\ll 1$  and
$\gamma\gg 1$, respectively. For weak coupling, therefore, there is only
a small repulsive correlation hole around each
particle. By contrast, in the strong coupling limit,  the probability
to find two bosons at the same point vanishes like $1/\gamma^2$.
In the Tonks-Girardeau limit,
the equivalence of density correlations to that of a free Fermi gas allows to
determine the full pair distribution function exactly as
$g^{(2)}(x)=1-\bigl(\sin(\pi n_1x)/\pi n_1x\bigr)^2$. For low densities, therefore,
the zero range repulsion strongly
suppresses configurations in which two bosons come closer than their mean
interparticle distance. Note that the pair correlation exhibits appreciable
oscillations with wave vector $2k_F=2\pi n_1$ even though the momentum
distribution is completely continuous at $k_F$. Experimentally, the local
value $g^{(2)}(0)$ of the pair correlation function has be determined
by \textcite{Kinoshita:2005}, using photo-association. As suggested by \textcite{Gangardt:2003},
the rate $K_1=K_3\cdot g^{(2)}(0)$ for stimulated transitions in which
two atoms in a continuum state are transferred
to a bound molecule is simply reduced by a factor $g^{(2)}(0)$ from the
corresponding value in 3D, provided the transfer occurs locally
on a length scale much less than the transverse oscillator length. The
photo-association rate in a one dimensional situation is thus strongly
reduced at $\gamma\gg 1$ due to the much smaller probability for two atoms to
be at the same point. From measurements of the atom loss after a
variable time of photo-association in an array of several thousand
1D traps with particle numbers in the range $40<N<240$,
\textcite{Kinoshita:2005} have thus extracted the value of
$g^{(2)}(0)$. As shown in Fig.~\ref{fig:g2gamma}  the results are in very good
agreement with theory over a wide range of interaction constants up to $\gamma\approx 10$.

The local value of three-body correlation function $g^{(3)}(0)$
has also been calculated by \textcite{Gangardt:2003}. It
behaves like $g^{(3)}(0)=1-6\sqrt{\gamma}/\pi+\ldots$  and
$g^{(3)}(0)\sim (\pi/\gamma)^6$ for small or large
$\gamma$ respectively. The predicted
suppression of three-body recombination losses has
been observed by \textcite{Tolra:2004}, using the strong confinement in
a 2D optical lattice. 

In-situ measurements of density fluctuations have been performed by
\textcite{Esteve:2006}. They observe a crossover from an
effectively high temperature regime at low densities
$n_1\ll (|a_1|/\lambda_T^4)^{1/3}$, where the number fluctuations
exceed the shot noise level due to bunching 
in an essentially ideal Bose gas. At high densities, a quasi-condensate
regime is reached with $n_1a\simeq 0.7$, close to the limit of
a single transverse mode. There, the number 
fluctuations are strongly suppressed and may be described 
by a 1D Bogoliubov description
of quasi-condensates \cite{Mora:2003}.

\paragraph*{Weak optical lattices and coupled 1D gases}
The problem of a 1D Bose gas in a {\it weak} optical lattice
has been discussed by \textcite{Buchler:2003a}.
Using the extension in 1D of the phase-density representation (\ref{eq:Bosefieldop})
of the field operator in a quantum hydrodynamic description
which accounts for the discrete nature of the particles \cite{Haldane:1981a},
a periodic potential commensurate with
the average particle density gives rise to an additional nonlinear
term $\cos{2\theta(x)}$ in (\ref{eq:1DQHD}). This is the well known
sine-Gordon model \cite{Giamarchi:2004}, which exhibits a
transition at a critical value $K_c=2$. For $K>2$ the
ground state of the Lieb-Liniger gas remains superfluid in a weak optical
lattice. For $(1<)K<2$, in turn,  the atoms are locked in an
incompressible Mott state even in an arbitrary weak
periodic lattice. From the exact Lieb-Liniger result for $K(\gamma)$,
the critical value $K_c=2$ is reached at $\gamma_{c}\approx 3.5$.

In a configuration using 2D optical lattices, a whole array
of typically several thousand parallel 1D gases are generated.
For a very large amplitude of the optical lattice $V_{\perp}\gtrsim 30\, E_r$,
hopping between different 1D gases is negligible and the system
decouples into independent 1D tubes. By continuously
lowering $V_{\perp}$, however, it is possible to study the
crossover from a 1D to a 3D situation.  
The equilibrium phase diagram of an array of 1D tubes with an
adjustable transverse hopping $J_{\perp}$ has been studied by \cite{Ho:2004b}.
It exhibits a fully phase coherent BEC for sufficiently large values of
$J_{\perp}$ and the 1D Luttinger parameter $K$. For a detailed 
discussion of  weakly coupled 1D gases see \textcite{Cazalilla:2006}.

\subsection{Repulsive and attractive fermions}
\label{subsec:1Dfermions}
As mentioned in subsection A, ultracold Fermi gases in a truly
1D regime $\varepsilon_F\ll\hbar\omega_{\perp}$ have been
realized using strong optical lattices. In the presence of an
additional axial confinement with frequency $\omega_0$, the
Fermi energy is $\varepsilon_F=N\hbar\omega_0$. The
requirement that only the lowest transverse mode is
populated therefore requires small particle numbers $N\ll\omega_{\perp}/\omega_0$.
Typical temperatures in these gases are around $k_BT\approx 0.2\varepsilon_F$
\cite{Moritz:2005}, which is not small enough, to observe non-trivial many-body effects.
We will therefore discuss the basic phenomena which may be studied with
ultracold fermions in one dimension only briefly.

In a spin-polarised 1D Fermi gas, only p-wave interactions are possible.
As shown by \textcite{Granger:2004}, the corresponding Feshbach resonances
are shifted due to the confinement in a similar way as in (\ref{eq:g_1exact}) above.
This was confirmed experimentally by \textcite{Guenter:2005}.
In the case of two different states, s-wave scattering dominates in the ultracold limit.
For repulsive interactions one obtains a
fermionic Luttinger liquid, which is basically a two component version
of the quantum hydrodynamic Hamiltonian (\ref{eq:1DQHD}).
It has a twofold linear excitation spectrum for fluctuations of the total density and the
density difference ('spin density')  respectively.
Generically, the velocities of  'charge' and spin
excitations are different. This is the most elementary form of 'spin-charge
separation', which has been verified experimentally in semiconductor quantum wires
\cite{Auslaender:2005}. In the context of ultracold Fermi gases in a harmonic trap,
spin-charge separation effects show up in collective excitation
frequencies \cite{Recati:2003} or in the propagation of wavepackets \cite{Kollath:2005}. A genuine observation of spin-charge separation, however, requires to study single particle correlations and cannot be inferred from collective excitations only.

For {\it attractive} interactions, a spin $1/2$ Fermi gas which is
described by the basic Hamiltonian Eq.~(\ref{eq:LiebLiniger})
is a so-called Luther-Emery liquid. Its
fluctuations in the total density still have a linear spectrum $\omega(q)=cq$,
however there is a finite gap for spin excitations \cite{Giamarchi:2004}.
The origin of this gap is the appearance of bound pairs of fermions with
opposite spin. The spectrum $\omega_s(q)=\sqrt{(\Delta_s/2 \hbar)^2+(v_s q)^2}$
for small oscillations of the spin density is similar to that of
quasiparticles in the BCS theory.
In analogy to Eq.~(\ref{eq:gamma_1}) for the bosonic problem,
the dimensionless coupling constant
$\gamma=-2/n_1a_1<0$ is inversely proportional to both the
1D scattering length $a_1=-\ell_{\perp}^2/a+A\ell_{\perp}$ (now for fermions
in different states, note that attractive interactions $a<0$ imply a {\it positive}
1D scattering length $a_1$) and to the total 1D density $n_1=n_{1\uparrow}+n_{1\downarrow}$.
As shown by \textcite{Gaudin:1967} and \textcite{Yang:1967},
the model is exactly soluble by the Bethe ansatz. For weak coupling,
$|\gamma|\lesssim 1$, the spin gap $\Delta_s\equiv 2\Delta_{\rm BCS}$
\begin{equation}
\label{eq:spingap}
\Delta_s(\gamma)=\frac{16\varepsilon_F}{\pi}\sqrt{\frac{|\gamma|}{\pi}}e^{-\pi^2/2|\gamma|}\, .
\end{equation}
has a form similar to that in BCS theory, except for an interaction dependent
prefactor $\sim \sqrt{|\gamma|}$. Note, however, that
the weak coupling regime is reached at {\it high} densities $n_1|a_1|\gg 1$,
in contrast to 3D, where $k_F|a|\ll 1$ in the BCS limit.  At low densities,
where $1/\gamma\to 0^{-}$, the spin gap approaches the two-body
bound state energy $\Delta_s\to\varepsilon_b$, which was measured
by rf-spectroscopy \cite{Moritz:2005}, as discussed above.
In this regime, the tightly bound fermion pairs behave like a hard
core Bose gas.  The strong coupling BEC limit of attractive fermions in
1D thus appears to be a Tonks-Girardeau gas, very different from the
nearly ideal Bose gas expected in a 3D situation (see section VIII). However,
in the presence of a harmonic waveguide, the associated transverse oscillator length
$\ell_{\perp}\equiv\sqrt{\hbar/M\omega_{\perp}}$ defines
an additional length, not present in a strictly 1D description.
As shown in section V.A, the exact solution of the
scattering problem for two particles in such a waveguide always exhibits
a two-body bound state, whatever the sign and magnitude of the
scattering length $a$.  Its binding energy right at the
confinement induced resonance is equal to $\varepsilon_b= 2\hbar\omega_{\perp}$.
Since $\hbar\omega_{\perp}\gg \varepsilon_F\/$ in the limit of
a singly occupied transverse channel, the two-particle bound state energy
$\epsilon_b$ is the largest energy scale in the problem beyond this point. 
In the regime after the confinement induced resonance,
where $\gamma\/$ becomes positive,
the appropriate  degrees of freedom are therefore no longer the single
atoms, but instead are strongly bound fermion pairs.  An exact solution
of the four-body problem in a quasi 1D geometry with tight harmonic
confinement shows, that these dimers have a \emph{repulsive}
interaction \cite{Mora:2005}.
Attractive fermions in 1D thus continuously evolve
from a fermionic Luther-Emery liquid to a gas of repulsive bosons.
As realized by \textcite{Fuchs:2004} and \textcite{Tokatly:2004},
the 1D BCS-BEC crossover problem can be solved exactly by the
Bethe ansatz, connecting the Gaudin-Yang model on the fermionic
side with the Lieb-Liniger model on the bosonic side.

The problem of attractive Fermi gases in one dimension at different
densities $n_{\uparrow}\ne n_{\downarrow}$ of the two components
has recently been solved by \textcite{Hu:2007} and \textcite{Orso:2007}.
The superfluid ground state with equal densities becomes
unstable above a critical field $h_c=\Delta_s/2$. This is the
analog of the Clogston-Chandrasekhar limit
\cite{Clogston:1962, Chandrasekhar:1962}, where
the paired ground state is destroyed by the paramagnetic, or Pauli, mechanism. 
In 3D this has been observed by \textcite{Zwierlein:2006}.
In contrast to the 3D case, however, the transition is continuous in 1D
and the result $h_c=\Delta_s/2$ holds for arbitrary coupling strengths.
Moreover, since the gap becomes large at {\it low} densities, the SF phase
with zero density imbalance appears at the trap {\it edge}.
The phase in the trap center, in turn,
is a partially polarized phase which still has superfluid correlations.
As argued by \textcite{Yang:2001} using Bosonization,
this phase exhibits an oscillating superfluid order parameter
similar to that predicted by \textcite{Fulde:1964,Larkin:1965} in a narrow range above
the Clogston-Chandrasekhar limit. In contrast to the 3D situation, non-conventional
superfluid order is thus expected in a rather wide range of parameters.

\section{TWO-DIMENSIONAL BOSE GASES}
\label{sec:twodim}

The two-dimensional Bose gas is a system which presents many
interesting features from a many-body physics perspective. The first
question that arises concerns the possibility to reach Bose-Einstein
condensation in a uniform system. The answer to this question is
negative, both for the ideal and the interacting gas. Indeed in
reduced dimensionality long range order is destroyed by thermal
fluctuations at any finite temperature. However in an interacting 2D
gas the destruction of order is only marginal and superfluidity can
still occur below a finite critical temperature $T_c$. Above $T_c$,
the quasi-long range order is destroyed via the mechanism that was
elucidated by \textcite{Berezinskii:1971}, and
\textcite{Kosterlitz:1973} and which consists in the breaking of
pairs of vortices with opposite circulations. As shown by
\textcite{Nelson:1977} this scenario implies a jump in the
superfluid   density\footnote{Note that, unless   explicitly
indicated, all densities in this section are {\it areal} and not
volume densities} from a finite and universal value
$n_s(T_c)/(k_BT_c)=2M/(\pi \hbar^2)$ to zero, right at $T_c$. Equivalently, the thermal wavelength $\lambda_T$ obeys $n_s(T_c)\lambda_T^2=4$. This
prediction has been experimentally tested using helium films
\cite{Bishop:1978,Bishop:1980}.

Quantum atomic gases provide a new system where this concept of a
quasi-long range order can be experimentally tested. However the
addition of a harmonic potential to confine the gas in the plane
changes the problem significantly. For example conventional
Bose-Einstein condensation of an ideal gas is possible in a 2D
harmonic potential. For an interacting gas, the situation is more
involved; a true BEC is still expected at extremely low
temperature. At slightly higher temperature phase fluctuations may
destabilize it and turn it into a quasi-condensate phase, which is
turned into a normal gas above the degeneracy temperature. We
review in this section the main features of atomic 2D gases and we
discuss the experimental results obtained so far.

We start with an ideal gas of $N$ bosons at temperature $T$,
confined in a square box of size $L^2$. Using the Bose-Einstein
distribution and assuming a smooth variation of the population of
the various energy states, we can take the thermodynamic limit
$N,L\to \infty$ in such a way that the density $n=N/L^2$   stays
constant. We then find a relation between the density $n$, the
thermal wavelength $\lambda_T=h/(2\pi M k_B T)^{1/2}$ and the
chemical potential $\mu$:
 \begin{equation}
 n \lambda_T^2 = - \ln \left( 1-e^{\mu/(k_BT)}  \right)\ .
 \label{2D_ideal}
  \end{equation}
This relation allows one to derive the value of $\mu$ for any value
of the degeneracy parameter $n \lambda_T^2$. It indicates that no
condensation takes place in 2D, contrarily to the 3D case. In the
latter case the relation between   $n_{\rm 3D} \lambda_T^3$ and
$\mu$ ceases to admit a solution above the critical value $n_{\rm
3D} \lambda_T^3=2.612$, which is the signature for BEC.

Consider now $N$ bosons confined by the potential
$V(r)=M\omega^2 r^2/2$ in the $xy$ plane. The presence of a trap
modifies the density of states and BEC is predicted to occur for an
ideal gas when the temperature is below the critical temperature
$T_0$ \cite{Bagnato:1991}:
 \begin{equation}
N=\frac{\pi^2}{6}\; \left( \frac{k_BT_0}{\hbar \omega} \right)^2\
.
 \label{threshold2D}
 \end{equation}
However it should be pointed out that the condensation remains a
very fragile phenomenon in a 2D harmonic potential. To show this
point we calculate the spatial density $n(\mathbf{x})$ using the
local density and semiclassical approximations, which amounts to replacing $\mu$ by
$\mu-V(\mathbf{x})$ in Eq.~(\ref{2D_ideal})
\begin{equation}
\label{eq:n_2semiclassical}
n(\mathbf{x})\lambda_T^2=-\ln(1-e^{(\mu-V(\mathbf{x}))/k_BT})\ .
 \end{equation}
Taking $\mu \to 0$ to reach the condensation threshold and
integrating over $\mathbf{x}$, we recover the   result
(\ref{threshold2D}). But we also note that $n_{\rm max}(0)=\infty$,
which means that the condensation in a 2D harmonic potential occurs
only when the 2D spatial density at the center of the trap is
infinite. This should be contrasted with the result for a 3D
harmonically trapped Bose gas, where condensation occurs at a
central density   $n_{\rm 3D, max}(0)=\zeta(3/2)/\lambda_T^3$, which
is equal (in the semi-classical approximation) to the threshold
density in a homogenous system  (see e.g.
\textcite{Pitaevskii:2003}).

\subsection{The uniform Bose gas in two dimensions}
\label{subsec:uniform2D}

We now turn to the more realistic case of a system with repulsive
interactions and we consider in this section the case of a uniform
gas. We will restrict here to well established results, the main
goal being to prepare the discussion of the trapped gas case, that
will be addressed in the next section. The first task is to model
the atom interaction in a convenient way. As it has been done for
a 1D system, it is tempting to use a contact term
$g_2\;\delta(\mathbf{x})$, which leads to a chemical potential
$\mu=g_2n$ in a mean-field approximation. However two-dimensional
scattering has very peculiar properties and we will see that in general it is
not possible to describe interactions in 2D by a coupling constant
$g_2$, and that one has to turn to an energy dependent
coefficient. We will then discuss the many-body state expected at
low temperature, and present briefly the
Berezinskii-Kosterlitz-Thouless transition.

We start by some considerations concerning quantum scattering in
two dimensions. We consider two particles of mass $M$ moving in
the $xy$ plane and we restrict here to low energy motion where the
scattering is isotropic. The scattering state can be written
\cite{Adhikari:1986}
 \begin{equation}
\psi_{\bs k}(\mathbf{x}) \sim e^{i\bs k\cdot\mathbf{x}}
-\sqrt{\frac{i}{8\pi}}\;f(k)\;\frac{e^{ikr}}{\sqrt{kr}}
 \label{scattstate}
 \end{equation}
where $\bs k$ is the incident wave vector and $f(k)$ the
dimensionless scattering amplitude for the relative energy
$E=\hbar^2k^2/M$. At low energy, one gets for the 
scattering amplitude the following variation
\begin{equation}
f(k)=\frac{4}{-\cot{\delta_0(k)}+i}\,\to\,
\frac{4\pi}{2\ln(1/k a_2)+i\pi},
\label{eq:sqwell2D}
\end{equation}
which defines the 2D scattering length $a_2$. Taking for instance 
a square well interaction potential of depth $V_0$ and diameter
$b$ it is equal to 
$a_2\! =\! bF(k_0 b)$, with $F(x)=\exp\left[J_0(x)/(x J_1(x)) \right]$ 
and $k_0=\sqrt{2M V_0}/\hbar$. Note that,
in contrast to the situation in 3D, where $\lim_{k\to 0}f(k)=-a$, in 2D
$f(k)$ tends to 0 when $k\to 0$. The total cross-section 
$\lambda=|f(k)|^2/4k$ (dimension of a length), however, tends to infinity.

Since the coupling coefficient $g_2$ is directly related to the
scattering amplitude, it appears that 2D systems are peculiar in
the sense that the coupling coefficient is intrinsically energy
dependent, by contrast to 1D and 3D systems. In addition to the
scattering amplitude $f(k)$, one may need the off-shell $T-$ matrix
when addressing many-body problems. It has been calculated for 2D
hard disks by \textcite{Morgan:2002}. The extension of the notion
of a zero-range interaction potential to the two-dimensional case
is discussed by \textcite{Olshanii:2002a}.

We now turn to a macroscopic assembly of bosonic particles and we
first address the $T=0$ situation. The case of a gas of hard disks
of diameter $b$ and surface density $n$ has been studied by
\textcite{Schick:1971}. The conclusion is that Bose-Einstein
condensation is reached with a large condensate fraction, provided
$\left(\ln(1/n b^2)\right)^{-1}\ll 1$. This constitutes the small
parameter of the problem, to be compared with $\sqrt{n a^3}$ in
3D. The chemical potential is then $\mu \simeq 4\pi \hbar^2
n/\left[ M\ln(1/n b^2)\right]$, indicating that the proper choice
for $g_2$ is (within logarithmic accuracy) $g_2= \hbar^2\tilde g_2 /M$,
where the dimensionless number $\tilde g_2$ is equal to the scattering
amplitude $f(k)$ taken for the energy $E=2\mu$. Corrections to the
result of \textcite{Schick:1971} for more realistic densities have been calculated by
\textcite{Andersen:2002b,Pricoupenko:2004,Pilati:2005}.

In the finite temperature case, the impossibility for 2D BEC
already mentioned for an ideal gas remains valid for an
interacting gas with repulsive interactions. This was anticipated
by \textcite{Peierls:1935} in the general context of long range
order in low dimensional systems. It was shown rigorously for
interacting bosons by \textcite{Hohenberg:1967}, based on
arguments by \textcite{Bogoliubov:1960}. A completely equivalent 
argument was given for lattice spin systems by \textcite{Mermin:1966}. 
To prove this result one can make a
\emph{reductio ad absurdum}. Suppose that the temperature is
small, but finite ($T\neq 0$) and that a condensate is present in
the mode $\bs k=0$, with a density $n_0$. By the Bogoliubov
inequality the number of particles $\tilde n_{\bs k}$ in state
$\bs k \neq 0$ satisfies
 \begin{equation}
\tilde n_{\bs k}+\frac{1}{2} \geq \frac{k_B T}{\hbar^2
k^2/M}\;\frac{n_0}{n}\ .
 \end{equation}
In the thermodynamic limit, the number of particles $N'$ in the
excited states is
 \begin{equation}
N'=\sum_{\bs k}\tilde n_{\bs k}=\frac{L^2}{4\pi^2}\int\tilde
n_{\bs k} \;d^2k\ .
 \label{Hohenberg}
 \end{equation}
When $k$ tends to zero the dominant term in the lower bound given
above varies as $1/k^2$. In 2D, this leads to a logarithmically
diverging contribution of the integral originating from low
momenta. This means that the starting hypothesis 
(existence of a condensate in $\bs k=0$) is wrong in 2D.

Even though there is no BEC for a homogeneous, infinite 2D Bose
gas, the system at low temperature can be viewed as a
quasi-condensate, i.e. a condensate with a fluctuating phase
\cite{Kagan:1987a,Popov:1987}. The state of the system is well
described by a wave function
$\psi(\mathbf{x})=\sqrt{\tilde{n}_0(\mathbf{x})}\;
e^{i\phi(\mathbf{x})}$ and the two dimensional character is
revealed by the specific statistical behavior of the spatial
correlation functions of the phase $\phi(\mathbf{x})$ and the
quasi-condensate density $\tilde{n}_0(\mathbf{x})$. Actually
repulsive interactions tend to reduce the density fluctuations and
one can in first approximation focus on phase fluctuations only.
The energy arising from these phase fluctuations has two
contributions. The first one originates from phonon-type
excitations, where the phase varies smoothly in space. The second
one is due to quantized vortices, i.e. points where the density is
zero and around which the phase varies by a multiple of $2\pi$.
For our purpose it is sufficient to consider only singly charged
vortices, where the phase varies by $\pm 2 \pi$ around the vortex
core.

\begin{figure}
 \centerline{\includegraphics[width=0.8\columnwidth]{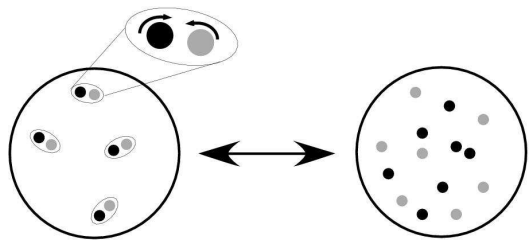}}\vskip 5mm
 \caption{Microscopic mechanism at the origin of the superfluid transition
 in the uniform 2D Bose gas. Below the transition temperature vortices only exist
 in the form of bound pairs formed by two vortices with opposite circulation.
 Above the transition temperature free vortices proliferate, causing an exponential
 decay of the one-body correlation function $g_1(r)$.}
 \label{fig:KT_principle}
\end{figure}

\textcite{Berezinskii:1971} and \textcite{Kosterlitz:1973} have
identified how a phase transition can occur in this system, when
the temperature is varied (Fig. \ref{fig:KT_principle}). At low
temperature the gas has a finite superfluid density $n_s$. The
one-body correlation function decays algebraically at large
distance:
 \begin{equation}
 T<T_c\ : \qquad n g^{(1)}(r)=\langle \hat \psi(\mathbf{x})\;\hat \psi(0)\rangle
\propto r^{-\eta}
 \label{eq:powerlawdecay}
 \end{equation}
with $\eta=(n_s \lambda_T^2)^{-1}$. The remarkable fact that there
is an exact relation between the coherence properties of the
system and the superfluid density is explained in the appendix.
Free vortices are absent in this low temperature phase and
vortices exist only in the form of bound pairs, formed by two
vortices with opposite circulations. At very low temperatures the
contribution of these vortex pairs to the correlation function
$g^{(1)}$ is negligible, and the algebraic decay of $g^{(1)}$ is
dominated by phonons. When $T$ increases, bound vortex pairs lead
to a renormalisation of $n_s$, which remains finite as long as $T$
is lower than the critical temperature $T_c$ defined by
 \begin{equation}
 T = T_c\ :\qquad n_s \lambda_T^2=4\ .
 \label{eq:BKT}
 \end{equation}
Above $T_c$ the decay of $g^{(1)}(r)$ is exponential or even Gaussian,
once the temperature is large enough that the gas is close to an
ideal system. With increasing temperature, therefore,
the superfluid density undergoes a
jump at the critical point from $4/\lambda_T^2(T_c)$ to 0 \cite{Nelson:1977}.
Note that Eq.~(\ref{eq:BKT}) is actually an implicit equation for the
temperature since the relation between the superfluid density $n_s
(T)$ and the total density $n$ remains to be determined. The
physical phenomenon at the origin of the
Berezinskii-Kosterlitz-Thouless phase transition is related to the
breaking of the pairs of vortices with opposite circulation. For
$T$ slightly above $T_c$, free vortices proliferate and form a disordered gas of
phase defects, which are responsible for the exponential decay of
$g^{(1)}$. For higher temperatures the gas eventually exhibits strong
density fluctuations and the notion of vortices becomes irrelevant.

The value given above for the transition temperature can be
recovered rather simply by evaluating the likelihood to have a
free vortex appearing in a superfluid occupying a disk of radius
$R$ \cite{Kosterlitz:1973}. One needs to calculate the free energy
$F=E-TS$ of this state. The energy $E$ corresponds to the kinetic
energy of the superfluid; assuming that the vortex is at the
center of the disk, the velocity field is $v=\hbar/(Mr)$, hence
$E=\pi n_s \int v^2(r)\,r\,dr=\pi \hbar^2/M \; \ln(R/\xi)$, where
we set the lower bound of the integral equal to the healing length
$\xi$, since it gives approximately the size of the vortex core.
The entropy associated with positioning the vortex core of area
$\pi \xi^2$ in the superfluid disk of area $\pi R^2$ is $k_B \ln
(R^2/\xi^2)$, hence the expression of the free energy:
 \begin{equation}
\frac{F}{k_B T}=\frac{1}{2}\left( n_s\lambda_T^2 -4\right) \;
\ln(R/\xi)\ .
 \end{equation}
For $n_s\lambda_T^2 > 4$ the free energy is large and positive for
a large system ($R\gg \xi$), indicating that the appearance of a
free vortex is very unlikely. On the opposite for $n_s\lambda_T^2
< 4$, the large and negative free energy signals the proliferation
of free vortices. The critical temperature estimated above from a single
vortex picture turns out to coincide with the temperature where pairs of
vortices with opposite circulation dissociate. Such pairs have a finite energy 
even in an infinite system \cite{Kosterlitz:1973}. 

The question remains how to relate the various spatial densities
appearing in this description, such as the total density $n$, and
the superfluid density $n_s$.   In the Coulomb gas analogy where
positive and negative charges correspond to clockwise and
counterclockwise vortices (see e.g. \textcite{Minnhagen:1987}) these
two quantities are related by $n_s/n=1/\varepsilon (T)$, where
$\varepsilon(T)$ is the dielectric constant of the 2D Coulomb gas.
For an extremely dilute Bose gas the relation between $n$ and
$n_s$ has been addressed by \textcite{Fisher:1988}. Their treatment
is valid in the limit of ultra weak interactions
$\epsilon=1/\ln(\ln(1/ (n a_2^2)))\ll 1$, where $a_2$ is the 2D
scattering length. They obtain the result $n_s/n \sim \epsilon$ on
the low temperature side of the transition point. Using
Monte-Carlo calculation, \textcite{Prokofev:2001} have
studied the case of weak, but more realistic interactions. Denoting
$\hbar^2 \tilde g_2/m$ the effective long wavelength interaction
constant, they obtain the following result for the total density at
the critical point: $n\lambda_T^2=\ln(C/\tilde g_2)$ where the
dimensionless number $C=380 \pm 3$. A typical value for $\tilde g_2$
in cold atom experiments is in the range $0.01$--$0.2$, which leads
to a total phase space density at the critical point $n\lambda_T^2$ in
the range $7.5$--$10.5$. \textcite{Prokofev:2001} also evaluate the
reduction of density fluctuation with respect to the expected result
$\langle n^2\rangle=2\langle n\rangle^2$ for an ideal gas. They
observe that these fluctuations are strongly reduced at the
transition point for the domain of coupling parameters relevant for
atomic gases. These high precision Monte-Carlo methods also allow
one to study the fluctuation region around the transition point
\cite{Prokofev:2002}.

The BKT mechanism has been the subject of several studies and
confirmations in various branches of condensed matter physics (for
a review see \textcite{Minnhagen:1987}). In the context of Bose
fluids, \textcite{Bishop:1978} performed an experiment with helium
films adsorbed on an oscillating substrate. The change in the
moment of inertia of the system gave access to the superfluid
fraction and showed a clear evidence for the BKT transition. In an
experiment performed with atomic hydrogen adsorbed on superfluid
helium, \textcite{Safonov:1998} observed a rapid variation of the
recombination rate of the 2D hydrogen gas when the phase space
density was approaching the critical value Eq.~(\ref{eq:BKT}).
However it is still a matter of debate whether one can reach a
quantitative agreement between these experimental observations and
the theoretical models \cite{Kagan:2000,Stoof:1994,Andersen:2002a}.

\subsection{The trapped Bose gas in 2D}

The recent progress concerning the manipulation, cooling and
trapping of neutral atomic gases with electromagnetic fields has
naturally opened the way to the study of planar Bose gases. In order
to prepare 2D atomic gases one freezes the motion along the $z$
direction using either light induced forces or magnetic forces. This
confining potential $V(z)$ has to be strong enough so that all
relevant energies for the gas (chemical potential, temperature) are
well below the excitation energy from the ground state to the first
excited state in $V(z)$. The two other directions $x,y$ are much
more weakly confined. The potential in the $xy$ plane is harmonic in
all experiments so far. Here we first review the main
experimental schemes that have been implemented. We then discuss the
new features that appear because of the harmonic confinement in
the $xy$ plane and we present the current status of experimental
investigations concerning the coherence properties of these trapped
2D gases.

\bigskip
\noindent {\it Experimental realizations of a 2D gas}
The conceptually simplest scheme to produce a 2D gas is to use a
sheet of light with a red detuning with respect to the atomic
resonance. The dipole potential then attracts the atoms towards the
locations of high light intensity, and ensures a strong
confinement in the direction perpendicular to the light sheet. This
technique has been implemented at MIT by \textcite{Gorlitz:2001b}
for sodium atoms. A 1064~nm laser was focused using cylindrical
lenses, and provided a trapping frequency $\omega_z/(2\pi)$ around
1000~Hz along the $z$ direction. The red-detuned light sheet also
ensured harmonic trapping in the $xy$ plane, with much smaller
frequencies (30 and 10~Hz along the $x$ and $y$ directions,
respectively). An adjustable number of atoms, varying between
$2\times 10^4$ and $2\times 10^6$, was loaded in the dipole trap
starting with a 3D condensate. The measurements were essentially
devoted to the size of the atom cloud after free ballistic
expansion. For small numbers of atoms (below $10^5$) it was observed
that the $z$ motion was indeed frozen, with a release energy
essentially equal to the kinetic energy of the ground state $\hbar
\omega_z/4$. For larger atom numbers, the interaction energy
exceeded $\hbar \omega_z$ and the gas was approaching the 3D
Thomas-Fermi limit.

Another way of implementing a 2D trap consists in using an
evanescent wave propagating at the surface of a glass prism.   In
2004 the group of R. Grimm in Innsbruck loaded such a trap with a
condensate of cesium atoms \cite{Rychtarik:2004}. The light was blue
detuned from resonance, so that the atoms levitated above the light
sheet, at a distance $\sim\;4\;\mu$m from the horizontal glass
surface ($\omega_z/(2\pi)\sim 500$~Hz). The confinement in the
horizontal $xy$ plane was provided by an additional hollow laser
beam, which was blue detuned from the atomic resonance and
propagating vertically. This provided an isotropic trapping with a
frequency $\omega_\bot/(2\pi) \sim 10$~Hz. As in the MIT experiment,
a time-of-flight technique revealed that for small atom numbers the
vertical expansion energy was approximately equal to $\hbar
\omega_z/4$, meaning that the $z$ motion was frozen. The number of
atoms was decreased together with temperature, and a signature of a
rapid increase of the spatial density, causing an increase of losses
due to 3-body recombination, was observed when the gas approached
quantum degeneracy. These data were consistent with the formation of
a condensate or a quasi-condensate at the bottom of the trap. 

A hybrid trap has been investigated in Oxford, where a blue
detuned, single node, Hermite Gaussian laser beam trapped Rb atoms
along the $z$ direction, whereas the confinement in the $xy$ plane
was provided by a magnetic trap \cite{Smith:2005}. This allowed to
achieve a very large anisotropy factor ($\sim 700$) between the
$z$ axis and the transverse plane. Here also the 2D regime was
reached for a degenerate gas with $\sim 10^5$ atoms.

Trapping potentials that are not based on light beams have also been
investigated. One possibility discussed by \textcite{Hinds:1998}
consists in trapping paramagnetic atoms just above the surface of a
magnetized material, producing an exponentially decaying field. The
advantage of this technique lies in the very large achievable
frequency $\omega_z$, typically in the MHz range. One drawback is
that the optical access in the vicinity of the magnetic material is
not as good as with optically generated trapping potentials. Another
appealing technique to produce a single 2D sheet of atoms uses the
so-called radio-frequency dressed state potentials
\cite{Zobay:2001}. The atoms are placed in an inhomogeneous static
magnetic field, superimposed with a radio-frequency field, whose
frequency is of the order of the energy splitting between two
consecutive Zeeman sublevels.  The dressed states are the
eigenstates of the atomic magnetic moment coupled to the static and
radio-frequency fields. Since the magnetic field is not homogeneous,
the exact resonance occurs on a 2D surface. There, one dressed state
(or possibly several, depending on the atom spin) has an energy
minimum, and the atoms prepared in this dressed state can form a 2D
gas. This method was implemented experimentally for a thermal gas by
\textcite{Colombe:2004}, but no experiment has yet been performed in
the degenerate regime.

Finally a 1D optical lattice setup, formed by the superposition of
two running laser waves, is a very convenient way to prepare stacks
of 2D gases   \cite{Orzel:2001,Burger:2002,Kohl:2005b,
Morsch:2006,Spielman:2007}. The 1D lattice provides a periodic
potential along $z$ with an oscillation frequency $\omega_z$ that
can easily exceed the typical scale for chemical potential and
temperature (a few kHz). The simplest lattice geometry is formed by
two counter-propagating laser waves, and it provides the largest
$\omega_z$ for a given laser intensity. One drawback of this
geometry is that it provides a small lattice period ($\lambda/2$
where $\lambda$ is the laser wavelength) so that many planes are
simultaneously populated. Therefore practical measurements only
provide averaged quantities. Such a setup has been successfully
used to explore the transition between a superfluid and a Mott
insulator in a 2D geometry \cite{Kohl:2005b,Spielman:2007}. Another
interesting geometry consists in forming a lattice with two beams
crossing at an angle $\theta$ smaller than 180$^\circ$
\cite{Hadzibabic:2004}. In this case the distance
$\lambda/(2\sin(\theta/2))$ between adjacent planes is adjustable, and
each plane can be individually addressable if this distance is large
enough \cite{Schrader:2004,Stock:2005}. Furthermore the tunneling
matrix element between planes can be made completely negligible,
which is important if one wants to achieve a true 2D geometry and
not a modulated 3D situation.
\bigskip

\noindent {\it From 3D to 2D scattering}
In section \ref{subsec:uniform2D} we discussed the properties of a
2D gas consisting of hard disks. Cold atomic gases, however,
interact through van der Waals forces, and one has to understand how
to switch from the 3D coupling constant to the 2D case. The
confining potential along $z$ is $V(z)=M\omega_z^2 z^2/2$, and we
assume that $\mu, k_BT \ll \hbar \omega_z$ such that the single atom
motion along the $z$ direction is frozen into the gaussian ground
state.

The scattering amplitude in this regime was calculated by
\textcite{Petrov:2000a}, and \textcite{Petrov:2001}. Quite
generally, low energy scattering in 2D is described by a scattering
amplitude of the form Eq.~(\ref{eq:sqwell2D}). Since $f(k)$ has a
pole at $k=i/a_2$, the relation between $a_2$ and the basic
scattering length $a$ of the 3D pseudopotential may be determined
from the bound state energy $\varepsilon_b=\hbar^2/(2M_ra_2^2)$ in a
2D confined geometry. This has been calculated in section V.A for
arbitrary values of the ratio between the 3D scattering length $a$
and the confinement length $\ell_{z}$. Using
Eq.~(\ref{eq:2Dboundstate}) in the limit of small binding energies,
the 2D scattering length is related to its 3D counterpart and the
confinement scale   $\ell_z=[\hbar/(M\omega_z)]^{1/2}$ by
\begin{equation}
\label{eq:a_2(a)} a_2(a)=\ell_z\sqrt{\frac{\pi}{B}}\exp{\left(
-\sqrt{\frac{\pi}{2}}\frac{\ell_z}{a}\right)}\, .
\end{equation}
with $B=0.905$ \cite{Petrov:2001}. As in 1D, the scattering
length for particles in the continuum is determined uniquely by the
two-particle binding energy in the limit
$\varepsilon_b\ll\hbar\omega_z$. The fact that $a_2(a)$ is positive,
independent of the sign of $a$, shows that for a 3D interaction
described by a pseudo potential, a two-particle bound state exists
for an arbitrary sign and strength of the ratio $a/\ell_z$ as
discussed in section V.A. Note that for realistic parameters 
$\ell_z\sim 100\, $nm and $a$ of the order of a few nm, the 2D scattering length is 
incredibly small. This is compensated for by the logarithmic 
dependence of the scattering amplitude on $a_2(a)$. Indeed, from Eqs.~(\ref{eq:sqwell2D}) and
(\ref{eq:a_2(a)}), the effective low energy scattering amplitude of
a strongly confined 2D gas is given by
\begin{equation}
f(k)=\frac{4\pi}{\sqrt{2\pi} \;\ell_z/a+ \ln(B/(\pi k^2
\ell_z^2))+i\pi}\ .
 \label{eq:2D_scat_amplitude}
 \end{equation}
When the binding along $z$ is not very strong, $\ell_z$ is much
larger than $a$ so that the logarithm and the imaginary term in
(\ref{eq:2D_scat_amplitude}) are negligible. This weak confinement
limit corresponds to the relevant regime for the experiments
performed up to now. The resulting scattering amplitude
 \begin{equation}
 f(k) \simeq\sqrt{8\pi}\; \frac{a}{\ell_z} \equiv  \tilde g_2 \ll 1
 \label{eq:2D_scat_amplitude_simple}
 \end{equation}
is independent of energy and the dimensionless coupling parameter 
$\tilde g_2=Mg_2/\hbar^2$ is much smaller than one. This implies that the gas is in
the weakly interacting regime in the sense that, at the degeneracy
point where $n\lambda_T^2=1$, the chemical potential $\mu\sim g_2n$ is
much smaller that the temperature ($\mu /(k_B T) = \tilde g_2/(2\pi)$).
An important feature of the $D=2$ dimensional gas is that the criterion for 
distinguishing the weakly and strongly interacting regime does
not depend on density. Indeed the analog of the ratio $\gamma$
given in Eq.~(\ref{eq:gamma_1}) is simply equal to $\tilde g_2$. 
In analogy to Eq.~(\ref{eq:g_1}) in the 1D case, the result
(\ref{eq:2D_scat_amplitude_simple}) can be recovered simply by
integrating the 3D pseudopotential over the $z$ oscillator ground
state. One often refers to a gas in this collisional regime as a
quasi-2D system in the sense that it can be considered as a 2D
system from the statistical physics point of view, but the
dynamics of binary collision remains governed by 3D properties; in
particular the 3D scattering length $a$ remains a relevant
parameter.

More generally, since the relevant energy for relative motion is
twice the chemical potential, the momentum $k=\sqrt{2M\mu}/\hbar$
in Eq.~(\ref{eq:2D_scat_amplitude}) is just the inverse healing
length $\xi$. At very low energies therefore, the effective
interaction in 2D is always repulsive, independent of the sign of
the 3D scattering length \cite{Petrov:2001}. This result, however,
is restricted to a regime, where $\ln{(\xi/\ell_z})\gg\ell_z/a$.
The logarithmic correction in (\ref{eq:2D_scat_amplitude}) is
therefore significant in the case of a strongly confining
potential, when $\ell_z$ and $a$ are comparable. One then recovers
a variation for $f(k)$ which is formally similar to that of a pure
2D square well (\ref{eq:sqwell2D}) with scattering length 
$a_2\simeq \ell_z$. This regime could be relevant
in a situation where the 3D scattering length $a$ is enhanced by a
Feshbach resonance
\cite{Wouters:2003,Rajagopal:2004,Kestner:2006}. If the 3D
scattering length $a$ is positive, the logarithmic correction in
(\ref{eq:2D_scat_amplitude}) is a mere reduction of the scattering
amplitude. On the other hand for a negative $a$, this correction
can lead to a strong increase of the amplitude for a particular
value of $\ell_z$  \cite{Petrov:2000a}, leading to a
confinement-induced resonance similar to those that we encountered
in the 1D case.

\noindent {\it Is there a true condensation in a trapped 2D Bose gas?}
This question has been strongly debated over the last decade as two
opposing lines of reasoning could be proposed. On the one hand we
recall that for an ideal gas the presence of a trap modifies the
density of states so that Bose-Einstein condensation becomes
possible in 2D. One could thus expect that this remains valid in the
presence of weak interactions. On the other hand in the presence of
repulsive interactions, the extension of the (quasi-)condensate in
the trap must increase with the number of atoms $N$. When $N$ is
large, a local density approximation entails that the correlation
function $g^{(1)}(r)$ decays algebraically as in
(\ref{eq:powerlawdecay}) over a domain where the density is
approximately uniform. This prevents from obtaining long range
order except for extremely low temperatures. A related reasoning uses the fact
that for the ideal gas, condensation is reached when the spatial
density calculated semi-classically becomes infinite (see
Eq.~(\ref{eq:n_2semiclassical})), which cannot occur in presence of
repulsive interactions.   The fragility of the condensation of the
ideal Bose gas in 2D is further illustrated by the existence at any
temperature of a non-condensed Hartree-Fock solution, for
arbitrarily small repulsive interactions \cite{Bhaduri:2000}.
However for very low temperature this solution is not the absolute
minimizer of the free energy, as shown using the
Hartree-Fock-Bogoliubov method by \textcite{Fernandez:2002},
\textcite{Gies:2004a} and \textcite{Gies:2004b}.

Currently the converging answer, though not yet fully tested
experimentally, is the following: at ultra-low temperature one
expects a true BEC, i.e. a system that is phase coherent over its
full extension. The ground state energy and density of a 2D Bose
gas in the limit $T=0$ can be obtained using the Gross-Pitaevskii
equation, as shown rigorously by \textcite{Lieb:2001} (see also
\textcite{Lee:2002,Kim:2000,Cherny:2001} and \textcite{Posazhennikova:2006}). 
The cross-over from a three
dimensional gas to a two dimensional gas at $T=0$ has been
addressed by \textcite{Tanatar:2002} and by
\textcite{Hechenblaikner:2005}.

When the temperature increases one meets the quasi-condensate,
superfluid regime, where phase fluctuations due to phonons dominate.
The scenario is then very reminiscent of the uniform case and it has
been thoroughly analyzed by \textcite{Petrov:2004a}. The function
$g^{(1)}(r)$ decays algebraically and vortices are found only in the
form of bound pairs. Finally at larger temperature these vortex
pairs break and the system becomes normal. A BKT transition is still
expected in the thermodynamic limit $N\to \infty$, $\omega\to 0$,
$N\omega^2$ constant, but the jump in the total superfluid mass in
suppressed because of the inhomogeneity of the atomic density
profile \cite{Holzmann:2005}. Indeed the energy for breaking a
vortex pair depends on the local density, and superfluidity will
probably be lost gradually from the edges of the quasi-condensate to
the center as the temperature increases. Assuming that the
atomic distribution is well approximated by the Hartree-Fock
solution at the transition point, \textcite{Holzmann:2005} predict
that the BKT transition temperature for a trapped gas is slightly
lower than the ideal BEC transition temperature (\ref{threshold2D}),
by an amount related to the (small) dimensionless coupling parameter
$\tilde g_2=Mg_2/\hbar^2$.

We focus for a moment on the quasi-condensate regime. It is
described by a macroscopic wave function
$\psi(\mathbf{x})=\sqrt{\tilde{n}_0 (\mathbf{x})}\;
\exp{i\phi(\mathbf{x})}$, and the density and phase fluctuations can
be analyzed using a Bogoliubov analysis. We refer the reader to the
work of \textcite{Mora:2003} and \textcite{Castin:2004} for a
thorough discussion of the extension of Bogoliubov theory to
quasi-condensates. As for the uniform gas \cite{Prokofev:2001},
repulsive interactions strongly reduce the density fluctuations for
$k_BT\lesssim \mu$ and $n \lambda_T^2 \gg 1$, so that   $\langle
\tilde{n}_0^2 (\mathbf{x})\rangle \simeq \left( \langle \tilde{n}_0
(\bs x)\rangle\right)^2$. For large atom numbers ($N\tilde g_2 \gg 1$) the
equilibrium shape of the gas can be derived using a Thomas-Fermi
approximation, as for a true condensate. The kinetic energy plays a
negligible role, and the density profile results from the balance
between the trapping potential and the repulsive interatomic
potential. It varies as an inverted parabola
 \begin{equation}
\tilde{n}_0(\mathbf{x})= \tilde{n}_0(0) \, \left(1-\frac{r^2}{R^2}
\right)\ , \quad \frac{\hbar^2}{M}\;\tilde g_2\;\tilde{n}_0(0)=\mu\ ,
 \label{TF2D}
 \end{equation}
where the chemical potential $\mu$ and the radius of the clouds
$R$ are:
 \begin{equation}
 \mu=\hbar \omega \, ( N\tilde g_2/\pi)^{1/2}
 \ , \quad
 R=\sqrt{2}\,a_\bot\,( N\tilde g_2/\pi)^{1/4}\ ,
 \end{equation}
with $a_\bot=\sqrt{\hbar/(M\omega)}$. 

The parabolic Thomas-Fermi profile appears on the top of a broader
background formed by the atoms out of the (quasi-)condensate. Such a
profile has first been experimentally observed by
\textcite{Gorlitz:2001b} and \textcite{Rychtarik:2004}. A precise
measurement of the onset at which a pure thermal distribution turns
into a bimodal (Thomas-Fermi + thermal) profile has recently been
performed by \textcite{Kruger:2007}. The experiment was performed
with a rubidium gas confined in a 1D optical lattice, such that
$\tilde g_2=0.13$. The phase space density at which bimodality arises
was found in good agreement with the prediction of
\textcite{Prokofev:2001} for the BKT threshold 
$n(0)\lambda_T^2=\ln(C/\tilde g_2)\simeq 8.0$, which is relevant
here if the local density approximation is valid at the center of
the trap. At the critical point, the total number of atoms in each
plane significantly exceeded the result (\ref{threshold2D}) expected
in the ideal case. In this experiment two to three planar gases were
actually produced simultaneously, and they could interfere with each
other when overlapping during time-of-flight, provided their spatial
coherence was large enough. It was observed that the onset of
bimodality coincides (within experimental accuracy) with the onset
of clearly visible interferences.

It is important to stress that since the expected Thomas-Fermi
profile is identical for a true and a quasi-condensate, its
observation cannot be used to discriminate between the two
situations. The phase fluctuations have been calculated by
\textcite{Petrov:2001} and \textcite{Petrov:2004a} in the regime
$\mu \lesssim k_B T$ and $n\lambda_T^2\gg 1$ (see
Eq.~(\ref{eq:phasefluct})   in the appendix)
 \begin{equation}
\delta\phi^2(\mathbf{x})=\langle (\phi(0)-\phi(\mathbf{x}))^2
\rangle \simeq \frac{2}{\tilde n_0(0)\lambda_T^2} \ln(r/\xi)\ .
 \end{equation}
This expression, which is reminiscent of the uniform result
(\ref{eq:powerlawdecay}), is valid for points $\mathbf{x}$ inside
the condensate. The healing length $\xi=\hbar/\sqrt{2M\mu}$
satisfies $\xi R=a_\bot^2$. Therefore it is only at a temperature
much below the degeneracy temperature, such that $\Delta
\phi(R)\lesssim \pi$, that one recovers a quasi uniform phase over
the whole sample, hence a true condensate.

\begin{figure}
 \vskip 5mm\centerline{\includegraphics[width=0.8\columnwidth]{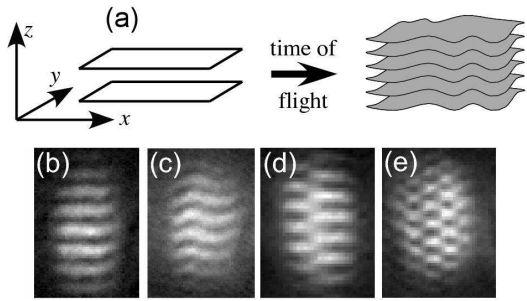}}\vskip 5mm
 \caption{Matter-wave heterodyning of 2D gases. (a) Principle of
 the method: two planar Bose gases are released from the trap,
 expand and overlap, giving rise to an interference pattern that
 is probed by absorption imaging. (b-e) Examples on experimental
 interference patterns obtained well below (b) and in the vicinity (c)
 of the degeneracy temperature. Some patterns show one (d) or
 several (e) dislocations, revealing the presence of vortices in
 one of the gases. Reprinted with permission from \textcite{Hadzibabic:2006}.
 }
 \label{fig:interf_pattern}
\end{figure}

\noindent {\it Experimental investigations of phase fluctuations}
A convenient way to access experimentally the phase coherence of
quasi-2D gases is the matter-wave heterodyning technique. It
consists in studying the statistical properties of the matter wave
interference pattern which forms when two independent, parallel 2D
Bose gases are released from the trap and overlap (figure
\ref{fig:interf_pattern}a). A detailed analysis of these patterns
has been given by \textcite{Polkovnikov:2006a} (see section III.C for
the 1D case). Assume that the two
gases have the same uniform amplitude $\psi_0$ and fluctuating
phases $\varphi_a (x,y)$ and $\varphi_b(x,y)$. The interference
signal $S(x,z)$ is recorded by sending an imaging beam along the
$y$ direction, which integrates the atomic density over a length
$L_y$:

\begin{equation}
S(x,z)\propto  2\psi_0^2+e^{2i\pi z/D}\;c(x)+e^{-2i\pi
z/D}\;c^*(x)  \label{modeling_interf}
\end{equation}

with

\begin{equation}
 c(x)=\frac{\psi_0^2}{L_y}\;\int_{-L_y/2}^{L_y/2}
 e^{i (\varphi_a(x,y)-\varphi_b(x,y))}\;dy .
\end{equation}

The period $D$ of the interference pattern is $D=2\pi \hbar
t/(Md)$, where $d$ is the initial distance between the two planes
and $t$ the expansion time. We now integrate the coefficient
$c(x)$ appearing in (\ref{modeling_interf}) over a variable length
$L_x$:
 \begin{equation}
C(L_x)=\frac{1}{L_x}\int_{-L_x/2}^{L_x/2} c(x)\; dx
 \end{equation}
and average $|C(L_x)|^2$ over many images recorded in the same
conditions. Using the fact that the phases $\varphi_a$ and
$\varphi_b$ are uncorrelated, we obtain for $L_x\gg L_y$
 \begin{eqnarray}
\langle |C(L_x)|^2 \rangle &=& \frac{1}{L_x^2}\int \hskip -2mm\int
\langle c(x)\;c^*(x') \rangle\;dx\;dx' \label{decay_C} \\
&=& \frac{1}{L_x} \int_{-L_x/2}^{L_x/2} |g^{(1)}(x,0)|^2\;dx
 \propto\ \left( \frac{1}{L_x} \right)^{2\alpha} \nonumber
 \end{eqnarray}
where we have assumed that the two gases have the same statistical
properties. The long-range physics is then captured in a single
parameter, the exponent $\alpha$. It is straightforward to
understand the expected values of $\alpha$ in some simple cases. In
a system with true long-range order, $g^{(1)}$ would be constant and
the interference fringes would be perfectly straight. In this case
$\alpha = 0$, corresponding to no decay of the contrast upon
integration. In the low temperature regime, where $g^{(1)}$ decays
algebraically (see Eq.~(\ref{eq:powerlawdecay})) the exponent
$\alpha$ coincides with the exponent $\eta (T)$ which describes the
quasi-long range order in $g^{(1)}$. In the high temperature case,
where $g^{(1)}$ decays exponentially on a length scale much shorter
than $L_x$, the integral in (\ref{decay_C}) is independent of $L_x$.
In this case $\alpha = 0.5$, corresponding to adding up local
interference fringes with random phases. The BKT mechanism
corresponds to a transition between a power law with exponent
$1/(n_s\lambda_T^2)\leq 0.25$ to an exponential decay of $g^{(1)}$.
It should thus manifest itself as a sudden jump of $\alpha$   from
0.25 to 0.5 when the temperature varies around $T_c$.

This method has been implemented at ENS with two rubidium planar
gases, forming two parallel, elongated strips ($L_x=120\mu$m,
$L_y=10\mu$m) \cite{Hadzibabic:2006} (figure
\ref{fig:interf_pattern}b-e). The experimental results confirm the
expected behavior, at least qualitatively. At relatively large
temperature the fitted exponent $\alpha$ is close to $0.5$. When the
temperature decreases a rapid transition occurs and $\alpha$ drops
to $\sim 0.25$. At the transition the estimated phase space density
of the quasi-condensate is $\tilde{n}_{0}(0)\lambda_T^2 \sim 6$.
Note that for a quantitative comparison between experiments and
theory, one should account for density fluctuations which are likely
to play an important point near the transition, in contrast to the
situation in superfluid liquid helium. Also the geometry effects in
these elongated samples ($R_x\sim 12 R_y$) may be significant.

In addition to the rapid variation of the exponent $\eta$
characterizing the decay of $g^{(1)}$, these experiments also gave
evidence for isolated vortices \cite{Stock:2005,Hadzibabic:2006}
((figure \ref{fig:interf_pattern}d-e). A vortex appears as a
dislocation of the fringes \cite{Chevy:2001,Inouye:2001}, and these
dislocations indeed proliferate on the high temperature side of the
transition. Using a theoretical analysis based on classical field
method, \textcite{Simula:2006}   obtained phase patterns of
quasi-condensates close to the critical temperature that indeed
exhibit an isolated, free vortex, in good agreement with
experimental observation. The probability for observing a vortex
pair in a similar configuration has been calculated by
\textcite{Simula:2005a}. 

The Berezinskii-Kosterlitz-Thouless mechanism has also been recently
investigated using a two-dimensional periodic array of $\sim 200$
Josephson coupled Bose-Einstein condensates \cite{Schweikhard:2007}.
Each tube-like condensate contains a few thousands atoms, and has a
length $\sim 35\;\mu$m along the $z$ direction. The condensates are
localized at the sites of a 2D hexagonal optical lattice of period
4.7$\mu$m in the $xy$ plane, and the coupling $J$ between adjacent
sites can be tuned by varying the optical lattice intensity. The
phase properties of the ensemble are probed by ramping down the
lattice, and recording the density profile in the $xy$ plane when
the wavefunctions from the various sites overlap. Vortices appear as
holes in the atomic density distribution, and the vortex surface
density is measured as a function of the Josephson coupling $J$ and
the temperature $T$. A universal vortex activation curve is obtained
as a function of the parameter $J/T$, showing vortex proliferation
for $J/T\lesssim 1$ in good agreement with the predictions of the 
BKT-mechanism.

\noindent {\it Breathing mode of a 2D gas}
In the previous subsection we have been mostly interested in the
static properties of 2D Bose gases. Here we point out a remarkable
dynamical property of these systems in an isotropic harmonic
potential, when the interaction potential between particles is
such that $V(\lambda r)=V(r)/\lambda^2$.
\textcite{Pitaevskii:1997a} showed that when the gas is prepared in
an arbitrary out-of-equilibrium state, the quantity $\langle
r^2\rangle$ oscillates at the frequency $2\omega$ without any
damping, irrespectively of the strength of the interaction. They
also proved that this property originates from the presence of a
hidden symmetry, described by the two-dimensional Lorentz group
SO(2,1). In fact precisely the same symmetry shows up in the case
of a unitary gas in 3D, as will be discussed in section VIII.B.

The Dirac distribution in 2D, $\delta^{(2)}(r)$, belongs to the
class of functions satisfying $V(\lambda r)=V(r)/\lambda^2$. It
makes this SO(2,1) symmetry relevant for trapped neutral atoms at
low energies, when the range of interaction is small compared to
all other scales. However a true contact interaction is singular
in 2D, and leads to logarithmic ultraviolet divergences that are
cut off by the finite range of the real interatomic potential.
Therefore one cannot hope to observe a fully undamped breathing
mode in atomic systems, but rather a very weakly damped dynamics.
It was pointed out by \textcite{Fedichev:2003} that vortex pair
nucleation could actually play a role in the residual expected
damping of this breathing mode. Note that the difficulties with
the contact interaction do not arise at the level of the
Gross-Pitaevskii equation, where the same property has been
predicted \cite{Pitaevskii:1996,Kagan:1996a}.

A precursor of this long lived breathing mode has been observed in a
3D, quasi cylindrical geometry by \textcite{Chevy:2002}. The
transverse breathing mode of the cylinder was found to oscillate at
a frequency very close to $2\omega$ with an extremely small damping
(quality factor of the mode $>2000$). The damping and shift of the
oscillation frequency could be calculated theoretically with a good
precision by \textcite{Jackson:2002b} (see also
\textcite{Guilleumas:2003}). In this case part of the damping is due
to the nucleation of pairs of phonons propagating along $\pm z$
\cite{Kagan:2003}, a mechanism that is of course absent in a pure 2D
geometry. This breathing mode has also been observed in a fast
rotating gas by \textcite{Stock:2004}. Its frequency was also $\sim
2\omega$, with a small correction due to the non-harmonicity of the
trapping potential that was necessary to stabilize the
center-of-mass motion of the atom cloud in the fast rotating regime
(see section \ref{subsec:fastrotationexp}).


\section{BOSE GASES IN FAST ROTATION}
\label{sec:fastrotation}

The investigation of rotating gases or liquids is a central issue
in the study of superfluidity \cite{Donnelly:1991}. It is relevant
for the study of liquid helium, rotating nuclei, neutron stars and
pulsars, and for the behavior of superconductors in a magnetic
field. During the recent years, several experiments using rotating
Bose-Einstein condensates have provided a spectacular illustration
of the notion of quantized vortices
\cite{Matthews:1999,Madison:2000a,AboShaeer:2001,Hodby:2001a}.
Depending on the rotation frequency $\Omega$ of the gas, a single
vortex or several vortices can be observed experimentally. When
the number of vortices is large compared to 1, they form a Abrikosov lattice,
i.e. a triangular array with a surface density $n_v= M\Omega/(\pi
\hbar)$. Since the circulation of the velocity around a single
charged vortex is $h/M$, this ensures that the velocity field of
the condensate, when calculated after coarse graining over
adjacent vortices, is equal to the orthoradial, rigid body
velocity field $v=\Omega r$ \cite{Feynman:1955}.

For a gas confined in a harmonic potential, the fast rotation
regime corresponds to stirring frequencies $\Omega$ of the order
of the trapping frequency $\omega$ in the plane perpendicular to
the rotation axis (hereafter denoted $z$). From a classical point
of view the transverse trapping and centrifugal forces then
compensate each other, and the motion of the particles in the $xy$
plane is only driven by Coriolis and interatomic forces. This
situation is similar to that of an electron gas in a magnetic
field, since Lorentz and Coriolis forces have the same
mathematical structure. The single particle energy levels are
macroscopically degenerate, as the celebrated Landau levels
obtained for the quantum motion of a single charge in a magnetic
field. When interactions between atoms are taken into account the
fast rotation regime presents a strong analogy with Quantum Hall
physics. One can distinguish two limiting cases in this fast
rotation regime. Firstly, when the number of vortices inside the
fluid $N_v$ remains small compared to the number of atoms $N$, the
ground state of the system is still a Bose-Einstein condensate
described by a macroscopic wave function $\psi(\mathbf{x})$. This
situation has been referred to as `mean field Quantum Hall regime'
\cite{Ho:2001,Fischer:2003}. Secondly, when $\Omega$ tends to
$\omega$, the number of vortices reaches values comparable to the
total number of atoms $N$. The description by a single macroscopic
wave function breaks down, and one expects a strongly correlated
ground state, such as that of an electron gas in the fractional
quantum Hall regime \cite{Cooper:2001}.

In this section we start by setting the Lowest Landau Level (LLL)
framework for the discussion of the fast rotation regime, and
discuss the main properties of a fast rotating condensate, when the
mean-field description remains valid. We then present recent
experimental results where the LLL regime has indeed been reached.
Finally we review some theoretical proposals to reach beyond mean
field physics, that present a close analogy with the physics of the
fractional quantum Hall effect. We do not discuss here the physics
of slowly rotating system, where one or a few vortices are involved.
We refer the interested reader to the review article of
\textcite{Fetter:2001} and to the recent book by
\textcite{Aftalion:2006Book}. Note that a rigorous derivation of
the Gross-Pitaevskii energy functional in the slowly rotating case
has been given by \textcite{Lieb:2006}. 

\subsection{The Lowest Landau Level formalism}
\label{subsec:LLLformalism}

\begin{figure}
\centerline{\includegraphics[width=0.8\columnwidth]{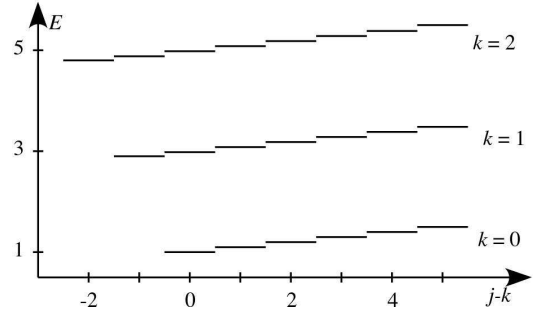}}
 \caption{Single particle energy spectrum for $\Omega=0.9\omega$. The index $k$
labels the Landau levels. The energy is expressed in units of
$\hbar \omega$. For $\Omega=\omega$ the Landau levels are
infinitely degenerate.}
 \label{fig:LL}
\end{figure}

\paragraph{The Landau Levels.}
We consider first a single particle confined in a two-dimensional
isotropic harmonic potential of frequency $\omega$ in the $xy$
plane. We are interested here in the energy level structure in the
frame rotating at angular frequency $\Omega$ ($>0$) around the $z$
axis, perpendicular to the $xy$ plane. The hamiltonian of the
particle is
 \begin{equation}
H^{(1)} =\frac{p^2}{2M}  + \frac{M\omega^2r^2}{2} -\Omega L_z
=\frac{\left(\bs p -\bs A \right)^2}{2M} + \frac{1}{2}
M(\omega^2-\Omega^2) r^2
 \label{singlepartH}
 \end{equation}
with $r^2=x^2+y^2$, $\bs A=M\bs \Omega \wedge \mathbf{x}$; $L_z$ is
the $z$ component of the angular momentum. Eq.~(\ref{singlepartH})
is formally identical to the hamiltonian of a particle of unit
charge placed in a uniform magnetic field $2m\Omega \bs {\hat z}$,
and confined in a potential with a spring constant
$M(\omega^2-\Omega^2)$. A common eigenbasis of $L_z$ and $H$ is the
set of (not normalized) Hermite functions
  \begin{equation}
\phi_{j,k}(\mathbf{x})=e^{r^2/(2a_\perp^2)}\;
(\partial_x+i\partial_y)^j\;(\partial_x-i\partial_y)^k
\left(e^{-r^2/a_\perp^2}\right)
 \label{Hermite}
  \end{equation}
where $j$ and $k$ are non-negative integers and
$a_\perp=\sqrt{\hbar/M\omega}$. The eigenvalues are $\hbar(j-k)$
for $L_z$ and
 \begin{equation}
 E_{j,k}=\hbar \omega+ \hbar(\omega-\Omega)j+ \hbar(\omega+\Omega)k
\label{spectrum}
 \end{equation}
for $H$. For $\Omega=\omega$, these energy levels group in series
of states with a given $k$, corresponding to the well known,
infinitely degenerate, Landau levels. For $\Omega$ slightly
smaller than $\omega$, this structure in terms of Landau levels
labeled by the index $k$ remains relevant, as shown in Fig.
\ref{fig:LL}. Two adjacent Landau levels are separated by $\sim 2
\hbar\omega$, whereas the distance between two adjacent states in
a given Landau level is $\hbar(\omega-\Omega) \ll \hbar\omega$. It
is clear from these considerations that the rotation frequency
$\Omega$ must be chosen smaller than the trapping frequency in the
$xy$ plane. Otherwise the single particle spectrum
(\ref{spectrum}) is not bounded from below. Physically, this
corresponds to the requirement that the expelling centrifugal
force $M\Omega^2 r$ must not exceed the trapping force in the $xy$
plane $-M\omega^2 r$.

We now consider an assembly of cold identical bosons rotating at a
frequency $\Omega$ close to $\omega$. Since the effective trapping
potential in (\ref{singlepartH}) becomes weaker as $\Omega$
increases, we expect that as $\Omega \to \omega$ the equilibrium
size of the atom cloud increases indefinitely, and the interaction
energy and the chemical potential $\mu$ tend to zero. We define
the Lowest Landau Level regime as the situation where $\mu, k_B T
\ll \hbar \omega$, so that the state of the system can be
accurately described in terms of Hermite functions $\phi_{j,k}$
with $k=0$ only.  Each basis function $\phi_{j,0}(\mathbf{x})$ is
proportional to $(x+iy)^j \; e^{-r^2/(2a_\bot^2)}$ and takes
significant values on a ring centered on $0$ with an average
radius $\sqrt{j}\, a_\bot$ and a width $\sim a_\bot$. Any function
$\psi(\mathbf{x})$ of the LLL is a linear combination of the
$\phi_{j,0}$'s and can be cast in the form:
 \begin{equation}
\psi(\mathbf{x})=e^{-r^2/(2a_\bot^2)}\; P(u)
 \label{formLLL1}
 \end{equation}
where $u=x+iy$ and $P(u)$ is a polynomial (or an analytic
function) of $u$. When $P(u)$ is a polynomial of degree $n$, an
alternative form of $\psi(\mathbf{x})$ is
 \begin{equation}
\psi(\mathbf{x})= e^{-r^2/(2a_\bot^2)}\;\prod_{j=1}^n (u-\zeta_j)
 \label{formLLL2}
 \end{equation}
where the $\zeta_j$ ($j=1 \ldots n$) are the $n$ complex zeroes of
$P(u)$. Each $\zeta_j$ is the position of a single-charged,
positive vortex, since the phase of $\psi(\mathbf{x})$ changes by
$+2\pi$ along a closed contour encircling $\zeta_j$. Therefore in
the LLL, there is a one-to-one correspondence between atom and
vortex distributions, contrarily to what happens for slower
rotation frequencies. This has interesting consequences on the
hydrodynamics of the gas, which cannot be described by
conventional Bernoulli and continuity equations
\cite{Bourne:2006}.

\paragraph{Equilibrium shape of a fast rotating BEC.}

We now address the question of the distribution of particles and
vortices in the case of fast rotation, assuming for the moment
that a mean field description is valid. We suppose that the motion
along the rotation axis $z$ is frozen in a way similar to what we
considered in the previous section devoted to static 2D gases.

Consider first the case of an ideal gas. At zero temperature all
atoms accumulate in the $j=k=0$ ground state. At low but finite
temperature ($k_BT \ll 2\hbar \omega$) the occupied states belong
to the LLL. The gas can be described at any time by a Hartree wave
function of the type (\ref{formLLL1}), where the coefficients
$c_m$ of the polynomial $P(u)=\sum c_m u^m$ are random independent
variables. This fast rotating ideal gas can thus be viewed as a
physical realization of a random polynomial \cite{Castin:2006}. A
measurement of the density distribution of the gas will reveal the
presence of the vortices, i.e. the roots of $P(u)$. Although the
gas is ideal, one can show that the positions of the vortices are
correlated and exhibit a strong anti-bunching phenomenon (see
\textcite{Castin:2006} and refs. in).

The case of a fast rotating condensate with repulsive interactions
has been analyzed by several authors
\cite{Ho:2001,Watanabe:2004,Aftalion:2005,Cooper:2004} and we will
now review the main results. In all this section we will assume that
the pair-wise interaction between atoms $i$ and $j$ can be described
by the contact term $g_2\,\delta(\mathbf{x}_i-\mathbf{x}_j)$. We
will furthermore assume that the 3D scattering length $a$ is much
smaller that the extension $\ell_z$ of the ground state of the
motion along $z$, so that $g_2\simeq\hbar^2 \tilde g_2/M$, with $\tilde g_2=
\sqrt{8\pi}\,a/\ell_z \ll 1$ (see Eq.~(\ref{eq:2D_scat_amplitude_simple})). 
Note that the restriction of the
contact interaction to the LLL subspace is a regular operator: it
does not lead to the same mathematical difficulties
as the ones encountered by considering the contact interaction
in the whole Hilbert space of 2D wave functions. In fact,
quite generally, interactions in the LLL are described by the
Haldane pseudopotentials $V_m$ \cite{Haldane:1983}. 
For a pseudopotential with scattering length $a$, the resulting 2D
contact interaction has $V_m=\sqrt{2/\pi}\hbar\omega\cdot a/\ell_z$
for $m=0$ and zero otherwise. 
In the fermionic case, where only odd values of $m$ are
allowed, the analog of this interaction is a hard core model, where
$V_m\ne 0$ only for $m=1$. The fact that the Laughlin states, to be
discussed in subsection C below, are
exact eigenstates for such pseudopotentials has been realized
by \textcite{Trugman:1985}.

\begin{figure}
 \centerline{\includegraphics[width=0.8\columnwidth]{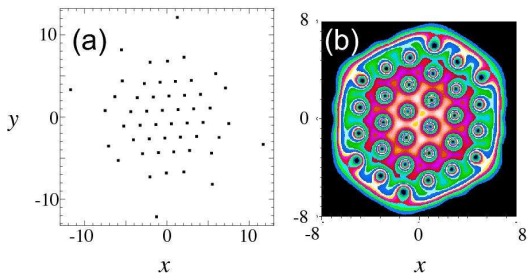}}
\caption{Calculated structure of the ground state
of a rotating Bose-Einstein condensate described by an LLL wave
function, showing vortex locations (a) and atomic density profile
(b). The parameters of the calculation correspond to 1000 rubidium
atoms confined in a trap with frequency $\omega/(2\pi)=150$~Hz and
rotating at a frequency $\Omega=0.99\,\omega$. The unit for the
positions x and y is $[\hbar/(m\omega)]^{1/2}$. Reprinted with permission from
from \cite{Aftalion:2005}.}
 \label{fig:vortex_lattice}
\end{figure}

We start with a gas rotating exactly at the trap
frequency ($\Omega=\omega$), with an infinite number of particles,
but a finite spatial density. In this case the numerical
minimization of the Gross-Pitaevskii energy functional indicates
that the vortices form an infinite regular triangular lattice. We turn now
to a gas with a finite number of particles, rotating at a frequency
$\Omega$ slightly below $\omega$. The initial treatment of
\textcite{Ho:2001} assumed an infinite, regular triangular vortex
lattice also in this case. The total energy of the system was
minimized by varying the spacing of the vortex lattice. When
injected in (\ref{formLLL2}) this led to the prediction of a
Gaussian atom distribution after coarse-graining over the vortex
lattice spacing. A more detailed analysis has recently been
performed, where the position $\zeta_j$ of each vortex is taken as a
variational parameter \cite{Watanabe:2004,Aftalion:2005,Cooper:2004}
(see also the work of \textcite{Anglin:2002} and
\textcite{Sheehy:2004a,Sheehy:2004b} for the case of a slower
rotation). One spans in this way the whole LLL subspace. These
studies have shown that the vortex distribution that minimizes the
total energy is nearly regular with the density $M\Omega/(\pi
\hbar)$ close to the center of the condensate, but it is strongly
deformed on the edges, with a rarefaction of vortices. For large
atom numbers, the predicted coarse-grained density distribution is
not gaussian as for a uniform vortex lattice, but it approaches a
Thomas-Fermi distribution $n_2(\mathbf{x}) \propto R^2-r^2$ similar
to (\ref{TF2D}), for the effective trapping potential
$M(\omega^2-\Omega^2)r^2/2$. This Thomas-Fermi prediction is in good
agreement with the results obtained in the experiments described
later in this section. The cloud radius is
 \begin{equation}
R\simeq a_\bot \left( \frac{2b}{\pi}\;\frac{N
\tilde g_2}{1-\Omega/\omega}\right)^{1/4}
 \label{eq:radius}
 \end{equation}
and diverges for $\Omega\to \omega$, as expected from the
compensation of the trapping force by the centrifugal one. The
dimensionless coefficient $b$ is the Abrikosov parameter
$b=1.1596$ \cite{Kleiner:1964} for a triangular lattice. It expresses the fact that due to
the restriction to the LLL and to the presence of vortices, the
energy and the size of the condensate are actually slightly larger
than what one would expect for a static trap with spring constant
$m(\omega^2-\Omega^2)$ and a smooth equilibrium distribution. The
chemical potential is
$$
 \mu \simeq \hbar \omega \left( \frac{2b}{\pi}\; N \tilde g_2
(1-\Omega/\omega)\right)^{1/2}
 $$
so that the condition $\mu \ll \hbar \omega$ for the validity of
the LLL approach reads $1-\Omega/\omega \ll 1/(N\tilde g_2)$. It is also
instructive to calculate the number $N_v$ of `visible' vortices,
i.e. those which sit in the disk of area $\pi R^2$. Using
$\Omega\simeq \omega$ so that $n_v \simeq m\omega/(\pi \hbar)$, we
get
 $$
 \frac{N_v}{N} \simeq
\left(\frac{\tilde g_2}{N(1-\Omega/\omega)} \right)^{1/2}\ .
 $$
As we will see further, the mean field approach is valid only if
$N_v\ll N$ so that the validity domain of the mean field LLL
approach corresponds to the interval:
 \begin{equation}
\mbox{mean field LLL:}\qquad \frac{\tilde g_2}{N}\ll
1-\frac{\Omega}{\omega} \ll \frac{1}{N\tilde g_2}
 \end{equation}
Note that the total number of vortices, including those sitting
outside the Thomas-Fermi radius, can be shown to be infinite for
the wave function that minimizes the energy in the LLL subspace
\cite{Aftalion:2006a}.

It is interesting to compare the behavior of a fast rotating BEC
with that of a fast rotating bucket of superfluid liquid helium,
or a type II superconductor in a large magnetic field
\cite{Fischer:2003}. In the latter cases, the size of the sample
is constant and the vortex density increases as the rotation
frequency (or the magnetic field) increases. Since the size of a
vortex core $\ell_c$ depends only on the spatial density of the
fluid ($\ell_c \sim$ the healing length $\xi$), it stays constant
as $\Omega$ increases and one reaches eventually a point where the
cores of adjacent vortices overlap. This corresponds to a loss of
superfluidity or superconductivity. For superfluid liquid helium,
the rotation frequency $\Omega_{c2}$ where this phenomenon should
happen is out of reach for realistic experiments. For
superconductors on the contrary, the critical field $H_{c2}$ where
the superconductivity is lost is a relevant experimental
parameter. For  fast rotating, harmonically trapped gases, the
scenario is very different: (i) the vortex density saturates to a
constant value $n_v=M\omega/(\pi \hbar)=1/(\pi a_\perp^2)$ when
$\Omega$ approaches $\omega$; (ii) the size $\ell_c$ of the vortex
core for a wave function of the type (\ref{formLLL2}) is no longer
dictated by interactions that would lead to $\ell_c\sim\xi$ as for
an incompressible fluid, but it is on the order of the vortex
spacing $a_\perp$. Therefore the fractional area
$\tilde{n}_0\ell_c^2$ occupied by vortices tends to a finite
value, as the trapped BEC rotates faster and faster. The
cross-over between the standard to the LLL regime is studied in
detail by \textcite{Baym:2003a} and \textcite{Cozzini:2006}.

\subsection{Experiments with fast rotating gases}
\label{subsec:fastrotationexp}

The most intuitive way to rotate a trapped atomic gas is to
superpose a rotating anisotropic potential to the axi-symmetric
trapping potential $V(r)=M\omega^2r^2/2$. The stirring anisotropy
can be written   $\delta V(\mathbf{x}, t)=\epsilon
M\omega^2(X^2-Y^2)/2$ where the coordinates $(X,Y)$ are deduced from
the static ones $(x,y)$ by a rotation of angle $\Omega t$. The
dimensionless parameter $\epsilon$ characterizes the strength of the
stirring potential with respect to the trapping one. In practice,
because of experimental limitations, 
$\epsilon$ has to be on the order of at least a few percents. Indeed it
must overcome by a significant factor the residual static anisotropy
of the trapping potential, that is typically in the
$10^{-3}-10^{-2}$ range \cite{GueryOdelin:2000}.

The stirring potential can be created by a modulated laser beam
\cite{Madison:2000a,AboShaeer:2001} or by a rotating magnetic field
\cite{Hodby:2001a,Haljan:2001}. The stirring method has been
successfully used to nucleate single vortices as well as large
vortex arrays in rotating BECs. However it is not fully
appropriate to approach the fast rotating regime of a harmonically
trapped gas. Indeed the center-of-mass motion of the atom cloud is
dynamically unstable when the rotation frequency $\Omega$ is set
in the interval $[\omega \sqrt{1-\epsilon},\omega
\sqrt{1+\epsilon}]$ \cite{Rosenbusch:2002}. A precise description
of the rotating system at the edge of the instability region
$\Omega=\omega \sqrt{1-\epsilon}$ has been given by
\textcite{Sinha:2005} (see also \textcite{Fetter:2007}) who showed
that the gas forms in this case a novel elongated quantum fluid,
with a roton-maxon excitation spectrum. Excitation modes with a
zero energy appear above a critical interaction strength, leading
to the creation of rows of vortices.

A possible way to circumvent the center-of-mass expulsion
occurring at $\Omega\sim \omega$ consists in adding a extra
trapping potential, that provides a stronger than quadratic
confinement. This method has been explored experimentally by
\textcite{Bretin:2004}. In this experiment the dipole potential
created by a strongly focused laser beam provided a quartic
confinement, in addition to the usual quadratic one. It was then
possible to explore the critical region $\Omega\sim \omega$ and to
approach the LLL regime $\mu \sim 2 \hbar \omega$. A striking
observation was a strong decrease of the visibility of the vortex
pattern in this region. Its origin is not fully understood yet,
but it may be related to the fact that the rotating gas was not in
the 2D regime. The shape of the rotating cloud was close to
spherical, and the vortex lines may have undergone a strong
bending with respect to the trap axis, which made them hardly
visible in the imaging process. This explanation is favored by the
theoretical study by \textcite{Aftalion:2004}: when looking for
the ground state of the system using imaginary time evolution of
the Gross-Pitaevskii equation, it was found that much longer times
were required for $\Omega\simeq \omega$ to reach a well-ordered
vortex lattice.

Note that the addition of a quartic potential brings some
interesting and novel aspects to the vortex dynamics in the trap,
with the possibility to nucleate `giant' vortices. This was
initially explored by \textcite{Fetter:2001b},
\textcite{Lundh:2002}, \textcite{Kasamatsu:2002}. The mean-field
description of the dynamics of a BEC in non harmonic potentials
has been recently the subject of an important theoretical
activity, and we refer the interested reader to the work of
\textcite{Cozzini:2006} and references in.

\begin{figure}
 \null
 \vskip5mm
 \centerline{\includegraphics[width=0.8\columnwidth]{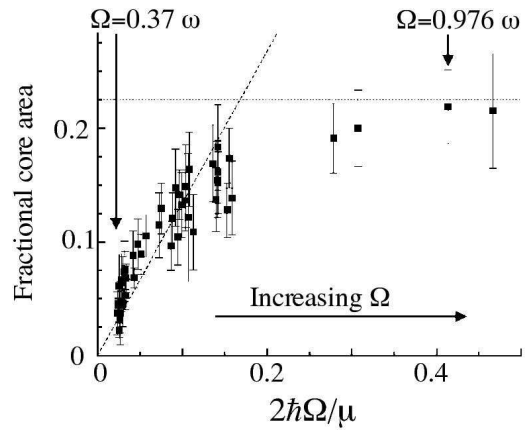}}
\caption{Fraction of the condensate surface area occupied by the
vortex cores, as a function of $2\hbar \Omega/\mu$. The vortex radius $r_v$ is defined as the
\emph{r.m.s.} radius of the Gaussian function giving the best fit
to the density dip at the vortex location. The dashed line is the
predicted 3D bulk value $r_v=1.94\,\xi$, where $\xi$ is the
healing length. For fast rotation the vortex core area deviates
from this prediction and it saturates at a value close to the
prediction by \textcite{Baym:2003b} (continuous line). Reprinted with permission from \textcite{Schweikhard:2004a}.}
 \label{fig:core_area}
\end{figure}

Another successful method to reach the fast rotation regime is the
evaporative spinup technique, developed by \textcite{Engels:2002}.
The cloud is first set in rotation at a frequency $\Omega$ notably
below $\omega$ using a magnetic stirrer, that is subsequently
switched off. Then a nearly one dimensional radio-frequency
evaporation along the axis of rotation cools the cloud.
Simultaneously the rotation speed of the gas increases since the
evaporated atoms carry less angular momentum than average. With this
tool the Boulder group has succeeded in producing a gas rotating at
$\Omega>0.99 \omega$ with a purely harmonic confinement. Thanks to
the centrifugal deformation, the radius of the gas in the $xy$ plane
increases whereas the thickness along $z$ shrinks to the size of the
ground state $\sqrt{\hbar/m\omega_z}$, setting the gas well inside
the 2D regime. As the total volume of the gas increases,
interactions are reduced: the chemical potential $\mu\sim 10$~Hz
drops below the splitting between two Landau levels $2\hbar\omega$
(17~Hz), and the LLL regime is reached \cite{Schweikhard:2004a}. With
this setup, the Boulder group has been able to test the prediction
that the fractional core area of the vortices saturates to a value
of the order of 0.2 \cite{Coddington:2004,Schweikhard:2004a}, as
predicted theoretically \cite{Baym:2003b} (figure
\ref{fig:core_area}). In addition the expected distortion of the
vortex lattice with respect to an ideal triangular array could be
detected experimentally on the edge of the rotating condensate
\cite{Coddington:2004}.   Following a suggestion by
\textcite{Anglin:2002a}, another interesting investigation performed
on this system dealt with the Tkachenko oscillations \cite{Tkachenko:1966b} (for a review see \textcite{Sonin:1987}), i.e. the
long-wavelength transverse excitations of the vortex lattice
\cite{Coddington:2003}. The Tkachenko waves could be directly imaged
and their frequency could be measured with a good precision. The
theoretical analysis of these oscillations has recently been
performed within the mean-field approximation by several authors
(\textcite{Baym:2003b,Baym:2004}, \textcite{Gifford:2004},
 \textcite{Choi:2003}, \textcite{Baksmaty:2004},
 \textcite{Woo:2004},
 \textcite{Mizushima:2004},
 \textcite{Sonin:2005a,Sonin:2005b},
 \textcite{Cozzini:2004} and \textcite{Chevy:2006}).

Fast rotation of a BEC can also be achieved by stirring the gas
with a potential that is more elaborate than a quadratic one. One
can use in particular a rotating optical lattice that creates a
rotating, spatially periodic pattern on the gas. This has been
explored recently by \textcite{Tung:2006}, who superimposed to a
rotating BEC a set of columnar pinning sites created by a
two-dimensional, co-rotating optical lattice. For a sufficiently
large laser intensity the optical lattice can impose its structure
to the vortex lattice; \textcite{Tung:2006} studied in particular
the transition from the usual triangular Abrikosov lattice to a
square configuration imposed by light. Theoretical investigations
of this problem were carried out by \textcite{Pu:2005} and
\textcite{Reijnders:2004a,Reijnders:2005}, who found that a rich
variety of structural phases can emerge in this geometry, from the
competition between vortex-vortex and vortex-optical lattice
interactions.

\subsection{Beyond the mean field regime}

In the mean-field description of a fast rotating gas, the
macroscopic wave function $\psi(\mathbf{x})$ is a solution of the
non-linear Gross-Pitaesvkii equation and corresponds to a vortex
lattice. The radius of the atom cloud, given in (\ref{eq:radius}),
is a measure of the number $j_{\rm max}$ of single particle LLL
states $\phi_{j,0}$ that have a significant population. Recalling
that $\phi_{j,0}$ is maximum for a radius $\sqrt{j}\; a_\bot$, we
find
 \begin{equation}
j_{\rm max} \simeq (R/a_\bot)^2 \simeq \left(
\frac{N\tilde g_2}{1-\Omega/\omega} \right)^{1/2}\ .
 \end{equation}
The filling factor $\nu=N/j_{\rm max}$ gives the average number of
particles in each occupied single particle state. When $\nu \gg 1$
on expects the mean field treatment to be valid, and $j_{\rm max}$
is equal to the number $N_v$ of visible vortices sitting in the
atom disk. On the opposite when $\Omega$ tends to $\omega$, $\nu$
becomes on the order of unity or below, the number of vortices
$N_v$ exceeds the number of atoms $N$, and one has to turn to a
full many-body treatment of the problem. This breakdown of
mean-field approximation occurs when
 \begin{equation}
\mbox{non mean-field:}\qquad 1-\frac{\Omega}{\omega}\lesssim
\frac{\tilde g_2}{N}\ .
 \label{eq:beyondmF}
 \end{equation}
The analysis of this ultra-fast rotating regime presents strong
analogies with the studies of the fractional quantum Hall effect
(FQH). In the latter case one is interested in the correlated
state of a 2D electron gas with Coulomb interaction when it is
placed in a strong magnetic field. In both cases the states of
interest are restricted to the LLL and one looks for specific
filling factors where ground states with specific properties can
emerge.  Note, that for bosons the filling factor can be
arbitrarily large within the LLL, while for fermions it is restricted
to $\nu<1$.  One remarkable feature is incompressibility, meaning that
the ground state is separated from all excited states by a
`macroscopic' energy gap scaling as $g_2$. Moreover it leads to
the appearance of edge states near the boundary of the system,
similar to the wedding-cake structure of the Mott-insulating state
of the Bose-Hubbard model. These edge states are crucial for
understanding the quantization of the Hall conductance, see
\textcite{MacDonald:1994} and - on a more mathematical level -
\textcite{Frohlich:1993}.

\paragraph*{Numerical studies.}

So far the ultra-fast rotation regime has not been reached
experimentally. Even for the fastest rotations realized in the
laboratory, the filling factor $\nu$ is $\sim 10^3$ ($N\sim 10^5$,
$N_v\sim 10^2$), well inside the mean field regime. Therefore the
results obtained so far originate from an exact numerical
diagonalization of the many-body hamiltonian and from the
connection with know features of the fermionic FQH.

Most studies are performed considering states with a given total
angular momentum $L_z$, so that the problem essentially consists
in finding the eigenstates of the interaction energy
 \begin{equation}
 V_{\rm int}=\frac{\hbar^2 \tilde g_2}{M} \sum_{i<j} \delta(\mathbf{x}_i-\mathbf{x}_j)
  \ .
 \end{equation}
All states considered hereafter belong to the LLL subspace, so
their functional form in the $xy$ plane is
 \begin{equation}
\Psi(\mathbf{x}_1,\ldots,\mathbf{x}_N)= P(u_1, \ldots,
u_N)\;\exp\left( -\sum_{j=1}^N r_j^2/2a_\bot^2 \right)
 \end{equation}
where $u_j=x_j+iy_j$, $r_j^2=x_j^2+y_j^2$, and where
$P(u_1,\ldots,u_N)$ is a symmetric polynomial.  The $z$ motion is
expected to be `frozen' to its ground state and it is not
explicitly written in what follows. If one is interested only in
the bulk properties of the vortex liquid state, it is convenient
to replace the inhomogeneous disk geometry of a real experimental
setup by a compact, homogeneous geometry. Both torus
\cite{Cooper:2001} and spherical
\cite{Regnault:2003,Nakajima:2003} manifolds have been considered.
The LLL is then a space of finite dimension $d_{\rm LLL}$,
proportional to the area ${\cal A}$ of the torus or the sphere:
$d_{\rm LLL}={\cal A}/(\pi a_\bot^2)$. This allows to define in a
non ambiguous way the filling factor $\nu$ ($\nu=N/d_{\rm LLL}$)
even for values on the order of 1 or below, where the notion of
visible vortices becomes dubious.

\paragraph*{Melting of the vortex lattice.}

When increasing the rotation speed of the gas, the first expected
deviation from the mean-field regime is the quantum melting of the
vortex lattice. This has been observed by \textcite{Cooper:2001}
in exact numerical calculations for filling factors $\nu=N/N_v\sim
6$ to $10$. This value can be recovered by calculating the quantum
fluctuations $\Delta$ of vortex positions and applying the
Lindemann criterion $\Delta_{\rm melt} \sim \ell/10$, where $\ell$
is the vortex spacing \cite{Sinova:2002}. For $\nu$ smaller than
the melting threshold, one meets for the ground state of the
many-body system a series of strongly correlated ground states
that we now briefly discuss.

\paragraph*{The Laughlin state and its daughter states.}
Since increasing the angular momentum spreads out the atoms in
space, one expects the interaction energy $E(L_z)$ of the ground
state for a given $L_z$ to decrease as $L_z$ increases (see e.g.
\textcite{Jackson:2000}). This decrease stops when one reaches the
celebrated Laughlin wave function, adapted here for bosonic
particles:
 \begin{equation}
P_{\rm Lau.}(u_1,u_2, \ldots, u_N)=\prod_{i<j} (u_i-u_j)^2
 \label{Laughlin}
 \end{equation}
Indeed since the probability to get two particles at the same
point vanishes, this state has the remarkable property to be an
eigenstate of the contact interaction potential with eigenvalue 0
\cite{Trugman:1985}. Increasing $L_z$ beyond this point cannot
reduce further $E(L_z)$. The total angular momentum $L_z/\hbar$ of
this state is equal to the degree $N(N-1)$ of each term of the
polynomial $P_{\rm Lau.}$ and this state expands over all LLL
single particle wavefunctions $\phi_{j,0}$ from $j=0$ to $j_{\rm
max}=2(N-1)$, i.e. a filling factor $\nu=1/2$. The Laughlin state
is incompressible and the gap to the first excited state is $\sim
0.1\, \tilde g_2\,\hbar \omega$ \cite{Regnault:2003,Regnault:2004}. The
Laughlin state is characterized by a quasi-uniform density of
particles over the circle of radius $a_\bot \sqrt{2N}$. The
two-body correlation function for this state shows a strong
anti-bunching $g^{(2)}(r\to 0)\sim r^2$. This correlation function has been
calculated numerically for a number of bosons $N$ up to 8 by
\textcite{Barberan:2006}.

For $L_z/\hbar > N(N-1)$, any state $P(u_1,\ldots,u_N)=P_{\rm
Lau.}(u_1, \ldots, u_N)\;Q(u_1, \ldots, u_N)$, where $Q$ is a
arbitrary symmetric polynomial, is a ground state of the system
with interaction energy 0. Depending on the total degree $d_Q$ of
$Q$, the physical interpretation of the state can be: (i) for
$d_Q\sim 1$, edge excitations of the Laughlin state
\cite{Cazalilla:2003,Cazalilla:2005}, (ii) for $d_Q= N$,
quasi-holes at a given point $U_0$ are obtained by taking
$Q=\prod_j (u_j-U_0)$ \cite{Paredes:2001,Paredes:2002}, (iii) for
$d_Q\sim N^2$, Laughlin-type wave functions with smaller filling
factors, by replacing the exponent $2$ by $4,6,\ldots$ in
(\ref{Laughlin}).

\paragraph*{The composite Fermion sequence.}
For filling factors between the melting point $\nu \sim 10$ and
the Laughlin state $\nu=1/2$, it is not possible to give an exact
analytical expression of the ground state at any given $\nu$. One
finds however a strong overlap between the numerically determined
ground states and some relevant states for the physics of
electronic FQH. An example is the composite fermion sequence which
presents strong analogies with the Jain principal sequence for
fermions \cite{Jain:1989}. A first evidence for this sequence was
found by \textcite{Cooper:1999}, and it has been studied in detail
by \textcite{Regnault:2003,Regnault:2004}. The physics at the
origin of the states of this sequence is reminiscent of that
explored in the section on 1D gases, where the problem of bosonic
particles with repulsive interaction is mapped onto the properties
of an assembly of non-interacting fermionic particles. Here one
considers that the gas is formed with fermionic composite
entities, each resulting from the attachment of a fermionic
particle with a vortex carrying one unit of statistical flux.
These composite fermions can be viewed as independent particles
which occupy various Landau levels. When they occupy exactly $n$
Landau levels they form an incompressible state. This occurs when
the filling fraction in the initial state is $\nu=n/(n+1)$. From a
more quantitative point of view, the composite fermion ansatz
corresponds to a wave function of the type:
 \begin{equation}
P(u_1,\ldots,u_N)= {\cal P}_{\rm LLL} \left[ Q_n(u_1,\ldots, u_N)
\prod_{i<j} (u_i-u_j)\;
 \right]
 \end{equation}
${\cal P}_{\rm LLL}$ describe the projector onto the LLL subspace
(for its precise definition see e.g. \textcite{Chang:2005}). The
first term in the bracket $Q_n(u_1,\ldots,u_N)$ is a Slater
determinant giving the state of $N$ fictitious fermions filling
exactly $n$ Landau levels. The second term involving products of
$u_i-u_j$ (Jastrow factor) corresponds to the attachment of a
vortex to each fermion. Since both terms in the bracket are
antisymmetric in the exchange of two particles, their product is a
symmetric wave function, suitable for the description of our $N$
identical bosons. Numerical evidence for such states was obtained
for $\nu=2/3$ and $3/4$ by \textcite{Regnault:2003,Regnault:2004}
and \textcite{Chang:2005}. The surface waves of the vortex liquids
whose wave functions can be described by the composite fermion
ansatz have been studied by \textcite{Cazalilla:2003},
\textcite{Regnault:2004} and \textcite{Cazalilla:2005}.

\paragraph*{The Read-Moore state and the Read-Rezayi series.}

For the filling factor $\nu=1$ yet another type of approximate
ground state have been identified \cite{Cooper:2001}, the
Moore-Read state, or Pfaffian. Assuming that $N$ is even the
expression of this state is
 \begin{equation}
P(u_1,\ldots,u_N)={\cal S} \left[
 \prod_{i<j\leq N/2}\hskip -1.5mm
 (u_i-u_j)^2 \prod_{N/2<l<n}
 \hskip -1.5mm (u_l-u_n)^2
 \right]
 \end{equation}
where ${\cal S}$ indicates symmetrization over all indices. The
total degree of each term of this polynomial is $N(N-2)/2$ and the
state expands over single particle LLL wavefunctions from $k=0$ up
to $k_{\rm max}=N-2$, corresponding to a filling factor $\nu=1$. As
for $\nu=1/2$ the ground state is incompressible with a gap $\sim
0.05\,\tilde g_2\, \hbar \omega$   \cite{Chang:2005}. It is noteworthy that
for this state, the probability is zero to have three particle at
the same location in space.

For even larger filling factors ($\nu$ between 1 and 10) an
analysis performed in a torus geometry suggested that the system
has incompressible ground states belonging to a family containing
clusters of $k$ particles \cite{Cooper:2001} and corresponding to
integer or half integer filling factors. This so-called
Read-Rezayi series \cite{Read:1999} is constructed by taking
symmetrized products of $k$ Laughlin states of the type
$\prod_{i<j\leq N/k}(u_i-u_j)^2$ (assuming that $N$ is a multiple
of $k$). The Laughlin and the Moore-Read states corresponds to
$k=1$ and $k=2$, respectively. The filling factor associated to
these states is $k/2$ and the total angular momentum is $L_z/\hbar
\sim N^2/k$. Further calculations performed in the spherical
geometry could not draw any conclusion concerning the survival of
the incompressibility of such states at the thermodynamic limit
\cite{Regnault:2004}.

\paragraph*{Possible detection schemes for fractional quantum Hall
effects.}

We now review some possible ways for observing experimentally
effects related to fractional quantum Hall physics. We first note
that it is unlikely that condensed-matter-type techniques, based
on transport properties with specific conductance plateaus can be
implemented with rotating atomic gases, at least in the near
future. We also point out that the condition (\ref{eq:beyondmF})
giving the threshold for the observation of mean-field effects
imposes \emph{de facto} to work with very small atomic samples.
Consider for example a purely harmonic trap. Due to residual trap
imperfections it is seems unlikely that one can achieve rotation
frequencies $\Omega$ larger than $0.999\;\omega$ (rotation
frequencies $\sim 0.99\;\omega$ have been achieved by
\textcite{Schweikhard:2004a}). Taking $\tilde g_2\sim 0.1$, this sets an
upper bound of $\sim 100$ on the atom number. Working with such
small atom numbers is not untractable, but it immediately makes
this type of experiment quite challenging from a technical point
of view.

A first possible experimental signature of the Laughlin and
Read-more states could lie in the fact that 3 particles can never
sit at the same location in space for these wave functions. As
3-body recombination is often the main source of atom losses one
can expect that the achievement of these states could be revealed
by a spectacular increase of the lifetime of the gas, as it has
been the case for 1D physics \cite{Tolra:2004}.

We now turn to more quantitative studies of these incompressible
states. A `simple' experimental evidence for a state such as the
Laughlin wave function could come from its specific density
profile, which is uniform over a disk of radius $a_\perp
\sqrt{2N}$, and zero elsewhere. This flat profile with density
$1/(2\pi a_\perp^2)$ is notably different from the parabolic
density profile expected in the mean-field regime, and its
observation would constitute a clear signature of a beyond
mean-field effect. Usually one does not measure directly the
in-trap density profile, because the relevant distances are on the
order of a few micrometers only, which is too small to be detected
with a good accuracy by optical means. The standard procedure to
circumvent this problem is to use a time-of-flight technique,
where the potential confining the atoms is suddenly switched off
so that the atoms fly away ballistically during an adjustable time
before being detected. For an arbitrary initial state the
functional form of the density distribution is modified during the
time-of-flight. For an LLL wavefunction this modification is a
mere scaling, at least when interactions between atoms are
negligible during the time-of-flight \cite{Read:2003}. In
particular the disk shape structure associated with the Laughlin
state should remain invariant in ballistic expansion.

If the number of atoms is larger than the one required to form a
Laughlin wave function at the particular frequency $\Omega$ that
is used, one expects a `wedding cake' structure for the atomic
density \cite{Cooper:2005}. This result is obtained within a local
density approximation and is very reminiscent of the structure
appearing in an atom Mott insulator confined in a harmonic trap.
At the center of the trap where the density is the largest, the
atoms may form an incompressible fluid corresponding for example
to the filling factor $2/3$, which is one of the composite fermion
states identified above. The atomic density is then expected to be
constant and equal to $2/(3\pi a_\perp^2)$ over a central disk
of radius $R_1$. For $r=R_1$ the gas switches abruptly (over a
distance $\sim a_\perp$) to the Laughlin state with filling
factor $1/2$ and the density drops to the value $1/(2\pi
a_\perp^2)$. It then stays constant over a ring of outer radius
$R_2$ ($R_2>R_1$) and for $r>R_2$ the density drops to zero. The
values of $R_1$ and $R_2$ can be obtained from a simple energy
minimization \cite{Cooper:2005}. For larger atom numbers, several
plateaus, with decreasing densities corresponding to the various
filling factors of the incompressible states, are expected.

The possibility to add an optical lattice along the $z$ direction
adds an interesting degree of freedom in the problem, and brings
experimentalists the hope to be allowed to work with a notably
larger atom number. In this configuration one deals with a stack
of $N_d$ parallel disks, all rotating at the same frequency
$\Omega$ along the $z$ axis. Each disk is coupled to its neighbors
by tunnelling across the lattice barrier, with a strength that can
be adjusted. For a large coupling the situation is similar to a
bulk 3D problem; when the coupling is reduced the system evolves
to the quasi-2D regime. This raises interesting questions even at
the level of a single vortex motion, as pointed out by
\textcite{Martikainen:2003}. The melting of the vortex lattice in
a stack of $N_d$ layers has been investigated by
\textcite{Cooper:2005} and \textcite{Snoek:2006a,Snoek:2006b}. For
smaller filling factors both the density profiles along $z$ and in
the $xy$ plane should show the weeding cake structure
characteristic of incompressible states \cite{Cooper:2005}.

More elaborate techniques have been proposed to test the anyonic
nature of the excitations of incompressible states.
\textcite{Paredes:2001} have investigated the possibility of
creating anyons in a Laughlin state by digging a hole in the atom
gas with a tightly focused laser. By moving adiabatically the hole
inside the cloud it should accumulate a phase, that could
subsequently be measured by an interference experiment. The
accumulated phase should then reveal the anyonic structure of the
hole-type excitation (see also \textcite{Paredes:2002}).

\subsection{Artificial gauge fields for atomic gases}

As shown in Eq.~(\ref{singlepartH}), rotating a neutral particle
system is equivalent to giving these particles a unit charge and
placing them in a magnetic field proportional to the rotation
vector $\bs \Omega$ (see Eq.~(\ref{singlepartH})). Other
possibilities have been suggested to apply an artificial gauge
field on a neutral gas. The common idea to these proposals is to
exploit the Berry' s phase \cite{Berry:1984} that arises when the
atomic ground level is split (e.g. by an electromagnetic field) in
several space-dependent sublevels, and the atoms follow
adiabatically one of them. We consider the one-body problem, and
label by $\{|n_{\bf x}\rangle\}$ the local energy basis of the
atomic ground level, with the associated energies $E_n({\bf x})$.
The most general state of the atom is a spinor $\sum_n \psi_n({\bf
x})|n_{\bf x}\rangle$ which evolves under the Hamiltonian
$p^2/(2m)+\sum_n E_n({\bf x})\;|n_{\bf x}\rangle\langle n_{\bf
x}|$. Suppose now that the atom is prepared in a given sublevel
$n$, and that the motion of the atomic center of mass is slow
enough to neglect transitions to other internal sublevels $n'$.
This is in particular the case in a magnetic trap where the index
$n$ simply labels the various Zeeman substates. One can then write
a Schr\"odinger equation for the component $\psi_n({\bf x})$ and
the corresponding hamiltonian reads $H=({\bf p}-{\bf A_n}({\bf
x}))^2/(2m)+V_n({\bf x})$. The vector potential ${\bf A_n}$ is
related to the spatial variation of the sublevel $|n\rangle$
 \begin{equation}
{\bf A}_n({\bf x})=i\hbar\, \langle n_{\bf x}|{\bs \nabla}n_{\bf
x}\rangle
 \end{equation}
and the scalar potential $V_n$ is
 \begin{equation}
V_n({\bf x})=E_n({\bf x})+\frac{\hbar^2}{2m}\left( \langle {\bs
\nabla} n_{\bf x}|{\bs \nabla}n_{\bf x}\rangle
 -
|\langle n_{\bf x}|{\bs \nabla}n_{\bf x}\rangle|^2 \right)\ .
 \end{equation}
If the spatial variation of the sublevels $|n_{\bf x}\rangle$ is
such that $\bs \nabla \wedge {\bf A}_n \neq 0$, this gauge field
can in principle have the same effect as rotating the atomic gas.

The simplest occurrence for an artificial gauge field happens in a
Ioffe-Pritchard magnetic trap \cite{Ho:1996a}. The sublevels
$|n_{\bf x}\rangle$ are the various Zeeman substates, which are
space-dependent because the direction of the trapping magnetic
field is not constant over the trap volume. However the gauge
field ${\bf A}_n$ that is generated in this configuration is too
small to initiate the formation of vortices. The addition of a
strong electric field could increase the magnitude of the
artificial gauge field, as shown by \textcite{Kailasvuori:2002}.
An alternative and promising line of research takes advantage of
the concept of \emph{dark states}, where two sublevels of the
atomic ground state are coupled by two laser waves to the same
excited state. If the laser frequencies are properly chosen, there
exists a linear combination of the two sublevels that is not
coupled to the light, and the spatial evolution of atoms prepared
in this dark state indeed involves the vector and scalar
potentials given above \cite{Dum:1996,Visser:1998,Dutta:1999}.
Possible spatial profile of the laser waves optimizing the
resulting artificial rotation field have been discussed by
\textcite{Juzeliunas:2004},
\textcite{Juzeliunas:2005,Juzeliunas:2006} and by
\textcite{Zhang:2005}. These proposals have not yet been
implemented experimentally.

Similar effects have also been predicted in a lattice geometry by
\textcite{Jaksch:2003}, where atoms with two distinct internal
ground state sublevels are trapped in different columns of the
lattice. Using a two-photon transition between the sublevels, one
can induce a non-vanishing phase of particles moving along a
closed path on the lattice. \textcite{Jaksch:2003} showed that one
can reach in this way a `high magnetic field' regime that is not
experimentally accessible for electrons in metals, characterized
by a fractal band structure (Hofstadter butterfly). The connection
between quantum Hall effect and the lattice geometry in presence
of an artificial gauge potential has been analyzed by
\textcite{Mueller:2004} and \textcite{Soerensen:2005}. One can
also generalize the Berry's phase approach to the case where
several energy states are degenerate \cite{Wilczek:1984}. Non
Abelian gauge fields emerge in this case, and possible
implementations on cold atom systems have been investigated
theoretically by \textcite{Osterloh:2005} and
\textcite{Ruseckas:2005}.

\section{BCS-BEC CROSSOVER}
\label{sec:crossover}

One of the basic many-body problems which has been brought
into focus by the study of ultracold atoms is that of a two component attractive
Fermi gas near a resonance of the $s$-wave scattering length.
The ability of tuning the interaction through a
Feshbach resonance allows to explore the
crossover from a BCS superfluid, when the attraction is weak and
pairing only shows up in momentum space, to a Bose-Einstein condensate
of tightly bound pairs in real space.
Here we will discuss the problem in the spin-balanced case,
which - in contrast to the situation at finite imbalance - is now
well understood.

\subsection{Molecular condensates and collisional stability}

The experimental study of the BCS-BEC crossover problem with 
ultracold atoms started with the realization of Fermi gases
in the regime of resonant interactions $k_F|a|\gg 1$ by \textcite{OHara:2002a}.
They observed an anisotropic expansion, characteristic for
hydrodynamic behavior. Typically, this is associated with
superfluidity because ultracold gases above the condensation
temperature are in the collisionless regime. Near a Feshbach resonance,
however, a hydrodynamic expansion is observed both
above and below the transition temperature. It is only through
the observation of stable vortices that superfluid and collision dominated
hydrodynamics can be distinguished. The BEC side of the 
crossover was first reached by creating ultracold molecules. This 
may be done either by direct evaporative cooling on the positive 
$a$ side \cite{Jochim:2003a}, where the weakly bound molecules 
are formed by inelastic three-body collisions. Alternatively, 
molecules can be generated in a perfectly reversible manner by
using a slow ramp of the magnetic field through a
Feshbach resonance \cite{Regal:2003b,Cubizolles:2003}.
This allows to convert a quasi-bound
state of two fermions at $a<0$ into a true bound state at $a>0$
(for a review of this technique see \textcite{Koehler:2006}).
Subsequently, a BEC of those molecules has been realized
both by direct evaporative cooling \cite{Jochim:2003b, Zwierlein:2003a} 
for $a>0$, or by converting a sufficiently cold attractive
Fermi gas at $a<0$ to a molecular condensate, using an
adiabatic ramp across the Feshbach resonance \cite{Greiner:2003b}.
The experiments are done with an equal mixture of the two
lowest hyperfine states of $^6$Li or of $^{40}$K confined optically
in a dipole trap. This allows to change the scattering length by a
magnetically tunable Feshbach resonance at $B_0=835\,$G or
$B_0=202\,$G respectively. 
On the BEC side, the fact that
the molecules are condensed can be verified experimentally by
observing a bimodal distribution in a time-of-flight experiment.
Probing superfluidity in Fermi gases on the BCS side
of the crossover, however, is much more difficult.  In particular, a time-of-flight
analysis of the expanding cloud does not work here.
Indeed, due to the factor $\exp{(-\pi/(2k_F|a|))}$ in the critical
temperature (see Eq.~(\ref{eq:GorkovT_c}) below),
superfluidity is lost upon expansion at constant
phase space density in contrast to the situation in BEC's
\footnote{ In this respect, the situation in two
dimensions, where pair binding appears for arbitrary 
values of the scattering length $a$, is much more favorable 
because the two-particle binding energy (\ref{eq:2Dboundstate})
is obviously density independent. Since
$T_c\sim\sqrt{\varepsilon_b\varepsilon_F}\sim n^{1/2}$, the superfluid
transition can thus be reached by an adiabatic expansion at constant
$T/T_F$, see \textcite{Petrov:2003}.}.
As discussed below, this problem may be circumvented by a
rapid ramp back into the BEC regime before the expansion.
A major surprise in the study of strongly interacting Fermi gases was the long lifetime
of the molecules near a Feshbach resonance 
\cite{Cubizolles:2003, Jochim:2003a, Strecker:2003}, 
in stark contrast to the situation encountered with bosonic atoms \cite{Herbig:2003, Durr:2004}.
The physics behind this was clarified by \textcite{Petrov:2004b} who
have solved the problem of scattering and relaxation into deeply
bound states of two fermions in the regime
where the scattering length $a$ is much larger than the characteristic
range of the interaction $r_e$.  As shown in section I.A, this range is
essentially the van der Waals length Eq.~(\ref{eq:vdwlength}), which is
much smaller than $a$ in the vicinity of a Feshbach resonance.
The basic physics which underlies the stability of fermionic dimers
in contrast to their bosonic counterparts is the fact that relaxation into
deep bound states is strongly suppressed by the Pauli-principle.
Indeed, the size of the weakly bound dimer states is just the scattering
length $a$, while that of the deep bound states is $r_e\ll a$.
By energy and momentum conservation,
a relaxation into a deep bound state requires that at least
three fermions are at a distance of order $r_e$. Since two of them are
necessarily in the same internal state and their typical momenta are of order $k\approx 1/a$,
the probability of a close three (or four) body encounter
is suppressed by a factor $(kr_e)^2\sim(r_e/a)^2$ due to
the antisymmetrization of the corresponding wave function.  From a detailed
calculation \cite{Petrov:2005}, the relaxation into deeply bound states
has a rate constant (in units cm$^3$/sec)
\begin{equation}
\label{eq:relaxation}
\alpha_{rel}=C\,\frac{\hbar r_e}{m}\cdot\Bigl(\frac{r_e}{a}\Bigr)^s\, ,
\end{equation}
which vanishes near a Feshbach resonance with a nontrivial power law.
The exponent $s=2.55$ or $s=3.33$
and the dimensionless prefactor $C$ depend on whether the
relaxation proceeds via dimer-dimer or dimer-atom collisions.
From experimental data the coefficient of the dominant dimer-dimer
relaxation is $C\approx 20$ \cite{Bourdel:2004, Regal:2004b}. Its value depends on short
range physics on the scale $r_e$ and thus cannot
be calculated within a pseudopotential approximation.
At a finite density, the power law dependence $a^{-s}$ holds only as
long as the scattering length is smaller than the average interparticle
distance. The actual relaxation rate $n\alpha_{rel}$ in fact 
stays finite near a Feshbach resonance and is essentially 
given by replacing the factor $r_e/a$ in Eq.~(\ref{eq:relaxation}) 
by $k_Fr_e\ll 1$ \cite{Petrov:2004b}. In practice, the measured 
lifetimes are on the order of $0.1\,$s for $^{40}$K and up to 
about $30\,$s for $^6$Li. This long lifetime of  fermionic atoms
near a Feshbach resonance is essential for the 
possibility to study the BCS-BEC crossover, because it
allows to reduce the physics near the 
resonance to an idealized, conservative many-body problem
in which relaxational processes are negligible.

The issue of dimer-dimer collisions has an additional aspect, which is
important for the stability of the strongly attractive Fermi gas.  
Indeed, molecules
consisting of two bound fermions also undergo purely elastic scattering.
It is obvious, that a molecular condensate will only be stable if the
associated interaction of these effectively bosonic dimers is repulsive.
From an exact solution of the four-particle Schr\"odinger equation with
pseudopotential interactions, \textcite{Petrov:2004b}
have shown that in the limit where
the distance $R$ (denoted by $R/\sqrt{2}$ in their paper) between the
centers of mass of two dimers is much larger than the
dimer size $a$ and at collision energies much
smaller than their respective binding energies $\hbar^2/2M_ra^2$,
the wave function has the asymptotic form
\begin{equation}
\label{eq:a_dd}
\Psi(\mathbf{x_1},\mathbf{x_2},\mathbf{R})=
\varphi_0(r_1)\varphi_0(r_2)\bigl( 1-a_{dd}/R\bigr)
\end{equation}
with $a_{dd}\simeq 0.60\, a$.
Here $\varphi_0(r)\sim\exp{(-r/a)}$ is the bound state wave function
of an individual dimer and $\mathbf{x}_{1,2}$ are the respective
interparticle distances between the two distinguishable fermions
which they are composed of.  It follows from Eq.~(\ref{eq:a_dd})
that the effective dimer-dimer interaction at low energies is
characterized by a positive scattering length, which is
proportional to the scattering length between its
fermionic constituents. This guarantees
the stability of molecular condensates and also implies that there
are no four-particle bound states for zero-range interactions
\footnote{Such states are discussed in nuclear physics, where
alpha particles in a nucleus may appear due to pairing correlations, see
e.g. \textcite{Roepke:1998}}.
Experimentally, the dimer-dimer scattering length can be inferred
from the radius $R=\ell_0(15Na_{dd}/\ell_0)^{1/5}$ of a molecular condensate
with $N$ dimers in a trap. The value found in fact agrees well
with the prediction $a_{dd}=0.60\, a$ \cite{Bartenstein:2004b, Bourdel:2004}.
Physically, the repulsion between dimers
can be understood as a statistical interaction due to the Pauli-principle
of its constituents. Within a phenomenological Ginzburg-Landau
description of the molecular condensate
by a complex order parameter $\psi(\mathbf{x})$, it is
simply related to a positive coefficient of the $\vert\psi\vert^4\,$-term.
In fact, the repulsive interaction between dimers was first derived from a
coherent state functional integral representation of the crossover problem
\cite{Drechsler:1992, Randeria:1993}. These results, however, were
restricted to a Born approximation of the scattering problem,
where $a_{dd}^{(B)}=2\, a$ \cite{Randeria:1993}. A derivation
of the exact result $a_{dd}=0.60\, a$ from diagrammatic
many-body theory has been given by \textcite{Brodsky:2006} and \textcite{Levinsen:2006}.
It is important to note that the stability of attractive fermions
along the BCS-BEC crossover relies crucially on the fact
that the range of the attractive interaction is much smaller than
the interparticle spacing. For more general interactions, where
this is not the case, instabilities may arise, as discussed
by \textcite{Fregoso:2006}.

\subsection{Crossover theory and Universality}

For a description of the many-body physics of the BCS-BEC
crossover, a natural starting point is a two-channel picture
in which fermions in an open channel couple resonantly to a
closed channel bound state. The resulting Hamiltonian
\begin{eqnarray}
\label{eq:BFM}
&\hat{H}_{BF}=\int d^3x \Bigg[ \sum_{\sigma}\hat{\psi}_{\sigma}^{\dagger}
\big(-\frac{\hbar^2}{2M}\nabla^2\big)\hat{\psi}_{\sigma}\; +&\nonumber \\
&\hat{\psi}_{B}^{\dagger}\big(-\frac{\hbar^2}{4M}\nabla^2+\nu\big)
\hat{\psi}_{B} + \tilde{g} \Big( \hat{\psi}_{B}^{\dagger}\hat{\psi}_{\uparrow}
\hat{\psi}_{\downarrow}+\rm{h.c.}\Big)\Bigg]&
\end{eqnarray}
defines the Bose-Fermi  resonance model. It was introduced
in this context by \textcite{Holland:2001} and by \textcite{Timmermans:2001}
and has been used subsequently e.g. by \textcite{Ohashi:2002} and 
\textcite{Drummond:2004}.
Here $\hat{\psi}_{\sigma}(\mathbf{x})$ are fermionic
field operators describing atoms in the open channel.
The two different hyperfine states are labelled by a formal
spin variable $\sigma=\uparrow,\downarrow$.  The bound
state in the closed channel is denoted by the bosonic field operator $\hat{\psi}_B$.
Its energy is detuned by $\nu$ with respect to the open channel 
continuum and $\tilde{g}$ is the coupling constant for the conversion of two atoms
into a closed channel state and vice-versa. It is caused by
the off-diagonal potential $W(r)$ in Eq.~(\ref{eq:2channel}) whose range
is of order the atomic dimension $r_c$. As a result,  the conversion is
pointlike on scales beyond $r_e$ where a pseudopotential description
applies. The magnitude of $\tilde{g}= \langle \phi_{\rm res}| W |\phi_{0}\rangle$
is determined by the matrix element of the off-diagonal potential between
the closed and open channel states. Using Eq.~(\ref{eq:rstar2}), its value
is directly connected with the characteristic scale $r^{\star}$ introduced in
Eq.~(\ref{eq:rstar}), such that $(2M_r\tilde{g}/\hbar^2)^2=4\pi/r^{\star}$
\cite{Bruun:2004a}. For simplicity, the background scattering
between fermions is neglected, i.e. there is no direct term quartic in the
fermionic fields. This is justified close enough to resonance $|B-B_0|\ll
|\Delta B|$, where the scattering
length is dominated by its resonant contribution.

\paragraph*{Broad and narrow Feshbach resonances}
As was discussed in section I.C, the weakly bound state
which appears at negative detuning, has always
a vanishing closed channel admixture near resonance.
For the experimentally relevant case $|a_{bg}|\gg r^{\star}$,
the virtual or real bound states within the range $|\nu|<\mu|\Delta B|$ of the detuning
may therefore be effectively described as a {\it single channel} zero-energy
resonance.
This criterion is based on two-body parameters only. In order
to justify a single channel model for describing the physics of the
crossover at a {\it finite} density $n=k_F^3/3\pi^2$ of fermions, it is
necessary that the potential resonance description is valid in the
relevant regime $k_F|a|\gtrsim 1$  of the many-body problem.
Now, the range in the detuning where $k_F|a|\gtrsim 1$ is given by $|\nu|\lesssim\sqrt{\varepsilon_F\varepsilon^{\star}}$. Since the closed
channel contribution is negligible as long as  $\nu\ll\varepsilon^{\star}$,
a single-channel description applies if $\varepsilon_F\ll\varepsilon^{\star}$
or $k_Fr^{\star}\ll 1$ \cite{Diener:2004, Bruun:2004a}. This is the
condition for a 'broad' Feshbach resonance, which only involves the
{\it many-body} parameter $k_Fr^{\star}$. In quantitative terms,
the Fermi wavelength $\lambda_F=2\pi/k_F$ of dilute gases
is of order $\mu$m while $r^{\star}$ is typically on the order of
or even smaller than the effective range $r_e$ of the interaction.
The condition $k_Fr^{\star}\ll 1$ is therefore very well
obeyed unless one is dealing with exceptionally narrow
Feshbach resonances. Physically, the assumption of
a broad resonance implies that the bosonic field in Eq.~(\ref{eq:BFM}),
which gives rise to the resonant scattering, is so strongly coupled to the
open channel that the relative phase between both fields is perfectly
locked, i.e. the closed channel molecules condense simultaneously
with the particles in the open channel.  In contrast to the two-particle
problem, therefore, there is a finite $Z\,$-factor precisely on resonance,
as verified experimentally by \textcite{Partridge:2005}.
An important point to realize is that this situation is
precisely opposite to that encountered
in conventional superconductors, where the role of $\varepsilon^{\star}$
is played by the Debye energy $\hbar\omega_D$. The ratio
$\hbar\omega_D/\varepsilon_F$ is
very small in this case, on the order of the sound velocity divided
by the Fermi velocity. Effectively, this corresponds to
the case of {\it narrow} resonances, where $k_Fr^{\star}\gg 1$.
The effective Fermi-Fermi interaction
is then retarded and the Bose field in Eq.~(\ref{eq:BFM})
is basically unaffected by the condensation
of the fermions. On a formal level, this case can be treated by replacing
the closed channel field by a c-number, giving rise to a reduced BCS model
with a mean-field order parameter $\Delta=\tilde{g}\langle\hat{\psi}_B\rangle$
\cite{DePalo:2005, Sheehy:2006}.

There is an essential simplification in describing the crossover problem
in the limit $k_Fr^{\star}\ll 1$. This is related to the
fact that the parameter $r^{\star}$ can be understood as an effective range
for interactions of fermions at energies below the continuum,
i.e. at $k=i\kappa$.  Indeed, consider the resonant  phase shift for
two-body scattering as given in Eq.~(\ref{eq:Breit-Wigner}).
At zero detuning $\nu=0$ and small $k$ the associated
scattering amplitude can be shown to be precisely of the form (\ref{eq:s-ampl})
with an effective range $r_e=-2r^{\star}$. Therefore, in the limit $k_Fr^{\star}\ll 1$,
the two-body interaction near resonance is described by
the scattering amplitude Eq.~(\ref{eq:p-potampl}) of
an ideal pseudopotential even at $k=k_F$. As a result,
the Fermi energy is the only energy scale in the problem
right at unitarity.  As pointed out by \textcite{Ho:2004a}, the thermodynamics
of the unitary Fermi gas is then universal, depending only on the dimensionless
temperature $\theta=T/T_F$. In fact, as found by \textcite{Nikolic:2007}, 
the universality is much more general and is tied to the 
existence of an unstable fixed point describing the unitary,
balanced gas at zero density.  As a result, by a proper rescaling, the 
complete thermodynamics and phase diagram of low density Fermi gases with 
short range attractive interactions is a universal function of temperature $T$,
detuning $\nu$, chemical potential $\mu$ and the external field $h$
conjugate to a possible density imbalance. 

\paragraph*{Universality}
The universality provides considerable insight into the problem
even without a specific solution of the relevant microscopic
Hamiltonian. For simplicity, we focus on the so-called
unitary Fermi gas right at the Feshbach resonance and the spin-balanced
case of an equal mixture of both hyperfine states which undergo pairing.
This problem has in fact first been discussed in nuclear physics
as a parameter free model of low density neutron matter \cite{Baker:1999, Heiselberg:2001}.
By dimensional arguments, at $a=\infty$, the particle density $n$ and
the temperature $T$ are the only
variables on which the thermodynamics depends. The free energy per particle,
which has $n$ and $T$ as its natural variables, thus acquires
a universal form
\begin{equation}
\label{eq:free energy}
F(T,V,N)=N\varepsilon_F\cdot f(\theta)
\end{equation}
with $\varepsilon_F\sim n^{2/3}$ the bare Fermi energy and
$\theta=T/T_F$ the dimensionless temperature.
The function $f(\theta)$ is monotonically {\it de}creasing,
because $s=-f'(\theta)$ is just the entropy per particle.
As will be shown below, the fact that the ground state is superfluid implies
that $f(0)-f(\theta)$ vanishes proportional to $\theta^4$ as the temperature
approaches zero, in contrast to a Fermi gas (or liquid), where the
behavior is $\sim\theta^2$. Physically, this is due to the fact that the low lying
excitations are sound modes and not fermionic quasiparticles.
By standard thermodynamic derivatives, the function $f$
determines both the dimensionless chemical potential
according to
\begin{equation}
\label{eq:mu}
\frac{\mu}{\varepsilon_F}=\frac{5}{3}\,f(\theta)-\frac{2}{3}\,\theta f'(\theta)= :\xi(\theta)
\end{equation}
and the pressure via $p/n\varepsilon_F=\mu/\varepsilon_F-f(\theta)$, consistent with
the Gibbs-Duhem relation $\mu N=F+pV$ for a homogeneous system. Moreover,  
the fact that $-f'(\theta)$ is the entropy per particle, immediately implies
that $3pV=2(F+TS)$. The internal energy $u$ per volume is therefore 
connected with pressure and density by the
simple identity $p=2u/3$, valid at all temperatures \cite{Ho:2004a}.
Naively, this appears like
the connection between pressure and energy density in
a non-interacting quantum gas. In the present case, however,
the internal energy has a nonvanishing contribution $\langle\hat H'\rangle$ from
interactions. A proper way of understanding the relation $p=2u/3$ is obtained by
considering the quantum virial theorem $2\langle\hat H_0\rangle-k\langle\hat H'\rangle=
3pV$ for a two-body interaction $V(\mathbf{x}_i-\mathbf{x}_j)\sim |\mathbf{x}_i-\mathbf{x}_j|^k$,
which is a homogeneous function of the interparticle distance. It implies
that $p=2u/3$ is valid for an interacting system if $k=-2$.
The pressure of fermions at unitarity is thus related to
the energy density as {\it if} the particles had a purely inverse square interaction.  
An important consequence of this is the virial theorem \cite{Thomas:2005}
\begin{equation}
\label{eq:virial}
\langle \hat H_{\rm tot}\rangle=2\langle \hat H_{\rm trap}\rangle
=2\int d^3x\, U_{\rm trap}(\mathbf{x})n(\mathbf{x})
\end{equation}
for a harmonically trapped unitary gas, which allows to determine the 
thermodynamics of the unitary gas from its equilibrium 
density profile $n(\mathbf{x})$.
The relation (\ref{eq:virial}) follows quite generally from the 
quantum virial theorem with $k=-2$ and the fact that 
the contribution $3pV$ of the external forces to the virial in the 
case of a box with volume $V$ is replaced by 
$2\langle \hat H_{\rm trap}\rangle$ 
in the presence of an external harmonic potential. It is 
therefore valid for finite temperature and arbitrary 
trap anisotropy, An alternative derivation
of Eq.~(\ref{eq:virial}) has been given by \textcite{Werner:2006}. They have noted
that the unitary Fermi gas in 3D exhibits a scale invariance which 
is related to a hidden $SO(2,1)$ symmetry. In fact, since the interaction 
potential at unitarity effectively obeys $V(\lambda r)=V(r)/\lambda^2$, the situation 
is precisely analogous to that discussed at the end of section VI for the 2D Bose gas 
with a pseudopotential interaction. In particular, 
the scale invariance implies a simple evolution of arbitrary initial states
in a time dependent trap and the existence of undamped breathing
modes with frequency $2\omega$ \cite{Werner:2006}. 
 
At zero temperature, the ground state properties of the unitary
gas are characterized by a {\it single} universal number $\xi(0)=5\, f(0)/3$,
which is often denoted by $1+\beta$ in this context.
It is smaller than one (i.e. $\beta<0$), because the attractive interaction leads to a 
reduction of the chemical potential at unitarity from its non-interacting value 
$\mu^{(0)}=\varepsilon_F$ to $\mu=\xi(0)\varepsilon_F$
\footnote{In a trap, the chemical potential $\mu_{\rm trap}\sim R^2$ is
reduced by a factor $\sqrt{\xi(0)}$ and not $\xi(0)$ as in the homogeneous case, because
the density in the trap center is increased by the attractive interaction.}. 
Experimentally, the most direct way of measuring the universal number $\xi(0)$ 
is obtained from in-situ, absorption imaging of the density distribution $n(\mathbf{x})$ 
in a trap. Indeed, within the local density approximation Eq.~(\ref{eq:LDA}), 
free fermions in an isotropic trap exhibit a density profile
$n(\mathbf{x})=n(0)\left(1-r^2/R_{TF}^2\right)^{3/2}$ with a Thomas-Fermi radius
$R_{TF}^{(0)}=(24\, N)^{1/6}\ell_0$. Since $\mu\sim n^{2/3}$ at unitarity
has the same dependence on density than for non-interacting fermions,
with a prefactor reduced just by $\xi(0)<1$,
the profile at unitarity is that of a free Fermi gas with a rescaled size.
For a given total particle number $N$ and mean trap frequency $\bar{\omega}$,
the resulting Thomas-Fermi radius at zero temperature, is therefore
reduced by a factor $\xi^{1/4}(0)$. Ideally, the value $R_{TF}^{(0)}$
would be measured by sweeping the magnetic field to the zero crossing 
of the scattering length at $B=B_0+\Delta B$, where an ideal Fermi gas
is realized. In practice, e.g. for $^6$Li, there is appreciable molecule
formation and subsequent decay processes at this field and it is more
convenient to ramp the field to values far on the BCS side, where the 
thermodynamics is again essentially that of an ideal Fermi gas.  
Results for the universal parameter $\xi(0)$ at the lowest attainable 
temperatures of around $\theta\approx 0.04$ have been obtained
in this way by \textcite{Bartenstein:2004b}, with the result 
$\xi(0)=0.32\pm\, 0.1$.  This value is considerably smaller than that inferred from
more recent in situ measurements of the trap radius by \textcite{Partridge:2006},
where $\xi(0)=0.46\pm 0.05$ is found. Alternatively, the parameter $\xi$ may be
determined by measuring the release energy of an expanding cloud
\cite{Bourdel:2004, Stewart:2006}.
In this case, however, an appreciable temperature dependence was found \cite{Stewart:2006},
which makes extrapolations to $T=0$ difficult. In particular, at
finite temperature, the relation between the density distributions at $a=0$ and
at $a=\infty$ involves the complete function $\xi(\theta)$ because the
Fermi temperature continuously decreases as one moves away from the trap center.

On the theoretical side, the ground state properties of a resonantly interacting
Fermi gas have been obtained numerically by
fixed-node Green function Monte Carlo calculations. They  provide quantitative
results for the equation of state \cite{Carlson:2003, Astrakharchik:2004a}
at arbitrary values of $a$ and in particular at unitarity. The resulting values
for $\xi(0)$ are $0.43$ \cite{Carlson:2003} or $0.41$ \cite{Astrakharchik:2004a}.
Very recently, the chemical potential and the gap of the unitary Fermi gas
at zero temperature have been calculated analytically from an effective
field theory using an $\epsilon=4-d$ expansion \cite{Nishida:2006}.
The possibility of such an expansion is based on an observation
made by \textcite{Nussinov:2004}, that
a unitary Fermi gas in four dimensions is in fact an ideal Bose gas.
Indeed, in $d=4$,  a two-particle bound state in a zero range potential only appears
at infinitely strong attraction. Thus, already at $\varepsilon_b=0^+$,
the resulting dimer size vanishes. At finite density, therefore, one ends up
with a non-interacting BEC, similar to the situation as $a\!\to\! 0^+$ in three
dimensions.
The expansion about the upper critical dimension $d=4$ may be complemented
by an expansion around the lower critical dimension, which is two for the
present problem \cite{Nishida:2007a}. Indeed, for $d\leq 2$ a bound state at zero binding energy
appears for an arbitrary weak attractive interaction, as shown explicitly
in section V.A. A unitary Fermi gas in $d\leq 2$ thus coincides with the
non-interacting gas and $\xi(0)\equiv 1$ for all $d\leq 2$ \cite{Nussinov:2004}.
The $d-2\,$ expansion, however, only captures
non-superfluid properties like the equation of state while all effects associated
with superfluidity are nonperturbative.
Combining these two expansions within
a Borel-Pad{\'e} method, the field-theoretical results 
for the universal parameter $\xi(0)$ give values in the range
$\xi(0)=0.36\, -\, 0.39$ for different Pad{\'e}-approximants \cite{Nishida:2007a}. 

\paragraph*{Critical temperature and pseudogap}
Within a single-channel description, a zero range interaction
$V(\mathbf{x}-\mathbf{x}')=g_0\,\delta(\mathbf{x}-\mathbf{x}')$ between
fermions of opposite spin $\sigma$ gives rise to an interaction
Hamiltonian in momentum space
\begin{equation}
\label{eq:Gorkov}
\hat{H}'=\frac{g_0}{2V}\sum_{\sigma}\sum_{k,k',Q}c_{k+Q,\sigma}^{\dagger}
c_{-k,-\sigma}^{\dagger}c_{-k',-\sigma}c_{k'+Q,\sigma}\, .
\end{equation}
Here $c_{k,\sigma}^{\dagger}$ are fermion creation operators
with momentum $\mathbf{k}$ and spin $\sigma$
and $V$ is the volume of the system. Moreover, $\mathbf{k}-\mathbf{k}'$ is the
momentum transfer due to the interaction and $\mathbf{Q}$
the conserved total momentum in the two-particle scattering process.
The bare coupling strength $g_0$ is determined by the
s-wave scattering length $a$ after a regularization, in which the
delta potential is replaced by the proper pseudopotential with finite strength $g$ 
(see below). For attractive interactions $g<0$,
\footnote{Note that the model (\ref{eq:Gorkov}) does not make sense
in the regime $g>0$, where it describes {\it repulsive} fermions.
However, with a proper pseudopotential, the two-particle interaction
has a bound state for positive scattering length. The Hamiltonian (\ref{eq:Gorkov})
is then understood to describe fermions along this branch and not in
their continuum states, where the interaction would be repulsive},
the Hamiltonian (\ref{eq:Gorkov}) was first discussed by
\textcite{Gorkov:1961}. In the weak-coupling regime $k_F|a|\ll 1$, where the
magnitude of the scattering length is much less than the average interparticle
spacing, they showed that a BCS-instability to state
with bound pairs appears at a temperature
\begin{equation}
\label{eq:GorkovT_c}
T_c=\frac{8e^C}{(4e)^{1/3}\pi e^2}\, T_F\,\exp\bigl(-\pi/(2k_F|a|)\bigr)
\end{equation}
($C= 0.577$ is Euler's constant). As expected for a weak-coupling
BCS-instability, the critical temperature vanishes with an essential
singularity. The absence of an energy cutoff in the interaction
leaves the Fermi temperature as the characteristic scale.
For typical densities and off-resonant scattering lengths in cold gases,
the parameter $k_F|a|\approx 0.02$ is very small, so (\ref{eq:GorkovT_c})
is applicable in principle. In practice, however,  fermionic
superfluidity in dilute gases, where $T_F$ is only of order
micro-Kelvin, is unobservable unless $k_F|a|$ becomes of order one.
In fact, the range of accessible coupling strengths on the BCS side
of the crossover is limited by the finite level spacing in the trap
or, alternatively,  by the trap size $R$, which must be larger than the
size $\xi_b\approx \hbar v_F/(k_{\rm B}T_c)$ of a Cooper pair \cite{Tinkham:1996}. 
Using the local density approximation, the condition
$k_{\rm B}T_c\gtrsim\hbar\omega$ on the BCS side
is equivalent to $\xi_b\lesssim R$ and implies particle
numbers $N\gtrsim N_{\star}=\exp(3\pi/(2k_F|a|))$. Since
$N_{\star}=10^5$ at $k_F|a|=0.4$ this shows that with
typical values for the particle numbers in a trap, the
regime $k_F|a|\ll 1$ is no longer described by the theory
of a locally homogeneous system. Instead, for $N<N_{\star}$
one reaches a regime which is similar to that of pairing in nuclei,
where the resulting energy gap obeys $\Delta_0\ll\hbar\omega$,
see e.g. \textcite{Heiselberg:2002}.

In the strong-coupling regime $k_F|a|\gtrsim 1$ near the
unitarity limit, where the critical temperature lies in
an accessible range of order $T_F$ itself,
no analytical solution of the problem is available.
In particular, the singular nature of the two-particle
scattering amplitude $f(k)=i/k$ right at unitarity rules out any obvious
perturbative approach. It is only far out on the
BEC side of the problem, where $k_Fa\ll 1$ again provides a small
parameter. In this regime, the
binding energy $\varepsilon_b$ is much larger than the Fermi energy $\varepsilon_F$.
At temperatures $k_BT\ll\varepsilon_b$ therefore,
a purely bosonic description applies for a
dilute gas of strongly bound pairs
\footnote{Note that the pseudopotential bound state is
strongly bound in the BEC limit of the crossover only
as far as the scales relevant for the BCS-BEC
crossover are concerned, while it is a very weakly bound state on the
scale of the actual interatomic poential, being
the highest, so-called rovibrational state of a total number $N_b\gg 1$
of bound states, see section I.A}
with density $n/2$ and a repulsive
interaction described by the dimer-dimer scattering length
of Eq.~(\ref{eq:a_dd}). Its dimensionless coupling constant
$(n/2)^{1/3}a_{dd}=0.16\,k_Fa$
is much smaller than one in the regime $1/k_Fa\gtrsim 2$.
Since the dimers eventually approach an ideal Bose gas,
with density $n/2$ and mass $2M$, the critical temperature
in the BEC limit is obtained by converting the associated ideal BEC
condensation temperature into the original Fermi energy.
In the homogeneous case this gives $T_c(a\to 0)=0.218\,T_F$
while in a trap the numerical factor is $0.518$.
The fact that $T_c$ is completely
independent of the coupling constant in the BEC limit is simple
to understand: On the BCS side, superfluidity is destroyed by
fermionic excitations, namely the breakup of pairs.
The critical temperature
is therefore of the same order as the pairing gap at zero temperature,
consistent with the well known BCS relation $2\Delta_0/k_BT_c=3.52$.
A relation of this type is characteristic for a situation, in which
the transition to superfluidity is driven by the gain in {\it potential} energy 
associated with pair formation. In particular, the formation and condensation 
of fermion pairs occur at the same temperature.
By contrast, on the BEC side,
the superfluid transition is driven by a gain in {\it kinetic} energy, 
associated with the condensation of pre-formed pairs.
The critical temperature is then on the order of the degeneracy
temperature of the gas, which is completely unrelated to the pair binding energy.

\begin{figure}
\includegraphics[width=0.8\columnwidth]{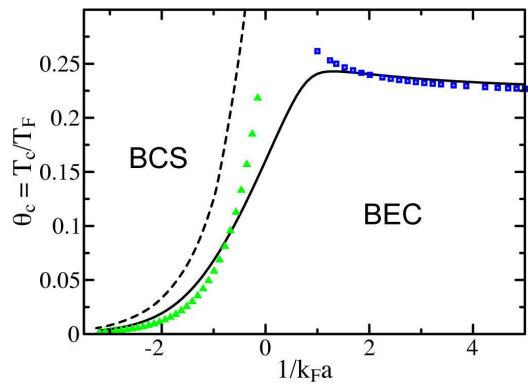}
\caption{Critical temperature of the homogeneous gas as a function of
the coupling strength. The exact asymptotic results Eq.~(\ref{eq:GorkovT_c})
and Eq.~(\ref{eq:Tc-shift}) in the BCS and BEC limits are indicated by
green triangles and blue squares, respectively. At unitarity $T_c= 0.16\, T_F$. The dashed line schematically denotes the evolution of $T^*$. Reprinted with permission from \cite{Haussmann:2007}. \label{fig:TcBECBCS}}
\end{figure}

To lowest order in $k_Fa$ in this regime, the shift Eq.~(\ref{eq:Tc-shift})
in the critical temperature due to the repulsive interaction
between dimers is positive and linear in $k_Fa$.
The critical temperature in the homogeneous case therefore
has a maximum as a function of the dimensionless inverse coupling
constant $v=1/k_Fa$, as found in the earliest calculation
of $T_c$ along the BCS-BEC crossover by \textcite{Nozieres:1985}.
More recent calculations of the universal curve $\theta_c(v)$
\cite{Haussmann:2007} indeed show a maximum around
$v\approx 1$, which is rather small, however (see Fig.~\ref{fig:TcBECBCS}).
The associated universal ratio $T_c/T_F=0.16$ at the unitarity
point $v=0$ agrees well with the value $0.152(7)$ obtained from
precise Quantum Monte Carlo calculations for the negative-U
Hubbard model at low filling by \textcite{Burovski:2006}.
Considerably larger values $0.23$ and $0.25$ for the ratio
$T_c/T_F$ at unitarity have been found by \textcite{Bulgac:2006}
from auxiliary field Quantum Monte Carlo calculations and by
\textcite{Akkineni:2006} from restricted path integral Monte
Carlo methods, the latter working directly with the continuum model.
In the presence of a trap, the critical temperature has been calculated
by \textcite{Perali:2004}. In this case, no maximum is found
as a function of $1/k_Fa$ because the repulsive interaction between
dimers on the BEC side leads to a density reduction in the trap center,
which eliminates the $T_c\,$-maximum at fixed density.

The increasing separation between the pair {\it formation} and the pair
{\it condensation} temperature as $v$ varies between the BCS- and the
BEC-limit implies that in the regime $-2\lesssim v\lesssim +2$ near 
unitarity, there is a substantial range of
temperatures above $T_c$, where preformed pairs exist, but do not form a 
superfluid. From recent path integral Monte Carlo calculations,
the characteristic temperature $T^{\star}$ below which strong
pair correlations appear has been found to be of order
$T^{\star}\approx 0.7\, T_F$ at unitarity \cite{Akkineni:2006},
which is at least three times the condensation temperature $T_c$ at this point.
It has been shown by \textcite{Randeria:1992} and \textcite{Trivedi:1995}, that
the existence of preformed pairs in the regime $T_c<T\lesssim T^{\star}$
leads to a normal state very different from a conventional Fermi liquid. 
For instance, the spin susceptibility 
is strongly suppressed due to singlet formation above 
the superfluid transition temperature 
\footnote{For a proposal to measure the spin susceptibility in trapped 
Fermi gases see \textcite{Recati:2006}.}. This is caused by the strong
attractive interactions near unitarity, which leads to pairs in the superfluid, whose size
is of the same order than the interparticle spacing. The temperature
range between $T_c$ and $T^{\star}$ may be considered as a
regime of strong superconducting fluctuations.  Such a regime is present 
also in high-temperature superconductors., where it is called the Nernst
region of the pseudogap phase \cite{Lee:2006}. Its characteristic temperature
$T^{\star}$ approaches $T_c$ in the regime of weak coupling
(see Fig.~\ref{fig:TcBECBCS}). It disappears in underdoped
cuprates, where $T_c$ vanishes. Remarkably, in these systems, the temperature 
below which the spin susceptiblity is suppressed, however, becomes {\it larger} 
at small doping \cite{Lee:2006}. 
Apart from the different nature of the pairing in both cases
(s- versus d-wave), the nature of the pseudogap in the cuprates, 
which appears in the proximity of a Mott-insulator with antiferromagnetic 
order, is thus a rather complex set of phenomena, which still lack a 
proper microscopic understanding \cite{Lee:2006}.  For a discussion of
the relevance and limitations of the analogy between the 
pseudogap phase in the BCS-BEC crossover to that in high $T_c$ 
cuprates see e.g. the reviews by \textcite{Randeria:1998} and by \textcite{Chen:2005}.

\paragraph*{Extended BCS description of the crossover}
A simple approximation, which covers the complete
range of coupling strengths analytically, is obtained by assuming
that, at least for the ground state, only zero momentum pairs are
relevant. In the subspace of states with only zero momentum pairs,
all contributions in Eq.~(\ref{eq:Gorkov}) with $\mathbf{Q}\ne 0$
vanish. The resulting Hamiltonian
\begin{equation}
\label{eq:BCSH}
\hat{H}'_{BCS}=\frac{g_0}{2V}\sum_{\sigma}\sum_{k,k'}c_{k,\sigma}^{\dagger}
c_{-k,-\sigma}^{\dagger}c_{-k',-\sigma}c_{k',\sigma}
\end{equation}
thus involves only {\it two} momentum sums. Eq.~(\ref{eq:BCSH})
is in fact just the reduced BCS-Hamiltonian, which is
a standard model Hamiltonian to describe the phenomenon
of superconductivity.
It is usually solved by a variational Ansatz
\begin{equation}
\label{eq:BCSN}
\Psi_{BCS}\left( 1,2,\ldots N\right)\,=\,
\hat{\cal A}\left[\phi(1,2)\phi(3,4)\cdots \phi(N-1,N)\right]
\end{equation}
in which an identical two-particle state $\phi(1,2)$ is assumed for each pair.
Here the arguments $1=(\mathbf{x}_1,\sigma_1)$ etc. denote
position and spin, $\hat{\cal A}$ is the operator which antisymmetrizes
the many-body wavefunction and we have assumed an even number
of fermions for simplicity. The wave function (\ref{eq:BCSN}) is  
a simple example of a so-called Pfaffian state (see section VII.C)
with $(N-1)\, !\, !$ terms, which is just the square root of the determinant 
of the completely antisymmetric  $N\times N$ matrix $\phi(i,j)$.
In second quantization, it can be written in the form
$\vert\Psi_{BCS}\left( 1,2,\ldots N\right)\rangle\,=\bigl(\hat{b}_0^{\dagger}\bigr)^{N/2}\vert 0\rangle$
of a Gross-Pitaevskii like state. The operator
$\hat{b}_0^{\dagger}=\sum_k\phi_k c_{k,\uparrow}^{\dagger}c_{-k,\downarrow}^{\dagger}$ 
creates a pair with zero total momentum, with  
$\phi_k=V^{-1/2}\int\phi(\mathbf{x})\exp{-i\mathbf{k}\mathbf{x}}$ 
the Fourier transform of the spatial part of the two particle wave 
function $\phi(1,2)$ in Eq.~(\ref{eq:BCSN}). It is important to note, however, 
that $\hat{b}_0^{\dagger}$ is {\it not} a Bose operator. It develops this character 
only in the limit, where the two-particle wave function $\phi(1,2)$ has a size 
much smaller than the interparticle spacing (see below). 
To avoid the difficult task of working with a
fixed particle number, it is standard practice to use a coherent state
\begin{equation}
\label{eq:BCS}
\vert\text{BCS}\rangle=C_{BCS}\,\exp\left (\alpha\hat{b}_0^{\dagger} \right )\,\vert 0\rangle=
\prod_k\bigl( u_k+v_kc_{k,\uparrow}^{\dagger}c_{-k,\downarrow}^{\dagger}\bigr)\vert 0\rangle\, .
\end{equation}
Since $\langle\hat{b}_0^{\dagger}\hat{b}_0\rangle=
\vert\sum_k\phi_k u_kv_k\vert^2= |\alpha|^2=N/2$ by
the number equation (see below), this state is characterized by a macroscopic
occupation of a single state, which is a bound fermion pair with zero total momentum.
The amplitudes $u_k,v_k$ are connected to the two-particle wavefunction via
$v_k/u_k=\alpha\phi_k$.  Since $u_k^2$ or $v_k^2$ are the probabilities
of a pair $\mathbf{k}\uparrow,-\mathbf{k}\downarrow$
being empty or occupied, they obey the normalization $u_k^2+v_k^2=1$.
The overall normalization constant
$C_{BCS}=\exp{-\frac{1}{2}\sum_k\ln\bigl(1+|\alpha\phi_k|^2\bigr)}\to \exp{-|\alpha|^2/2}$
approaches the standard result of a coherent state of bosons
in the strong coupling limit, where $|\alpha\phi_k|^2\approx v_k^2\ll 1$ for all $k$. In this limit,
$\hat{b}_0^{\dagger}$ is indeed a Bose operator and the wave function
Eq.~(\ref{eq:BCSN}) is that of an ideal BEC of dimers. In fact, 
antisymmetrization becomes irrelevant in the limit where the
occupation $v_k^2$ of all fermion states is much less than one
\footnote{Note that the wavefunction (\ref{eq:BCSN}) still contains
$(N-1)\, !\, !$ terms even in the BEC limit. In practice, however, only
a single term is relevant, unless one is probing correlations between
fermions in different pairs.}.
The BCS wave function has the gap $\Delta$ as a single variational
parameter, which appears in the fermion momentum distribution
\begin{equation}
\label{eq:v_k}
v_k^2=
\frac{1}{2}\left( 1-\frac{\varepsilon_k-\mu}{\sqrt{(\varepsilon_k-\mu)^2+\Delta^2}}\right)\, .
\end{equation}
With increasing strength of the attractive interaction, this
evolves continuously from a slightly smeared Fermi distribution
to a rather broad distribution $v_k^2\to\Delta^2/4(\varepsilon_k-\mu)^2\sim (1+(k\xi_b)^2)^{-2}$
in the BEC limit, where the chemical potential is large and negative (see below).
Its width $\xi_b^{-1}$ increases as the pair size $\xi_b=\hbar/\sqrt{2M|\mu|}$ approaches zero.
Experimentally the fermionic momentum distribution near the Feshbach resonance
has been determined from time-of-flight measurements  by \textcite{Regal:2005}.
Accounting for the additional smearing due to the trap, the results
are in good agreement with Monte-Carlo calculations of the momentum distribution
for the model (\ref{eq:Gorkov}) \cite{Astrakharchik:2005b}.  
An analysis of the distribution at finite temperature allows to 
determine the decrease of the average fermionic excitation gap
with temperature \cite{Chen:2006b}. 

\begin{figure}
\includegraphics[width=0.8\columnwidth]{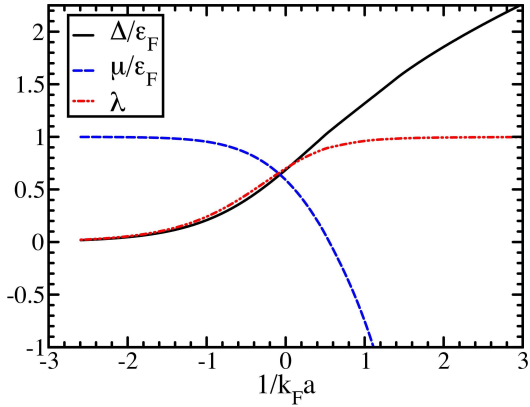}
\caption{Solution of the gap and number equations ~(\ref{eq:gaprenormalized})
and ~(\ref{eq:number}) for the reduced BCS Hamiltonian Eq.~(\ref{eq:BCSH}).
The dimensionless gap parameter, chemical potential and condensate fraction
(\ref{eq:BCSn_0}) of the ground state are shown as a function of the
dimensionless interaction parameter $1/k_Fa$. \label{fig:BCSgap}}
\end{figure}

Within the extended BCS description,  the magnitude of $\Delta$ is 
determined by the standard gap equation
\begin{equation}
\label{eq:gap}
-\frac{1}{g_0}=\frac{1}{2V}\sum_k\frac{1}{\sqrt{(\varepsilon_k-\mu)^2+\Delta^2}}
\end{equation}
where $E_k=\sqrt{(\varepsilon_k-\mu)^2+\Delta^2}$ is the BCS quasiparticle energy.
In conventional superconductors the momentum sum in Eq.~(\ref{eq:gap}) is restricted
to a thin shell around the Fermi energy and the solution $\Delta\sim\exp{-1/|g_0|N(0)}$
for $g_0<0$ depends only on the density of states per spin $N(0)$ in the normal state right
{\it at} the Fermi energy. In cold gases, however, there is no such cutoff as long as
$\varepsilon_F\ll\varepsilon^{\star}$. Moreover, the true dimensionless
coupling constant $N(0)|g|=2k_F|a|/\pi$ is far from small, approaching infinity
at the Feshbach resonance. The pairing interaction thus affects
fermions deep in the Fermi sea and eventually completely melts the Fermi
sphere. Within the pseudopotential approximation the apparent divergence in
Eq.~(\ref{eq:gap}) can be regularized by the formal replacement
$1/g_0\to 1/g\,-1/2V\,\sum_k(1/\varepsilon_k)$. Physically, this amounts to integrating
out the high energy contributions in Eq.~(\ref{eq:gap}) where the spectrum is
unaffected by the pairing. A general procedure for doing this, including the case
of strong pairing in nonzero angular momentum states, has been given by
\textcite{Randeria:1990}. Converting the sum over $k$ to an integral over the
free particle density of states, the renormalized gap equation at zero temperature
can then be written in the form
\begin{equation}
\label{eq:gaprenormalized}
\frac{1}{k_Fa}=\bigl(\tilde{\mu}^2+\tilde{\Delta}^2\bigr)^{1/4}\, P_{1/2}(x)\, .
\end{equation}
Here $\tilde{\mu}=\mu/\varepsilon_F$ and $\tilde{\Delta}=\Delta/\varepsilon_F$
are the dimensionless chemical potential and gap respectively, while
$P_{1/2}(x)$ is a Legendre function of the first kind. The parameter
$x=-\mu/\bigl(\mu^2+\Delta^2\bigr)^{1/2}$ varies between $-1$ in the BCS-
and $+1$ in the BEC-limit because the fermion chemical potential continously drops
from $\mu=\varepsilon_F$ in weak-coupling to
$\mu\to-\varepsilon_b/2$ for strongly bound pairs and $|\mu|\gg\Delta$ in both limits.
Physically the behavior of the chemical potential
in the BEC-limit can be understood by noting that the energy
gained by adding {\it two} fermions is just the molecular binding energy.
The detailed evolution of $\mu$ as a function of the dimensionless
coupling strength $1/k_Fa$ follows from the equation $N=2\sum_k v_k^2$
for the average particle number. In dimensionless form this gives
\begin{equation}
\label{eq:number}
\frac{4}{\pi}=\tilde{\mu}\bigl(\tilde{\mu}^2+\tilde{\Delta}^2\bigr)^{1/4}\, P_{1/2}(x)
+\bigl(\tilde{\mu}^2+\tilde{\Delta}^2\bigr)^{3/4}\, P_{-1/2}(x)\, .
\end{equation}
The equations (\ref{eq:gaprenormalized}, \ref{eq:number}), 
originally discussed by \textcite{Eagles:1969}
and \textcite{Leggett:1980}, determine the gap, the chemical potential
and related quantities like the condensate fraction \cite{Ortiz:2005}
\begin{equation}
\label{eq:BCSn_0}
\lambda_{\rm BCS}=\frac{n_0}{n}\vert_{\rm BCS}=\frac{3\pi}{16}\frac{\tilde{\Delta}^2}{{\rm  Im}\,\bigl(\tilde{\mu}+i\tilde{\Delta}\bigr)^{1/2}}
\end{equation}
for arbitrary coupling (see Figure \ref{fig:BCSgap}).

\begin{figure}
\includegraphics[width=0.9\columnwidth]{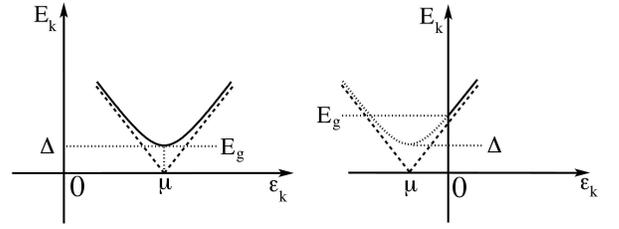}
\caption{Change in the fermionic excitation spectrum of the extended BCS
description as the chemical potential changes from positive to negative
values.\label{fig:BCSspectrum}}
\end{figure}

They provide a simple approximation
for the crossover between weak-coupling $1/k_Fa\to -\infty$ and the
BEC-limit  $1/k_Fa\to \infty$ within
the variational Ansatz Eq.~(\ref{eq:BCS}) for the ground state wave function.
In fact, as realized long ago by Richardson and Gaudin, the results are {\it exact}
for the reduced BCS Hamiltonian Eq.~(\ref{eq:BCSH}) and not just of a
{\it variational} nature as usually presented in textbooks
\footnote{For a review see \textcite{Dukelsky:2004}. It is interesting to note
that although the BCS wave function (\ref{eq:BCS}) gives the
exact thermodynamics of the model,
its number projected form does not seem to be exact
beyond the trivial weak coupling or BEC limit, see \textcite{Ortiz:2005}.}.
Both the gap and the condensate fraction increase
continuously with coupling strength, while the chemical potential
becomes negative for $1/k_Fa>0.55$. The values $\xi_{\rm BCS}(0)=0.59$,
$\tilde{\Delta}_{\rm BCS}=0.69$ and $\lambda_{\rm BCS}=0.70$ for the
chemical potential, the gap and the condensate fraction at unitarity
differ, however,  considerably from the corresponding
results $\xi(0)\approx 0.4$ and $\tilde{\Delta}\approx\lambda\approx 0.5$
obtained by both numerical and field-theoretic methods
for the physically relevant model ~(\ref{eq:Gorkov}).
For strong coupling, the gap increases
like $\tilde{\Delta}=4(3\pi k_Fa)^{-1/2}$. Apparently, this is much smaller than the
two-particle binding energy. To explain why $2\Delta$ differs from the energy
$\varepsilon_b$ of a strongly bound dimer
even in the BEC limit, it is necessary to determine the minimum value of the
energy $E_k=\sqrt{(\varepsilon_k-\mu)^2+\Delta^2}$ for single fermion excitations.
For negative chemical potentials, this minimum
is not at $\Delta$ as in the usual situation $\mu>0$, but at $\sqrt{\Delta^2+\mu^2}$
(see Fig.~\ref{fig:BCSspectrum}). Since $|\mu|\gg\Delta$ in the BEC limit, the minimum
energy for a single fermionic excitation is therefore
$|\mu|=\varepsilon_b/2$ and not $\Delta$ \cite{Randeria:1990}.

The size of the pairs $\xi_b$
continuously shrinks from an exponentially large value
$k_F\xi_b\simeq\varepsilon_F/\Delta$ in the BCS-limit to essentially
zero $\xi_b\simeq a$ in the BEC-limit of tightly bound pairs. For weak
coupling, the size of the pair coincides with the coherence length $\xi$.
This is no longer the case on the BEC-side of the crossover, however.
Indeed, as shown by \textcite{Pistolesi:1996} and \textcite{Randeria:1997},
the coherence length reaches a minimum value on the order of the 
interparticle spacing around
the unitary limit and then increases slowly to approach the value
$\xi=(4\pi na_{dd})^{-1/2}$ of a weakly interacting Bose gas of dimers
with density $n/2$. This minimum is closely related to a maximum
in the critical velocity around the unitary point. Indeed, as discussed
in section IV.D, the critical momentum for superfluid flow within
a mean-field description is simply $k_c\approx 1/\xi$. More precisely, 
the critical velocity $v_c$ on the BEC side of the crossover coincides with 
the sound velocity according to the Landau criterion. Near unitarity,
this velocity reaches $c=v_F\sqrt{\xi(0)/3}\approx 0.36\, v_F$ (see below). 
On the BCS side, the destruction of superfluidity does not involve the excitation
of phonons but is due to pair-breaking. As shown by \textcite{Combescot:2006}
and \textcite{Sensarma:2006}, the resulting critical velocity exhibits a maximum 
$v_c^{max}\approx 0.36 v_F$ around the unitarity point which is close to the value
of the local sound velocity there. Using a moving optical lattice near the trap center, 
this prediction has been verified recently by \textcite{Miller:2007}.

\paragraph*{Failure of extended BCS-theory}
In the regime of weak-coupling,
the gap $\tilde{\Delta}_{BCS}=8/e^2\,\exp{-\pi/2k_F|a|}$
is exponentially small. Using Eq.~(\ref{eq:GorkovT_c})
for the critical temperature, however, the ratio $2\Delta/k_BT_c$
differs from the well known BCS-value $2\pi/e^C=3.52$ by a factor $(4e)^{1/3}$.
The reason for this discrepancy is subtle and important from a basic point of
view. It has to do with the fact that the reduced BCS-model contains only
zero momentum pairs and thus no density fluctuations are possible.
By contrast, the original model (\ref{eq:Gorkov}) includes
such fluctuations, which are present in any neutral system. The innocent looking
BCS assumption of pairs with zero momentum only thus eliminates
an important part of the physics. To understand why
the reduced BCS-model fails to account for the correct
low energy excitations, it is useful to rewrite the interaction
Eq.~(\ref{eq:BCSH}) in real space, where
\begin{equation}
\label{eq:BCSHrealspace}
\hat{H}_{BCS}'=\frac{g_0}{2V}\sum_{\sigma}\int_x\int_{x'}
\hat{\psi}_{\sigma}^{\dagger}(\mathbf{x})
\hat{\psi}_{-\sigma}^{\dagger}(\mathbf{x})
 \hat{\psi}_{-\sigma}(\mathbf{x}')\hat{\psi}_{\sigma}(\mathbf{x}')\, .
\end{equation}
The associated non-standard (note the order of $\mathbf{x}$ and $\mathbf{x}'$ !)
'interaction' $V_{BCS}(\mathbf{x}-\mathbf{x}')=g_0/V$ has {\it infinite} range, but
is scaled with $1/V$ to give a proper thermodynamic limit. It is well known,
that models of this type are exactly soluble and that their phase
transitions are described by mean-field theory. Moreover, while
Eq.~(\ref{eq:BCSHrealspace}) is invariant under a {\it global}
gauge transformation
$\psi^{\ }_\sigma(\mathbf{x}) \rightarrow
e^{i\phi} \psi^{\ }_\sigma(\mathbf{x})$ (particle number is conserved),
this is no longer the case if $\phi(\mathbf{x})$ varies spatially.
In the exactly soluble reduced BCS model therefore,
the change in energy associated with a slowly varying phase does not
vanish like $(\nabla\phi(\mathbf{x}))^2$ as it should for a superfluid (see
Appendix).
The omission of pairs with finite momentum thus eliminates the
well known Bogoliubov-Anderson mode  $\omega(q)=cq$
of a neutral superfluid \cite{Anderson:1958b}.
A proper description of the crossover problem in a neutral system like
cold gases requires to account for both the bosonic as well as the
fermionic excitations for an arbitrary strength of the coupling.
By contrast, for charged superconductors,
the Bogoliubov-Anderson mode is pushed up to a high frequency plasmon
mode,  which is irrelevant for the description of superconductivity.
A reduced BCS model, in which they are omitted anyway, is thus appropriate.

The absence
of the collective Bogoliubov-Anderson mode implies that already the
leading corrections to the ground state in powers of $k_F|a|$ are incorrect
in a description which is based on just solving the
gap- and number-equation for arbitrary coupling.
For weak interactions, one misses the Gorkov, Melik-Barkhudarov
reduction of the gap. Indeed, as shown by \textcite{Heiselberg:2000},
density fluctuations give rise to a screening of the attractive interaction at
finite density, changing the dimensionless
coupling constant of the two-particle problem to
$g_{\rm eff}=g+g^2N(0)(1+2\ln{2})/3+\ldots$.
Since the additional contribution to the two-body scattering amplitude
$g<0$ is positive, the bare attraction between two fermions is weakened.
Due to the nonanalytic dependence of
the weak-coupling gap or transition temperature on the coupling constant
$k_Fa$, the renormalization $g\to g_{\rm eff}$ gives rise
to a reduction of both the gap and the critical temperature by a finite factor
$(4e)^{-1/3}\approx 0.45$.
The universal ratio $2\Delta/k_BT_c=3.52$, valid in weak coupling,
is thus unaffected. It is interesting to note that a naive extrapolation
of the Gorkov Melik-Barkhudarov result $\Delta/\varepsilon_F=(2/e)^{7/3}\exp{-\pi/2k_F|a|}$
for the zero temperature gap to infinite coupling $k_Fa=\infty$ gives
a value $\Delta(v=0)=0.49\varepsilon_F$ which is close to the
one obtained from Quantum Monte Carlo calculations at unitarity \cite{Carlson:2003}.
In the BEC limit, the condensate fraction Eq.~(\ref{eq:BCSn_0})
reaches the trivial limit one of the ideal Bose gas like $\lambda_{BCS}=1-2\pi na^3+\ldots$.
The correct behavior, however, is described by the Bogoliubov result
Eq.~(\ref{eq:depletion}) with density $n/2$ and scattering length $a\to a_{dd}$,
i.e. it involves the {\it square root} of the gas parameter $na^3$.
As pointed out by \textcite{Lamacraft:2006}, the presence of pairs
with nonzero momentum is also important for measurements of
noise correlations in time-of-flight pictures near the BCS-BEC crossover.
As discussed in section III.B, such measurements provide information about
the truncated density correlation function
\begin{equation}
\label{eq:BCSg_2}
G_{\uparrow\downarrow}(\mathbf{k}_1,\mathbf{k}_2)=
\langle\hat{n}_{\uparrow}(\mathbf{k}_1)\hat{n}_{\downarrow}(\mathbf{k}_2)\rangle-
\langle\hat{n}_{\uparrow}(\mathbf{k}_1)\rangle\,\langle\hat{n}_{\downarrow}(\mathbf{k}_2)\rangle\, .
\end{equation}
The momenta $\mathbf{k}_{1,2}=M\mathbf{x}_{1,2}/\hbar t$ are simply
related to the positions $\mathbf{x}_{1,2}$ at which the correlations are determined
after a free flight expansion for time $t$. While the formation of bound states
shows up as a peak at $\mathbf{k}_1=-\mathbf{k}_2$ as observed by
\textcite{Greiner:2005b} in a molecular gas above condensation,
the presence of pairs with nonzero momenta give rise to an additional
contribution proportional to $1/|\mathbf{k}_1+\mathbf{k}_2|$.
This reflects the depletion of the condensate due to bosons at finite momentum
as found in the Bogoliubov theory of an interacting Bose gas.

\paragraph*{Crossover thermodynamics}
As pointed out above, a complete description of the BCS-BEC crossover
involves both bosonic and fermionic degrees of freedom. Since the fermionic
spectrum has a gap, only the bosonic Bogoliubov-Anderson mode is 
relevant at low temperatures \cite{Liu:2006}.
Similar to the situation in superfluid $^4$He, these
sound modes determine the low temperature thermodynamics along the full
BCS-BEC crossover. They give rise to an entropy
\begin{equation}
\label{eq:entropy}
S(T)=V \frac{2 \pi^2}{45}k_B\left(\frac{k_B T}{\hbar c}\right)^3+\ldots
\end{equation}
which vanishes like $T^3$ for arbitrary coupling strength.
The associated sound velocity $c$ at zero temperature follows from the
ground state pressure or the chemical potential via the thermodynamic relation
$Mc^2=\partial p/\partial n=n\partial\mu/\partial n$. 
It was shown by \textcite{Randeria:1997}, that the sound velocity decreases
monotonically from $c=v_F/\sqrt{3}$ to zero on the BEC side. There, to lowest
order in $k_Fa$, it is given by the Gross-Pitaevskii result
$(c/v_F)^2=k_F a_{dd}/(6\pi)$ for the sound velocity of a dilute gas of dimers
with a repulsive interaction obtained from Eq.~(\ref{eq:a_dd}).
At unitarity, $c=v_F\sqrt{\xi(0)/3}\approx 0.36\, v_F$ is related to the
Fermi velocity by the universal constant $\xi(0)$. This has been used 
in recent measurements of the sound velocity along the axial $z$ direction
direction of an anisotropic trap by \textcite{Joseph:2007}. The velocity 
$c_0=c(z=0)$ measured near the trap center is given by
$c_0^2=\sqrt{\xi(0)}(v_F^{(0)})^2/5$. Here, $v_F^{(0)}$ is the Fermi 
velocity of a non-interacting gas, which is fixed by the associated
Fermi energy in the trap by $\varepsilon_F(N)=M(v_F^{(0)})^2/2$. 
A factor $3/5$ arises from averaging over the transverse 
density profile \cite{Capuzzi:2006}.
An additional factor $1/\sqrt{\xi(0)}$ is due to the fact that at unitarity, 
the attractive interactions increase the local density $n(0)$ in 
the trap center compared to that of a non-interacting Fermi gas
by a factor $\xi(0)^{-3/4}$. The local Fermi velocity $v_F[n(0)]$ is thus
enhanced by  $\xi(0)^{-1/4}$ compared to $v_F^{(0)}$. 
Precise measurements of the ratio $c_0/v_F^{(0)}=0.362\pm 0.006$ \cite{Joseph:2007}
at the lowest available temperatures thus give $\xi(0)=0.43\pm 0.03$,
in good agreement with the Quantum Monte Carlo results.
In addition, the independence of this ratio on the density has been 
tested over a wide range, thus confirming the universality at the unitary 
point. 

\begin{figure}
	\includegraphics[width=0.9\columnwidth]{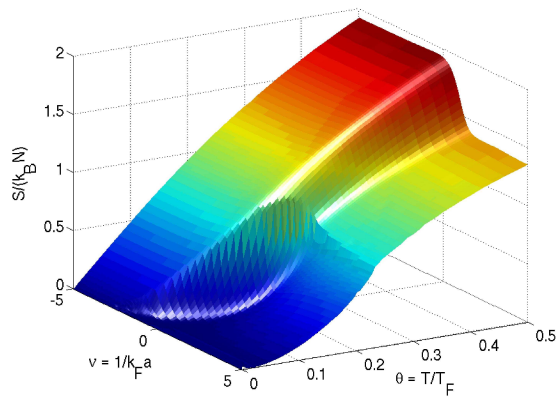}
\caption{Entropy per particle in units of $k_{\rm B}$ as a function
of the dimensionless coupling $1/k_Fa$ and temperature
$\theta=T/T_F$. The crossover from a Fermi gas with a
tiny superfluid regime as $T\to 0$ to a Bose gas which is
superfluid below $T\approx 0.2\, T_F$ occurs in a narrow range
of coupling strengths. An isentropic ramp, starting in the normal phase on the negative
$a$ side at sufficiently low $T$ leads to a molecular condensate. Reprinted with permission from \textcite{Haussmann:2007}.
\label{fig:entropy}}
\end{figure}

Numerical results on the crossover problem at finite temperatures
have been obtained by \textcite{Bulgac:2006}, by \textcite{Burovski:2006} and
by \textcite{Akkineni:2006} at the unitarity point. More recently, they have been
complemented by analytical methods, using expansions around
the upper and lower critical dimension \cite{Nishida:2007b}
or in the inverse number $1/N$
of a strongly attractive Fermi gas with $2N$ components \cite{Nikolic:2007}.
A rather complete picture of the crossover thermodynamics
for arbitrary couplings and temperatures in the spin-balanced case was
given by \textcite{Haussmann:2007} on the basis of a variational approach to the
many-body problem developed by
Luttinger-Ward and DeDominicis-Martin. Although the theory does not capture
correctly the critical behavior near the superfluid transition, which is
a continuous transition of the 3D XY-type for arbitrary coupling, the results obey
standard thermodynamic relations and the specific relation $p=2u/3$ 
at unitarity at the level of a few percent.  In addition, the resulting
value $T_c/T_F\approx 0.16$ for the critical temperature at unitarity
agrees rather well with the presently most precise numerical
results by \textcite{Burovski:2006}. 
As an example, a 3D plot of the entropy per particle is shown in Fig.~\ref{fig:entropy}
Apparently, the freezing out of fermionic excitations with increasing
coupling $v$ leads to a strong suppression of the low temperature entropy.
An adiabatic ramp across the Feshbach resonance from the BEC to the BCS
side is thus associated with a lowering of the temperature, as emphasized by \textcite{Carr:2004}.
In particular, it is evident from this picture that a molecular condensate
can be reached by going isentropically from negative to positive scattering
lengths even if the initial state on the fermionic side is above the
transition to superfluidity, as was the case in the experiments by \textcite{Greiner:2003b}.
The plot provides a quantitative picture of how an attractive gas of fermions
gradually evolves into a repulsive gas of bosons. Apparently,
most of the quantitative change happens in the
range $-2\lesssim v\lesssim +2$, which is precisely the regime accessible
experimentally with cold atoms. Note, that  the exact entropy and its first 
derivative are continuous near the critical temperature.
Indeed, the singular contribution to $S(T)$ is proportional to 
$|T-T_c|^{(1-\alpha)}$ and $\alpha<0$ for the 
3D XY-transition. Moreover, the value $S(T_c)/N\approx 0.7\, k_B$ of the
entropy per particle at $T_c$ at the unitarity point \cite{Haussmann:2007}, 
provides a limit on the entropy of any initial state which is is required to reach 
the superfluid regime $T<T_c$ near unitarity by an adiabatic process.

\subsection{Experiments near the unitarity limit}

\paragraph*{Thermodynamic properties} 
The thermodynamics of a strongly interacting Fermi gas near unitarity
has been studied experimentally by the Duke group \cite{Kinast:2005}.
It has been found, that the spatial profiles of both the trapped and released
gas are rather close to a Thomas-Fermi profile with a size parameter
determined from hydrodynamic scaling. This is consistent with the fact that
the equation of state of  the unitary gas is identical with that of a
non-interacting gas up to a scale factor. Strictly speaking, however, 
this is true only at zero temperature. At finite temperature, the function 
$\xi(\theta)$ defined in Eq.~(\ref{eq:mu}) is clearly different from that 
of an ideal Fermi gas.  Assuming a Thomas-Fermi profile, 
the effective temperature of a near unitary gas can be 
inferred from fitting the observed cloud profiles with a Thomas-Fermi distribution 
at finite temperature. The temperature dependent internal energy $E(T)$ 
and specific heat of the gas could then be determined by adding a well 
defined energy to the gas through a process, where the cloud is released
abruptly and recaptured after a varying expansion time \cite{Kinast:2005}.
The resulting $E(T)$ curve essentially follows that of a non-interacting gas
above a characteristic temperature $\tilde{T}_c\approx 0.27\,T_F^{(0)}$. The
data below $\tilde{T}_c$ follow a power law $E(T)-E_0\sim T^a$ with an 
exponent $a\simeq 3.73$, which is not far from the exact low $T$ result
$E(T)-E_0=V\cdot (\pi^2/30) k_BT(k_BT/\hbar c)^3$ of a uniform superfluid
in the phonon-dominated regime $k_BT\ll Mc^2$. Quantitatively, the data 
are well described by a finite temperature generalization of the extended BCS
description of the crossover \cite{Chen:2005}. In particular, the   
characteristic temperature $\tilde{T}_c$ is close to the theoretically
expected transition temperature $T_c^{trap}\approx 0.25\, T_F^{(0)}$ 
to the superfluid state. Indeed, within the 
local density approximation (LDA), the superfluid transition in a trap 
occurs when the local Fermi temperature $T_F[n(0)]$ in the trap center reaches the
critical value $T_c$ of the homogeneous gas at density $n(0)$.
At the unitarity point, the latter is obtained from the universal ratio
$T_c/T_F\approx 0.16$. As noted above, the attractive interactions
increase the local density in the trap center compared to the non-interacting 
gas with (bare) Fermi energy  $\varepsilon_F(N)=
\hbar\bar{\omega}\cdot(3N)^{1/3}=k_BT_F^{(0)}$. Thus $T_F[n(0)]$ is 
enhanced by a factor $1/\sqrt{\xi}$. Neglecting for simplicity the
variation of the chemical potential over the size of the trap due to
the local variation of $T_F$, this translates into a critical temperature
$T_c^{trap}/T_F^{(0)}=0.16\cdot\xi^{-1/2}(T_c)\approx 0.25$,
where $\xi(T_c)\approx 0.39$ is again a universal number at unitarity
\cite{Haussmann:2007}.  
 
Experimental results on thermodynamic properties which do not rely
on the difficult issue of a proper temperature calibration are possible 
by using the virial theorem Eq.~(\ref{eq:virial}). It allows to measure
the energy of the strongly interacting gas directly from its density profile. 
Within LDA, the contributions to $\langle \hat H_{\rm trap}\rangle$ from the 
three spatial directions are identical, even in an anisotropic trap. 
The total energy per particle thus follows directly from the 
average mean square radius $E/N=3M\omega_z^2\langle z^2\rangle$
along (say) the axial direction. The predicted linear increase of 
$\langle z^2\rangle$ at unitarity  with the energy input has been
verified experimentally by \textcite{Thomas:2005}. More generally, since the 
internal energy per particle is equal to $f(\theta)-\theta f'(\theta)$ in units of the 
bare Fermi energy, the universal function $f(\theta)$ is 
in principle accessible by measuring the density profile. 
A possible way to do thermometry for such measurements has
recently been developed by \textcite{Luo:2007}. They have measured the 
dependence $S(E)$ of the entropy on energy by determining the energy 
from the mean square radii and the entropy by adiabatically ramping the 
magnetic field far into the BCS regime.  There, the gas is essentially
an ideal Fermi gas in a trap and its entropy may be inferred again from
$\langle z^2\rangle$. At low energies, $S(E)\sim (E-E_0)^b$ is fit 
to a power law  with an exponent $b\approx 0.59$
(note that in the phonon dominated regime of a uniform gas one 
expects $b=3/4$ exactly at {\it arbitrary} coupling). The associated ground 
state energy has been determined by \textcite{Drummond:2007}
from a fit to the theory of \textcite{Nozieres:1985}, generalized 
to the symmetry broken state (see \cite{Randeria:1997}). The resulting
value $(E_0/N\varepsilon_F)_{\rm trap}=0.48\pm 0.03$ implies 
$\xi(0)=0.40\pm 0.03$, because the total energy in the trap is just twice of 
the result $(E_0/N\varepsilon_F)=f(0)=3\xi(0)/5$ obtained in a uniform
system. At higher energies, a different power law for $S(E)$ is found.
The crossover energy is used to provide a characteristic
temperature scale $\tilde{T}_c/T_F^{(0)}=0.29\pm 0.03$, which agrees 
well with the critical temperature inferred from the earlier measurements.   

\begin{figure}
	\includegraphics[width=0.9\columnwidth]{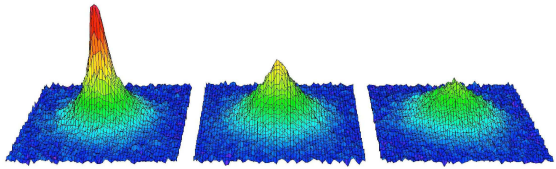}
\caption{Time-of-flight images showing a fermionic
	condensate after projection onto a molecular gas. The images
	show condensates at different detunings $\Delta B=0.12, 0.25,$ and
	$0.55.\,$G (left to right) from the Feshbach resonance (BCS side), starting
	with initial temperatures $T/T_F=0.07$. Reprinted with permission from \textcite{Regal:2004a}.
\label{fig:FermionCondensation}}
\end{figure}

\paragraph*{Condensate fraction}
The standard signature of BEC in ultracold gases is the
appearance of a bimodal density distribution below the
condensation temperature. Fitting the density profile from
the absorption image to a superposition of a Thomas Fermi
profile and a Gaussian background from the thermal atoms
allows a rather precise measurement of the condensate fraction
$n_0/n$. In the case of fermionic superfluidity this does not work
because pairs break in time-of-flight.
A way out is the rapid transfer technique \cite{Regal:2004a, Zwierlein:2004}
in which the fragile pairs on the BCS side of the crossover are
preserved by sweeping the magnetic field towards the BEC side of the
resonance, transforming them to stable molecules. In an adiabatic situation,
each fermionic pair is thereby transformed into a tightly bound dimer and
a time-of-flight analysis of the molecular condensate allows to infer the momentum
distribution of the original pairs on the BCS side of the crossover. 
The resulting absorption images are shown in Fig.~\ref{fig:FermionCondensation}.
 
\begin{figure}
	\includegraphics[width=0.9\columnwidth]{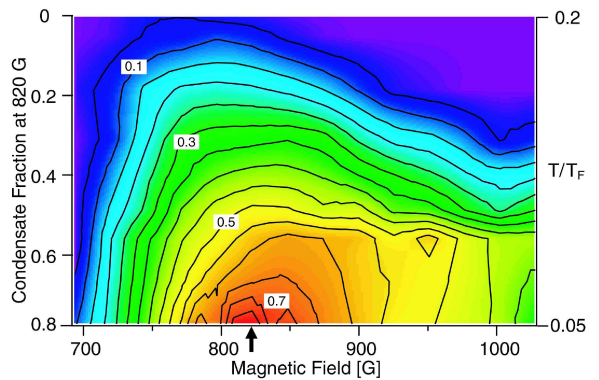}
\caption{Condensate fraction as a function of magnetic field and 
	temperature. The highest condensate fractions and onset temperatures
	are obtained on the BEC side, close to the resonance at 
	$B_0=834\,$G. As a measure of temperature, the 
	condensate fraction at $820\,$G (see arrow) is used as the vertical 
	axis. Reprinted with permission from \textcite{Zwierlein:2004}.
\label{fig:ResonanceCondensation}}
\end{figure}

In practice, the process is non-adiabatic and essentially projects
the initial many-body state onto that of a molecular condensate.
A theoretical analysis of the condensate fraction extracted from the
rapid transfer technique has been given by \textcite{Perali:2005} and by
\textcite{Altman:2005}. The fraction of molecules depends on the sweep
rate and, in a strongly non-adiabatic situation, provides information
about pair correlations in the initial state even in the absence of
a condensate \cite{Altman:2005}. Experimentally,
the observed condensate fractions in $^{40}$K were at most around
$0.14$ \cite{Regal:2004a}, much smaller than the expected equilibrium values
$n_0/n\approx 0.5$ at unitarity. A possible origin of this discrepancy
may be the rather short lifetime
of $^{40}$K dimers near resonance on the order of $100$ ms.
Yet, the qualitative features of the phase diagram agree with
an analysis based on an equilibrium theory \cite{Chen:2006a}.    
For $^6$Li, in turn, the condensate fractions determined via the
rapid transfer technique turned out to be much larger,
with a maximum value $0.8$ at $B\approx 820\, $G, on the BEC
side of the resonance \cite{Zwierlein:2004} (see Fig.~\ref{fig:ResonanceCondensation}). 
Apparently, in this case, there were no problems with the lifetime,
however the experimental definition of $n_0/n$ via the ratio of the
particle numbers in the central peak to the total number
overestimates the true condensate fraction. This is probably due to the
fact that the central peak contains an appreciable contribution
from particles which are removed from the condensate by the strong
interactions even at zero temperature.
The observation \cite{Zwierlein:2004} that
the condensate fraction decreases to zero on the BEC side (see Fig.~\ref{fig:ResonanceCondensation})
instead of approaching one is probably caused by fast vibrational relaxation
into deeply bound states further away from the resonance.

\begin{figure}
	\includegraphics[width=0.8\columnwidth]{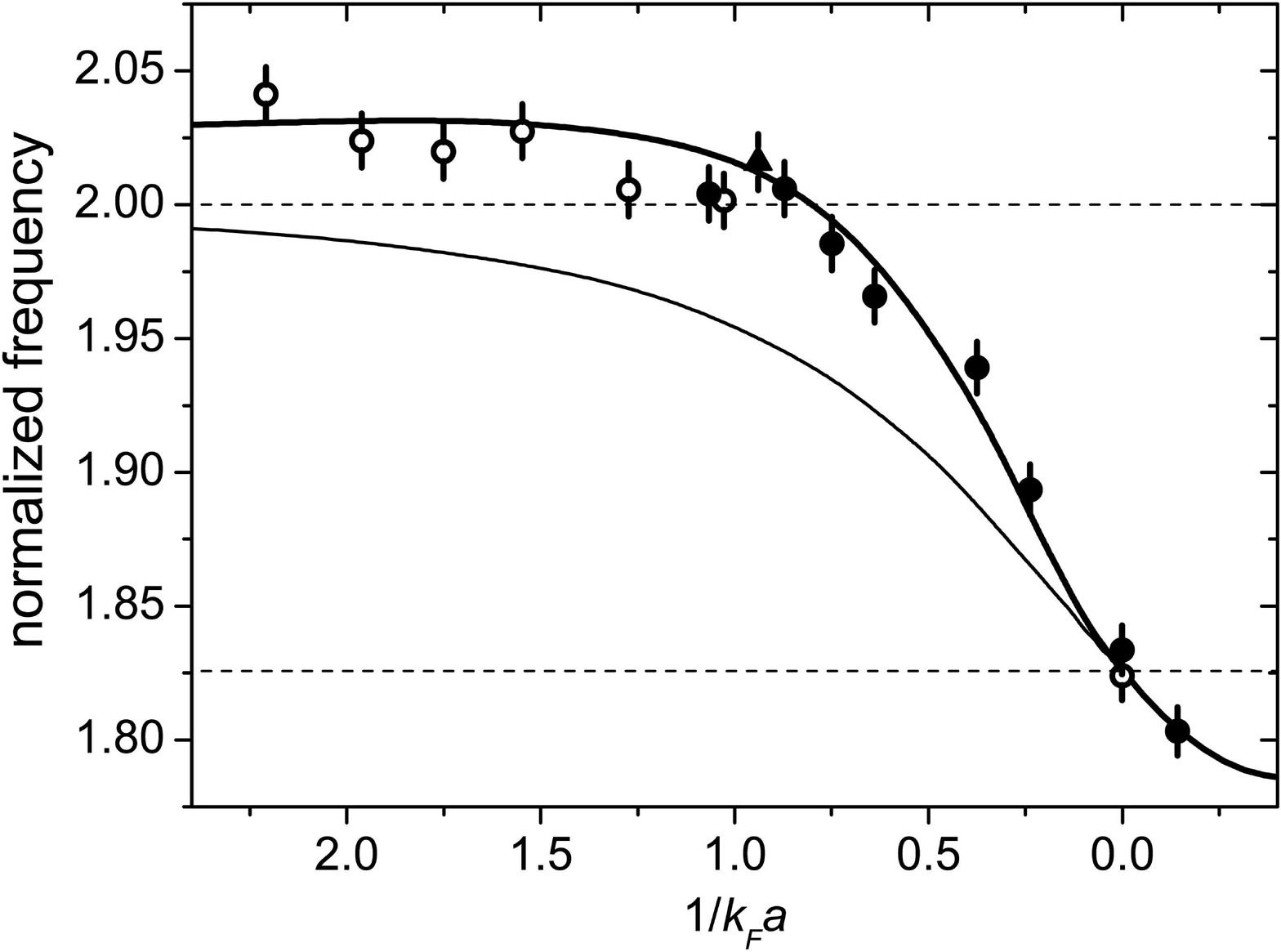}
\caption{Normalized frequency $\omega_r/\omega_{\perp}$ of the
radial compression mode as a function of dimensionless interaction
parameter $1/k_Fa$ on the BEC
side of the crossover. The result at unitarity agrees well with
the predicted value $\sqrt{10/3}=1.826$.  The BEC limit
$\omega_r/\omega_{\perp}=2$ is approached from {\it above}
due to the positive correction to the ground state chemical potential
beyond mean field, predicted by Lee, Huang and Yang. The monotonic
line is the result of an extended BCS mean field theory. Reprinted with permission from \textcite{Altmeyer:2007}. \label{fig:LHY}}
\end{figure}

\paragraph*{Collective modes}
Collective modes in a harmonic trap have been a major tool to
study cold gases in the BEC regime, where the dynamics
is determined by superfluid hydrodynamics. For a mixture
of fermions with an adjustable attractive interaction, the corresponding
eigenfrequencies have been worked out by \textcite{Stringari:2004} and
by \textcite{Heiselberg:2004}.
For a highly elongated trap, where the axial trap frequency $\omega_z$ is
much smaller the radial confinement frequency $\omega_{\perp}$, 
two important eigenmodes are the axial and radial compression modes, 
with respective frequencies \cite{Heiselberg:2004, Stringari:2004}
\begin{equation}
\label{eq:breathingmode}
\omega_B=\omega_z\,\sqrt{3-\frac{1}{\gamma +1}}\;\;\;\text{and}\;\;\;
\omega_r=\omega_{\perp}\,\sqrt{2(\gamma +1)}\, .
\end{equation}
They are completely determined by the effective polytropic index
$\gamma=d\ln{p}/d\ln{n} -1$ and thus give information about the equation of state $p(n)$
along the crossover. Since $\gamma=2/3$ exactly for the unitary Fermi gas, one obtains
universal numbers for these frequencies precisely at the Feshbach resonance,
as pointed out by \textcite{Stringari:2004}. In particular the radial compression
mode frequency $\omega_r$ is equal to $\omega_r=\sqrt{10/3}\,\omega_{\perp}$
in the BCS-limit and at unitarity, while it approaches $\omega_r=2\,\omega_{\perp}$
in the BEC-limit, where $\gamma=1$. The fact that the chemical potential
Eq.~(\ref{eq:mu_B}) of a dilute repulsive Bose gas is larger than its
mean-field limit $\mu_{\rm Bose}\sim n$, implies $\gamma>1$, i.e.  the BEC-limit is
reached from {\it above}. The expected nonmonotonic behavior of $\omega_r$
as a function of $1/k_Fa$ has recently been observed by \textcite{Altmeyer:2007}.
Remarkably, it provides the first quantitative test of the
Lee-Huang-Yang correction to the chemical potential Eq.~(\ref{eq:mu_B})
of a dilute repulsive Bose gas (of dimers) (see Fig.~\ref{fig:LHY}).

The issue of damping of the collective modes
in the BCS-BEC crossover has raised a number of 
questions, which - remarkably - are connected with 
recent developments in QCD and field theory. As  
pointed out by \textcite{Gelman:2004}, 
not only the thermodynamics but also dynamical properties like the
kinetic coefficients are described by universal scaling functions 
at the unitarity point. An example is the shear viscosity $\eta$
which determines the damping of sound and collective oscillations
in trapped gases.  At unitarity, its dependence on
density $n$ and temperature $T$ is fixed by dimensional arguments to be
$\eta=\hbar n\alpha(T/\mu)$, where $\mu$ is the chemical potential
and $\alpha(x)$ a dimensionless universal function.
At zero temperature, in particular, $\eta(T=0)=\alpha_{\eta}\hbar n$
is linear in the density, which defines a 'quantum viscosity' coefficient 
$\alpha_{\eta}$. From a simple fluctuation-dissipation type argument 
in the {\it normal} phase, a lower bound $\alpha_{\eta}\geq 1/6\pi$ 
was derived by \textcite{Gelman:2004}.  Assuming that a hydrodynamic 
description applies, the effective shear viscosity can be inferred from
the damping rate of the collective modes. In particular, the damping of
the axial mode near the unitarity limit in the experiments by
\textcite{Bartenstein:2004b} gives $\alpha_{\eta}\approx 0.3$ at the lowest attainable temperatures.
Based on these results, it has been speculated that ultracold atoms near
a Feshbach resonance are a nearly perfect liquid \cite{Gelman:2004}.
A proper definition of viscosity coefficients like $\eta$, however,
requires hydrodynamic damping. At zero temperature, this is usually not
the case.  It is valid at $T=0$ in one dimension, where exact results on 
$\alpha_{\eta}$ have been obtained for arbitrary coupling 
along the BCS-BEC crossover \cite{Punk:2006}.

At finite temperature, damping is typically associated with the 
presence of thermally excited quasiparticles, which also give rise to a 
nonzero entropy density $s(T)$. It has been shown by
\textcite{Kovtun:2005} that in a rather special class of 
relativistic field theories,  which are dual to some string theory,
the ratio $\eta(T)/s(T)=\hbar/(4\pi k_B)$ 
has a universal value. For more general models, this value 
is conjectured to provide a lower bound on $\eta/s$ for
a vanishing chemical potential, i.e. effectively in the 
zero density limit.  Since the ratio does not involve the 
velocity of light, the string theory bound on $\eta/s$ may 
apply also to non-relativistic
systems like the unitary Fermi gas, which lack an intrinsic
scale beyond temperature and density.  From Eq.~(\ref{eq:entropy}),
both entropy density and shear viscosity should therefore vanish like
$T^3$ in the low temperature, superfluid regime. Recent 
measurements of this ratio, using the damping of the radial 
breathing mode in a strongly aniosotropic trap have been 
performed by \textcite{Turlapov:2007} as a function of 
the energy of the gas. The results indicate a very low viscosity
of the unitary gas
in both the superfluid and normal regime, and a ratio
$\eta/s$ which is indeed of order $\hbar/k_B$. Exact 
results on {\it bulk} rather than shear viscosities 
have been obtained by \textcite{Son:2007}.
He noted that the unitary Fermi gas exhibits a conformal symmetry which constrains
the phenomenological coefficients in the dissipative part of the stress tensor. In particular, the bulk viscosity vanishes identically in the normal state and thus
no entropy is generated in a uniform expansion. 
In the superfluid phase, which is quite generally characterized by 
the shear plus three different bulk viscosities \cite{Forster:1975}, this result implies
that two of the bulk viscosity coefficients vanish at unitarity, while
one of them may still be finite.

\begin{figure}
	\includegraphics[width=0.9\columnwidth]{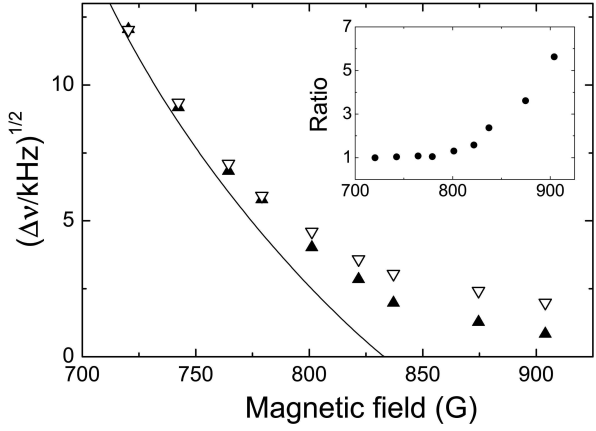}
\caption{Effective pairing gap in $^6$Li from rf-spectroscopy as a function of
magnetic field. The solid line is the two-particle binding
energy, which vanishes at the Feshbach resonance at $B_0=834\,$G
coming from the BEC side. The open and closed symbols
are for Fermi temperatures $T_F=3.6\,\mu$K and $T_F=1.2\,\mu$K
respectively. The ratio of the effective pairing gaps
strongly depends on $T_F$ on the BCS side (inset). Reprinted with permission from \textcite{Chin:2004}. \label{fig:gap}}
\end{figure}

\paragraph*{Rf-spectroscopy}
A microscopic signature of pairing between fermions is provided by
rf-spectroscopy of the gap on the BCS-side of the crossover,
where no bound states exist in the absence of a Fermi sea. 
This was first suggested by \textcite{Torma:2000}. 
Following earlier work by \textcite{Regal:2003a} and \textcite{Gupta:2003}
in the non-superfluid regime,
such an experiment has been performed by \textcite{Chin:2004}.
An rf-field with frequency $\omega_L$ is used to drive transitions between 
one of the hyperfine states $|2\rangle=|\!\!\downarrow\rangle$ which is 
involved in the pairing and an empty hyperfine state $|3\rangle$ which lies 
above it by an energy $\hbar\omega_{23}$  due to the magnetic field splitting 
of the bare atom hyperfine levels. In the absence of any interactions, the spectrum
exhibits a sharp peak at $\omega_L=\omega_{23}$. The presence of attractive interactions
between the two lowest hyperfine states $|1\rangle$ and $|2\rangle$ will
lead to an upward shift of this resonance. In a molecular picture, which is valid
far on the BEC-side of the crossover, this shift is expected to coincide with
the two-particle binding energy $\varepsilon_b$. Indeed, it was shown by 
\textcite{Julienne:2005}, that both scattering lengths and molecular binding 
energies may be extracted from rf-spectra of weakly bound molecules.
For a quantitative analysis, however, it is important even at the level of
single molecules, to properly account 
for the nonvanishing interaction $a_{13,23}\ne 0$ of atoms in state 
$|3\rangle$ with those in the initial states $|1\rangle$ and $|2\rangle$
forming the molecule \cite{Gupta:2003}.
In the experiments of \textcite{Chin:2004}, the rf-spectrum exhibits
a dominant free atom peak centered at $\omega_L=\omega_{23}$ for temperatures
$T\approx T_F$. At low temperatures, $T\lesssim 0.2\, T_F$ where the gas is superfluid,
an additional peak is observed, which is shifted with respect to the free transition. 
As shown in Figure~\ref{fig:gap},
the shift essentially follows the two-particle binding energy on the BEC-side of the crossover
but stays finite on the BCS-side $a<0$. In particular, the size of the shift near unitarity
increases with the Fermi energy because the formation of bound pairs
is a many-body effect. A theoretical analysis of these observations, which is 
based on an extended BCS description of pairing generalized to finite
temperature within a T-matrix formalism, has been given
by \textcite{Kinnunen:2004} and \textcite{He:2005, Ohashi:2005}. 
Including the necessary average over the inhomogeneous gap parameter 
$\Delta(\mathbf{x})$ in a harmonic trap,
reasonable agreement with the experimentally observed spectra has
been obtained. An important point to realize is that the strong attractive 
interactions near unitarity lead to an effective 'pairing gap' already in the normal state above 
$T_c$. The RF-shift is therefore not a direct measure of the superfluid 
order parameter \cite{Kinnunen:2004, He:2005}.

The models discussed above rely on the assumption that 
interactions with the final state $|3\rangle$ are negligible.
When this is not the case, the results can change significantly.
For instance,  when the interaction constants $g_{\sigma\sigma'}$
between the states $|\sigma=1,2,3\rangle$ which are involved are identical,
neither mean field ('clock')
shifts nor the effects of pairing show up in the long wavelength rf-spectrum
\cite{Yu:2006}. A theory for the average frequency shift of the rf-spectra of homogeneous
gases at zero temperature, which takes into account both many-body
correlations and the interactions with the final state $|3\rangle$ has recently been
given by \textcite{Baym:2007} and \textcite{Punk:2007}. Near $T=0$, where only a 
single peak is observed in the rf-spectrum, its position can be determined from 
a sum rule approach. Introducing the detuning $\omega=\omega_L-\omega_{23}$
of the RF field from the bare $|2\rangle-|3\rangle$ transition, the average clock shift 
\begin{equation}
\label{eq:rf-shift}
\hbar\bar{\omega}=\left(\frac{1}{g_{12}}-\frac{1}{g_{13}}\right)
\frac{1}{n_2} \frac{\partial (-u)}{\partial g_{12}^{-1}}
\end{equation}
can be expressed in terms of the derivative of the ground state energy density
$u$ with respect to the inverse of the renormalized coupling constant   
$g_{12}= 4 \pi \hbar^2 a_{12}/M$ \cite{Baym:2007}. The expression is finite for all coupling strengths 
$g_{12}$ and evolves smoothly from the BCS- to the BEC-limit. In particular,
the average clock shift vanishes, if $g_{12}=g_{13}$ \cite{Zwierlein:2003b}. 
This is directly connected with the result mentioned above,
because the interaction between states $|2\rangle$  and $|3\rangle$ drops out
for the average shift $\bar\omega$.  
For negligible populations of the state $|3\rangle$,   
the derivative $\partial (-u)/\partial g_{12}^{-1}=\hbar^4\, C/M^2$ can be
expressed in terms of the constant $C$ which characterizes the asymptotic behavior 
$\lim n_{\mathbf{k}}=C/k^4$ of the momentum distribution at large momenta
of the crossover problem in the $|1\rangle -|2\rangle$ channel \cite{Punk:2007}.
Within an extended BCS-description of the ground state wavefunction, the 
constant $\hbar^4\, C_{\rm BCS}/M^2\equiv\Delta^2$ is precisely the square of 
the gap parameter.  In the BCS-limit, Eq.~(\ref{eq:rf-shift}) thus reproduces 
the weak coupling result obtained by \textcite{Yu:2006}. In the BEC-limit, 
where the BCS ground state becomes exact again, the asymptotic behavior 
$\Delta_{BEC}=4\varepsilon_F/\sqrt{3\pi k_Fa_{12}}$ gives 
$\hbar\bar{\omega}=2\varepsilon_b(1-a_{12}/a_{13})$,
with $\varepsilon_b=\hbar^2/Ma_{12}^2$ the two-particle binding energy.  
This agrees precisely with the first moment of the spectrum of a bound-free transition
in the molecular limit \cite{Julienne:2005}. The dependence on $k_F$ drops 
out, as it must. The most interesting regime is that around the unitarity point  
$1/g_{12}=0$, where the average rf-shift is simply
given by $\bar{\omega}=-0.46\, v_F/a_{13}$. The prefactor is obtained from 
a numerical evaluation of the derivative in Eq.~(\ref{eq:rf-shift}) \cite{Baym:2007}
or - equivalently - the constant $C$ in the momentum distribution at the unitarity 
point \cite{Punk:2007} (note that the dependence $\bar{\omega}\sim -\, v_F/a_{13}$ 
at unitarity also holds in an extended BCS description, however the numerical 
factor $0.56$ differs from the exact value). Compared
with locally resolved rf-spectra, which were measured recently by \textcite{Shin:2007},
the predicted average shift in the case of $^6$Li is almost twice
the observed peak position. This is probably due to the fact that $\bar{\omega}$ 
has a  considerable contribution from the higher frequency part of the spectrum.  
An unexpected prediction of Eq.~(\ref{eq:rf-shift})
is the {\it linear}  dependence of the shift on the Fermi 
wavevector $k_F$ at unitarity. Experimentally, the spatial resolution necessary
to distinguish this from the naive scaling proportional to $\varepsilon_F$
has not yet been achieved \cite{Shin:2007}. 

\paragraph*{Vortices}

\begin{figure}
	\includegraphics[width=0.95\columnwidth]{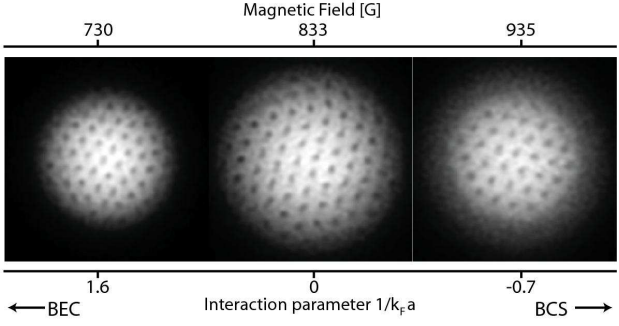}
\caption{Vortex lattice in a rotating gas of $^6$Li precisely at the Feshbach resonance
and on the BEC and BCS side. Reprinted with permission from \textcite{Zwierlein:2005}. \label{fig:vortexlattice}}
\end{figure}

While the appearance of a pairing gap in the rf-spectrum
is a strong indication for superfluidity, it is not a conclusive proof.
Indeed, pairing effects appear in the normal state above $T_c$
as a precursor to superfluidity or may be present even at zero
temperature in unbalanced Fermi gases above the Clogston-Chandrasekhar
limit, as observed very recently by \textcite{Schunck:2007}. 
A crucial step, which verifies the existence of a paired
fermionic superfluid was thus the observation of perfect triangular
vortex lattices in rotating Fermi gases near unitarity by \textcite{Zwierlein:2005}
(see Fig.~\ref{fig:vortexlattice}). 
Vortex lattices require conservation of vorticity,
which is a consequence of superfluid hydrodynamics.
The regularity of the lattice shows that all vortices have the same
vorticity. Since it is a superfluid of fermionic {\it pairs}, the expected
circulation per vortex is $h/2M$. This is indeed the value found from
equating the total circulation at a given stirring frequency
with the number of vortices and the transverse area of the cloud.

\section{PERSPECTIVES}
\label{sec:perspectives}

From the examples discussed in this review, it is evident that cold atoms provide a novel tool
to study the physics of strong correlations in a widely tunable range and in
unprecedentedly clean sytems. Basic models in many-body physics
like the Hubbard-model with on-site interaction or the Haldane pseudopotentials
for physics in the lowest Landau level,
which were originally introduced in a condensed matter context as an
idealized description of strong correlation effects in real materials
can now be applied on a quantitative level. In the 
following, a brief outline of possible future directions is given.

\subsection{Quantum magnetism}

One of the major challenges with ultracold atoms in optical lattices in the near future lies in the
realization and study of configurations, which can serve as tunable model systems for basic problems in quantum magnetism. For two component mixtures of bosonic or fermionic atoms in an optical lattice
which are labelled by a pseudo-spin variable $\spinupstate,\spindownstate$ and dominating interactions $U\gg J$ between them, second order perturbation theory in the kinetic energy term allows one to map the corresponding (Bose-)Hubbard Hamiltonian to an anisotropic Heisenberg model (XXZ model) with effective spin-spin interactions between atoms on neighboring lattice sites:

\begin{equation}
  \sum_{\Rvec,\Rvec'} \left [ J^{z}_{ex} S^z_\Rvec S^z_{\Rvec'} \pm J^{\perp}_{ex}\left( S^x_\Rvec S^x_{\Rvec'} + S^y_\Rvec S^y_{\Rvec'}\right)\right]
\label{eq:spin_hamiltonian}	
\end{equation}

where the +(-) sign holds for the case of fermions or bosons, respectively \cite{Duan:2003,Kuklov:2003}. By tuning the exchange terms $J^{z}_{ex}= 2(J_\uparrow^2+J_\downarrow^2)/U_{\uparrow\downarrow}-4J^2_\uparrow/U_{\uparrow\uparrow}-4J^2_\downarrow/U_{\downarrow\downarrow}$  and $J^{\perp}_{ex}= 4 J_{\uparrow}J_{\downarrow}/U_{\uparrow\downarrow}$ via spin-dependent tunneling amplitudes or spin-dependent interactions, one can change between various quantum phases. One intriguing possibility, pointed out by \textcite{Kuklov:2003} is the realization of supercounterflow, i.e. superfluidity in the relative motion of the two components, while the system is a Mott insulator, as far as the total density is concerned. Another possibility is the formation of topological quantum phases, arising for different exchange coupling along different lattice directions \textcite{Kitaev:2006}, which can be realized with optical lattices \textcite{Duan:2003}. The most natural case occurs, however, for equal tunneling amplitudes and onsite interactions, which yields an isotropic Heisenberg type spin-hamiltonian:

\begin{equation}
	H=J_{ex} \sum_{\Rvec,\Rvec'} \mathbf{S_\Rvec} \cdot \mathbf{S_{\Rvec'}}.
\end{equation}

Here the superexchange coupling $J_{ex}=\pm 4J^2/U_{\uparrow\downarrow}$ has a positive (negative) sign for the case of fermions (bosons) and thus favors antiferromagnetically (ferromagnetically) ordered phases. In the case of fermions, it can easily be understood why the antiferromagnetically ordered phase is preferred: an initial spin-triplet state cannot lower its energy via second order hopping processes, as the two spins pointing in the same direction can never be placed on a single lattice site. The spin-singlet term, however, is not subject to this restriction and can lower its energy via a second order exchange hopping process \cite{Auerbachbook}. Recently, the tunnelling dynamics based on second order hopping events has been observed in an array of tightly confining double wells \cite{Foelling:2007}. Remarkably, the observed second order coupling strengths $\sim J^2/hU$ can be almost on the order of one kHz and thus an order of magnitude larger than the direct magnetic dipole interaction between two alkali atoms on neighboring lattice sites. 

The observation of such magnetically ordered quantum phases requires to reach very low temperatures $k_B T< 4 J^2/U$ (for $J/U\ll 1$) in the experiment. \textcite{Werner:2005} have shown that such temperatures could be within reach experimentally, as a Pomeranchuk type cooling effect during the loading of the atoms in the lattice assists in a cooling of the atoms. In particular, it turns out that
for an initial temperature of a homogeneous two-component Fermi gas below
$T/T_F \simeq 0.08$, an antiferromagnetically ordered phase could be reached at unit filling, when the optical lattice potential is ramped up adiabatically.
A more robust implementation of spin Hamiltonians might be reached via ground state polar molecules, for which much larger spin-spin coupling strengths have been recently found to be present \cite{Micheli:2006}. Such nearest neighbor spin interactions would be mediated via a long range electric dipole-dipole interaction between the molecules. Several experimental groups are currently pursuing the goal of creating such ground state polar molecules out of ultracold atoms via Feshbach sweeps and subsequent photoassociation or by directly slowing and sympathetically cooling stable polar molecules. Noise correlation or Bragg spectroscopy could allow one to uniquely identify such antiferromagnetically ordered phases. 

From a condensed matter point of view, one of the most challenging
problems is the realization and study of the fermionic repulsive
Hubbard model in an optical lattice with adjustable interactions and
filling fraction, in particular in 2D. As discussed by
\textcite{Hofstetter:2002} such a system would constitute a cold atom version of one of the
most intensely studied models in condensed matter physics,
allowing to access unconventional normal and
$d$-wave superconducting phases as found in the high temperature superconductors
 \cite{Lee:2006}. Of course, realization of these models with cold atoms would not solve the
latter problem, however it would be an extremely valuable tool to test
some of the still open issues in this field.  In this context,
specific proposals have been made for realizing so-called
resonating valence bond states by adiabatically transforming spin
patterns in optical superlattices \cite{Trebst:2006}. More complex spin-liquid states might be created by  enforcing 'frustration'  in antiferromagnetically ordered phases through triangular or Kagome type lattices \cite{Santos:2004}.
For a detailed review of these models, see \textcite{Lewenstein:2007}. 

Quantum impurity problems, which have played an important role in
the study of magnetism, may be realized in the cold atom context by
confining single atoms in a tight optical trap or in a
deep optical lattice. For hard-core bosons or fermions, the 
effective pseudo-spin one half associated with the possible local
occupation numbers $n=0,1$ can be coupled 
to a reservoir of either a BEC or a degenerate Fermi gas, 
using Raman transitions \cite{Recati:2005}. 

\subsection{Disorder}
It was noted by \textcite{Anderson:1958a} that waves in a 
medium with a static ('quenched') randomness may become 
localized due to constructive interference between 
multiply reflected waves. Qualitatively, this happens 
below a 'mobility edge', where the mean free
path $\ell$ (calculated e.g. in a Born approximation treatment of
the scattering by the disorder potential) becomes smaller than
the wavelength $\lambda$. This so-called Ioffe-Regel
criterion applies in a 3D situation with short range disorder 
and in the absence of interactions
\footnote{In 1D and 2D, an arbitrary weak disorder
leads to localization, i.e. even waves with $\lambda\ll\ell$
are localized.}. In the interacting case, where both 
localization due to disorder (Anderson) or due to interactions 
(Mott) are possible, the problem is 
still not well understood (see e.g. the reviews by \textcite{Belitz:1994}
and \textcite{Basko:2006}). Cold atoms thus provide
a novel tool to investigate the localization problem, in 
particular since the interactions are 
tunable over a wide range.  

For non-interacting bosons, the ground state in the presence 
of disorder is trivially obtained by putting all the particles
in the lowest single-particle level of the random potential.
Adding weak repulsive interactions, a finite number of 
localized states in the so-called Lifschitz tails of the  
single-particle spectrum \cite{Lifschitz:1988} will be occupied.
As long as the chemical potential is in this low energy range,
these states have negligible spatial overlap. For very low 
densities, repulsive bosons are therefore expected to form a 
'Lifshitz glass' of fragmented, local condensates \cite{Lugan:2007}. 
With increasing densities, these local condensates will 
be coupled by Josephson tunneling and eventually form
a superfluid, where coherence is established over the whole
sample. A quantitative analysis of this transition was first given 
by \textcite{Giamarchi:1988} in the particular case of one dimension.  
Using a quantum hydrodynamic, Luttinger 
liquid description, they found that weak interactions tend to 
suppress the effect of Anderson localization (this effect is also 
present in the case of a commensurate filling in an optical lattice, 
where also a Mott-insulating phase appears, see \textcite{Rapsch:1999}). 
Specifically, weak disorder does not destroy the superfluid, 
provided the Luttinger exponent $K$
introduced in section V.B is larger than $3/2$. Within the Lieb-Liniger
model with an effective coupling constant $\gamma$ defined in Eq.~(\ref{eq:gamma_1}),
this requires $\gamma\lesssim 8$. In the opposite regime 
$K<3/2$ of low densities or strong interactions near the Tonks-Girardeau limit, 
even weak disorder destroys the superfluid.  The ground state then 
lacks long range phase coherence, consistent with the picture of a 
'Lifshitz glass' discussed above.  Exact results for the momentum
distribution and the local density of states have been obtained by 
\textcite{DeMartino:2005} 
in the special limit of the Tonks-Girardeau gas, 
where the problem is equivalent to non-interacting fermions 
(see section V.B).  Interacting bosons 
in higher dimensions were studied by \textcite{Fisher:1989},
using the Bose-Hubbard model Eq.~(\ref{eq:BHM}).
To account for disorder, the on-site energies $\epsilon_{\mathbf{R}}$,
are assumed to have a random component, with zero average and
finite variance
\begin{equation}
\label{eq:disorder}
\langle(\epsilon_{\mathbf{R}}-\langle\epsilon_{\mathbf{R}}\rangle)
(\epsilon_{\mathbf{R}'}-\langle\epsilon_{\mathbf{R}'}\rangle)\rangle=
\Delta\cdot\delta_{\mathbf{R},\mathbf{R}'}\, ,
\end{equation}
where $\Delta$ is a measure of the strength of the disorder. It has been shown
by \textcite{Fisher:1989}, that even at weak disorder $\Delta<U/2$  a novel
so-called Bose glass phase
appears, which separates the SF- and MI-states
\footnote{The 'Lifshitz glass' mentioned above may be thought of as the 
low density or strong disorder limit of the Bose glass.}. 
At strong disorder $\Delta>U/2$,
the MI-states are destroyed completely.  The Bose glass is characterized by a
vanishing superfluid and condensate density. It will thus show no
sharp interference peaks in a time-of-flight experiment after release 
from an optical lattice (see section IV.B). 
In contrast to the MI-phase, the Bose glass has both a finite 
compressibility (i.e. there is no shell structure in a trap) and a continuous 
excitation spectrum. 

A concrete proposal for studying localization effects with cold atoms
has been made by \textcite{Damski:2003a},
who suggested to use laser speckle patterns as a means to realize the 
frozen disorder. Such patterns have been employed in this 
context by the groups in Florence \cite{Lye:2005} 
and in Orsay \cite{Clement:2005}. They are produced by a laser beam, 
which is scattered from a ground glass diffuser
\cite{Clement:2006}. The speckle pattern has a random intensity $I$, 
which is exponentially distributed $P(I)\sim\exp{(-I/\langle I\rangle)}$.
The rms intensity fluctuation $\sigma_I$ is thus equal to the 
average intensity $\langle I\rangle$. The fluctuations in the
intensity give rise to a random value of the optical dipole potential
in Eq.~(\ref{eq:dippot}), which is experienced by the atoms. By varying 
the detuning $\Delta$, this random potential (purely repulsive for $\Delta>0$) 
can be tuned in the range between zero and around $h\cdot 4$ kHz
\cite{Clement:2006}. An important parameter characterizing the speckle pattern
is the spatial correlation length $\sigma_R$, which is the scale over which
$\langle I(\mathbf{x}) I(0)\rangle$ decays to $\langle I\rangle^2$. When the 
optical setup is diffraction limited, the smallest achievable $\sigma_R$'s can
take values down to one micrometer \cite{Clement:2006}, 
which is comparable to typical healing lengths $\xi$ in dilute gases.
Reaching this limit is important, since smooth disorder potentials with 
$\sigma_R\gg\xi$ are not suitable
to study Anderson localization in expanding BEC's.  Indeed, the typical
range of momenta after expansion reaches up to $k_{\rm max}\approx 1/\xi$.
For  $\sigma_R\gg\xi$ therefore, the spectral range of the disorder is 
much smaller than the momenta of the matter waves. Even
speckle patterns with long correlation lengths, however, 
can lead to a strong suppression 
of the axial expansion of an elongated
BEC. This was observed experimentally by \textcite{Clement:2005, Fort:2005, Schulte:2005}.
The effect is not due to Anderson localization, however. Instead, it is caused by
classical total reflection, because during expansion 
the density and chemical potential of the gas decrease. 
Eventually, therefore, the matter waves have energies below the typical
disorder potentials and the gas undergoes fragmentation.
A suggestion to realize Anderson localization of 
non-interacting particles in 1D has recently been made by \textcite{Sanchez:2007}.
It is based on a 1D BEC, which after expansion is transformed into 
a distribution of free matter waves with momenta up to $1/\xi$. 
For short range disorder with $\sigma_R<\xi$, these waves will all be localized, even for rather weak disorder. 

A rather direct approach to study the interplay between disorder and strong interactions has recently been followed by the group in Florence, using  bi-chromatic 
optical lattices \cite{Fallani:2007}. Such lattices provide a pseudo-random 
potential if the lattice periods are incommensurate. Adding the
incommensurate lattice in the superfluid state, 
a strong suppression of the interference pattern is found. 
Starting from a MI-phase, the sharp excitation spectrum is smeared 
out with increasing amplitude of the incommensurate lattice. Both 
observations are consistent with the presence of 
a Bose glass phase for a strong incommensurate lattice potential \cite{Fallani:2007}. 
A different method to realize short range disorder in cold gases has been suggested by 
\textcite{Gavish:2005}. It employs a two species mixture of atoms in an optical 
lattice. Due to a finite inter-species scattering length, the component 
which is free to move around in the lattice  
experiences an on-site random potential of the type ~(\ref{eq:disorder}),
provided the atoms of the different species or spin state are frozen at random sites. 
The interplay between disorder and interactions is, of course, also 
an intriguing problem for fermions. A quantitative phase
diagram for short range disorder plus repulsive interactions has been
determined in this case by \textcite{Byczuk:2005}. So far, however, no experiments have been done in this direction.

\subsection{Nonequilibrium Dynamics}
A unique feature of many-body physics with cold atoms is the 
possibility to modify both the interactions and the external potentials
dynamically. In the context of the SF-MI transition, this has been
discussed already in section IV.C. Below, we give a brief outline
of some of the recent developments in this area.

The basic question about the efficiency of collisions in 
establishing a new equilibrium from an initial out-of-equilibrium 
state has been adressed by \textcite{Kinoshita:2006} for 
Bose gases in one dimension. An array of several thousand
1D tubes created by a strong 2D optical lattice (see section V.B) 
was subject to a pulsed optical lattice along the axial direction.
The zero momentum state is thus depleted and essentially all 
the atoms are transferred to momenta $\pm 2\hbar k$, where 
$k$ is the wavevector of the pulsed axial optical lattice. The two wavepackets
in each tube separate and then recollide again after a time $\pi/\omega_0$, which is 
half the oscillation period in the harmonic axial trap with frequency 
$\omega_0$.  The associated collision energy $(2\hbar k)^2/M$ was 
around $0.45\,\hbar\omega_{\perp}$, i.e. much smaller than the
minimum energy $2\hbar\omega_{\perp}$ necessary to excite
higher transverse modes. The system thus remains strictly 1D 
during its time evolution. It was found, that even after 
several hundred oscillation periods, the initial non-equilibrium 
momentum distribution was preserved. 

This striking observation
raises a number of questions. Specifically, is the absence of a 
broadening of the momentum distribution connected 
with the integrability of the 1D Bose gas ? More generally, 
one can ask, whether and under which conditions,
the unitary time-evolution of a non-equilibrium initial state in a 
strongly interacting but non-integrable quantum system will evolve into 
a state, in which at least one- and two-particle correlations are stationary
and the information about the precise initial conditions is hidden
in some - in practice unobservable - high order correlations.
The non-equilibrium dynamics for the integrable case of bosons in 1D 
has recently been addressed by \textcite{Rigol:2007}. Taking as a model a
Tonks-Girardeau gas on a lattice, they have shown numerically 
that the momentum distribution at large times is well described by 
that of an 'equilibrium' state with 
a density matrix of the form $\hat{\rho}\sim\exp{(-\sum_m\lambda_m\hat{A}_m)}$.
Here, the $\hat{A}_m$ denote the full set of conserved quantities,
which are known explicitely for the Tonks-Girardeau gas because it 
is equivalent to free fermions. The Langrange multipliers $\lambda_m$ 
are fixed by the initial 
conditions of given expectation values $\langle\hat{A}_m\rangle$
at $t=0$. This is the standard procedure in statistical physics,
where equilibrium is described microscopically as a state of 
maximum entropy consistent with the given 'macroscopic' data
 $\langle\hat{A}_m\rangle$ \cite{Balian:1991}. In particular,
an initially double-peaked momentum distribution is
preserved in the stationary state described by the maximum
entropy density operator $\hat{\rho}$ \cite{Rigol:2007}. 
The hypothesis, that the absence of momentum relaxation 
is related to the integrability of the 1D Bose 
gas, could be tested by going to higher momenta $k$, where the 1D 
scattering amplitude is no longer given by the low energy form
of Eq.~(\ref{eq:1Ds-ampl}). For such a case, the pseudopotential 
approximation breaks down and three body collisions or longer ranged 
interactions can become relevant. For the extreme case of two free 3D 
colliding BECs, equilibration has indeed been observed to set in after
a few collisions \cite{Kinoshita:2006}. 

In non-integrable systems
like the 1D Bose-Hubbard model, where no conserved 
quantities exist beyond the energy, the standard reasoning of
statistical physics gives rise to a micro-canonical density 
operator, whose equivalent 'temperature'  is set by the energy of the initial state
\footnote{In the semiclassical limit, this 'eigenstate thermalization 
hypothesis' can be derived under relatively weak assumptions,
see \textcite{Srednicki:1994}.}. 
Note that the micro-state $\exp{(-i\hat{H}t/\hbar)}\vert\psi(0)\rangle$, 
which evolves from the initial state by the unitary time evolution
of a closed system, remains time dependent. Its statistical (von Neumann)
entropy vanishes. By contrast, the microcanonical density operator describes a 
stationary situation, with a nonzero {\it thermodynamic} entropy. It is determined
by the number of energy eigenstates near the exact initial energy which are
accessible in an energy range much smaller than the microscopic scale set by
the one- or two two-body terms in the Hamiltonian but much larger that 
the inverse of the recurrence time. For 1D problems, the hypothesis 
that simple macroscopic observables are eventually well desribed by such a 
stationary density operator can be tested
quantitatively by using the adaptive time dependent 
density matrix renormalization group \cite{White:2004, Daley:2004}.
In the case of the 1D BHM, an effectively 'thermal' stationary 
state indeed arises for long time dynamics after a quench 
from the SF to the MI \cite{Kollath:2007} (see section IV.C).  Apparently, however, the description by a stationary 'thermal' density operator is valid only for not too large values of the final repulsion $U_f$.

A different aspect of the non-equilibrium 
dynamics of 1D Bose gases was studied by \textcite{Hofferberth:2007}.
A single 1D condensate formed in a 
magnetic micro trap on an atom chip \cite{Folman:2002}
is split into two parts by applying rf-potentials \cite{Schumm:2005}
The splitting process is done in a phase coherent manner, such that 
at time $t=0$, the two condensates have  
a vanishing relative phase. They are kept in 
a double well potential for a time $t$ and then 
released from the trap. As discussed in section III.C,
the resulting interfence pattern provides information
about the statistics of the interference amplitude. In the 
non-equilibrium situation discussed here, the relevant 
observable analogous to Eq.~(\ref{eq:observable})
is the operator $\exp{(i\hat{\theta}(z,t))}$ integrated along the
axial $z$-direction of the two condensates. Here 
$\hat{\theta}(z,t)$ is the time-dependent phase {\it difference}
between the two independently fluctuating condensates. 
Using the quantum hydrodynamic Hamiltonian Eq.~(\ref{eq:1DQHD}),
it has been shown by \textcite{Burkov:2007}
that the expectation value of this operator decays sub-exponentially
for large times $t\gg\hbar/k_BT$ where the phase fluctuations can be 
described classically. This behavior is in rather good agreement with 
experiments \cite{Hofferberth:2007}. In particular, it allows to determine 
the temperature of the 1D gas very precisely.

The dynamics of the superfluid gap parameter in attractive Fermi gases
after a sudden change of the coupling constant has been 
investigated by \textcite{Barankov:2004, Barankov:2006}
and \textcite{Yuzbashyan:2006} using the exactly integrable 
BCS Hamiltonian. Such changes in the coupling constant are
experimentally feasible by simply changing the magnetic field in a Feshbach 
resonance. Depending on the initial conditions, different 
regimes have been found where the gap parameter may 
oscillate without damping,  approaches an 'equilibrium'
value different from that associated with the coupling 
constant after the quench, or decays to zero monotonically 
if the coupling constant is reduced to very small values.   

\section*{Acknowledgments}

Over the past several years, a great number of people have
contributed to our understanding of many-body phenomena in cold
gases, too numerous to be listed individually. In particular, however,
we acknowledge E. Demler, F. Gerbier, Z. Hadzibabic, N. Nygaard, 
A. Polkovnikov, D. Petrov, C. Salomon, R. Seiringer and M. Zwierlein for valuable 
comments on the manuscript. 

Labaratoire Kastler et Brossel is a reserach unit 
of Ecole Normale Superieure, Universite Pierre et Marie Curie and CNRS.

\section{APPENDIX: BEC AND SUPERFLUIDITY}
\label{sec:appendix}

Following the early suggestion of London that superfluidity (SF) in
$^{4}$He has its origin in Bose-Einstein-Condensation (BEC), the
relation between both phenomena has been a subject of
considerable debate. In the following, we
outline their basic definitions and show that SF is the more general
phenomenon, which is both necessary and
sufficient for the existence of either
standard BEC or of quasi-condensates in low dimensional systems.
For a detailed discussion of the connections between BEC and
superfluidity see also the book by \textcite{Leggett:2006}.

The definition of BEC in an interacting system of bosons is based on
the properties of the one-particle density matrix  $\hat{\rho}_1$,
which is conveniently defined by its matrix elements
\begin{equation}\label{eq:rho1}
G^{(1)}(\mathbf{x},\mathbf{x}')=\langle\mathbf{x}'\vert\hat{\rho}_1\vert\mathbf{x}\rangle =
\langle\hat{\psi}^{\dagger}(\mathbf{x}) \hat{\psi}(\mathbf{x}')\rangle
\end{equation}
in position space.\footnote{In an inhomogeneous situation, it is sometimes convenient
to define a reduced one-particle density matrix $g^{(1)}$ by
$G^{(1)}(\mathbf{x},\mathbf{x}')=\sqrt{n(\mathbf{x})n(\mathbf{x}')}g^{(1)}(\mathbf{x},\mathbf{x}')$.}
As a hermitean operator with $\text{Tr}\,\hat{\rho}_1=
\int n(\mathbf{x})=N$, $\hat{\rho}_1$ has a complete set $\vert n\rangle$
of eigenstates with positive eigenvalues $\lambda_n^{(1)}$ which sum up
to $N$. As realized by \textcite{Penrose:1956}, the criterion for
BEC is that there is precisely one
eigenvalue $\lambda_0^{(1)}=N_0$ of order $N$ while all other
eigenvalues are non-extensive
\footnote{In principle it is also possible that more than one eigenvalue
is extensive. This leads to so-called fragmented BEC's which may appear
e.g. in multicomponent spinor condensates as shown by \textcite{Ho:1998}}.
Of course, the separation between extensive and non-extensive
eigenvalues $\lambda_n^{(1)}$ is well defined only in the thermodynamic
limit $N\to\infty$. In practice, however, the distinction between the BEC
and the normal phase above $T_c$
is rather sharp even for the typical particle numbers $N\approx 10^4-10^7$ of
cold atoms in a trap. This is true in spite of the fact that the non-extensive
eigenvalues are still rather large (see below).
The macroscopic eigenvalue $N_0$ determines the
number of particles in the condensate. In terms of the single particle eigenfunctions
$\varphi_n(\mathbf{x})=\langle\mathbf{x}\vert n\rangle$ associated
with the eigenstates $\vert n\rangle$ of $\hat{\rho}_1$,
the existence of a condensate is equivalent to a
macroscopic occupation of a single state created by the operator
$\hat{b}_0^{\dagger}=\int\varphi_0(\mathbf{x})\hat{\psi}^{\dagger}(\mathbf{x})$.
In a translation invariant situation, the eigenfunctions
are plane waves. The eigenvalues $\lambda_k^{(1)}$ are then just
the occupation numbers $\langle\hat{b}_k^{\dagger}\hat{b}_k\rangle$ in momentum
space. In the thermodynamic limit $N,V\to\infty$ at constant density $n$,
the one particle density matrix
\begin{equation}\label{eq:ODLRO}
\langle\mathbf{x}'\vert \hat{\rho}_1\vert\mathbf{x}\rangle =
n_0+\int_k\tilde{n}(\mathbf{k})e^{-i\mathbf{k}(\mathbf{x}-\mathbf{x}')}\to n_0
\end{equation}
thus approaches a finite value $n_0=\lim N_0/V$ as $r=\vert\mathbf{x}-\mathbf{x}'\vert\to\infty$.
This property is called off-diagonal long range order (ODLRO). Physically,
the existence of ODLRO implies that the states
in which a boson is removed from an $N\/$-particle system at positions $\mathbf{x}$
and $\mathbf{x}'$ have a finite overlap even in the limit when the separation
between the two points is taken to infinity
\footnote{For a  discussion of the topological properties of the many-body
wave function which are required to give ODLRO see
\textcite{Leggett:1973}}.
For cold gases, the presence of ODLRO below $T_c$, at least over a scale
of order $\mu$m, was observed experimentally by measuring the decay of the
interference contrast between atomic beams outcoupled from two points at a
distance $r$ from a BEC in a trap \cite{Bloch:2000}.
The momentum distribution $\tilde{n}(\mathbf{k})$ of non-condensate
particles is singular at small momenta. In the limit $\mathbf{k}\to 0$,
it behaves like $\tilde{n}(\mathbf{k})\sim c/k$ at zero
and $\tilde{n}(\mathbf{k})\sim T/k^2$ at finite temperature, respectively
\cite{Lifshitz:1980}. Since $k_{min}\sim N^{-1/3}$ in a finite
system, the lowest non-extensive eigenvalues of the one
particle density matrix are still very large, scaling like $\lambda_n^{(1)}\sim N^{1/3}$
at zero and $\lambda_n^{(1)}\sim N^{2/3}$ at finite temperature.

 The definition of BEC via the existence of ODLRO ~(\ref{eq:ODLRO})
is closely related with Feynman's intuitive picture of Bose-Einstein condensation as a transition,
below which bosons are involved in existse cycles of infinite size \cite{Feynman:1953}.
In fact, a precise connection exists with the superfluid density defined in Eq.~(\ref{eq:n_s}) 
below, which can be expressed in terms of the square of the winding number
in a Feynman path representation of the equilibrium density matrix
\cite{Pollock:1987}. As discussed by \textcite{Ceperley:1995} and 
\textcite{Holzmann:1999}, the connection between the {\it condensate} density 
and infinite cycles is also suggestive in terms of the path integral representation 
of the one-particle density matrix. 
It is difficult, however,  to put this on a rigorous footing beyond the
simple case of an ideal Bose gas \cite{Ueltschi:2006}. 

While a microscopic definition of BEC is straightforward at least
in principle, the notion of superfluidity is more subtle. On a
phenomenological level, the basic properties of superfluids may
be explained by introducing a complex order parameter
$\psi(\mathbf{x})=|\psi(\mathbf{x})|
\exp{i\phi(\mathbf{x})}$, whose magnitude
squared gives the superfluid density $n_s$, while the  phase $\phi(\mathbf{x})$
determines the superfluid velocity via $\mathbf{v}_s=\hbar/M\cdot\nabla\phi(\mathbf{x})$
\cite{Pitaevskii:2003}.
The latter equation immediately implies that
superfluid flow is irrotational and that the circulation
$\Gamma=\oint\mathbf{v_s}d\mathbf{s}$ is quantized in an integer
number of circulation quanta $h/M$.
Note, that these conclusions require a finite value $n_s\ne 0$, yet are
completely independent of its magnitude. In order to connect this
phenomenological picture of SF with the microscopic definition of BEC,
the most obvious assumption is to identify the order parameter
$\psi(\mathbf{x})$  with the eigenfunction $\varphi_0(\mathbf{x})$ of
$\hat{\rho}_1$ associated with the single extensive eigenvalue,
usually choosing a normalization such that
$\psi(\mathbf{x})=\sqrt{N_0}\varphi_0(\mathbf{x})$.
Within this assumption, BEC and SF appear as essentially identical
phenomena. This simple identification is not valid, however, beyond a
Gross-Pitaevskii description or for low dimensional systems.
The basic idea of how to define superfluidity in quite general terms goes
back to \textcite{Leggett:1973} and \textcite{Fisher:1973}.
It is based on considering the sensitivity of the many-body
wave function with respect to a change in the boundary condition (bc).
Specifically, consider $N\/$ bosons in a volume $V=L^3\/$ and choose
boundary conditions where the many-body wave function
\begin{equation}\label{eq:bc}
\Psi_{\Theta}\left( \mathbf{x}_1, \mathbf{x}_2, \ldots\mathbf{x}_i, \ldots \mathbf{x}_N\right)
\end{equation}
is multiplied by a pure phase factor $e^{i\Theta}$ if $\mathbf{x}_i\to\mathbf{x}_i+L\mathbf{e}$
for all $i=1\ldots N$ and with $\mathbf{e}$ a unit vector in one of the directions
\footnote{For simplicity, we assume isotropy in space. More generally the
sensitivity with respect to changes in the bc's may depend on the direction, in which case
the superfluid density becomes a tensor.}.
The dependence of the many-body energy eigenvalues
$E_n(\Theta)$ on $\Theta$ leads to a phase
dependent equilibrium free energy $F(\Theta)$. The difference
$\Delta F(\Theta)=F(\Theta)-F(\Theta=0)$ is thus a
measure for the sensitivity of the many-body
system in a thermal equilibrium state to a change in the bc's.
Since the eigenstates of a time reversal invariant Hamiltonian
can always be choosen real, the energies $E_n(\Theta)$
and the resulting  $\Delta F(\Theta)$ must be even in $\Theta$.
For small deviations $\Theta\ll 1$ from periodic
bc's, the expected leading behavior is therefore quadratic.
The superfluid density $n_s(T)$ is then defined
by the free energy difference per volume
\begin{equation}\label{eq:n_s}
\frac{\Delta F(\Theta)}{V}=\frac{\hbar^2}{2M}\, n_s(T)\cdot\left(\frac{\Theta}{L}\right)^2+\ldots
\end{equation}
to leading order in $\Theta$. In a superfluid, therefore,
a small change in the bc's leads to a change in the free energy
per volume which scales like $\gamma/2\,(\Theta/L)^2$. The
associated proportionality constant $\gamma=\hbar^2 n_s/M$
is called the helicity modulus.
Clearly, the definition (\ref{eq:n_s}) for superfluidity, which is based only on
{\it equilibrium} properties and also applies to finite systems, is
quite different from that for the existence of BEC.  Yet, it turns out, that
the two phenomena are intimately connected in the sense that
a finite superfluid density is both necessary and sufficient for
either standard BEC or the existence of quasicondensates in lower dimensions.

Following \textcite{Leggett:1973}, the physical meaning of the phase
$\Theta$ and the associated
definition of $n_s$ can be understood by considering bosons (gas or liquid)
in a superfluid state, which are enclosed between two concentric cylinders
 with almost equal radii $R$, co-rotating
with an angular frequency $\mathbf{\Omega}=\Omega\mathbf{e}_z$
\footnote{The walls are assumed to violate perfect cylindrical symmetry,
to allow for the transfer of angular momentum to the fluid!}.
In the rotating frame, the problem is stationary, however it
acquires an effective gauge potential
$\bs A=M\bs \Omega \wedge \mathbf{x}$, as shown in Eq.~(\ref{singlepartH}).
Formally, $\bs A$ can be eliminated by a gauge transformation at the expense
of a many-body wavefunction, which is no longer
single-valued. It changes by a factor $e^{i\Theta}$ under changes $\theta_i\to\theta_i+2\pi$
of the angular coordinate of each particle $i$, precisely as in (\ref{eq:bc}),
where $\Theta=-2\pi MR^2\Omega/\hbar$ is linear in the angular
frequency. The presence of a phase dependent free energy increase
$\Delta F(\Theta)$ of the superfluid in the rotating frame
implies, that the equilibrium state in this frame carries
a nonzero (kinematic) angular momentum $L_z'=-\partial\Delta F(\Theta)/\partial\Omega
=-(n_s/n)\cdot L_z^{(0)}$, where $L_z^{(0)}=I_{cl}\Omega$ is the
rigid body angular momentum in the lab frame and $I_{cl}=NMR^2$
the classical moment of inertia. A fraction $n_s/n$
of the superfluid thus appears to stay at rest in the lab frame
for small angular frequencies $\Omega\ll\hbar/MR^2$, where $\Theta\ll 1$.
As a result, the apparent moment of inertia is smaller than that of classical
rigid body like rotation.  Superfluidity,
as defined by Eq.~(\ref{eq:n_s}), thus implies the appearance
of non-classical rotational inertia (NCRI) \cite{Leggett:1973}.
In the context of cold gases, this phenomenon has been observed experimentally
by \textcite{Madison:2000a}. They have shown that a trapped gas in the
presence of a small, non-symmetric perturbation remains at zero angular
momentum (i.e. no vortex enters), for sufficiently small angular frequencies.

The example of a rotating system shows, that a finite value of
$\Theta$ is associated with a non-vanishing current in the system.
Indeed, a finite superfluid density in the sense of Eq.~(\ref{eq:n_s})
implies the existence of long range current-current correlations,
which is the original microscopic definition of $n_s$ by \textcite{Hohenberg:1965}.
To describe states of a superfluid with finite currents, it is
useful to introduce a smoothly varying local phase $\phi(\mathbf{x})$
on scales much larger than the interparticle spacing, which is connected with
the total phase difference between two arbitrary points by
$\Theta=\int d\mathbf{s}\cdot\mathbf{\nabla}\phi(\mathbf{x})$.
This local phase variable is now precisely the phase of the
coarse grained complex order parameter $\psi(\mathbf{x})$
introduced by Landau. This identification becomes evident by
noting that a non-vanishing phase gradient gives rise to a finite
superfluid velocity $\mathbf{v}_s=\hbar/M\cdot\nabla\phi(\mathbf{x})$.
The free energy increase
\begin{equation}\label{eq:DeltaF}
\Delta F(\Theta)=\frac{\gamma(T)}{2}\,\int d^3x \left(\nabla\phi(\mathbf{x})\right)^2
\end{equation}
associated with a change in the bc's is therefore just
the kinetic energy $M/2\,\int n_s(\mathbf{v}_s)^2$ of superfluid flow with
velocity $\mathbf{v}_s$ and density $n_s$.  An immediate consequence
of Eq.~(\ref{eq:DeltaF}) is the quite general form of the excitation
spectrum of superfluids at low energies. Indeed, considering the fact that
phase and particle number are conjugate variables, the operator
$\delta\hat{n}(\mathbf{x})$ for small
fluctuations of the density obeys the canonical commutation relation
$[\delta\hat{n}(\mathbf{x}),\hat{\phi}(\mathbf{x}']=i\delta(\mathbf{x}-\mathbf{x}')$
with the quantized phase operator $\hat{\phi}(\mathbf{x})$.
For any Bose gas (or liquid) with a finite compressibility
$\kappa=\partial n/\partial\mu\ne 0$ at zero temperature
\footnote{This condition rules out the singular case of an
ideal Bose gas, where $\kappa=\infty$ and thus $c=0$.
The ideal gas is therefore {\it not} a SF, even though it has
a finite superfluid density which coincides with the condensate
density \cite{Fisher:1973}.},
the energy of small density fluctuations is
$\int (\delta n(\mathbf{x}))^2/2\kappa$. Combining this with
Eq.~(\ref{eq:DeltaF}), the effective Hamiltonian
for the low lying excitations of an arbitrary superfluid is
of the quantum hydrodynamic form
\begin{equation}
\label{eq:QHD}
\hat{H}=\int d^3x\left[ \frac{\hbar^2n_s}{2M}\left(\nabla\hat{\phi}(\mathbf{x})\right)^2
+\frac{1}{2\kappa}\left(\delta\hat{n}(\mathbf{x})\right)^2\right]\, .
\end{equation}
This Hamiltonian describes harmonic phonons with a linear spectrum
$\omega=cq$ and a velocity, which is determined by
$Mc^2=n_s/\kappa$. In a translation invariant situation,
where $n_s(T=0)\equiv n$ (see below), this velocity coincides with
that of a standard (first sound) compression mode in a gas or liquid.
This coincidence is misleading, however, since Eq.~(\ref{eq:QHD})
describes true elementary excitations and not a hydrodynamic,
collision dominated mode. States with a single phonon are thus
exact low-lying eigenstates of the strongly interacting many-body
system, whose ground state exhibits the property (\ref{eq:n_s}).
As the Lieb-Liniger solution of a 1D Bose gas shows, this mode exists
even in the absence of BEC. It only requires a finite value of
the superfluid density, as specified by Eq.~(\ref{eq:n_s}).
A mode of this type will therefore be present  in the SF phase of
bosons in an optical lattice in {\it any}
dimension and also in the presence of a
finite disorder, as long as the superfluid density is non-vanishing.

The connection between SF and BEC now follows
from the macroscopic representation \cite{Popov:1983}
\begin{equation}
\label{eq:Bosefieldop}
\hat{\psi}(\mathbf{x})=\exp{\Bigl(i\hat{\phi}(\mathbf{x})\Bigr)}\, \Bigl[
n+\delta\hat{n}(\mathbf{x})\Bigr]^{1/2}\approx \sqrt{\tilde{n}_0}\exp{i\hat{\phi}(\mathbf{x})}
\end{equation}
of the Bose field in terms of density and phase operators.
The parameter $\tilde{n}_0$ is a quasi-condensate density.
Its existence relies on the assumption that there is a broad
range $\xi\ll |\mathbf{x}| \ll\ell_{\phi}$ of intermediate distances,
where the one-particle density matrix is equal to a finite value
$\tilde{n}_0<n$,
\footnote{The short range decay from $G^{(1)}(\mathbf{0})=n$
to $\tilde{n}_0$ is due fluctuations at scales smaller than $\xi$
which require a microscopic calculation. The integral (\ref{eq:phasefluct})
is therefore cutoff at $q_{\rm max}\approx 1/\xi$.}
beyond which only phase fluctuations contribute.
Using the harmonic Hamiltonian Eq.~(\ref{eq:QHD}), the
asymptotic decay of $G^{(1)}(\mathbf{x})=\tilde{n}_0
\exp{\Bigl(-\delta\phi^2(\mathbf{x})/2\Bigr)}$ at large separations is thus
determined by the mean square fluctuations
\begin{equation}
\label{eq:phasefluct}
\delta\phi^2(\mathbf{x})=\frac{1}{\hbar\kappa}
\int\frac{d^dq}{(2\pi)^d}\frac{1-\cos{\mathbf{q}\mathbf{x}}}{cq}
\coth{\frac{\hbar cq}{2k_BT}}
\end{equation}
of the phase difference between points separated by $\mathbf{x}$.
The representation Eq.~(\ref{eq:Bosefieldop})
and the notion of a quasi-condensate require the
existence of a finite superfluid density at least at $T=0$, but {\it not} that of BEC.
From Eq.~(\ref{eq:phasefluct}), it is straightforward to see, that plain BEC
in the sense of a non-vanishing condensate density
$n_0=\tilde{n}_0\exp{-\Bigl(\delta\phi^2(\infty)/2\Bigr)}$
at non-zero temperatures only exists in 3D.
In two dimensions, the logarithmic divergence of
$\delta\phi^2(\mathbf{x})\to 2\eta\ln{|\mathbf{x}|}$ at large distances
leads to an algebraic decay $g^{(1)}(\mathbf{x})\sim |\mathbf{x}|^{-\eta}$,
consistent with the Mermin-Wagner-Hohenberg
theorem. Since $\kappa c^2=n_s(T)/M$,
the exponent $\eta(T)=(n_s(T)\lambda_T^2)^{-1}$ is related to the exact
value of the superfluid density as pointed out in Eq.~(\ref{eq:powerlawdecay}).
A similar behavior, due to {\it quantum} rather than thermal phase
fluctuations, applies in 1D at zero temperature, see section V.B.

The arguments above show that SF in the sense of Eq.~(\ref{eq:n_s})
{\it plus} the assumption of a finite compressibility $\kappa$ are
sufficient conditions for either plain BEC in 3D or
the existence of quasi-condensates in 2D at finite and in 1D
at zero temperature
\footnote{In a trap, a quasicondensate in 1D may exist even
at finite temperature, provided the cloud size is smaller than
the phase coherence length, see section V.B}.
For the reverse question, whether SF is also {\it necessary}
for the existence of BEC, the answer is again yes, in the
general sense that BEC is replaced by quasi-condensates
in low dimensions. Indeed, for a translation invariant system, \textcite{Leggett:1998}
has shown that  the existence of ODLRO in the
ground state with an arbitrary small condensate fraction $n_0/n$ implies
perfect superfluidity $n_s(T=0)\equiv n$ at zero temperature. Note that
this includes the case of 2D gases, where only a quasi-condensate survives
at non-zero temperature. In 1D, the presence of quasi long range
order in $g^{(1)}\sim |x|^{-1/2K}$ implies a finite value of $n_s$ by the
relation $K=\pi\hbar\kappa c$ for bosonic Luttinger liquids.
 In more general terms, the fact that
ODLRO in a BEC implies superfluidity follows from the Nambu-Goldstone
theorem. It states that the appearence of (quasi) long range order in the phase implies the existence of an elementary excitation, whose energy
vanishes in the limit of zero momentum.  As emphasized e.g. by \textcite{Weinberg:1986},
the order parameter phase $\phi(\mathbf{x})$ introduced above is just the
Nambu-Goldstone field associated with the
broken $U(1)$ 'gauge' symmetry (this notion has to be treated with care, see
e.g. \cite{Wen:2004}. In the present context, a broken gauge symmetry
is just the statement of Eq.~(\ref{eq:n_s}), that the free energy contains
a term quadratic in a phase twist imposed at the boundaries of the 
system). For systems with a finite compressibility, the dynamics of the 
Nambu-Goldstone mode is described by the quantum hydrodynamic
Hamiltonian Eq.~(\ref{eq:QHD}).
The definition (\ref{eq:n_s}) of superfluidity and the resulting generic quantum hydrodynamic Hamiltonian (\ref{eq:QHD}) which are connected to long (or quasi long) range order via Eqs.~(\ref{eq:Bosefieldop}, \ref{eq:phasefluct}), are thus completely general and are not tied e.g. to translation invariant systems. Despite their different definitions
based on long range {\it current} or long range {\it phase} correlations,
the two phenomena SF and BEC (in its generalized sense) are
therefore indeed just two sides of the same coin. 

\bibliographystyle{apsrmp}
\bibliography{RMP}

\end{document}